\documentclass{aastex63}

\usepackage{natbib}
\usepackage{times}
\usepackage{gensymb}
\usepackage{graphicx}
\usepackage{amsmath}

\newcommand{\Chimera}{{\sc Chimera}}
\newcommand{\chimera}{{\sc Chimera}}

\newcommand{\nue}{\ensuremath{\nu_{e}}}
\newcommand{\nuebar}{\ensuremath{\bar \nu_e}}
\newcommand{\nubar}{\ensuremath{\bar \nu}}
\newcommand{\numt}{\ensuremath{\nu_{\mu\tau}}}
\newcommand{\numtbar}{\ensuremath{\bar \nu_{\mu\tau}}}
\newcommand{\numu}{\ensuremath{\nu_{\mu}}}
\newcommand{\nutau}{\ensuremath{\nu_{\tau}}}
\newcommand{\numubar}{\ensuremath{\bar \nu_{\mu}}}
\newcommand{\nutaubar}{\ensuremath{\bar \nu_{\tau}}}
\newcommand{\nux}{\ensuremath{\nu_{x}}}

\newcommand{\Ye}{\ensuremath{Y_{\rm e}}}
\newcommand{\Yein}[2]{\ensuremath{Y_{{\rm e},#1}^{#2}}}

\newcommand{\alp}{\ensuremath{\alpha}}

\newcommand{\TNSE}{\ensuremath{T_{\mathrm{NSE}}}}
\renewcommand\Re{\operatorname{Re}}

\newcommand{\be}{\begin{eqnarray}}
\newcommand{\ee}{\end{eqnarray}}
\newcommand{\ds}{\displaystyle}

\newcommand{\rfrac}[2]{{}^{#1}\!/_{#2}}
\newcommand{\e}[1]{\ensuremath{\times 10^{#1}}}
\newcommand{\half}{\ensuremath{\frac{1}{2}}}

\newcommand{\epscm}{\ensuremath{\epsilon_{_{\! 0}}}}
\newcommand{\epscmi}[1]{\ensuremath{\epsilon_{_{\! 0} \, #1}}}
\newcommand{\psimomcm}[1]{\ensuremath{\psi_{_{\! 0}}^{(#1)}}}
\newcommand{\psimomcmi}[2]{\ensuremath{\psi_{_{\! 0} \, #2}^{(#1)}}}
\newcommand{\psimomcmin}[3]{\ensuremath{\psi_{_{\! 0} \, #2}^{(#1) \, #3}}}

\newcommand{\psibarmomcm}[1]{\ensuremath{\bar{\psi}_{_{\! 0}}^{(#1)}}}
\newcommand{\psibarmomcmi}[2]{\ensuremath{\bar{\psi}_{_{\! 0} \, #2}^{(#1)}}}
\newcommand{\psibarmomcmin}[3]{\ensuremath{\bar{\psi}_{_{\! 0} \, #2}^{(#1) \, #3}}}
\newcommand{\Psimomcm}[1]{\ensuremath{\Psi_{_{\! 0}}^{(#1)}}}
\newcommand{\Ecm}{\ensuremath{E_{_{\! 0}}}}
\newcommand{\Ecmi}[1]{\ensuremath{E_{_{\! 0} \, #1}}}
\newcommand{\mucm}{\ensuremath{\mu_{_{\! 0}}}}
\newcommand{\mucmi}[1]{\ensuremath{\mu_{_{\! 0} \, #1}}}
\newcommand{\Vcm}{\ensuremath{{\cal V}_{_{\! 0}}}}
\newcommand{\Vcmi}[1]{\ensuremath{{\cal V}_{_{\! 0} \, #1}}}
\newcommand{\fcm}{\ensuremath{f_{_{\! 0}}}}
\newcommand{\fcmi}[1]{\ensuremath{f_{_{\! 0} \, #1}}}
\newcommand{\Ncmi}[1]{\ensuremath{N_{_{\! 0} \, #1}}}
\newcommand{\Lcmi}[1]{\ensuremath{L_{_{\! 0} \, #1}}}
\newcommand{\pcm}{\ensuremath{p_{_{\! 0}}}}
\newcommand{\Omegacm}{\ensuremath{\Omega_{_{\! 0}}}}
\newcommand{\dtcmi}[1]{\ensuremath{dt_{_{\! 0} \, #1}}}

\newcommand{\isotope}[2]{\ensuremath{\mathrm {^{#2}#1}}}
\newcommand{\gcc}{\ensuremath{{\mbox{g~cm}}^{-3}}}
\newcommand{\den}[2]{\ensuremath{#1 \times 10^{#2}\,\textrm{g~cm}^{-3}}}

\newcommand{\cmps}{\ensuremath{\mbox{cm~s}^{-1}}}

\newcommand{\mev}{\mbox{MeV}}

\newcommand{\msun}{\ensuremath{M_\sun}}

\newcommand{\ccsn}{{\sc CCSN}}
\newcommand{\ccsne}{{\sc CCSN}e}
\newcommand{\vertex}{{\sc Vertex}}
\newcommand{\agileboltztran}{{\sc Agile-Boltztran}}
\newcommand{\prometheus}{{\sc Prometheus}}
\newcommand{\prometheusvertex}{{\sc Prometheus-Vertex}}

\newcommand{\eden}[1]{\ensuremath{e_{\rm #1}}}
\newcommand{\edenx}[2]{\ensuremath{e_{{\rm #1},#2}}}
\newcommand{\edint}{\ensuremath{\eden{int}}}
\newcommand{\edinti}{\ensuremath{e_{{\rm int},i}}}
\newcommand{\edth}{\ensuremath{\eden{th}}}
\newcommand{\edthi}{\ensuremath{e_{{\rm th}, i}}}
\newcommand{\edkin}{\ensuremath{\eden{kin}}}
\newcommand{\edkini}{\ensuremath{e_{{\rm kin}, i}}}
\newcommand{\edgrav}{\ensuremath{\eden{grav}}}
\newcommand{\edgravi}{\ensuremath{e_{{\rm grav}, i}}}
\newcommand{\edbind}{\ensuremath{\eden{bind}}}
\newcommand{\edbindi}{\ensuremath{e_{{\rm bind}, i}}}
\newcommand{\edtot}{\ensuremath{\eden{tot}}}
\newcommand{\edtoti}{\ensuremath{e_{{\rm tot}, i}}}
\newcommand{\edtotp}{\ensuremath{\eden{tot'}}}
\newcommand{\edtotpi}{\ensuremath{e_{{\rm tot'}, i}}}
\newcommand{\edtotpp}{\ensuremath{\eden{tot''}}}

\newcommand{\uvec}{\ensuremath{\mathbf{u}}}
\newcommand{\fvec}{\ensuremath{\mathbf{f}}}
\newcommand{\nvec}{\ensuremath{\mathbf{\hat{n}}}}
\newcommand{\Area}[1]{\ensuremath{{\rm Area}_{#1}}}
\newcommand{\Vol}[1]{\ensuremath{{\rm Vol}_{#1}}}
\newcommand{\Flx}[1]{\ensuremath{{\rm Flx}_{#1}}}
\newcommand{\pderiv}[2]{\ensuremath{\frac{\partial #1 }{\partial #2}}}
\newcommand{\pderivoo}[3]{ \left( \frac{\partial{#1}}{\partial{#2}} \right)_{#3}}
\newcommand{\pderivo}[2]{ \left( \frac{\partial{#1}}{\partial{#2}} \right)}

\newcommand{\epsp}[1]{\ensuremath{\varepsilon_{\bar{#1}}}}
\newcommand{\epspo}[1]{\ensuremath{\varepsilon_{\bar{#1}+1}}}
\newcommand{\eps}[1]{\ensuremath{\varepsilon_{#1}}}
\newcommand{\depsp}[1]{\ensuremath{\delta\varepsilon_{\bar{#1}}}}
\newcommand{\deps}[1]{\ensuremath{\delta\varepsilon_{#1}}}
\newcommand{\rtye}{\ensuremath{(\rho,T,Y_e)}}

\newcommand{\lrtye}{\ensuremath{(\log\rho,\log T,\Ye)}}
\newcommand{\aeps}{\ensuremath{\alpha\epsilon}}

\newcommand{\series}[2]{ \item \textbf{Series #1} {\begin{itemize} #2 \end{itemize}} }
\newcommand{\sitem}[2]{ \item \textbf{#1:} #2}

\newcommand{\UTphys}{Department of Physics and Astronomy, University of Tennessee, Knoxville, TN 37996-1200, USA}
\newcommand{\ORNLphys}{Physics Division, Oak Ridge National Laboratory, P.O. Box 2008, Oak Ridge, TN 37831-6354, USA}
\newcommand{\JICS}{Joint Institute for Computational Sciences, Oak Ridge National Laboratory, P.O. Box 2008, Oak Ridge, TN 37831-6173, USA}
\newcommand{\NCCS}{National Center for Computational Sciences, Oak Ridge National Laboratory, P.O. Box 2008, Oak Ridge, TN 37831-6164, USA}
\newcommand{\NCSU}{Department of Physics, North Carolina State University,  Raleigh, NC 27695-8202, USA}
\newcommand{\FAU}{Department of Physics, Florida Atlantic University, 777 Glades Road, Boca Raton, FL 33431-0991, USA}
\newcommand{\CSMD}{Computer Science and Mathematics Division, Oak Ridge National Laboratory, P.O. Box 2008, Oak Ridge, TN 37831-6164, USA}
\newcommand{\NSF}{Physics Division, National Science Foundation, Alexandria, VA 22314 USA}

\shorttitle{\Chimera\ methodology}
\shortauthors{Bruenn et al.}

\begin{document}

\title{\Chimera: A Massively Parallel Code for core collapse Supernova Simulation}

\author[0000-0003-0999-5297]{Stephen W. Bruenn}
\affiliation{\FAU}

\author[0000-0001-9691-6803]{John M. Blondin}
\affiliation{\NCSU}
\author[0000-0002-9481-9126]{W. Raphael Hix}
\affiliation{\ORNLphys}
\affiliation{\UTphys}
\author[0000-0002-5231-0532]{Eric J. Lentz}
\affiliation{\UTphys}
\affiliation{\ORNLphys}
\affiliation{\JICS}
\author[0000-0002-5358-5415]{O. E. Bronson Messer}
\affiliation{\NCCS}
\affiliation{\ORNLphys}
\affiliation{\UTphys}
\author[0000-0001-9816-9741]{Anthony Mezzacappa}
\affiliation{\UTphys}
\affiliation{\JICS}

\author[0000-0003-1251-9507]{Eirik Endeve}
\affiliation{\CSMD}
\affiliation{\UTphys}
\affiliation{\JICS}
\author[0000-0003-3023-7140]{J. Austin Harris}
\affiliation{\NCCS}
\affiliation{\ORNLphys}
\author[0000-0003-3070-0625]{Pedro Marronetti}
\affiliation{\NSF}

\author[0000-0003-0395-8532]{Reuben D. Budiardja}
\affiliation{\UTphys}
\affiliation{\NCCS}
\author{Merek A. Chertkow}
\affiliation{\UTphys}
\author{Ching-Tsai Lee}
\affiliation{\UTphys}

\correspondingauthor{Stephen W. Bruenn}
\email{bruenn@fau.edu}

\accepted{February 24, 2020, ApJS}

\begin{abstract}
We provide a detailed description of the \chimera\ code, a code developed to model core collapse supernovae in multiple spatial dimensions. 
The core collapse supernova explosion mechanism remains the subject of intense research. 
Progress to date demonstrates that it involves a complex interplay of neutrino production, transport, 
and interaction in the stellar core, three-dimensional stellar core fluid dynamics and its associated 
instabilities, nuclear burning, and the foundational physics of the neutrino--stellar core weak interactions
and the equations of state of all stellar core constituents -- particularly, the nuclear equation of state 
associated with core nucleons, both free and bound in nuclei.
\Chimera, by incorporating detailed neutrino transport, realistic neutrino--matter interactions, three-dimensional hydrodynamics, realistic nuclear, leptonic, 
and photonic equations of state, and a nuclear reaction network, along with other refinements, can be used to study the role of neutrino radiation, hydrodynamic 
instabilities, and a variety of input physics in the explosion mechanism itself.
It can also be used 
to compute observables such as neutrino signatures, gravitational radiation, and the products of nucleosynthesis
associated with core collapse supernovae.
The code contains modules for neutrino transport, multidimensional compressible hydrodynamics, nuclear reactions, a variety of neutrino interactions, equations of state, and modules to provide data for post-processing observables such as the products of nucleosynthesis, and gravitational radiation. 
\chimera\ is an evolving code, being updated periodically with improved input physics and numerical refinements.
We detail here the current version of the code, from which future improvements will stem, which can in turn be described 
as needed in future publications.
\end{abstract}


\tableofcontents

\section{Introduction}

Modeling core collapse supernovae (\ccsne) has become over the years an extremely demanding computational problem, requiring realistic multidimensional, general relativistic, multigroup, neutrino radiation hydrodynamics; sophisticated nuclear equations of state; and extensive nuclear reaction networks to elucidate the CCSN explosion mechanism and capture some of the important observables.
The sheer breadth of interdependent physics, over a vast range of density and energy scales, is one aspect of the computational challenge.
This includes general relativistic gravity; matter velocities at non-negligible fractions of the speed of light; the production, transport, and interaction of neutrinos and anti-neutrinos of all flavors  across three regimes: the tight coupling of neutrinos and matter at high densities in the core, weaker coupling far from the core, and intermediate coupling in between; the evolution of the nuclear composition both in and out of nuclear statistical equilibrium as mediated by both strong and weak nuclear interactions; and an equation of state that spans a density range that can exceed fourteen orders of magnitude. 
And there is a second notable aspect of the computational challenge. The \ccsn\ explosion mechanism also appears to be marginal in the sense that rather modest changes in the numerical modeling of the neutrino transport and/or the neutrino interactions, and/or the use of Newtonian versus general relativistic gravity, for example, can change the outcome of a simulation not only quantitatively, but qualitatively as well.
Therefore, the physical treatment and numerical implementation of the above-described physics has to be realistic and highly accurate (in the numerical sense) if meaningful results are to be obtained. 
Finally, a \ccsn\ simulation may require millions of time steps to integrate a model forward in time through a sufficiently long period in order to determine outcomes such as explosion energies and other observables. This requires that the various numerical algorithms implemented in the code be highly optimized.

Our development of the radiation hydrodynamics code \Chimera\ to model \ccsne\ has drawn on previous codes that have successfully modeled one or another physical process relevant to \ccsne. 
The hydrodynamics module has been built on the dimensionally-split, Lagrangian-plus-remap scheme with piecewise parabolic reconstruction as formulated by \citet{CoWo84} and implemented in VH1 as described by \citet{HaBlLi12} and \citet{BlLu93a}, but extended to include multi-species advection, energy absorbed or released due to compositional changes, nuclei coming in or out of nuclear statistical equilibrium, multidimensional gravity, momentum and energy exchange with neutrinos, and a sliding radial grid algorithm that continually adjusts the radial grid to resolve structures that arise during the course of a simulation. 
The neutrino transport stems from the original formulation in \citet{Brue85} but modified and improved, and with a number of neutrino source terms either refined or added to the \chimera\ suite of neutrino--matter interactions.
The nuclear composition, when material is not under conditions appropriate for nuclear statistical equilibrium, is evolved by the thermonuclear reaction network code XNet developed by \citet{HiMe06,TrHi13}.

In this paper we present a detailed description of our \Chimera\ code, which has been developed with the aim of realistically modeling \ccsn\ and the associated observables. 
In the sections that follow, we describe each of our algorithms in detail. 
\chimera\ is under continuous development. When it is deemed that significant improvements or refinements have been implemented and tested, the version of the code current at that time is frozen, removed from further development, and used to execute and follow new simulations in our ongoing investigations of \ccsne.
These frozen code versions, along with common sets of microphysics inputs are designated as a lettered `Series' of \chimera\ simulations, where simulations within a `Series' share not only the code base and algorithm choices, but the default microphysics and control parameters as well.

The first test results from \chimera\ for low-resolution 2D (axisymmetric) simulations were reported in \citet{BrDiMe06}, as was an early description of the code \citep{MeBrBl08}.
The first attempt at production simulations in 2D and 3D, with a reasonably complete set of physics and with reasonable resolution were reported in \citet{BrMeHi09b} and retroactively designated as `Series-A.'
Though some flaws emerged late in the Series-A simulations, the earlier portions of the 2D simulations were used for gravitational wave extraction analysis \citep{YaMaMe10}.
Subsequent to the Series-A simulations, we made several improvements to and enhanced \chimera, as well as updated some input physics.
The `Series-B' simulations \citep{BrMeHi13, BrLeHi16} used the updated code to recompute the four 2D simulations of Series-A, with enhanced resolution, from which we analyzed the gravitational wave signals \citep{YaMeMa15} and nucleosynthesis of the ejecta \citep{HaHiCh18}.
The methods and algorithms described in this paper largely reflect those used in the \chimera\ Series~B.
The next group of models, Series-C, focused on the 3D modeling \citep{LeBrHi15,LeKeCa18} using the same input microphysics as Series~B.
Direct analyses of the gravitational wave \citep{YaMeMa17} and neutrino detector signals \citep{MeDeLe18} have been performed for these simulations.
The primary differences between Series-B and Series-C were code consolidation and optimizations, but a few improvements are described herein, including,
an improved treatment of neutrino transport through the shock (Section~\ref{trans_shock}), a more efficient interpolation of neutrino opacities (Section~\ref{opac_interp}), and parallel IO (Section~\ref{sec:IO}).
Further series will follow, including a general `Series D' consisting of several 3D models and 2D studies, and a `Series E' focused on the nuclear equation of state and related code improvements.
All of the above mentioned series have employed the so-called ray-by-ray plus approximation to neutrino transport in which spherically symmetric multi-energy neutrino transport is performed along each radial ray with lateral advection of neutrinos with matter performed in optically thick regions. 
Fully multi-energy, multi-angle neutrino transport has recently been implemented in core-collapse simulations with various levels of microphysics \citet{KuTaKo16, KuKoTa18, OtRodSS18, VaBuRa18, SkDoBu19}
Multi-angle transport will be a future upgrade of \chimera.
Descriptions of the modifications and enhancements for these, and subsequent, series will appear with the simulation results, as needed.

Section~\ref{sec:overview} gives a general overview of \chimera\ including the domain decomposition used in implementing parallel computing architectures, the directional splitting used, and the sequence of computational steps in a complete time cycle.
Section~\ref{app:Eos} details the implementation of the nuclear reaction network and equations of state, including the technique for juxtaposing more than one equation of state in adjacent density regimes.
Section~\ref{app:hydro} presents a detailed description of our hydrodynamics algorithms, with test problems in Section~\ref{sec:Hydro_Test}.
The neutrino transport method and numerical methods are described in Section~\ref{app:transport}, with a detailed description of the neutrino transport source terms given in Section~\ref{app:source}.
Finally, Section~\ref{trans_tests} presents results of static neutrino transport tests, and Section~\ref{comparisons} presents comparisons of 1D core collapse supernova simulations with the Boltzmann code \agileboltztran\ \citep{LiMeMe04,LiRaJa05}. 
Additional specific test results verifying a number of our algorithms are presented at the end of the relevant sections.

\newpage
\section{General Overview}
\label{sec:overview}

To drive the component pieces integrated from other codes, \chimera\ uses a master--data model, where data is stored in a master copy on each computing element, transposed as needed to appropriate dimensional sweeps, checked-out to constituent codes, processed, and checked back to the master copy.
At the master level, general program control, data input and output, and monitoring are also managed.

\subsection{Directional Splitting}
\label{app:Dir_Splt}

\begin{figure}
\gridline{
               \fig{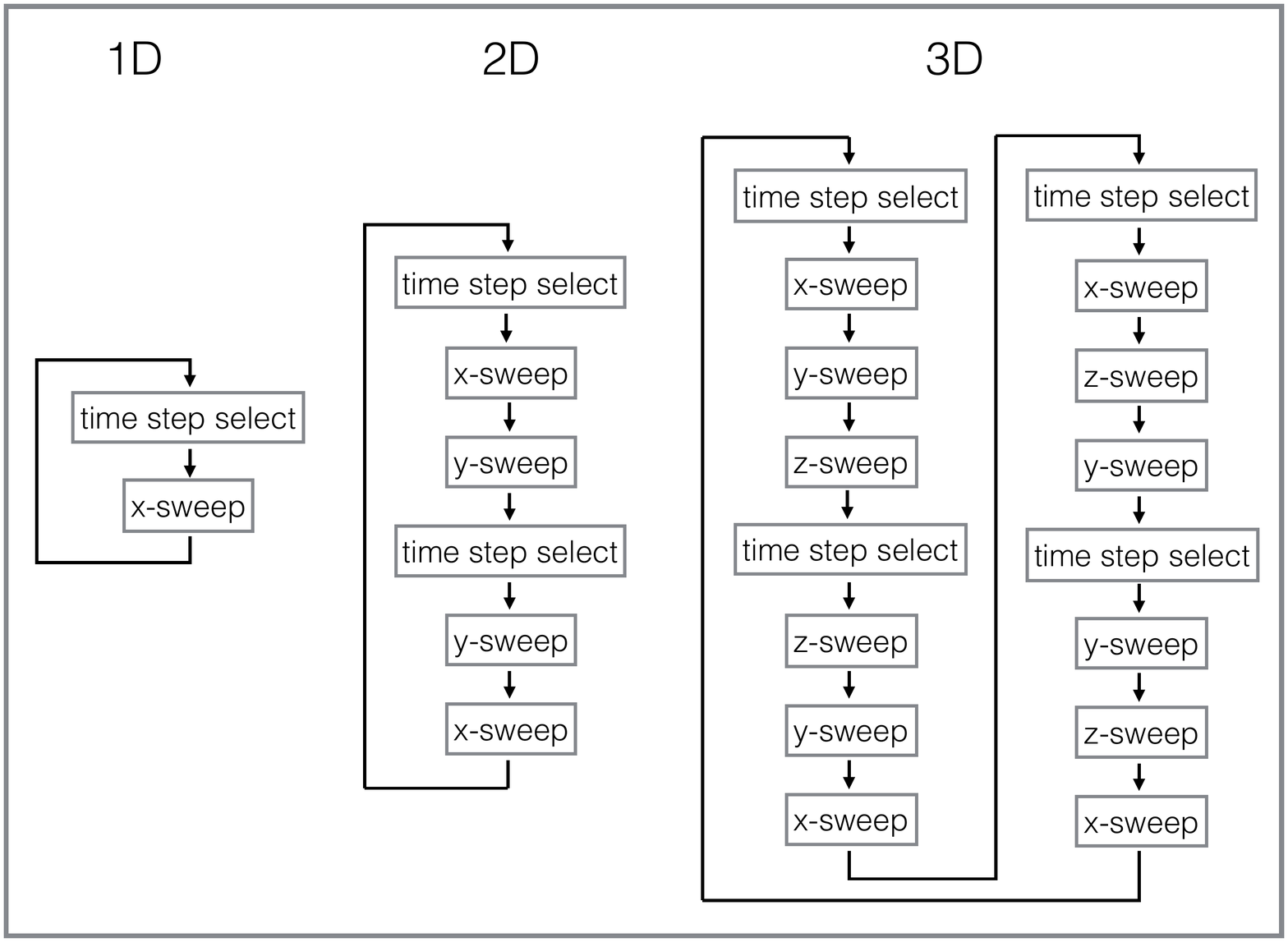}{0.5\textwidth}{(a)}
               }
\gridline{
              \fig{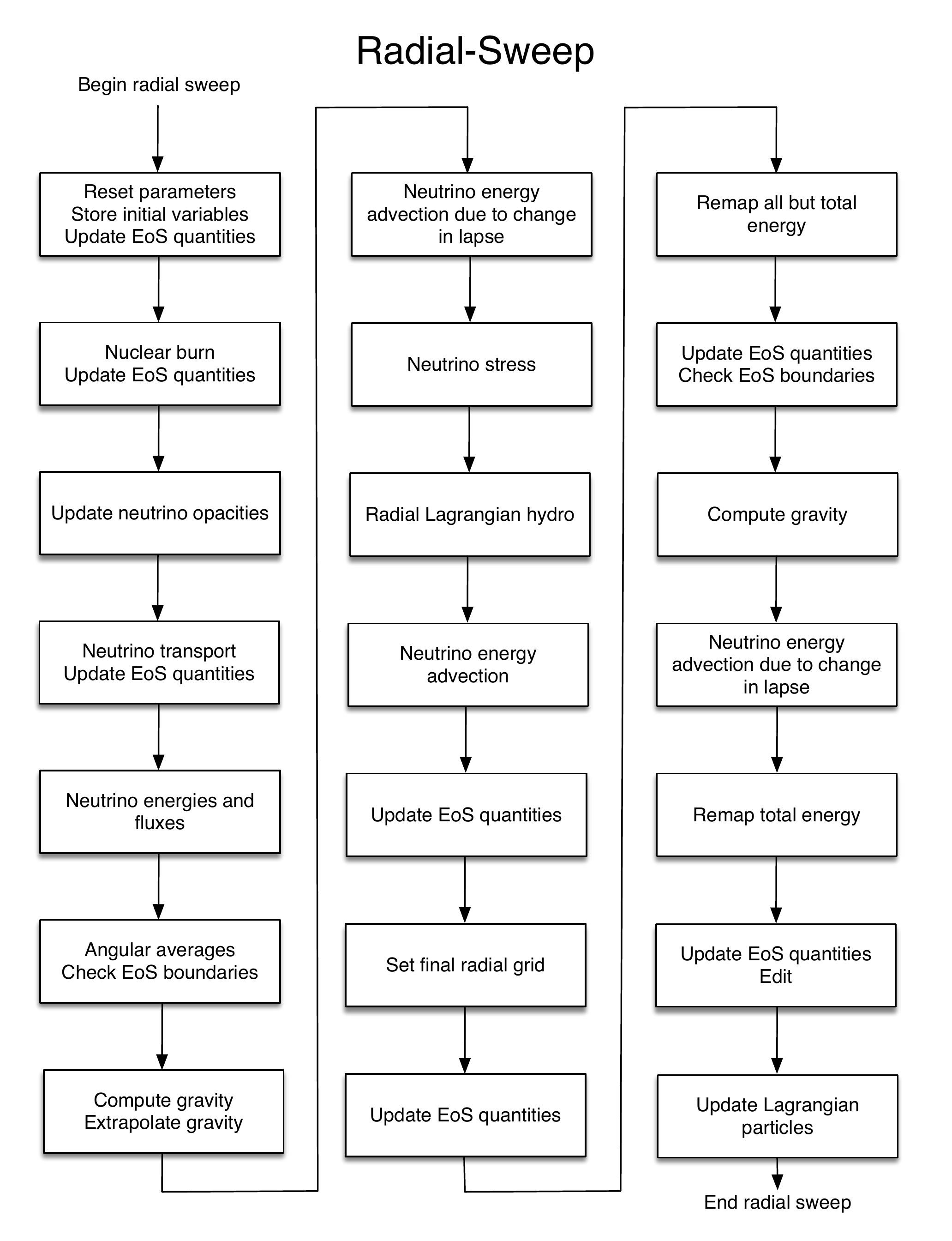}{0.5\textwidth}{(b)}
              \fig{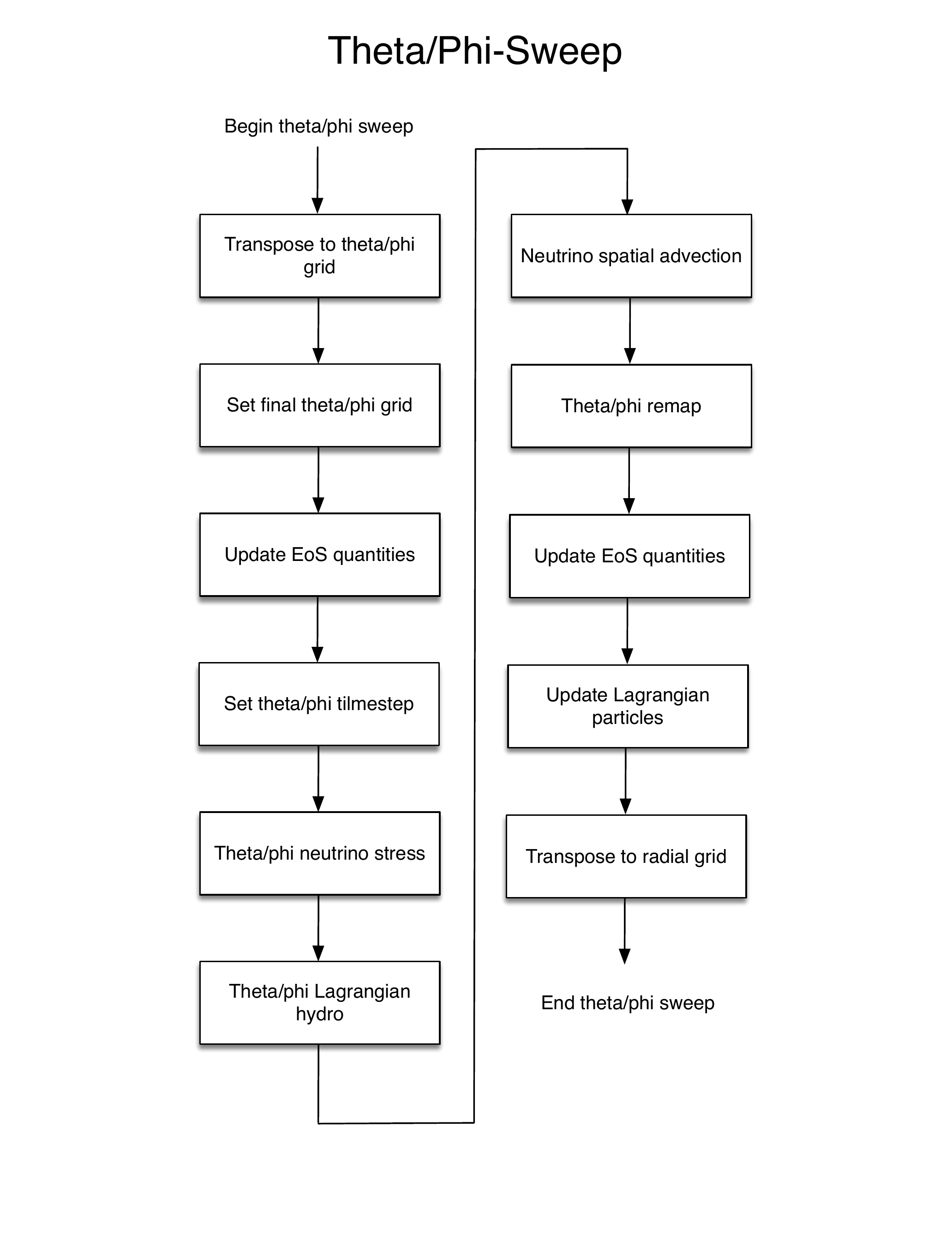}{0.5\textwidth}{(c)}
               }
\caption{\label{fig:sweep}
Panel (a): Sweep sequence used in \chimera\ for numerically solving 1-, 2-, and 3-dimensional problems. 
Panels (b) and (c): Sequence of operations performed during the (b) radial sweep and (c) theta or phi sweep.
}
\end{figure}

For multidimensional applications \chimera\ uses the method of dimensional splitting \citep{Stra68}. 
In this method the numerical solution of the multidimensional problem proceeds by a series of  one-dimensional steps or ``sweeps'' to build up the full solution. 
The updated variables after each sweep are used as initial conditions for the next sweep. 
The order of the sweeps is shown in Figure~\ref{fig:sweep}. 
For second-order accuracy in time, the time step, $\Delta t$, should be selected at the beginning of the sweep sequences, and for the 2D case, $\frac{1}{2}\Delta t$ should be used in the x-y subsequence of the sweep sequence and again in the y-x subsequence. Likewise, in the 3D case $\frac{1}{4} \Delta t$ should be used for each of the four x-y-z transposition subsequences. 
In practice, the full time step computed before each subsequence is used for that subsequence. 
Since the time steps computed for each of the subsequences are nearly the same, approximate time-centering is maintained. 
Furthermore, the larger time steps permit a more refined grid to be used for the same amount of computer time. 

\subsection{Domain Decomposition}
\label{domain}

\chimera\ is designed to numerically evolve CCSNe on spherical polar grids and to run on multiprocessor machines.
The domain decomposition in \chimera\ is dictated, in part, by the present implementation of neutrino transport, which is in the radial direction. 
The decomposition is along rays, or bundles of rays, each ray being one zone wide in two dimensions and spanning the entire set of zones from boundary to boundary along the third. 
Thus, an x-ray, or radial ray refers to a set of zones along a radial line from the center to the outer edge of the grid. 
A y-ray, or angular ray, refers to a set of zones at a given radius and azimuth which, for a 180\degree\ angular grid, spans the arc from the ``north" pole to the ``south" pole.
Finally, a z-ray, or azimuthal ray, refers to a set of zones at a given radius and polar angle that, in the case of a 360\degree\ azimuthal grid, completely encircles the polar axis.

\begin{figure}
\fig{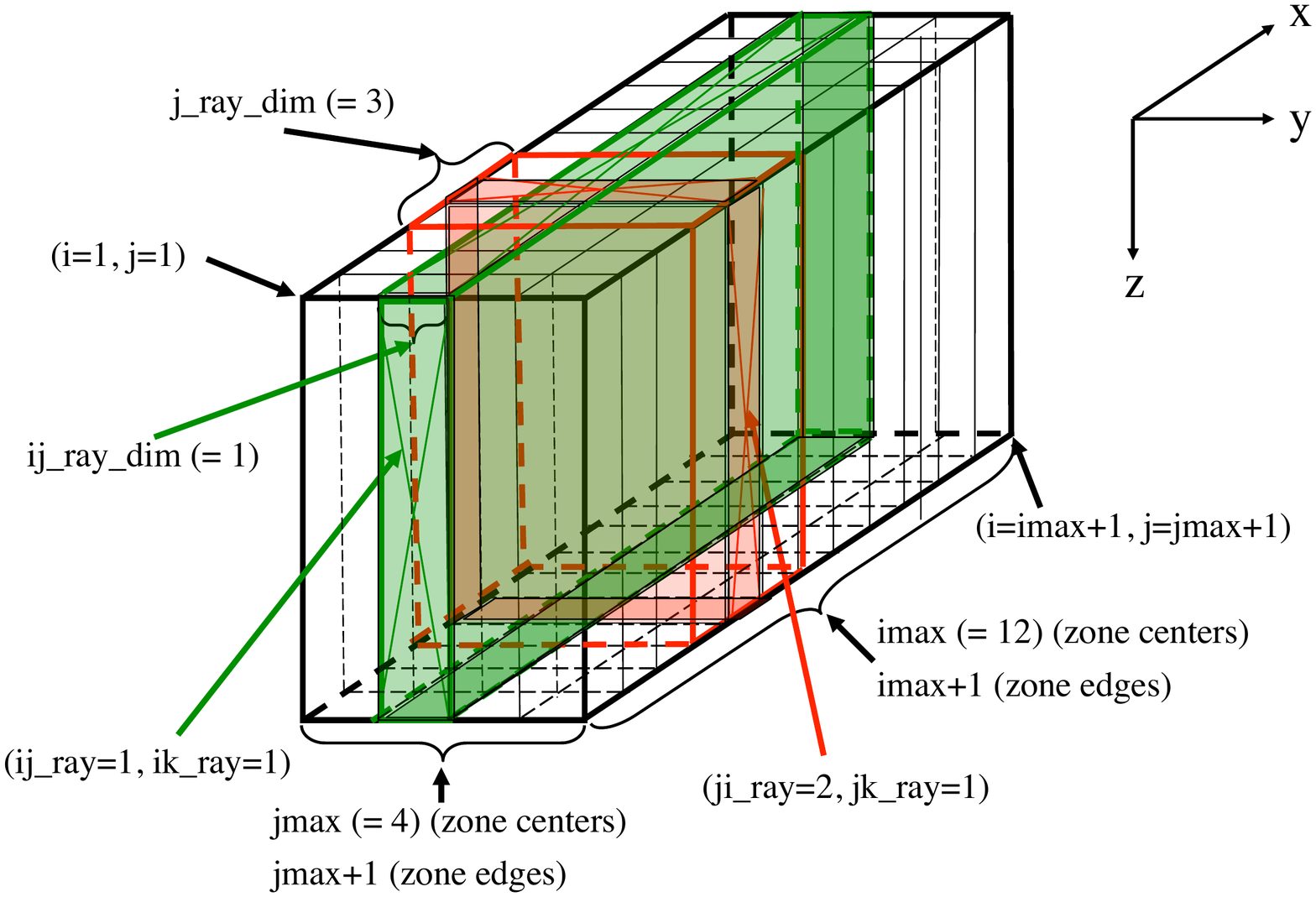}{0.7\textwidth}{}
\caption{\label{fig:slicing1}
Example \chimera\ domain decomposition for 2-dimensional models with one x ray per MPI rank.}
\end{figure}

As an extremely simple example of the domain decomposition in \chimera, consider a 2D grid consisting of four radial rays, each radial ray consisting of twelve radial zones with one radial ray per MPI rank, four MPI ranks in all. 
Figure~\ref{fig:slicing1} shows the logical structure of this grid. 
The large rectangular block bordered in thick black lines encompasses the logical grid with the x (radial), y (angular), and z (azimuthal) directions shown at the upper right. 
Because the grid is two-dimensional, the third or z-dimension is superfluous, but is shown as an elongated cell wall in the z-direction for the sake of comparison with a 3D example below.
Each zone is shown in the figure as a square-sided vertical rectangular volume.
The total number of radial, angular, and azimuthal zone-centers are denoted by \texttt{imax}, \texttt{jmax}, and \texttt{kmax}, respectively, and the total number of zone edges by $\texttt{imax}+1$, $\texttt{jmax}+1$, and $\texttt{kmax}+1$.
In this case $\texttt{imax} = 12$, $\texttt{jmax} = 4$, and $\texttt{kmax} = 1$.

The computation in \chimera\ is directionally split, and the (x-, y-, z-) sweeps -- i.e., the (radial-, angular-, azimuthal-) sweeps -- refer to the direction of computation.
During the x-sweep, a set number, or bundle, of radial rays (in this example just one) is assigned to each MPI rank.
In Figure~\ref{fig:slicing1} a radial ray is shown in green for a particular, but otherwise arbitrary, MPI rank. 
The dimensions \texttt{ij\_ray\_dim} and \texttt{ik\_ray\_dim} denote the dimensions of a bundle of x-rays in the y- and z-directions, respectively, so that $\texttt{ij\_ray\_dim} \times \texttt{ik\_ray\_dim}$ is the number of x-rays in the bundle. 
In this example, where the bundle of rays per MPI rank consists of just one ray, both \texttt{ij\_ray\_dim} and \texttt{ik\_ray\_dim} are unity.
The local indices \texttt{ij\_ray} and \texttt{ik\_ray} locate a particular ray within a bundle relative to the upper left corner of the bundle, so that $\texttt{ij\_ray} \in [1,\texttt{ij\_ray\_dim}]$ and $\texttt{ik\_ray} \in [1,\texttt{ik\_ray\_dim}]$. 
In this example both of these indices are unity as there is only one ray per bundle.
During the x-sweep, the radial hydrodynamics, radial ray-by-ray transport, and nuclear reactions are evolved along with global gravity solves.

Following the x-sweep, a transpose to the y- or z-oriented rays is performed.
In this 2D example, the transpose is just to the y-oriented rays.
Because there are twelve radial zones and only four angular zones in this example, each MPI rank now consists of a bundle of 3 y-rays in order that the total number of MPI ranks remain the same.
This is delineated in Figure~\ref{fig:slicing1} by the rays enclosed by red for a particular, but arbitrary, bundle.
Now (\texttt{ji\_ray}, \texttt{jk\_ray}) takes the place of (\texttt{ij\_ray}, \texttt{ik\_ray}), and locate a particular angular ray in the x-z plane of the bundle.
The widths in the x-z plane of each bundle is given by \texttt{j\_ray\_dim} and \texttt{ik\_ray\_dim}, which in this example are equal to 3 and 1, respectively.
\texttt{j\_ray\_dim} and \texttt{k\_ray\_dim} are the number of radial zones on each MPI rank after transposing to the y-oriented rays and z-oriented rays, respectively.

\begin{figure}
\fig{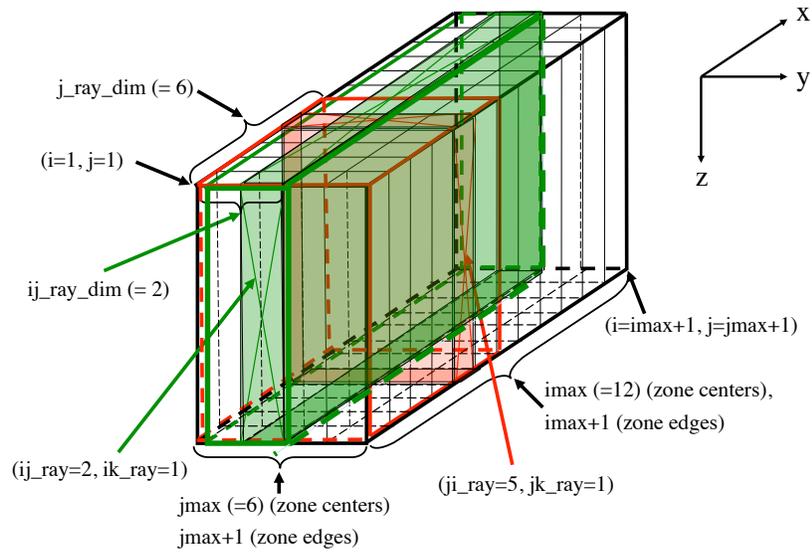}{0.7\textwidth}{}
\caption{\label{fig:slicing2}
Example \chimera\ domain decomposition for 2-dimensional models with two x rays per MPI rank. Some lines have been slightly offset to render them visible.}
\end{figure}

Figure \ref{fig:slicing2} illustrates a slight variation of the previous domain decomposition example using the same logical grid. 
The total number of radial, angular, and azimuthal zones are the same as before, but but in this example bundles of 2 x-rays are associated with each MPI rank, two MPI ranks in all.
In Figure \ref{fig:slicing2} a particular, but arbitrary, bundle of x-rays is shown bounded by thick green lines.
In this case the y-z dimensions of the bundle are given by $\texttt{ij\_ray\_dim} = 2$ and $\texttt{ik\_ray\_dim} = 1$, respectively, and the particular, but arbitrary, radial ray designated by the narrow-lined green X at the front and top has the local indices in the bundle ($\texttt{ij\_ray} = 2, \texttt{jk\_ray} = 1$).
Following the x-sweep, a transpose to the y-oriented rays is made, and because there are only 2 MPI ranks in this case, the number of y-rays associated with each MPI rank is 6, as there are 12 y-rays in all.
This is delineated in Figure~\ref{fig:slicing2} by the rays enclosed by the thick red lines for a particular, but arbitrary, y-bundle,
The x-z dimensions of the y-bundles are $\texttt{j\_ray\_dim} = 6$ and $\texttt{ik\_ray\_dim} = 1$, respectively, and the particular, but arbitrary, y-ray designated by the narrow-lined red X at the front and top has the local indices in the bundle ($\texttt{ji\_ray} = 5, \texttt{jk\_ray} = 1$).

\begin{figure}
\fig{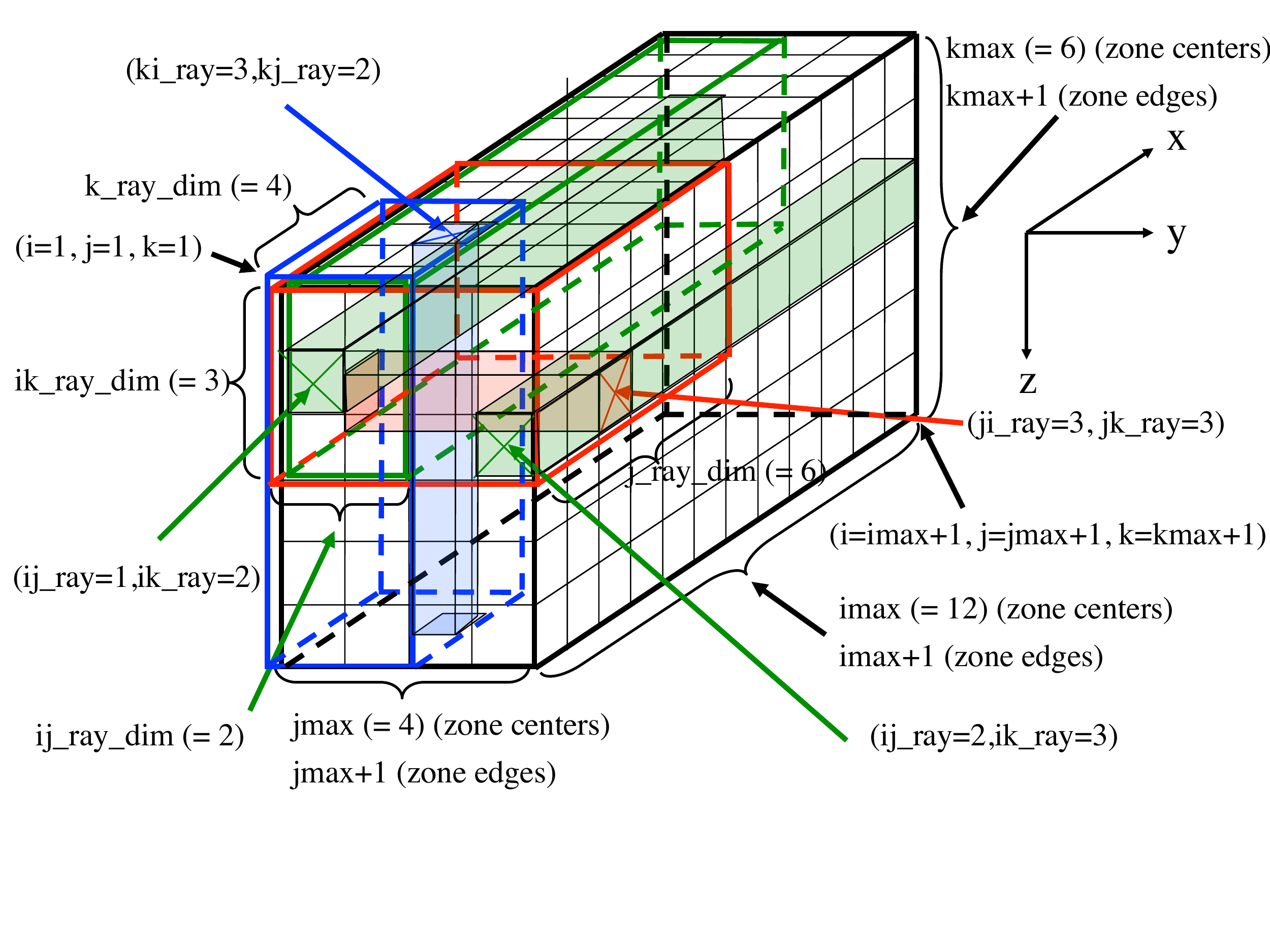}{0.7\textwidth}{}
\caption{\label{fig:dlicing2}
Example \chimera\ domain decomposition for a 3-dimensional model. Some of the lines have been slightly offset to render them visible.}
\end{figure}

Figure~\ref{fig:dlicing2} illustrates a general domain decomposition example using the same logical grid for the x- and y-rays, but adding a z-ray consisting of 6 zones. 
Now there are $\texttt{imax} \times \texttt{jmax} \times \texttt{kmax} = 12 \times 4 \times 6 = 288$ zones in all, and $\texttt{jmax} \times \texttt{kmax} = 4 \times 6 = 24$ x-rays.
Let us suppose we wish to compute with 4 MPI ranks.
We must then use 4 bundles of x-rays, each bundle assigned to one MPI rank and consisting of 6 x-rays.
One such particular, but arbitrary, bundle is shown in Figure~\ref{fig:dlicing2}, outlined in the thick green lines.
The y-z dimensions of the bundle are $\texttt{ij\_ray\_dim} = 2$ and $\texttt{ik\_ray\_dim}= 3$, respectively.
A particular, but arbitrary, x-ray is shown in this bundle by a thin-lined green X on its front face, having the local indices ($\texttt{ij\_ray}=1, \texttt{ik\_ray} = 2$); another x-ray located in an adjacent bundle is delineated by another thin-lined green X on its front face having the local indices ($\texttt{ij\_ray}=2, \texttt{ik\_ray} = 3$).
On executing the x-sweep, each MPI rank performs the computation required to complete the individual x-sweep for each x-ray in its bundle.

Following the x-sweep, a transpose to the y- and z-oriented rays is performed, either in that order, or reversed order, as shown above in Figure~\ref{fig:sweep}.
Consider the transpose to the y-oriented rays. 
There are $\texttt{imax} \times \texttt{kmax} = 12 \times 6 = 72$ y-rays. With 4 MPI ranks, we assign to each MPI rank 4 bundles of y-rays with 18 y-rays each, having the x-z dimensions $\texttt{j\_ray\_dim} = 6$ and $\texttt{ik\_ray\_dim} = 3$, respectively.
A particular, but arbitrary, bundle of y-rays is shown bounded by thick red lines in Figure~\ref{fig:dlicing2}, and a particular, but arbitrary, y-ray in that bundle is singled out by a thin-lined red X, having the local indices ($\texttt{ji\_ray} = 4, \texttt{jk\_ray} = 3$).

Following or preceding the y-sweep, a transpose to the z-oriented rays is performed.
There are $\texttt{imax} \times \texttt{jmax} = 12 \times 4 = 48$ z-rays. With 4 MPI ranks, we assign to each MPI rank 4 bundles of z-rays with 12 z-rays each, having the x-y dimensions $\texttt{k\_ray\_dim} = 4$ and $\texttt{ij\_ray\_dim} = 2$, respectively.
A particular, but arbitrary, bundle of z-rays is shown bounded by thick blue lines in Figure~\ref{fig:dlicing2}, and a particular, but arbitrary, z-ray in that bundle is singled out by a thin-lined blue X, having the local indices ($\texttt{ki\_ray} = 3, \texttt{kj\_ray} = 2$).

In general, for a 2D or 3D grid with a domain decomposition consisting of x-rays, y-rays, and z-rays, to ensure load balancing the number number of rays in x-bundles, y-bundles, and z-bundles must satisfy
\begin{equation}
N_{\rm MPI} = \frac{ \texttt{jmax} \times \texttt{kmax} }{ \texttt{ij\_ray\_dim} \times \texttt{ik\_ray\_dim} } = \frac{ \texttt{imax} \times \texttt{kmax} }{ \texttt{j\_ray\_dim} \times \texttt{ik\_ray\_dim} } = \frac{ \texttt{imax} \times \texttt{jmax} }{ \texttt{ij\_ray\_dim} \times \texttt{k\_ray\_dim} }
\label{eq:g1}
\end{equation}
where $N_{\rm MPI}$ is the number of MPI ranks.

\subsection{IO Subsystem}
\label{sec:IO}

The input and output (IO) of data by simulation codes presents a significant challenge, and good IO is required to achieve acceptable computational performance.
\Chimera\ has several components to its IO subsystem.
A typical full explosion model requires $\mathcal{O}(1000)$ hours to complete.
This requires writing checkpoint files for restarting the simulation and data files for later analysis, plotting, and visualization.
We have implemented two schemes for general restart and analysis IO: (1) the original, serial method used for much of our early 2D work and (2) a parallel method required for effective computation in 3D.

\subsubsection{Serial IO Method}

The original serial IO scheme had separate components for restart and analysis output.
The restart IO consisted of one file written per MPI rank, containing only the necessary data to restart the simulation and to maintain tracking of conservation, including the positions of the Lagrangian tracer particles associated with the processors domain (see Sect.~\ref{sec:particles}).
These sets of files were written on typically 100-cycle intervals to alternating files in a pair.
At predetermined points in the computation, `permanent' restart files were saved, to be used to provide the initial data if a simulation had to be `rewound' to fix a problem.

These restart files were supplemented with plot files written on fixed time intervals, typically every 0.2~ms after core bounce.
Each file consisted of single variable (e.g., entropy, radial velocity, \nuebar\ luminosity, etc.) data for the entire grid and was assembled by gathering the data to the root processor before writing it out.
For 3D runs with more than $\sim$10000 radial rays, the memory available on the root node was typically insufficient.
These plot files did not contain all information needed for restart, but did contain derived quantities (like luminosity) useful for visualization and analysis.

The above binary files were also supplemented with {\it in situ} analysis output of global properties (shock radius information, explosion energies, radial traces along fixed angles, etc.) written to plain text files.
To generate the `trace files' for each Lagrangian tracer, the thermodynamic, abundance, and neutrino quantities were interpolated to the particle position and recorded in a binary tracer file. 

\subsubsection{Parallel IO Method}
\label{sec:parIO}

To achieve scaling to larger process counts in 3D and improved file performance, we implemented parallel IO with the \texttt{HDF5} library\footnote{\url{hdfgroup.org}}.
The \texttt{HDF5} library permits complex file structures, metadata, and file portability.
Initially, the \texttt{HDF5} implementation was a replacement for the restart IO, without the alternating file scheme, with the analysis and plotting data added from the binary one-variable files.
To improve the time resolution of the gravitational wave analysis from the $\Delta t = 0.2$~ms resolution used in Series-A \citep{YaMaMe10} and Series-B \citep{YaMeMa15} analyses, we added a finer-resolution sampling of the quantities needed to compute the matter contribution to the gravitational wave signal -- i.e., density and velocity -- to the \texttt{HDF5} `Restart' files.

For Series-D, we are moving toward a fully \texttt{HDF5}-based system for large IO.
We have implemented fixed-time-interval `Frame' \texttt{HDF5} files, without the extra data for gravitational wave analysis, to replace the single-variable equivalents.
The density and velocity information at finer intervals is retained in `GW' \texttt{HDF5} at intervals matching those of the `Restart' files from which the `GW' data have now been removed.
This separation of the data renders the `Restart' \texttt{HDF5} files unnecessary after serving their primary role as checkpoint files for restarting a simulation.
(The `Frames' files contain the exact same data as the `Restart' files and can be used to restart a simulation, as well, if needed.)
The combination of `Frames'+`GW' \texttt{HDF5} files retained for analysis is smaller than the set of `Restart' \texttt{HDF5} files retained for the C-series.

\newpage
\subsection{Tracer Particles}
\label{sec:particles}

In addition to the IO of grid-based data, \chimera\ also outputs data for passive Lagrangian tracer particles.  
These are used to record the thermodynamic and neutrino exposure histories of individual mass elements that are then used for post-processing analysis.
Following each directional sweep of the hydrodynamics, the position of a tracer particle in that direction at time $t^{n}$ and position $(r^{n},\theta^{n},\phi^{n})$ is advanced to $t^{n+1}$ according to the simple Euler method:
\begin{eqnarray}
	r^{n+1} & = & r^{n} + u_{r}^{n} \Delta t^{n}, \\
	\theta^{n+1} & = & \theta^{n} + \frac{ u_{\theta}^{n} }{r} \, \Delta t^{n}, \\
	\phi^{n+1} & = & \phi^{n} + \frac{u_{\phi}^{n}}{(r \sin\theta)^{n}} \, \Delta t^{n},
\label{eq:traceradvect}
\end{eqnarray}
assuming constant velocity $(u_{r}^{n},u_{\theta}^{n},u_{\phi}^{n})$ through the time interval $\Delta t^{n} = t^{n+1} - t^{n}$.
For the B-series models, which were all 2D and included 4000--8000 particles, individual output files were maintained for each tracer.
These contained the physical quantities of interest, linearly interpolated in radius to the tracer particle positions from the zone-center (cell-averaged) values of the computational grid; the lone exception being the interpolation of differential neutrino number fluxes, which are defined at radial zone edges.
Output for individual particles was performed whenever the physical quantities of interest changed by 10\% \citep[see ][for more details]{HaHiCh17}.

For the C-series and later models, which included 3D models with as much as 400,000 tracers, this approach proved impractical, thus \chimera\ simply records the tracer positions in `Frames' and `Restart' files, with interpolation of quantities of interest relegated to a post-processing step.
Nucleosynthesis tests by \citet{HaHiCh17} show that output intervals of $\Delta t \sim 1$~ms are sufficient in CCSN to resolve the features in the thermodynamic profiles needed for accurate post-processing of the Lagrangian tracers.
Thus, the computational time expended on the temporally detailed individual particle traces written out in the B-series models can now be better spent elsewhere.

\subsection{Scalability}
\label{Scale}

\begin{figure}
\fig{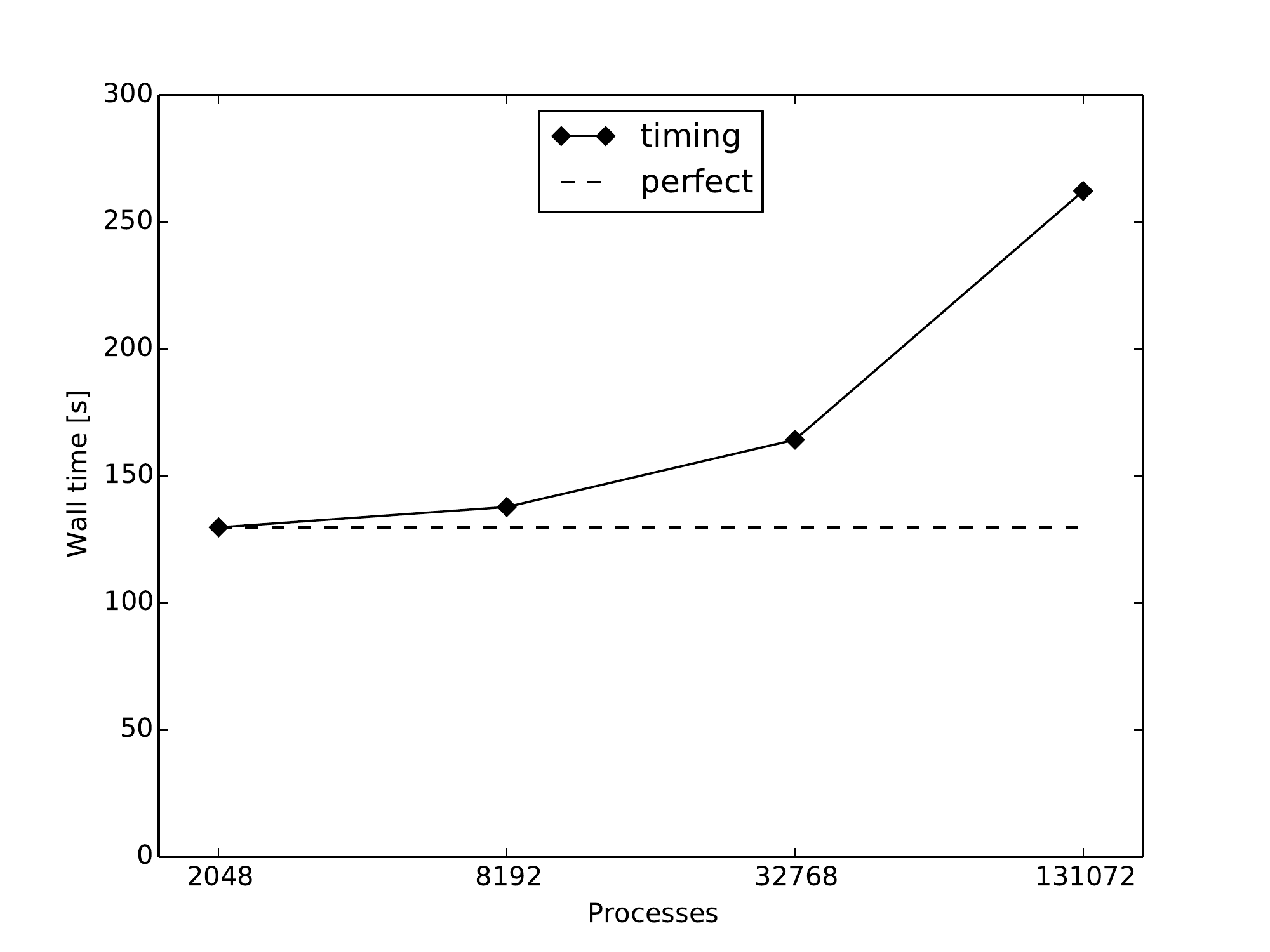}{0.7\textwidth}{}
\caption{\label{fig:scaling}
\chimera\ weak scaling for 100 cycles of full calculation on the Cray XK7 (OLCF Titan) for 32x64 (2048) rays to 256x512 (131072) rays,  with one ray per core}
\end{figure}

To test the scalability of \chimera\ we computed 3D models with all of the standard physics of the B- and C-series, started from the same 15-\msun\ progenitor used in the B15-WH07 run \citep{BrMeHi13,BrLeHi16}, with 512 radial zones.
Models were run from the onset of collapse for a couple of hours for $n_\theta$ of 32, 64, 128, and 256, with $n_\phi = 2 n_\theta$, on the Cray XK7 (`Titan') at OLCF, utilizing `maximal decomposition' with one radial ray per process (MPI rank).
The results plotted in Figure~\ref{fig:scaling} use the version of the code used for the C15-3D model \citep{LeBrHi15} and show slightly more than doubled wall time for the largest model relative to the smallest, but the typical size of our actual production 3D runs are closer to the 32768-ray (128x256) model, which takes about 25\% longer than the smallest model to complete 100 steps.

\subsection{\Chimera\ Series}
\label{Chimera_V}

The lettered series of \chimera\ simulations are not in the strict sense of software engineering and development practices `versions' of the code, but rather represent a combination of default physics, parameter choices, and a common code base used in those simulations.
As it has occurred, all of the series discussed in this paper have utilized different code bases, with progressive improvement with each subsequent series, but in the future we may use the series designation to separate groups of simulations with different defaults run with the same code.
The series label combined with a basic description of the progenitor and any deviations from the series defaults are used to construct unique simulation designations to identify a model across publications.

Initially there were separate 2D and 3D branches of the code and both were used for the simulations retroactively designated `Series~A' \citep{BrMeHi09b} with the 2D branch using only the binary IO for restart and plotting and the 3D branch using HDF5 for restarts.
The Series-B runs included many improvements to the code and simulation design, reflecting the acquired experience of Series~A.
While the 2D-only Series~B code was running, the 3D code was optimized and became the only branch of \chimera\ for all subsequent runs of any dimension and included changes to the code structure needed for the Yin-Yang grid. The Yin-Yang mesh was not yet satisfactory when Series~C was started and the major differences between that and Series~D are the finalized Yin-Yang grid and an improved radial-mesh motion routine.
Major improvements were made to the handling of the NSE state and the boundary with the non-NSE material handled by the network before Series-B, with geometric upgrades before Series~D permitting free islands of NSE or non-NSE material to embed, as necessary, in the other.
Some operational choices in the dual EoS scheme were made for Series~B, while Series~E added pre-tabulated equations of state (R.~E. Landfield et al., in prep.).
The temporal sub-cycling of the non-radial sweeps deep in the core generated entropy and was replaced for Series~B by a frozen core, wherein the non-radial hydrodynamics was skipped, which was in turn replaced by a spherical averaging scheme late in Series~B and used in all later series.

In the list that follows we describe major differences between the code revisions used in the series indicated.
For descriptions of the utilized physics and default choices for resolution, the associated simulation papers should be referenced. 
\begin{itemize}

\series{A}{ \sitem {EoS}{LS180, $\rho > \den{1.7}{8}$} \sitem {Hydrodynamics}{subcycle lateral sweep} \sitem{NSE}{spherical boundary} \sitem{IO}{binary restart and plot files} }
\series{B}{ \sitem {EoS}{LS220, $\rho > \den{1}{11}$} \sitem {Hydrodynamics}{frozen core, carbuncle update} \sitem {NSE}{17-species( $\alpha$, n, p, \isotope{Fe}{56}), boundary a function of latitude } }
\series{C}{ \sitem {Opacities}{shared points (Section~\ref{sec:oppool})} \sitem {Hydrodynamics}{averaged core} \sitem{Transport}{shock fix (Section~\ref{trans_shock}), Minerbo closure substituted fo geometric clisure}\sitem{IO}{merged restart and plot quatities into single HDF5 file} \sitem{Mesh}{Yin-Yang 3D grid related changes}    }
\end{itemize}

\section{Equation of State}
\label{app:Eos}

In order that the hydrodynamics, nuclear transmutations, and neutrino transport be tied closely to the thermodynamics, the equation of state (EoS) must be invoked several times each cycle (See Figure~\ref{fig:sweep}).
Furthermore, the EoS must provide not only the quantities needed for the hydrodynamics -- e.g., the pressure, internal energy, and entropy as a function of density, temperature, and electron fraction -- but the element composition and chemical potentials, as well, as these are needed for the computation of the opacities which, in turn, are needed in the neutrino transport. 

Furthermore, the derivatives of a number of these thermodynamic quantities are needed to compute, by a Newton-Raphson iteration, updates in one or more independent variables given updates in other independent and dependent variables. For example, we need the derivative of the internal energy with respect to temperature to update the temperature given updates in the internal energy, density, and electron fraction. And we need the derivatives of the internal energy with respect to the temperature and electron fraction (as well as derivatives of the neutrino opacities with respect to these variables) to build the Jacobian for the neutrino transport solve.

\subsection{General EoS Methods}

\begin{figure}
\fig{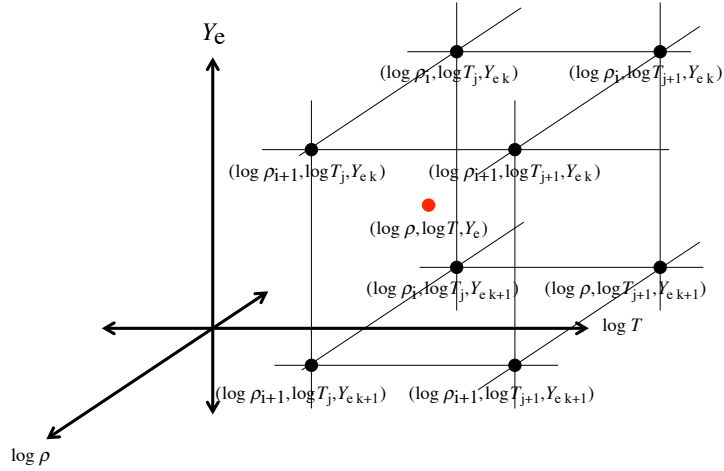}{0.6\textwidth}{}
\caption{\label{fig:Eos_cubes}
The thermodynamic state, $(\log \rho, \log T, \Ye)$, of a zone inside a grid element in $ \log \rho, \log T, and \Ye$ space. 
}
\end{figure}

\begin{figure}
\fig{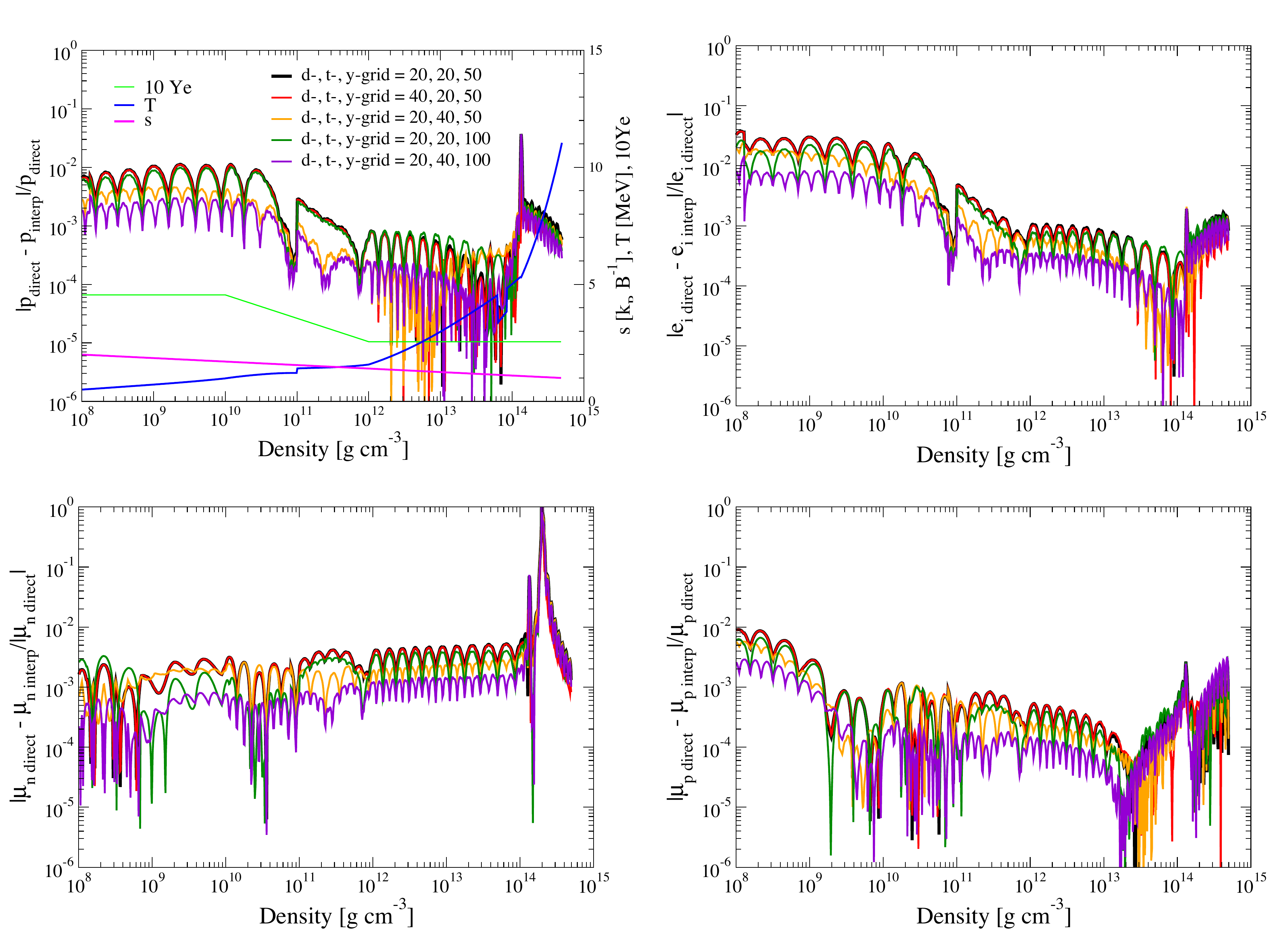}{1.0\textwidth}{}
\caption{\label{fig:s2_EoS_Prec}
Relative deviation of representative thermodynamic quantities obtained by direct output from the EoS versus 
by interpolation, for the listed EoS grid resolutions and for a `Near Bounce'  profile typical of a stellar core just prior to the formation of the bounce shock. 
Panels (a), (b), (c), and (d) show the relative deviation for the pressure, internal energy, neutron chemical potential, and proton chemical potential, respectively.
}
\end{figure}

Two general cases must be distinguished when considering the thermodynamic state of the fluid, nuclear statistical equilibrium (NSE), and non-NSE. 
In NSE, the thermodynamic state of the fluid in a given zone is specified by the values of its density, $\rho$, temperature, $T$, and electron fraction, $\Ye$. 
To accommodate the demand for frequent EoS interrogations and the need for derivatives of some of the dependent thermodynamic quantities, \Chimera\ constructs a thermodynamic grid in ($\log \rho, \log T, \Ye$)-space defined by a user specified number, $\texttt{dgrid}(\texttt{m})$, of evenly spaced points of $\log \rho$ per decade change in $\rho$, a user specified number, $\texttt{tgrid}(\texttt{m})$, of evenly spaced points of $\log T$ per decade change in $T$, and a user specified number, $\texttt{ygrid}(\texttt{m})$, of points in \Ye\ over the range [0,1]. 
The index $\texttt{m} = 1, 2, 3$ allows the user to select three different thermodynamic grid resolutions for the three density ranges $\rho < \rho_{\rm es}(1)$, $\rho_{\rm es}(1) < \rho < \rho_{\rm es}(2)$, and $\rho > \rho_{\rm es}(2)$, where $\rho _{\rm es}(1)$ and $\rho_{\rm es}(2)$ are user selected densities. 
A particular dependent thermodynamic function, corresponding to the thermodynamic state ($\log \rho, \log T, \Ye$), is computed by linear interpolation from its values at the eight surrounding grid points, which satisfy
\begin{equation}
\log \rho_{i} < \log \rho \le \log \rho_{i+1}, \quad \log T_{j} < \log T \le \log T_{j+1},  \quad Y_{{\rm e},k+1} < \Ye \le Y_{{\rm e},k},
\label{eq:e1}
\end{equation}
where the $\rho_{i}$, $T_{j}$, and $Y_{{\rm e},k}$ are the values of $\rho$, $T$, and \Ye\ at the grid points.
Figure~\ref{fig:Eos_cubes} shows a cell of the thermodynamic grid within which the thermodynamic state, ($\log \rho, \log T, \Ye$), is located.
We will refer to the eight grid points surrounding a given mass zone as the `surrounding grid points', and the cell itself simply as the `EoS cell'.
We emphasize that not all grid points of the thermodynamic grid have thermodynamic quantities evaluated and  stored there, but only those grid points surrounding the thermodynamic states of mass zones, with a cell of grid points tied to each zone. 
Initially, the thermodynamic state of each mass zone gets a suite of thermodynamic quantities computed and stored at the eight surrounding grid points.
During a simulation, when the changing thermodynamic state of a mass zone causes the state to enter a different EoS cell, the needed thermodynamic quantities are in turn computed and stored on the grid points of that cell.
Thus, thermodynamic quantities (and neutrino opacities) are computed on an `as needed' basis, keeping the thermodynamic state of each mass zone surrounded by the needed thermodynamic quantities on the eight nearest ($\log \rho, \log T, \Ye$)-grid points.
Lastly, to avoid involving an excessive number of quantities in internode communication when transposing from one set of rays (radial, angular, or azimuthal) to another, the EoS grids along these rays are maintained independently. 

A total of 14 dependent EoS variables comprise the thermodynamic vector that is computed and stored at each of the grid points of a cell surrounding the thermodynamic state of a mass zone. 
These quantities are the pressure, $p$; specific internal energy, $e$; specific entropy, $s$; neutron chemical potential, $\mu_{\rm n}$;  proton chemical potential, $\mu_{\rm p}$;  electron chemical potential, $\mu_{\rm e}$; neutron mass fraction, $X_{\rm n}$; proton mass fraction, $X_{\rm p}$; representative heavy nucleus mass fraction $X_{\rm H}$ (nuclei with mass numbers greater than helium), along with the mass number, $A$, charge number $Z$, and mean binding energy per particle $b_{\rm A}$ of the representative heavy nucleus; the adiabatic exponent $\Gamma_{s} = \left({\partial p}/{\partial \rho} \right)_{s,\Ye}$, and the specific internal energy, $e_{\rm int}$, with the particle rest masses and arbitrary constants subtracted out. 
The helium mass fraction, $X_{\alpha}$, is not stored, but computed as $X_{\alpha} = 1 - X_{\rm n} - X_{\rm p} - X_{\rm a}$. 
The specific internal energy, $e_{\rm int}$, is used to compute the quantity $\Gamma_{e} = {p}/{ e_{\rm int} \rho } + 1$, utilized by the Riemann solver to prevent unphysical $\Gamma_{e} < {4}/{3}$, which can cause post-shock oscillations in the some of the thermodynamic variables \citep{BuRaJa06}. 
The formula used to interpolate a thermodynamic quantity from its values at the eight surrounding grid points is differentiated (exactly) to obtain partial derivatives of that quantity with respect to either $\rho$, $T$, or \Ye, as needed.

\begin{figure}
\fig{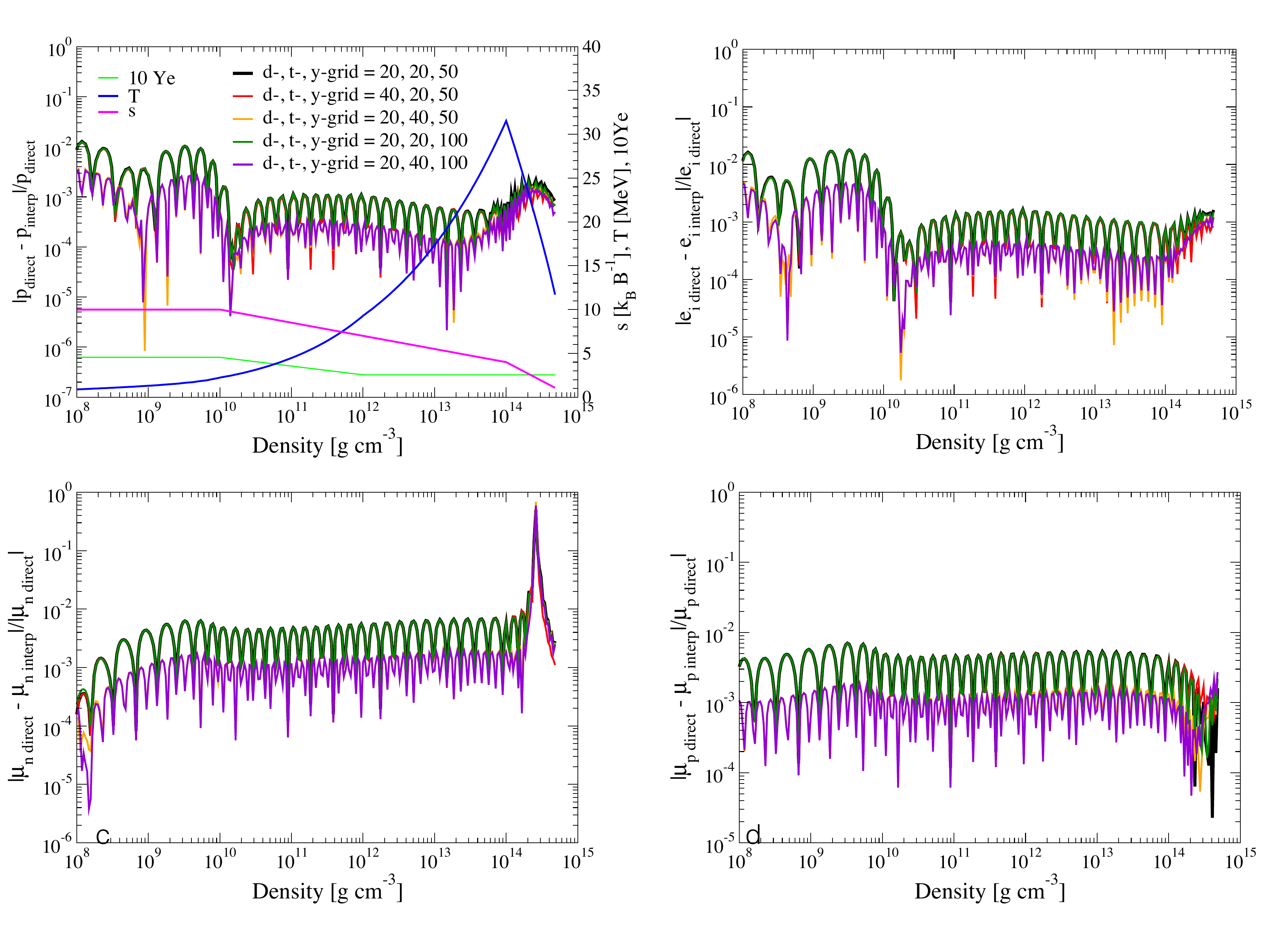}{1.0\textwidth}{}
\caption{\label{fig:s10_EoS_Prec}
As in Figure \ref{fig:s2_EoS_Prec}, but for a `Shock Stagnation' profile representative of a stellar core during the epoch of shock stagnation.
}
\end{figure}

To provide a sense of the accuracy of our EoS interpolation scheme, Figures \ref{fig:s2_EoS_Prec} and \ref{fig:s10_EoS_Prec} show the relative deviation of the pressure, specific internal energy, and the neutron and proton chemical potentials obtained by direct output from our stellar EoS (described below) versus interpolation in the EoS grid, for the grid resolution listed in the figures. 
The $\rho$, $s$, and \Ye\ profiles used for generating Figures~\ref{fig:s2_EoS_Prec} and \ref{fig:s10_EoS_Prec} are representative of models near bounce but before the formation of the shock, and many tens of ms after bounce during shock stagnation.
We will refer to these profiles as `Near Bounce' (Figure~\ref{fig:s2_EoS_Prec}) and `Shock Stagnation' (Figure~\ref{fig:s10_EoS_Prec}).
It is clear from the figures that increasing the density resolution from 10 to 20 grid points per decade in density does not decrease the relative deviation of these thermodynamic quantities except for densities above nuclear saturation. 
(The black and red curves lie on top of each other below the saturation density.) 
Increasing the temperature resolution from 20 to 40 points per decade in temperature (orange curve) reduces the relative deviation for all of the graphed quantities throughout most of the density range displayed. 
Increasing the electron fraction resolution from 50 to 100 over the range [0,1] in \Ye\ decreases the relative deviation for quantities (namely, abundances and chemical potentials) in regimes (low entropy) where there is a strong dependence on $\Ye$ -- e.g., where there is partial dissociation. 
The B-series models were performed with the EoS grid resolution of (\texttt{d}-, \texttt{t}-, \texttt{ygrid}) = 20, 20, 50, which, except for a few exceptions described below, typically obtains values of interpolated thermodynamic quantities within a percent or so of the values obtained directly from the EoS (in most cases less than a percent).
The D-series models are being performed with the higher grid resolution of (\texttt{d}-, \texttt{t}-, \texttt{ygrid}) = 20, 40, 100, which typically gives a relative deviation about five times smaller.

\begin{figure}
\fig{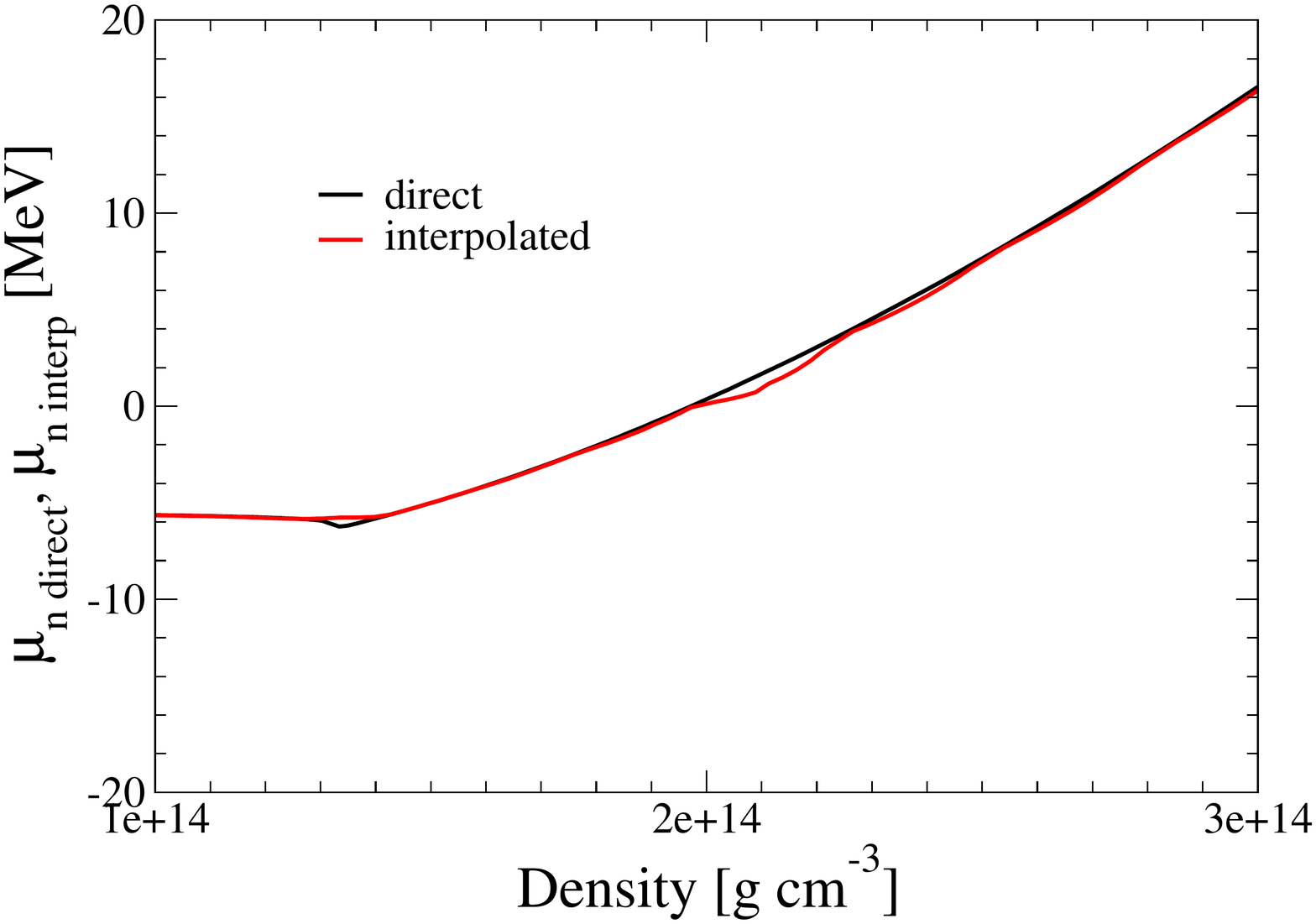}{0.5\textwidth}{}
\caption{\label{fig:s2_mu_n_comp}
Neutron chemical potential for the `Near Bounce' profile, obtained directly from the EoS (black line) versus by interpolation for (\texttt{d}-, \texttt{t}-, \texttt{ygrid}) = 20, 40, 100  (red line). 
Slight differences in the chemical potential while it passes through zero at densities 2--\den{3}{14}\ are the cause of the large relative errors there. 
The `Shock Stagnation' profile yields similar results and discrepancies.
}
\end{figure}

A few features of the graphs deserve comment. One feature is the slight kink in the temperature at $\rho =  10^{11}$ \gcc\ in the `Near Bounce' profile (Figure \ref{fig:s2_EoS_Prec}a), and is is due to the LS EoS -- C EoS (see Section~\ref{StellarEoS}) transition at that density. 
Another is the substantial relative error in the neutron chemical potential at a density of $2 \times 10^{14}$ \gcc\ for the `Near Bounce' profile, and at a density of $2.5 \times 10^{14}$ \gcc\ for the `Shock Stagnation' profile.
These are the densities for the respective profiles at which the neutron chemical potentials pass through zero, so any slight deviation in the interpolated versus the directly obtained values for these quantities will be amplified by their small absolute values when computing their relative deviations.
The region where the neutron chemical potentials change sign is shown in Figure \ref{fig:s2_mu_n_comp} for the `Near Bounce' profile, but is representative of the `Shock Stagnation' profile as well.
The spikes in the relative deviation of all quantities at $1.3 \times 10^{14}$ \gcc\, for the `Near Bounce' profile deserves mention, as well.
These are caused by the nuclei--nuclear matter phase transition at that density, and the abrupt change in composition there. 
Table interpolation smooths this transition across the width of the density grid, while direct calls to the EoS see this transition as a discontinuity.
These spikes do not appear in the `Shock Stagnation' profile because the higher entropy results in matter being completely dissociated at the above density, leading to a  smoother transition across the nuclei--nuclear matter phase transition.

\subsection{NSE EoSs}
\label{StellarEoS}

For \ccsn\ simulations, the pressure, specific internal energy, and specific entropy are taken as the sum of contributions from different species, namely
\begin{equation}
p = p_{\rm ion} + p_{\rm e^{-}+e^{+}} + p_{\rm rad}, \quad e = e_{\rm ion} + e_{\rm e^{-}+e^{+}} + e_{\rm rad}, \quad s = s_{\rm ion} + s_{\rm e^{-}+e^{+}} + s_{\rm rad} ,
\label{eq:e14}
\end{equation}
where the subscripts ``ion,'' ``e$^{-}$+e$^{+}$,'' and ``rad'' denote contributions from nuclei, electrons and positrons, and photons, respectively. 
For the B-, C-, and D-series runs, \Chimera\ employs, for densities above $10^{11}$ \gcc, the $K = 220$~\mev\ incompressibility version of the \citet{LaSw91} (LS) EoS  for the ion and photon components.
(The retroactively named A-series used the $K = 180$~\mev\ version of LS EoS.)
The LS EoS utilizes a compressible liquid drop model for nuclei modeled after the \citet{LaLaPe78} (LLPR) formalism, and considers an ion composition of free neutrons and protons, helium, and a representative heavy nucleus. 
For matter in NSE at densities below $10^{11}$ \gcc\ with $\Ye < \rfrac{26}{56}$, the \citet{Coop85} (C) EoS is used. 
The C~EoS does not treat the high-density parameters of the liquid drop model (nuclear incompressibility modulus, surface energy, symmetry energy) as consistently as the LS EoS, but computes the mass fraction of helium more accurately than the LS EoS in the regime below $10^{11}$ \gcc\, where it is being employed.
Advection of material across this EoS boundary requires consistent tracking of the specific internal energy. (See Section~\ref{remap3} for details.)
To improve the fidelity of the composition of matter that may eventually become part of the ejecta, in regions of NSE where $\Ye \geq \rfrac{26}{56}$ (the value of $Z/A$ for \isotope{Fe}{56}) the NSE calculation in C EoS has been upgraded to a 17-species representation of the composition, including free neutrons, free protons, the 14 even-Z and even-A nuclei between \isotope{He}{4} and \isotope{Zn}{60} plus  \isotope{Fe}{56}.
This NSE calculation functions similarly to others in the literature \citep[e.g.,][]{HaWoEl85,ClTa65} but is limited by the small isotope set.

\subsection{Nuclear Network and Non-NSE Region}
\label{app:nuc_network}

In zones where the timescale required to reach NSE is larger than other physical timescales (e.g., those associated with the evolution of stellar core fluid), the nuclear composition is evolved using the XNet thermonuclear reaction network code. In these regions, the thermodynamic state depends on the isotopic composition, as well as $\rho$ and $T$, and the electron fraction (\Ye) is calculated from $Y_e = \sum_i Z_i Y_i$, where $Z_i$ is the proton number of an isotope and $Y_i$ is the molar abundance of that isotope.

\begin{deluxetable*}{ll}
\tabletypesize{\scriptsize}
\tablecaption{Available Nuclear Networks \label{tab:networks}}
\tablecolumns{3}
\tablewidth{0pt}
\tablehead{
\colhead{Network} & \colhead{Nuclear Species} 
}
\startdata
{\tt alpha} & \isotope{He}{4}, \isotope{C}{12}, \isotope{O}{16}, \isotope{Ne}{20}, \isotope{Mg}{24}, \isotope{Si}{28}, \isotope{S}{32}, \isotope{Ar}{36}, \isotope{Ca}{40}, \isotope{Ti}{44}, \isotope{Cr}{48}, \isotope{Fe}{52}, \isotope{Ni}{56}, \isotope{Zn}{60} \\
{\tt alpnp} & \isotope{n}{}\tablenotemark{$\dag$}, \isotope{H}{1}\tablenotemark{$\dag$}, \isotope{He}{4}, \isotope{C}{12}, \isotope{O}{16}, \isotope{Ne}{20}, \isotope{Mg}{24}, \isotope{Si}{28}, \isotope{S}{32}, \isotope{Ar}{36}, \isotope{Ca}{40}, \isotope{Ti}{44}, \isotope{Cr}{48}, \isotope{Fe}{52}, \isotope{Ni}{56}, \isotope{Zn}{60}  \\
{\tt anp56} & \isotope{n}{}\tablenotemark{$\dag$}, \isotope{H}{1}\tablenotemark{$\dag$}, \isotope{He}{4}, \isotope{C}{12}, \isotope{O}{16}, \isotope{Ne}{20}, \isotope{Mg}{24}, \isotope{Si}{28}, \isotope{S}{32}, \isotope{Ar}{36}, \isotope{Ca}{40}, \isotope{Ti}{44}, \isotope{Cr}{48}, \isotope{Fe}{52}, \isotope{Fe}{56}\tablenotemark{$\dag$}, \isotope{Ni}{56}, \isotope{Zn}{60}  \\
\\
{         } & \isotope{n}{}, \isotope{H}{1\textrm{--}2}, \isotope{He}{3\textrm{--}4}, \isotope{Li}{6\textrm{--}7}, \isotope{Be}{7,9},\isotope{B}{8,10,11}, \isotope{C}{12\textrm{--}14}, \isotope{N}{13\textrm{--}15}, \isotope{O}{14\textrm{--}18}, \isotope{F}{17\textrm{--}19}, \isotope{Ne}{18\textrm{--}22}, \isotope{Na}{21\textrm{--}23}, \\
{\tt sn150} & \isotope{Mg}{23\textrm{--}26}, \isotope{Al}{24\textrm{--}27}, \isotope{Si}{28\textrm{--}32}, \isotope{P}{29\textrm{--}33}, \isotope{S}{32\textrm{--}36}, \isotope{Cl}{33\textrm{--}37}, \isotope{Ar}{36\textrm{--}40}, \isotope{K}{37\textrm{--}41}, \isotope{Ca}{40\textrm{--}48}, \isotope{Sc}{43\textrm{--}49},  \\
{         } & \isotope{Ti}{44\textrm{--}50}, \isotope{V}{46\textrm{--}51}, \isotope{Cr}{48\textrm{--}54}, \isotope{Mn}{50\textrm{--}55}, \isotope{Fe}{52\textrm{--}58}, \isotope{Co}{53\textrm{--}59}, \isotope{Ni}{56\textrm{--}62}, \isotope{Cu}{57\textrm{--}63}, \isotope{Zn}{59\textrm{--}66} \\
\\
{         } & \isotope{n}{}, \isotope{H}{1\textrm{--}2}, \isotope{He}{3\textrm{--}4}, \isotope{Li}{6\textrm{--}7}, \isotope{Be}{7,9},\isotope{B}{8,10,11}, \isotope{C}{12\textrm{--}14}, \isotope{N}{13\textrm{--}15}, \isotope{O}{14\textrm{--}18}, \isotope{F}{17\textrm{--}19}, \isotope{Ne}{18\textrm{--}22}, \isotope{Na}{21\textrm{--}23}, \\
{\tt sn160} & \isotope{Mg}{23\textrm{--}26}, \isotope{Al}{25\textrm{--}27}, \isotope{Si}{28\textrm{--}32}, \isotope{P}{29\textrm{--}33}, \isotope{S}{32\textrm{--}36}, \isotope{Cl}{33\textrm{--}37}, \isotope{Ar}{36\textrm{--}40}, \isotope{K}{37\textrm{--}41}, \isotope{Ca}{40\textrm{--}48}, \isotope{Sc}{43\textrm{--}49}, \isotope{Ti}{44\textrm{--}51},   \\
{         } & \isotope{V}{46\textrm{--}52}, \isotope{Cr}{48\textrm{--}54}, \isotope{Mn}{50\textrm{--}55}, \isotope{Fe}{52\textrm{--}58}, \isotope{Co}{53\textrm{--}59}, \isotope{Ni}{56\textrm{--}64}, \isotope{Cu}{57\textrm{--}65}, \isotope{Zn}{59\textrm{--}66}, \isotope{Ga}{62\textrm{--}64}, \isotope{Ge}{63\textrm{--}64} \\
\enddata
\tablenotetext{$\dag$}{inert species, which are advected but not reactive}
\end{deluxetable*}

The initial value problem presented by a nuclear reaction network for an isolated region (individual zone) can, in principle, be solved by a wide range of methods discussed in the literature.  
However the physical nature of the problem, reflected in the wide range of reaction timescales, renders these numerical systems \emph{stiff}.  
The challenges of solving such stiff astrophysical systems are detailed in a number of review articles on the subject \citep[see, e.g., ][]{HiMe06,TrHi13}.  
In A-, B- and C-series of \chimera\ models, XNet utilizes the fully-implicit Backward-Euler method, introduced to nuclear astrophysics by \citet{ArTr69}.
Data for these reactions is drawn from the REACLIB compilation \citep{RaTh00}.
Unfortunately, A- and B-series \chimera\ models neglect screening of nuclear reactions.
The nuclear state is updated for each non-NSE zone in each time step. As needed in each zone, the network is sub-cycled until the hydrodynamic time step is reached.

Several pre-built networks available in \chimera\ are shown in Table~\ref{tab:networks}.
The A- and B-series models all utilize the simple 14-species \alp-network \texttt{alpha}.
The active nuclear material evolved in XNet excludes free protons, free neutrons, and an auxiliary heavy nucleus that are advected with the nuclear composition.
In the C-series models, we switched to the \texttt{alpnp} network that adds protons and neutrons to the network, though the free nucleons are effectively inert, as their are no reactions included that connect them to other species.
The properties (mass and charge number, binding energy, and mass fraction) of the auxiliary heavy nucleus are taken initially from the part of the composition of the progenitor that cannot be mapped onto the network species. 
For material that has come out of NSE, the properties are taken from the sum of nuclei not included in the network composition vector.
For networks \texttt{alpha} and \texttt{alpnp} this consists of \isotope{Fe}{56} for $\Ye \geq \rfrac{26}{56}$, or the representative heavy nucleus from the nuclear EoS for $\Ye \le \rfrac{26}{56}$).
For the various D-series models underway, the base network has been updated to \texttt{anp56}, which adds \isotope{Fe}{56} as an additional unconnected, inert species and permits the network to map directly to the 17-species NSE used by the extended C-EoS and also reduces the mass fraction traced by the auxiliary heavy nucleus.
These modifications to the network infrastructure primarily serve the development of even larger networks (\texttt{sn150} and \texttt{sn160}) for CCSN simulations.
The D-series includes 2D and 3D simulations utilizing the \texttt{sn160} network.

When not in NSE, the thermodynamic quantities for the EoS cell are computed assuming the same composition fo nuclei on all 8 vertices. 
This gives rise to an apparent inconsistency, however, as the electron fraction, \Ye, at some or all of the \added{EoS} grid points will not correspond to the \Ye\ of the composition of nuclei.
Furthermore, the electron--positron contribution to the thermodynamic vector at each EoS grid point is computed from the values of $\rho, T,$ and $\Ye$ at that grid point, the result being that the \Ye\ of the electron--positron gas is not consistent with the \Ye\ of the composition of nuclei at a grid point.
This procedure, however, allows us to take finite derivatives with respect to the electron fraction, albeit with some approximation, but accurate enough to stabilize the neutrino transport in the non-NSE regions.
At the same time, when the thermodynamic state of a mass zone is interpolated from the EoS cell, the contribution of the ions will be based on a \Ye\ common to all EoS cell points, while the contributions of the electron--positron gas will be interpolated to the same \Ye, and the two contributions will reflect the same value of \Ye.
The ions used to compute the thermodynamic properties are those in the network, the  heavy nucleus advected with the active composition, and, in the case of the original \texttt{alpha} network used in Series A and B, the free nucleon mass fractions whose values are also stored with the thermodynamic vector.

\subsection{NSE Transition}

\chimera's treatment of the transition of matter into NSE is comparable to that used in other CCSN codes of similar capability \citep{BuJaRa06,MuJaMa12,NaTaKo14,SkDoBu19}.
The transition condition is motivated by the temperatures and densities at which complete silicon burning would occur within the current global time step.
For temperatures above this threshold, the use of the nuclear network is superfluous, as the network will achieve NSE every time step.
For transitioning into NSE, there will be a slight change in the nuclear binding energy, due to the change of the representative nucleus, the ensemble of nuclei, etc., and the extent of this change is one metric for gauging the accuracy of our assumption of NSE.
In order to maintain hydrodynamic stability across this transition, we adjust temperature to maintain constant pressure for a given density and electron fraction.
This may result in a small change in the specific internal energy (including binding energy), but the dynamical impact that results from the transition between inconsistent thermodynamic states is minimized.
The advection of material across an NSE/non-NSE interface in either direction, as well as the transition into NSE and freeze-out from NSE of entire zones, includes the appropriate gain or loss of nuclear binding energy (see Section~\ref{remap4} for details).
In the A-series simulations, the NSE interface was a sphere of fixed radius, while in the B-series simulations the NSE boundary was independent for each radial ray.
The C and later series allow multiple NSE/non-NSE interfaces along any coordinate direction.

To determine whether a zone is in NSE and may, therefore, be omitted from nuclear burning, \chimera\ applies an empirically determined linear relationship between the NSE transition temperature, \TNSE, and the density:
\begin{align}
	\TNSE(\rho) = 
	\begin{cases}
		C_{1} \rho + C_{2}	& \text{if } \rho < 2\e{8}~\gcc; \\
		6.5\e9~\mathrm{K}	& \text{otherwise},
	\end{cases}
	\label{eq:tnse}
\end{align}
where $C_{1} \equiv 5.333$~$\mathrm{K~g^{-1}~cm^{3}}$ and $C_{2} \equiv 5.433\e{9}~\mathrm{K}$.
At the beginning of a global timestep, any non-NSE zone for which $T \geq \TNSE$ is transitioned to NSE.
A zone which is in NSE at the beginning of a timestep will be transitioned out of NSE if $T < (\TNSE - 2 \times 10^{8}~\mathrm{K})$ and if the representative heavy nucleus, split into \isotope{Ni}{56} and \isotope{Fe}{56}, will result in less than half of the mass fraction being \isotope{Fe}{56}.
If this latter condition is not met, NSE is maintained as long as is practical, until the condition is met or until $T < 4.9\e{9}~\mathrm{K}$, as the NSE representation of neutron-rich composition is better than that allowed by this limited network.

For simplicity, the transition out of NSE occurs when the temperature drops below this NSE condition (Equation~(\ref{eq:tnse}) for \chimera).
However, for the rapidly changing conditions in expanding CCSN matter, the assumption of NSE has been shown to break down when the temperature falls below 6~GK \citep{MeKrCl98}.
For transitioning out of NSE using the alpha network (but before evolving the network), the nuclear binding energy does not change.
The NSE composition is evaluated using some analytically calculated state, but the temperature is then adjusted so that, for a given density and electron fraction, we will get the same specific internal energy (including binding energy) using a local EoS cube interpolation.

\subsection{Electron--Positron EoS}
\label{sec:epeos}

\begin{figure}
\fig{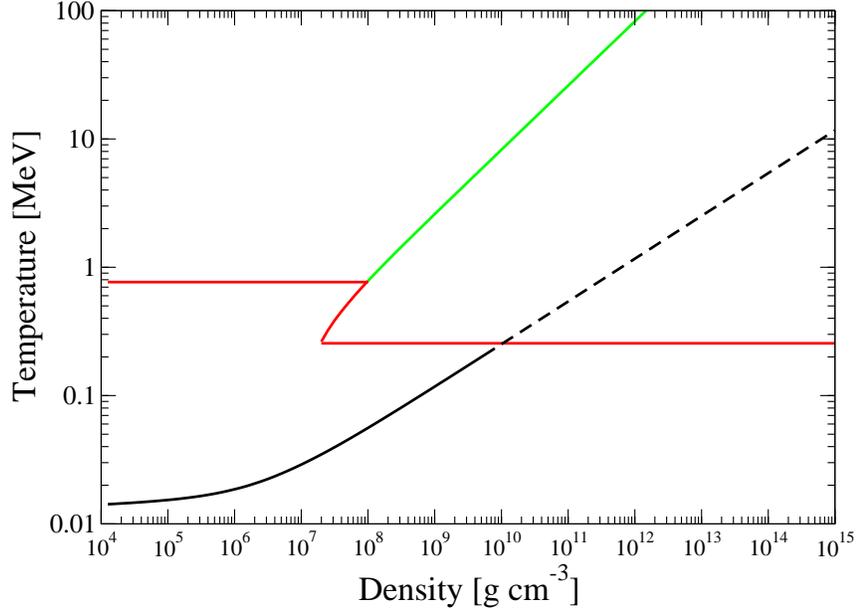}{0.7\textwidth}{}
\caption{\label{fig:EP_Eos_Regimes}
Regimes in the T-$\rho$ plane, showing the different schemes for computing the electron--positron EoS quantities at $\Ye = 0.5$.
}
\end{figure}

The computation of the electron--positron component of the EoS is divided into six regimes (Figure~\ref{fig:EP_Eos_Regimes}). 
The first major division is based on whether the electrons are relativistic or non-relativistic. 
The electron--positron gas is regarded as non-relativistic if
\begin{equation}
e_{\rm nr,Fermi} < 0.01 m_{\rm e}c^{2} \qquad kT < 0.01 \, m_{\rm e}c^{2} ,
\label{eq:e15}
\end{equation}
where the nonrelativistic electron Fermi energy, $e_{\rm nr \, Fermi}$, is given by
\begin{equation}
e_{\rm nr,Fermi} = \frac{ (\hbar c)^{2} }{ 2 m_{\rm e}c^{2} } \left( 3 \pi^{2} \right)^{2/3} n_{\rm e}^{2/3}, \quad n_{\rm e} = \frac{\rho \Ye }{m_{\rm B} },
\label{eq:e16}
\end{equation}
where $n_{\rm e} = n_{\rm e^{-}} - n_{\rm e^{+}}$ is the net electron number density, $n_{\rm e^{-}}$ is the electron density, and $n_{\rm e^{+}}$ is the positron density. Otherwise, the electron--positron gas is treated as relativistic. 

If the electron--positron gas is relativistic, there are three regimes where different approximations are used: high temperature, very degenerate, and intermediate. In the intermediate regime, the Fermi integrals for the electron--positron thermodynamic functions are integrated numerically. Since this is the most computationally intensive procedure, it is first ascertained whether the thermodynamic state can be considered to be in the high-temperature or the very degenerate regime. 
If the thermodynamic state is found to be in neither of those regimes, the calculation of the thermodynamic functions begins with net electron density, given by
\begin{eqnarray}
 n_{\rm e} &=&  \frac{1}{\pi^{2}\hbar^{3}} \left[ \int_{0}^{\infty} \frac{ p^{2} dp }{ \exp{[ (  \sqrt{ p^{2} c^{2} + m_{\rm e}^{2} c^{4} } - \mu_{\rm e} )/kT ]} + 1 } - \int_{0}^{\infty} \frac{ p^{2} dp }{ \exp{[ (  \sqrt{ p^{2} c^{2} + m_{\rm e}^{2} c^{4} } + \mu_{\rm e} )/kT ]} + 1 } \right] 
  \nonumber \\ 
  &=& \frac{k^{3}T^{3}}{\pi^{2}\hbar^{3}} \left[ \int_{0}^{\infty} \frac{ \left( x + \beta \right) dx \sqrt{ x ( x + 2 \beta ) } }{ e^{ x + \beta - \eta_{\rm e} } + 1 } - \int_{0}^{\infty} \frac{ \left( x + \beta \right) dx \sqrt{ x ( x + 2 \beta ) } }{ e^{ x + \beta + \eta_{\rm e} } + 1 } \right], 
\label{eq:e18} 
\end{eqnarray}
where in the last expression $\beta = m_{\rm e}c^{2}/kT$, $\eta_{\rm e} = \mu_{\rm e}/kT$, and the following substitutions have been made: $y = p/m_{\rm e}c$, $z^{2} = y^{2} + 1$, and $x = \beta ( z - 1 )$. 
To obtain $\eta_{\rm e}$, the right-hand side of Equation~({\ref{eq:e18}) is integrated numerically by means of a 48-point Gauss-Laguerre scheme using a guess for $\eta_{\rm e}$, and then iterated for the $\eta_{\rm e}$ until the right-hand side equals $n_{\rm e}$ to within one part in $10^{6}$. 
Once $\eta_{\rm e}$ is obtained, $p_{\rm e+p}$ and  $e_{\rm e+p}$, the latter of which includes the electron--positron rest mass energy, are obtained by 48-point Gauss-Laguerre numerical integration of the following Fermi integrals
\begin{equation}
p_{\rm e+p} = \frac{ \hbar c }{ 3 \pi^{2} } \left( \frac{ kT }{ \hbar c } \right)^{4} \int_{0}^{\infty} \left[ dx \frac{ ( x^{2} + 2 \beta x )^{3/2} }{ e^{ x + \beta - \eta_{\rm e} } + 1 } + dx \frac{ ( x^{2} + 2 \beta x )^{3/2} }{ e^{ x + \beta + \eta_{\rm e} } + 1 } \right],
\label{eq:e19} 
\end{equation}
\begin{equation}
e_{\rm e+p} = \frac{ \hbar c }{ \pi^{2} } \left( \frac{ kT }{ \hbar c } \right)^{4} \int_{0}^{\infty} \left[ dx \frac{ ( x + \beta )^{2} \sqrt{ x^{2} + 2 \beta x } }{ e^{ x + \beta - \eta_{\rm e} } + 1 } + dx \frac{ ( x + \beta )^{2} \sqrt{ x^{2} + 2 \beta x } }{ e^{ x + \beta + \eta_{\rm e} } + 1 } \right] .
\label{eq:e20} 
\end{equation}
The entropy is obtained from the pressure, specific internal energy, and chemical potential from the thermodynamic relation
\begin{equation}
s_{\rm e+p} =  \frac{1}{kT} \left[ \frac{ e_{\rm e+p} }{ n_{\rm B} } + \frac{ p_{\rm e+p} }{ n_{\rm B } } - \mu_{\rm e} \Ye \right] .
\label{eq:e21}
\end{equation}

To determine whether the high-temperature approximation may be applied, Equation (\ref{eq:e18}) is rewritten with the substitution $z = x+\beta$ to get
\begin{equation}
n_{\rm e} = \frac{1}{\pi^{2}} \left( \frac{kT}{ \hbar c} \right)^{3} \left[ \int_{\beta}^{\infty} \frac{ z dz \sqrt{ z^{2} - \beta^{2} } }{ e^{ z - \eta_{\rm e} } + 1 } - \int_{\beta}^{\infty} \frac{ z dz \sqrt{ z^{2} - \beta^{2} } }{ e^{ z + \eta_{\rm e} } + 1 } \right] .
\label{eq:e22} 
\end{equation}
The high-temperature approximation consists of setting $\beta = 0$ in Equation (\ref{eq:e22}) and performing the integrations analytically
\begin{eqnarray}
 n_{\rm e} &\simeq& \frac{1}{\pi^{2}} \left( \frac{kT}{ \hbar c} \right)^{3} \left[ \int_{0}^{\infty} \frac{ x^{2} dx }{ e^{ x - \eta_{\rm e} } + 1} - \int_{0}^{\infty} \frac{ x^{2} dx }{ e^{ x + \eta_{\rm e} } + 1} \right] \nonumber \\ 
&  =& \frac{1}{\pi^{2}} \left( \frac{kT}{ \hbar c} \right)^{3} \left[ F_{2}(\eta_{\rm e}) - F_{2}(-\eta_{\rm e}) \right] \nonumber \\ 
& =& \frac{1}{\pi^{2}} \left( \frac{kT}{ \hbar c} \right)^{3} \left[ \frac{1}{3} \eta_{\rm e} \left( \pi^{2} + \eta_{\rm e}^{2} \right) \right] ,
\label{eq:e23} 
\end{eqnarray}
with the solution for $\eta_{\rm e}$
\begin{equation}
\eta_{\rm e} = \sqrt[3]{ \frac{3 \chi}{2} + \sqrt{ \frac{ 9 \chi^{2} }{4} + \frac{ \pi^{6} }{27} } } + \sqrt[3]{ \frac{3 \chi}{2} - \sqrt{ \frac{ 9 \chi^{2} }{4} + \frac{ \pi^{6} }{27} } } ,
\label{eq:e24}
\end{equation}
where $\chi = \pi^{2} \left( { \hbar c}/{kT} \right)^{3} n_{\rm e}$. 
The high-temperature approximation is then applied if $\beta < 2/3$, or if $\beta < 2$ and $\eta_{\rm e}$ given by Equation~(\ref{eq:e24}) satisfies $\eta_{\rm e} > 2 \beta$.
The high-temperature approximation is applied if values of $\rho$ and $T$ lie above and to the right of the solid red line in Figure~\ref{fig:EP_Eos_Regimes}. 
The solid red horizontal line at the left is given by the first condition above, the rest of the red line is given by the second condition. 
The solid green line terminating in the curved red segment is given by the condition $\eta_{\rm e} > 2 \beta$. Once $\eta_{\rm e}$ is obtained, analytic expressions for $p_{\rm e+p}$ and  $e_{\rm e+p}$ in the high-temperature approximation are given by
\begin{eqnarray}
 p_{\rm e+p} &=& \frac{\hbar c}{ 3 \pi^{2}} \left( \frac{kT}{ \hbar c} \right)^{4} \left[ F_{3}(\eta_{\rm e}) + F_{3}(-\eta_{\rm e}) - \frac{3}{2} \beta^{2} \left[ F_{1}(\eta_{\rm e}) + F_{1}(-\eta_{\rm e}) \right] \right]  \nonumber \\ 
&  =& \frac{\hbar c}{ 3 \pi^{2}} \left( \frac{kT}{ \hbar c} \right)^{4} \left[ \frac{ 7 \pi^{4} }{60} + \frac{1}{2} \eta_{\rm e}^{2} \left( \pi^{2} + \frac{1}{2} \eta_{\rm e}^{2} \right) - \frac{3}{2} \frac{\beta^{2}}{6} \left( \pi^{2} + 3 \eta_{\rm e}^{2} \right) \right] , \label{eq:e25} \\
e_{\rm e+p} &=& \frac{\hbar c}{ \pi^{2}} \left( \frac{kT}{ \hbar c} \right)^{4} \left[ F_{3}(\eta_{\rm e}) + F_{3}(-\eta_{\rm e}) - \frac{1}{2} \beta^{2} \left[ F_{1}(\eta_{\rm e}) + F_{1}(-\eta_{\rm e}) \right] \right] , \label{eq:e26} 
\end{eqnarray}
so that
\begin{eqnarray}
 e_{\rm e+p} + p_{\rm e+p} &=& \frac{\hbar c}{ 3 \pi^{2}} \left( \frac{kT}{ \hbar c} \right)^{4} \left\{ 4 \left( F_{3}(\eta_{\rm e}) + F_{3}(-\eta_{\rm e}) \right) - 4 \times \frac{3}{2} \beta^{2} \left[ F_{1}(\eta_{\rm e}) + F_{1}(-\eta_{\rm e}) \right] \right. \nonumber \\ 
 && \left. + 2 \times \frac{3}{2} \beta^{2} \left[ F_{1}(\eta_{\rm e}) + F_{1}(-\eta_{\rm e}) \right] \right\} \nonumber \\ 
& =& 4 p_{\rm e+p} + \frac{\hbar c}{\pi^{2}} \left( \frac{kT}{ \hbar c} \right)^{4} \frac{\beta^{2}}{6} \left( \pi^{2} + 3 \eta_{\rm e}^{2} \right) .
\label{eq:e27} 
\end{eqnarray}
Substituting Equation~(\ref{eq:e27}) into Equation~(\ref{eq:e21}) for the entropy, gives
\begin{equation}
s_{\rm e+p} =  \frac{1}{kT n_{\rm B} } \left[ 4 p_{\rm e+p} + \frac{\hbar c}{\pi^{2}} \left( \frac{kT}{ \hbar c} \right)^{4} \frac{\beta^{2}}{6} \left( \pi^{2} + 3 \eta_{\rm e}^{2} \right)  \right] - \eta_{\rm e} \Ye .
\label{eq:e28}
\end{equation}
After computing the pressure from Equation~(\ref{eq:e25}), the entropy is computed from Equation~(\ref{eq:e28}), and then the specific internal energy is computed from the thermodynamic relation 
\begin{equation}
e_{\rm e+p} =  n_{\rm B} kT \left( s_{\rm e+p} + \eta_{\rm e} \Ye \right) - p_{\rm e+p} .
\label{eq:e29}
\end{equation}

The thermodynamic state is considered to be in the very degenerate regime if, as computed by Equation~(\ref{eq:e24}) or determined by iterating Equation~(\ref{eq:e18}),  $\eta_{\rm e} > 35$. 
In this case the relativistic Sommerfeld approximation is used for the thermodynamic functions, starting with
\begin{equation}
n_{\rm e^{-}} \simeq \frac{1}{3 \pi^{2} } \left( \frac{ m_{\rm e} c^{2} }{ \hbar c } \right)^{3} \left\{ x_{\eta}^{3}
+ \left( \frac{ kT }{ m_{\rm e} c^{2} } \right)^{2} \frac{ \pi^{2} }{2} \left[ \frac{ 1}{ x_{\eta}} \left( 2 x_{\eta}^{2} + 1 \right) \right]
+ \left( \frac{ kT }{ m_{\rm e} c^{2} } \right)^{4} \frac{ 7 \pi^{4} }{40} x_{\eta}^{-5} \right\} ,
\label{eq:e30} 
\end{equation}
which is iterated on $x_{\eta} ={pc}/{m_{\rm e}c^{2}}$, where $p$ is the electron momentum, until the right-hand side is equal to $n_{\rm e^{-}}$ to one part in $10^{6}$. 
Once $x_{\eta}$ is determined, the other thermodynamic functions are computed as follows:
\begin{eqnarray}
p_{\rm e} &=& \frac{ \left( m_{\rm e} c^{2} \right)^{4} }{ ( \hbar c )^{3} } \frac{1}{24 \pi^{2} } \left[  x_{\eta} \left( 2 x_{\eta}^{2} - 3 \right) \sqrt{ x_{\eta}^{2} + 1 } + 3 \log \left( x_{\eta} + \sqrt{ x_{\eta}^{2} + 1 } \right) \right. \nonumber \\ 
&& + \left. 4 \pi^{2} \left( \frac{ kT }{ m_{\rm e} c^{2} } \right)^{2} x_{\eta} \sqrt{ x_{\eta}^{2} + 1 }
+ \left( \frac{ kT }{ m_{\rm e} c^{2}  } \right)^{4} \frac{ 7 \pi^{4} }{15} \frac{ \sqrt{ x_{\eta}^{2} + 1 } \left( 2 x_{\eta}^{2} - 1 \right) }{ x_{\eta}^{3} } \right],
\label{eq:e31}
\end{eqnarray}
\begin{eqnarray}
e_{\rm e} &=& \frac{ \left( m_{\rm e} c^{2} \right)^{4} }{ ( \hbar c )^{3} } \frac{1}{24 \pi^{2} } \left[  8 x_{\eta}^{3} \left( \sqrt{ x_{\eta}^{2} + 1 } - 1 \right) - x_{\eta} \left( 2 x_{\eta}^{2} - 3 \right) \sqrt{ x_{\eta}^{2} + 1 } - 3 \log \left( x_{\eta} + \sqrt{ x_{\eta}^{2} + 1 } \right) \right. \nonumber \\ 
&& + \left. 4 \pi^{2} \left( \frac{ kT }{ m_{\rm e} c^{2} } \right)^{2} \left( \sqrt{ x_{\eta}^{2} + 1 } \left( 3 x_{\eta}^{2} + 1 \right)/x_{\eta} - \left( 2 x_{\eta}^{2} + 1 \right)/x_{\eta} \right) \right] + n_{\rm B} \Ye m_{\rm e} ,
\label{eq:e32}
\end{eqnarray}
\begin{equation}
\mu_{\rm e} =  m_{\rm e} c^{2} \sqrt{ x_{\eta}^{2} + 1 } ,
\label{eq:e33}
\end{equation}
and
\begin{equation}
s_{\rm e} =  \frac{1}{kT} \left( \frac{ e_{\rm e} + p_{\rm e} }{ n_{\rm B} } - \mu_{\rm e}  \Ye \right) .
\label{eq:e34}
\end{equation}

\begin{figure*}
\gridline{\fig{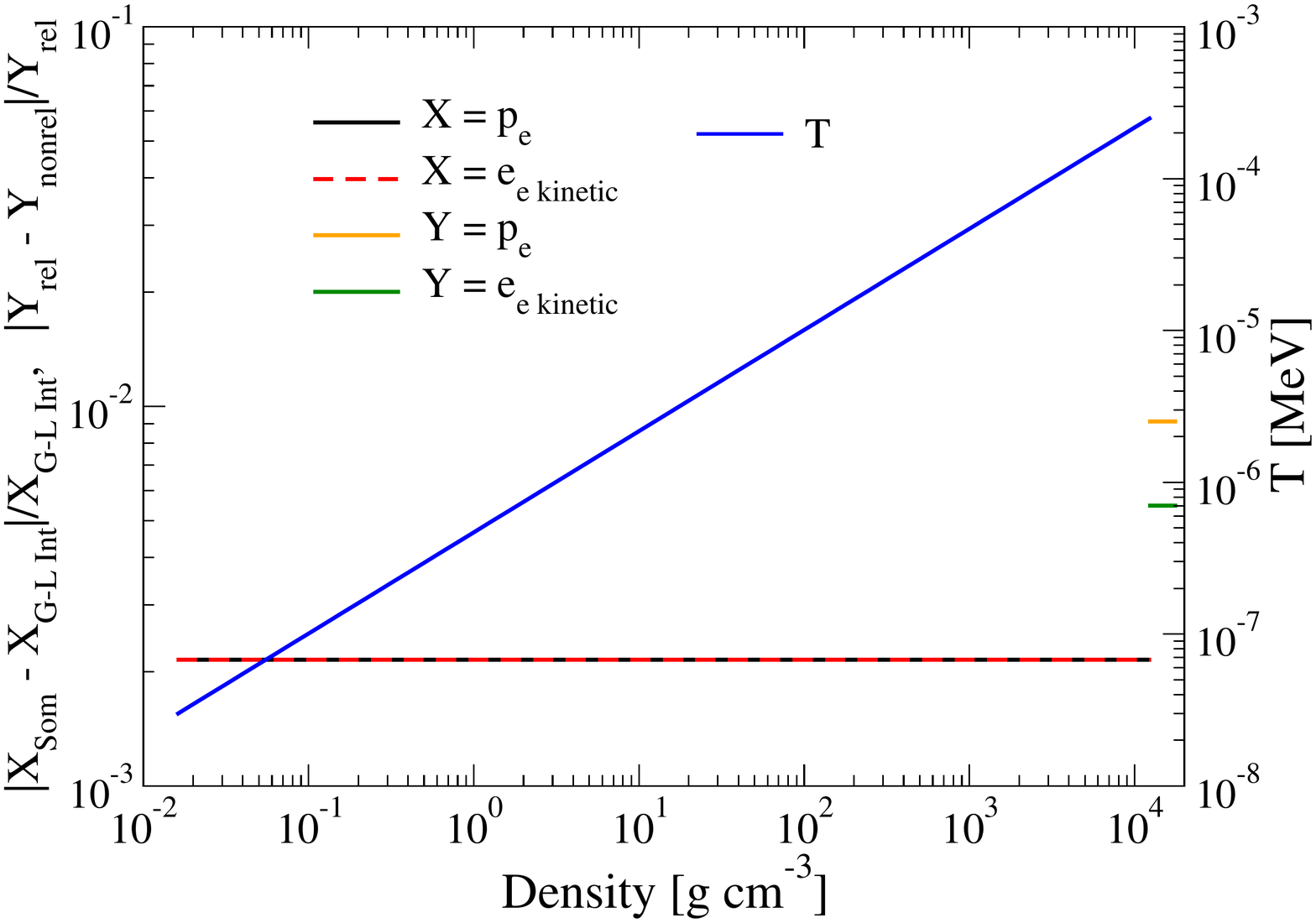}{0.5\textwidth}{(a)}
              \fig{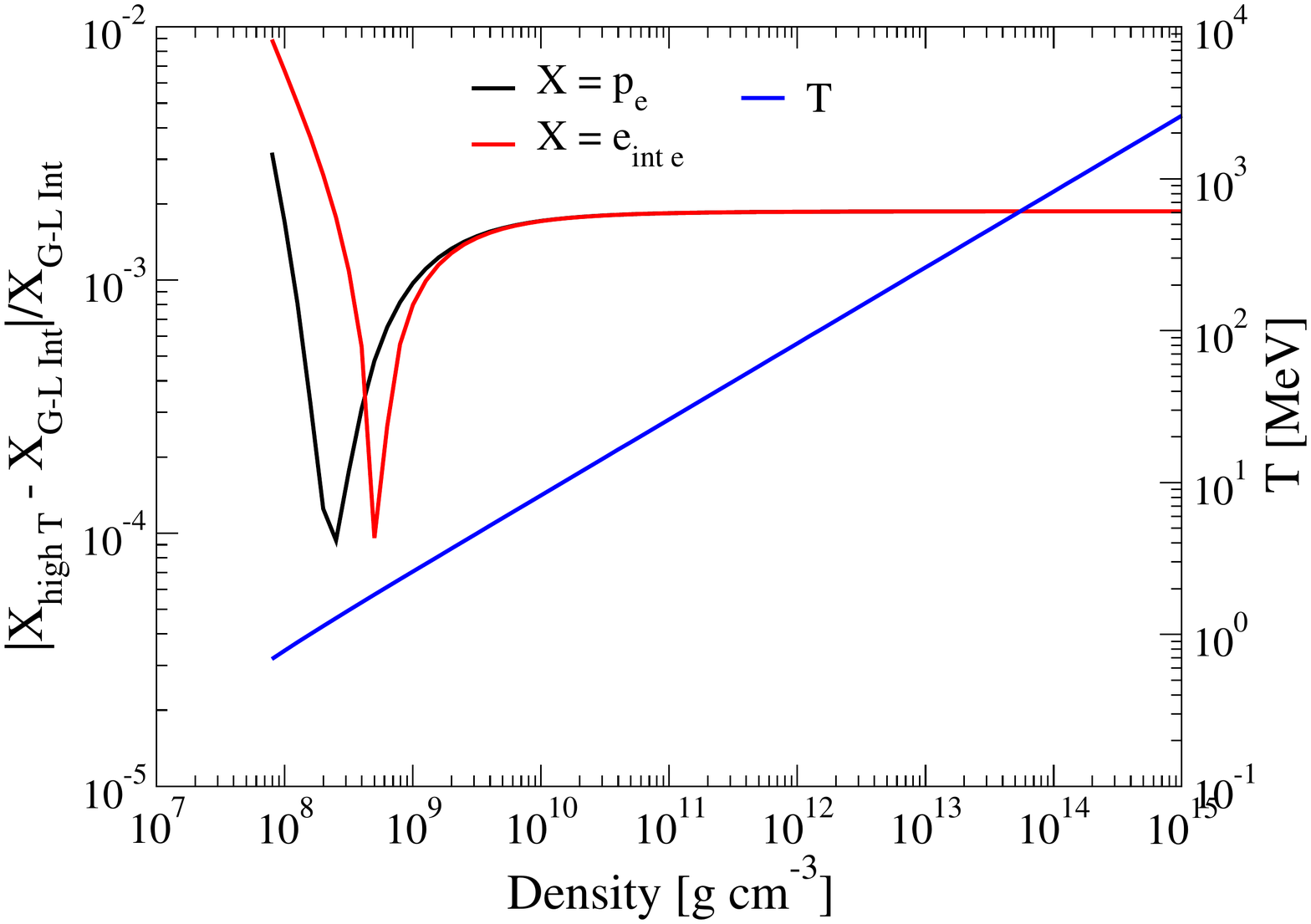}{0.5\textwidth}{(b)}
               }
\caption{\label{fig:e+p_prec}
Panel~(a): For the $T$ profile of the Gauss--Laguerre---Sommerfeld boundary (blue line; right axis), the normalized deviation ($\| X_{\rm A} - X_{\rm B}\|/X_{\rm B}$; left axis) for electron pressure ($p_{\rm e}$; black solid line) and electron specific kinetic energy ($e_{\rm e, kin}$; red line) between the Gauss--Laguerre integration and Sommerfeld approximation. 
The short segments on the right side show the relative deviation of the electron--positron pressure ($p_{\rm e^{-}+e^{+}}$; orange) and the electron--positron specific kinetic energy ($e_{\rm e^{-}+e^{+},kin}$; green) as computed with the relativistic and nonrelativistic formalisms at the relativistic--nonrelativistic transition density.  
Panel~(b): For the $T$ profile (blue) at the boundaries of (i) the high-temperature approximation and Gauss--Laguerre integration regions (solid lines) and (ii) the Relativistic Sommerfeld approximation and Gauss--Laguerre integration regions (dashed lines), we plot the normalized deviation of electron--positron pressure ($p_{\rm e^{-}+e^{+}}$; black) and specific internal energy ($e_{\rm int,e^{-}+e^{+}}$; red).
}
\end{figure*}

If the nonrelativistic criteria in Equations (\ref{eq:e15}) are satisfied, then there are two regimes in which different approximations are applied. 
If $\epsilon_{\rm F}/kT > 35$, the nonrelativistic Sommerfeld approximation is used, otherwise the nonrelativistic Fermi integrals for the electrons are integrated numerically. 
In the latter case, the Fermi integral for the electron number density 
\begin{equation}
n_{\rm e} = \frac{ 2^{1/2} (m_{\rm e} c^{2} )^{3/2} (kT)^{3/2}}{\pi^{2}(\hbar c)^{3}} \int_{0}^{\infty} \frac{ x^{1/2} dx }{ e^{ x - \eta_{\rm e} } + 1 } .
\label{eq:e35}
\end{equation}
is integrated by a 48-point Gauss-Laguerre scheme and iterated on $\eta_{\rm e}$ until the right and left sides of Equation~(\ref{eq:e35}) are equal to one part in $10^{6}$.
Once $\eta_{\rm e}$ has been obtained, $\mu_{\rm e}$ is computed by $\mu_{\rm e} = kT  \eta_{\rm e} + m_{\rm e} c^{2}$, and the electron pressure, specific internal energy, and entropy are obtained by
\begin{equation}
e_{\rm e,kin} = \frac{3}{2} \frac{( 2 m_{\rm e} )^{3/2} ( kT )^{5/2}}{ 3 \pi^{2} \hbar^{3}} \int_{0}^{\infty} x^{3/2} dx \frac{ 1 }{ e^{ x - \eta_{\rm e} } + 1 } ,
\label{eq:e36}
\end{equation}
and
\begin{equation}
e_{\rm e} = e_{\rm e,kin} + \frac{ \rho \Ye }{m _{\rm B}} m_{\rm e} c^{2}, \quad p_{\rm e} = \frac{2}{3} e_{\rm e,kin}, \quad s_{\rm e} =  \frac{1}{kT} \left( \frac{ e_{\rm e} + p_{\rm e} }{ n_{\rm B} } - \mu_{\rm e} \Ye \right) .
\label{eq:e37}
\end{equation}

In the case that $\epsilon_{\rm F}/kT > 35$, the Sommerfeld approximations for the nonrelativistic Fermi expressions are used. Thus, the electron density is given by
\begin{equation}
n_{\rm e} = \frac{ (2 m_{\rm e} c^{2} )^{3/2} (kT)^{3/2}}{ 3 \pi^{2}(\hbar c)^{3}} \eta_{\rm e}^{3/2} \left[ 1 + \frac{ \pi^{2} }{8} \eta_{\rm e}^{-2} + \frac{ 7 \pi^{4} }{ 640 } \eta_{\rm e}^{-4} \right] .
\label{eq:e38}
\end{equation}
Equation (\ref{eq:e38}) is iterated on $\eta_{\rm e}$ as described above, $\mu_{\rm e}$ is then computed also as described above, and the electron pressure, specific internal energy, and entropy are obtained by
\begin{equation}
p_{\rm e} = \frac{ 2 ( 2 m_{\rm e} )^{3/2} ( kT )^{5/2}}{ 15 \pi^{2} \hbar^{3}} \eta_{\rm e}^{5/2} \left[ 1 + \frac{ 5 \pi^{2} }{8} \eta_{\rm e}^{-2} + \frac{ 7 \pi^{4} }{ 384 } \eta_{\rm e}^{-4} \right] ,
\label{eq:e39}
\end{equation}
\begin{equation}
e_{\rm e} = \frac{3}{2} p_{\rm e} + \frac{ \rho \Ye }{m_{\rm B} } m_{\rm e} c^{2}, \quad s_{\rm e} =  \frac{1}{kT} \left( \frac{ e_{\rm e} + p_{\rm e} }{ n_{\rm B} } - \mu_{\rm e} \Ye \right) .
\label{eq:e40}
\end{equation}

The accuracies of the various approximations are indicated by the relative deviations of the electron--positron pressure and specific internal energy as computed by pairs of these approximations at the boundaries separating their respective regimes of applicability.
This is shown in Figure~\ref{fig:e+p_prec}, which indicates that relative deviations are at most a few tenths of a percent. 
The discontinuity between approximation regimes is smoothed by the finite resolution of the EoS table.

\subsection{Double-$\gamma$ EoS}
\label{sec:ggeos}

A ``2$\gamma$,'' thermodynamically consistent EoS has been developed, simple enough to be completely analytic, yet rich enough to be used for testing the hydrodynamics and transport modules of \Chimera. 
A system of completely degenerate and relativistic free electrons, partially degenerate interacting neutrons, and nondegenerate free protons, is modeled by the free energy 
\begin{eqnarray}
 F &=& F_{1} \left( \frac{ N_{e} }{V} \right)^{\gamma_{1}} \left( \frac{ V_{0} }{N_{0} } \right)^{\gamma_{1} - 4/3} V + F_{2} \left( \frac{N_{\rm n}}{V} \right)^{\gamma_{2}} V
 \nonumber \\ 
&& - kT \left[ N_{\rm n} \ln \left( g_{\rm n} V \right) + E_{\rm coef} N_{\rm n} \ln \left( \frac{ 2 \pi m_{\rm B} k T }{h^{2}} \right) - N_{\rm n} \ln N_{\rm n} + N_{\rm n} \right]
 \nonumber \\ 
&& - kT \left[ N_{\rm p} \ln \left( g_{\rm p} V \right) + E_{\rm coef} N_{\rm p} \ln \left( \frac{ 2 \pi m_{\rm B} k T }{h^{2}} \right) - N_{\rm p} \ln N_{\rm p} + N_{\rm p} \right],
\label{eq:e2}
\end{eqnarray}
where $m_{\rm B}$ is the baryon mass, $g_{\rm n} = g_{\rm p} = 2$ are the statistical weights of the neutrons and protons, $k$ and $h$ have their usual meanings, and $\gamma_{1}$, $\gamma_{2}$, $F_{1}$, $F_{2}$, $E_{\rm coef}$, and $X_{\rm p} \equiv N_{\rm p}/( \rho/m_{\rm B} ) = \Ye \equiv ( N_{\rm e^{-}} - N_{\rm e^{+}} )/( \rho/m_{\rm B}) = 1 - X_{\rm n} = N_{\rm n}/( \rho/m_{\rm B} )$ are free parameters. 
The mass fraction of free protons, $X_{\rm p}$, is typically taken to be 0.5, and the parameters $\gamma_{\rm 1}$ and $F_{1}$  are typically chosen to be
\begin{equation}
\gamma_{1} = \frac{4}{3} ,
\label{eq:e3}
\end{equation}
\begin{equation}
F_{1} = \frac{3}{4} \left( \frac{3}{8 \pi } \right)^{1/3} h c ,
\label{eq:e4}
\end{equation}
so that the first term on the right-had side of Equation~(\ref{eq:e2}) represents completely degenerate and extremely relativistic free electrons. 
Given the above expression for the free energy, the pressure is
\begin{equation}
p = - \left( \pderiv{F}{V} \right)_{T, N_{i}} = \left( \gamma_{1} - 1 \right) F_{1} \left( \frac{ \rho \Ye }{m_{\rm B}} \right)^{\gamma_{1}} \left( \frac{ m_{\rm B} }{ \rho_{0} } \right)^{\gamma_{1} - 4/3} + \left( \gamma_{2} - 1 \right) F_{2} \left( \frac{ \rho X_{\rm n} }{m_{\rm B}} \right)^{\gamma_{2}} + \frac{ \rho kT}{m_{\rm B}}.
\label{eq:e5}
\end{equation}
We choose $F_{2}$ so that the contributions of the first two terms for $p$ in Equation~(\ref{eq:e4}) become equal at $\rho_{\rm nuc}$, which, for the case in which $Y_{\rm p}$ is 0.5, requires that
\begin{equation}
\left( \gamma_{1} - 1 \right) F_{1} \left( \frac{ \rho_{\rm nuc} 0.5 }{m_{\rm B}} \right)^{\gamma_{1}} \left( \frac{ m_{\rm B} }{ \rho_{0} } \right)^{\gamma_{1} - 4/3} = \left( \gamma_{2} - 1 \right) F_{2} \left( \frac{ \rho_{\rm nuc} 0.5 }{m_{\rm B}} \right)^{\gamma_{2}}
\label{eq:e6}
\end{equation}
or
\begin{equation}
F_{2} = \frac{ \gamma_{1} - 1 }{ \gamma_{2} - 1 } F_{1}  \left( \frac{ \rho_{\rm nuc} 0.5 }{m_{\rm B}} \right)^{\gamma_{1} - \gamma_{2} } \left( \frac{ m_{\rm B} }{ \rho_{0} } \right)^{\gamma_{1} - 4/3} .
\label{eq:e7}
\end{equation}
The entropy of the system is given by
\begin{equation}
S = - \left( \pderiv{F}{T} \right)_{V, N_{i}} = S_{\rm n} + S_{\rm p} ,
\label{eq:e8}
\end{equation}
where $S_{\rm i}$, ${\rm i} = {\rm n}, {\rm p}$ is given by
\begin{equation}
S_{\rm i} = N_{\rm i} k \left[ \ln \left\{ \frac{ V g_{\rm i} }{N_{\rm n} } \left( \frac{ 2 \pi m_{\rm B} k T }{h^{2}} \right)^{ E_{\rm coef} } \right\} + E_{\rm coef} + 1 \right] .
\label{eq:e9}
\end{equation}
The internal energy of the system is given by
\begin{equation}
E = F + TS
= F_{1} \left( \frac{ \rho \Ye }{m_{\rm B}} \right)^{\gamma_{1}} \left( \frac{ m_{\rm B} }{ \rho_{0} } \right)^{\gamma_{1} - 4/3} V + F_{2} \left( \frac{ \rho X_{\rm n} }{m_{\rm B}} \right)^{\gamma_{2}} V + E_{\rm coef} N_{\rm B} kT .
\label{eq:e10}
\end{equation}
Lastly, the chemical potentials are given by
\begin{equation}
\mu_{\rm n} = - \left( \pderiv{F}{N_{\rm n}} \right)_{V, N_{\rm p}, N_{\rm e}}
= \gamma_{2}  F_{2} \left( \frac{ \rho X_{\rm n} }{m_{\rm B}} \right)^{\gamma_{2} - 1} + kT \ln \left[ \frac{ \rho X_{\rm n} }{ g_{\rm n} m_{\rm B}} \left( \frac{ h^{2} }{ 2 \pi m_{\rm B} kT } \right)^{E_{\rm coef}} \right] ,
\label{eq:e11}
\end{equation}
\begin{equation}
\mu_{\rm p} = - \left( \pderiv{F}{N_{\rm p}} \right)_{V, N_{\rm n}, N_{\rm e}}
=  kT \ln \left[ \frac{ \rho X_{\rm p} }{ g_{\rm p} m_{\rm B}} \left( \frac{ h^{2} }{ 2 \pi m_{\rm B} kT } \right)^{E_{\rm coef}} \right] ,
\label{eq:e12}
\end{equation}
and
\begin{equation}
\mu_{\rm e} = - \left( \pderiv{F}{N_{\rm e}} \right)_{V, N_{\rm n}, N_{\rm p}}
= \gamma_{1}  F_{1} \left( \frac{\rho \Ye}{m_{\rm B}} \right)^{\gamma_{1} - 1} \left( \frac{ m_{\rm B} }{\rho_{0} } \right)^{\gamma_{1} - 4/3} .
\label{eq:e13}
\end{equation}
That all of the thermodynamic functions are derived from an analytic thermodynamic potential, namely, the free energy, insures that the EoS is thermodynamically consistent. 
Furthermore, being completely analytic and simple, the interpolation scheme described above is not necessary when using this EoS.
Derivatives of thermodynamic functions can be obtained analytically, and expressions for the independent variables, $\rho$, $T$, and $\Ye$ can be solved for directly without having to invert a complicated expression by resorting to an iteration scheme.

\section{Hydrodynamics}
\label{app:hydro}

\Chimera's  hydrodynamics are evolved using a dimensionally-split, Lagrangian-plus-remap version of the Piecewise Parabolic Method \citep[PPMLR;][]{CoWo84} as implemented in VH1 \citep{HaBlLi12}, but extended to include multi-species advection, multidimensional gravity, neutrino coupling (energy and momentum) to the hydrodynamics, and radial grid movement.
PPM is a high-order Godunov-type scheme \citep{Go59, Va79} in which fluxes at zone interfaces are calculated from solutions of the Riemann problem and, therefore, is well suited for capturing shocks and contact discontinuities within one or two zones.
The PPMLR evolves the zone averages of the density, $\rho$,  fluid velocity, \uvec,  specific internal energy, \edint,  electron fraction, \Ye,  mass fractions, $X_{n}$, of nuclear ion species in regions that are not in nuclear statistical equilibrium (non-NSE regions), and zero-angular moments of the neutrino distribution functions $\psi_{\nu}^{(0)}$.

There are several advantages to a Lagrangian-plus-remap 
for CCSN simulations.
For explicit differencing, the time step constraints are less severe, 
as they are only applied during the Lagrangian step and depend only on the Lagrangian wave speeds rather than the sum of the Lagrangian wave plus advection speeds. 
In for example a quasi-Eulerian approach, the remap does not have to map the grid back to its placement prior to the Lagrangian step, but can allow the grid to evolve to accommodate itself to changing physical situations, such as moving with the fluid during core infall, thereby keeping the fluid well resolved, or ensuring adequate zoning in the vicinity of the neutrinosphere during the formation of the density cliff, or tracking the interface between two compositionally different fluids.

While the hydrodynamics  can be performed in either Cartesian, cylindrical or spherical coordinates, the neutrino radiation transport is performed in the ``ray-by-ray'' (RbR) approximation (Section~\ref{app:transport}) and requires a spherical coordinate system. 
We will therefore limit our discussion to spherical coordinate systems, as these are the coordinate systems that have been used for all \chimera\ CCSN simulations.

\subsection{General Overview}
\label{overview}

The method of solution of the hydrodynamics equations is a finite-volume method, wherein conserved quantities are represented as integrated over computational volumes, or zones, and changes to these variables occur by means of sources or sinks or by means of fluxes through zone boundaries due to the relative velocity between the fluid and these boundaries. 
\chimera\ uses a Lagrangian-plus-remap version of the PPM method, which can be described by considering the zone-integrated conserved quantity, $q({\bf x},t)$, integrated over a computational element, $\Delta V(t)$, whose boundaries may be time-dependent
\begin{equation}
Q({\bf x},t) =  \int_{\Delta V(t)} q({\bf x},t) \, dV.
\end{equation}
The time rate of change of $Q$ is
\begin{eqnarray}
 \frac{dQ}{dt} & = & \frac{d}{dt} \int_{\Delta V(t)} q({\bf x},t)\, dV \nonumber \\
 &=& \int_{\Delta V(t)} \pderiv{ q({\bf x},t) }{t}\, dV + \int_{\Delta S(t)} q({\bf x},t) {\bf u}_{g} \cdot {\bf n}\, dS \nonumber \\ 
 &=& - \int_{\Delta S(t)} q({\bf x},t) {\bf u} \cdot {\bf n}\, dS + \int_{\Delta V(t)} {\cal S}_{q}({\bf x},t)\, dV + \int_{\Delta S(t)} q({\bf x},t) {\bf u}_{g} \cdot {\bf n}\, dS \nonumber \\ 
 &=&  - \int_{\Delta S(t)} q({\bf x},t) \left( {\bf u} - {\bf u}_{g} \right) \cdot {\bf n}\, dS + \int_{\Delta V(t)} {\cal S}_{q}({\bf x},t) \, dV,
\end{eqnarray}
where the second line results from the use of the Reynolds transport theorem, with ${\bf u}_{g}$, which we will refer to as the grid velocity, being the velocity of the bounding surface $\Delta S(t)$ of the zone volume $\Delta V(t)$.
The first term in the third line is the flux of $q({\bf x},t)$ across the surface $\Delta S(t)$ due to the fluid velocity {\bf u}, assuming linear flux and the surface $\Delta S(t)$ to be fixed, and the second term represents a possible volume source ${\cal S}_{q}({\bf x},t)$ of $q({\bf x},t)$.
To calculate ${dQ}/{dt}$, \chimera\ splits the calculation into two steps,
\begin{equation}
\left( \frac{dQ}{dt} \right) = \left( \frac{dQ}{dt} \right)^{\rm step \; 1} + \left( \frac{dQ}{dt} \right)^{\rm step \; 2} .
\label{eq:g2}
\end{equation}
In `step 1' the grid velocity is set equal to the fluid velocity and the resulting equation
\begin{equation}
\left( \frac{dQ}{dt} \right)^{\rm step 1} =  \int_{\Delta V(t)} {\cal S}_{q}({\bf x},t)\, dV
\label{eq:g3}
\end{equation}
is solved. This is just the Lagrangian form of the equation for ${ \partial Q({\bf x},t) }/{\partial t}$. 
In `step 2' the fluid velocity is set to zero and the initial configuration of ${\cal S}_{q}({\bf x},t)$ is that of its final configuration after `step 1.'
The grid is then given a prescribed velocity so that 
\begin{equation}
\left( \frac{dQ}{dt} \right)^{\rm step 2} =  \int_{\Delta S(t)} q({\bf x},t)\, {\bf u}_{g} \cdot {\bf n}\, dS .
\label{eq:g4}
\end{equation}
If the grid velocity, ${\bf u}_{g}$, is chosen to be the negative of the fluid velocity, ${\bf u}$, then `step 1' plus `step 2' would be equivalent to
\begin{equation}
\frac{dQ}{dt} =  - \int_{\Delta S(t)} q({\bf x},t)\, {\bf u} \cdot {\bf n}\, dS + \int_{\Delta V(t)} {\cal S}_q({\bf x},t)\, dV,
\label{eq:g5}
\end{equation}
which is just the Eulerian form of the equation for ${ \partial Q({\bf x},t) }/{\partial t}$. 
For the $\theta$- and $\phi$-grid, we choose to set ${\bf u}_{g} = - {\bf u}$ to keep these grids stationary. 
For the radial grid, however, we use the freedom of choice for ${\bf u}_{g}$ to make the grid dynamically adaptive, allowing it to move in such a way as to maintain good resolution during such epochs as core collapse or the formation of a density cliff in the vicinity of the neutrinospheres.  

\subsection{PPM Interpolation Scheme}
\label{Interpolation}

\begin{figure}

\fig{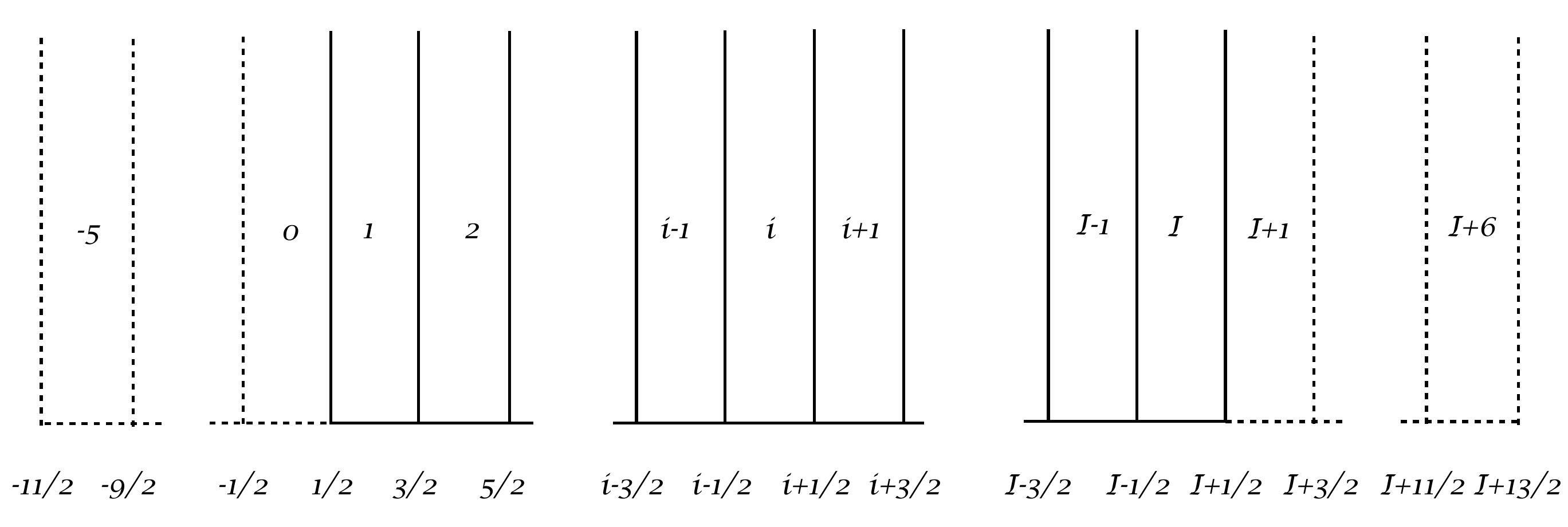}{0.7\textwidth}{}
\caption{\label{g}
Schematic representation of the grid used to construct the finite-difference equations for the hydrodynamics in any given dimension. Integer zone indices represent zones, and half-integer zone indices represent zone interfaces.
\label{fig:Zoning}
}
\end{figure}

The solution of the hydrodynamics equations proceeds in a dimensionally-split manner. We will describe the solution as it proceeds in a particular, but otherwise arbitrary, dimension and will refer to a specific dimension (e.g., $r$, $\theta$, or $\phi$) only when expressions specific to this dimension arise.
In order to construct the finite difference representations of the underlying partial differential equations that \Chimera\ solves, a discrete grid is set up dividing the interval $\xi_{\rm min}$ to $\xi_{\rm max}$ into a total of $I$ zones, where $\xi$ is the parameter  ($r$, $\theta$, or $\phi$) specifying the coordinate distance along a ray in a given dimension.
Figure~\ref{fig:Zoning} illustrates our indexing convection.
At each end of the grid, six ghost zones are appended to hold boundary values of the quantities stored in the real zones $1 \cdots I$.
In a given dimension, both the Lagrangian and the remap steps in PPMLR hydrodynamics begin by constructing zone interface values of primitive quantities, such as $\rho$, $p$, and the components of \uvec, from zone-average values of these quantities.
\citet{LuHa93} have shown that a differencing scheme that uses zone averages to represent fluid variables will not converge to the continuity equation (essential for conserving quantities during advection) unless the interpolation scheme, $a(\xi)$, for the zone-averaged quantities, $a_{i}$, satisfies 
\begin{equation}
a_{i} = \frac{1}{V_{i}}\int_{\xi_{i-\half}}^{\xi_{i+\half}} a(\xi) \frac{dV}{d\xi} d\xi ,
\label{eq:h1}
\end{equation}
where $V_{i}$ is the volume of zone $i$, and ${dV}/{d\xi}$ is the one-dimensional Jacobian determinant for $V$. 

The interpolation scheme used here is that described in \citet{CoWo84}, as modified by \citet{BlLu93a} to accommodate curvilinear coordinates while at the same time representing a linear velocity field accurately near coordinate boundaries -- e.g., $r = 0$ in the radial dimension.
The procedure for determining the zone interface value at $i-\half$ of the zone-averaged variables, $a_{i}$, is to construct a quartic polynomial, $A(\xi)$, for each zone interface $i - \half$ such that it takes on the respective values, $A_{i-\frac{5}{2}} \cdots A_{i+\frac{3}{2}}$ at the five points $\xi_{i-\frac{5}{2}} \cdots \xi_{i+\frac{3}{2}}$, where
\begin{equation}
A_{i-\half} = \sum_{k<i} a_{k} \Delta V_{k} .
\label{eq:h2}
\end{equation}
The desired interpolating polynomial, $a(\xi)$, is given by the integrand of the indefinite integral 
\begin{equation}
A(\xi) = \int^{\xi} a(\xi') dV(\xi') = \int^{\xi} a(\xi') \frac{dV}{d\xi'} d\xi';
\end{equation}
that is,
\begin{equation}
a(\xi) = \frac{dA(\xi)}{d\xi} \left(\frac{dV}{d\xi}\right)^{-1} .
\label{eq:h3}
\end{equation}
By construction, each cubic polynomial $a(\xi)$ so obtained has the desired property
\begin{equation}
a_{j} \Delta V_{j} = \int_{\xi_{j - \half}}^{\xi_{j + \half}} a(\xi) \frac{dV}{d\xi'} d\xi' \quad j = i - 2, \cdots, i+1 .
\label{eq:h4}
\end{equation}
The interface value, $a_{i-\half}$, is obtained from Equation~(\ref{eq:h3}) by evaluating it at $\xi = \xi_{i - \half}$:
\begin{equation}
a_{i-\half} = \left. \frac{dA(\xi)}{d\xi} \right|_{i-\half} \left( \left. \frac{dV}{d\xi} \right|_{i-\half} \right)^{-1},
\label{eq:h5}
\end{equation}
where the explicit expression for $A(\xi)$ is given by Equations~(12) and (13) in \citet{BlLu93a}.

The interpolation for $a_{i-\half}$ given by \citet{BlLu93a} differs from Equations~(1.6) and (1.7) of \citet{CoWo84}. In the former publication, the zone-averaged quantities $a_{i}$ are multiplied by the geometry-dependent correction factors ${\Delta V_{i}}/{ \Delta \xi_{i} }$.
Denoting these geometry corrected quantities by $a_{i}^{*}$, we have
\begin{equation}
a_{i}^{*} = a_{i} \frac{ \Delta V_{i}}{ \Delta \xi_{i} },
\label{eq:h6}
\end{equation}
where the geometry-dependent correction factor ${\Delta V_{i}}/{ \Delta \xi_{i} }$ is given by
\begin{equation}
\frac{ \Delta V_{i}}{ \Delta \xi_{i} } = \left\{ \begin{array}{cc} \frac{1}{3} \left( \xi_{i-\half}^{2} + \xi_{i-\half} \xi_{i+\half} + \xi_{i+\half}^{2} \right), & \mbox{$r$-direction} \\ \left( \cos \xi_{i-\half} - \cos \xi_{i+\half} \right)/\Delta \xi_{i}, & \mbox{$\theta$-direction} \\ 1, & \mbox{$\phi$-direction} \end{array} \right.
\label{eq:h7}
\end{equation}
and where
\begin{equation}
\Delta \xi_{i} = \xi_{i+\half} - \xi_{i-\half} .
\label{eq:h8}
\end{equation}
The average slope, $\delta a_{i}^{*}$, of the parabolas are then computed from Equation~(1.7) of \citet{CoWo84} but using the quantities $a_{i}^{*}$; e.g.,
\begin{equation}
\delta a_{i}^{*} = \frac{ \Delta \xi_{i} }{ \Delta \xi_{i-1} + \Delta \xi_{i} + \Delta \xi_{i+1} } \left[ \frac{ 2 \Delta \xi_{i-1} + \Delta \xi_{i} }{ \Delta \xi_{i+1} + \Delta_{i} } \Delta a_{i+\half}^{*} + \frac{ \Delta \xi_{i} + 2 \Delta \xi_{i+1} }{ \Delta \xi_{i-1} + \Delta_{i} } \Delta a_{i-\half}^{*} \right] .
\label{eq:h9}
\end{equation}
where
\begin{equation}
\Delta a_{i-\half}^{*} = a_{i}^{*} -  a_{i-1}^{*}.
\label{eq:h10}
\end{equation}
The $\delta a_{i}^{*}$ are then modified as follows \citep[cf.;][Equation 1.8]{CoWo84}:
\begin{equation}
\delta_{m} a_{i}^{*} = \left\{ \begin{array}{cc}\min\left( \left| \delta a_{i}^{*} \right|, 2 \left| a_{i+1}^{*} - a_{i}^{*} \right|, 2 \left| a_{i}^{*} - a_{i-1}^{*} \right| \right) \mbox{sign} \,( \delta a_{i}^{*}), & \mbox{if} \;  \Delta a_{i+\half}^{*} \Delta a_{i-\half}^{*} > 0 , \\ 0 & \mbox{if} \;  \Delta a_{i+\half}^{*} \Delta a_{i-\half}^{*} < 0 . \end{array} \right.
\label{eq:h11}
\end{equation}
The modifications for both cases implement monotonicity constraints, ensuring that no new maxima or minima appear (i.e., that $a_{i-\half}$ lies in the range of $a_{i}^{*}$ and $a_{i-1}^{*}$), and in the case of  $\Delta a_{i+\half}^{*} \Delta a_{i-\half}^{*} > 0$ lead to somewhat steeper representations of discontinuities.

Given $a_{i}^{*}$ and $\delta_{m} a_{i}^{*}$, the interface value $a_{i+\half}$ is now obtained from the cubic interpolating polynomial, Equation~(\ref{eq:h3}), which is Equation (1.6) of \citet{CoWo84} with the geometry-dependent corrections applied as above in Equation~(\ref{eq:h6}) and below in Equation~(\ref{eq:h13}):
\begin{eqnarray}
a_{i+\half}^{*} &=& a_{i}^{*} + \frac{ \Delta \xi_{i} }{ \Delta \xi_{i} + \Delta \xi_{i+1} } \left( a_{i+1}^{*} - a_{i}^{*} \right)
 \nonumber \\ 
 && + \frac{1}{ \sum_{k = -1}^{2} \Delta \xi_{i + k} } \left\{ \frac{ 2 \Delta \xi_{i+1} \Delta \xi_{i} }{ \Delta \xi_{i} \Delta \xi_{i+1} } \left[ \frac{ \Delta \xi_{i-1} + \Delta \xi_{i} }{ 2 \Delta \xi_{i} + \Delta \xi_{i+1} } -  \frac{ \Delta \xi_{i+1} + \Delta \xi_{i+2} }{ 2 \Delta \xi_{i} + \Delta \xi_{i+1} } \right] \left( a_{i+1}^{*} - a_{i}^{*} \right) \right. \nonumber \\ 
&&-  \left. \Delta \xi_{i} \frac{ \Delta \xi_{i-1} + \Delta \xi_{i} }{ 2 \Delta \xi_{i} + \Delta \xi_{i+1} } \delta a_{m, i+1}^{*} + \Delta \xi_{i+1} \frac{ \Delta \xi_{i+1} + \Delta \xi_{i+2} }{ \Delta \xi_{i} + 2 \Delta \xi_{i+1} } \delta a_{m, i}^{*}  \right\},\label{eq:h12} \\
a_{i+\half} &=& a_{i+\half}^{*} \left( \pderiv{V}{\xi} \right)_{i+\half}^{-1} ,\label{eq:h13}
\end{eqnarray}
where 
\begin{equation}
\left( \pderiv{V}{\xi} \right)_{i+\half} = \left\{ \begin{array}{cc} \xi_{i+\half}^{2} \quad (1 \mbox{ if } \xi_{\half} = 0 ), & \mbox{$r$-direction} \\ \sin \xi_{i+\half}  \quad (1 \mbox{ if } \xi_{\half} = 0 ), & \mbox{$\theta$-direction} \\ 1. & \mbox{$\phi$-direction} \end{array} \right.
\label{eq:h14}
\end{equation}
where the values at $\xi_{\half} = 0$ are to avoid singularities in Equation (\ref{eq:h13}). Finally, the range of $a_{R, i}$ is limited to be within the range of $a_{i}$ and $a_{i+1}$:
\begin{equation}
a_{R, i} = \max( a_{i+\half}, \min( a_{i}, a_{i+1} ) ) ,
\label{eq:h15}
\end{equation}
\begin{equation}
a_{R, i} = \min( a_{R, i}, \max( a_{i}, a_{i+1} ) ) ,
\label{eq:h16}
\end{equation}
and
\begin{equation}
a_{L, i+1} = a_{R, i} ,
\label{eq:h17}
\end{equation}
where $a_{L, i+1}$ is the value of $a(\xi)$ at the left interface of zone $i+1$, and $a_{R, i}$ is its value at the right interface of zone $i$. 
At zone boundaries, $a_{L, i}$ and $a_{R, i}$ must be modified in certain cases as follows. In the $r$-direction, if $\xi_{\half} = 0$, then $a_{L, i}$ must be modified in a manner depending on whether the variable is odd (e.g., velocity) or even (e.g., density, specific energy) at the origin:
\begin{equation}
\begin{array}{lll}
 a_{R, 0} = 0, & a_{L, 1} = 0, & \mbox{odd variables at r = 0} , \\
a_{R, 0} = \frac{5}{2} a_{1} - \frac{3}{2} a_{L, 2}, & a_{L, 1} = a_{R, 0} & \mbox{even variables at r = 0} . \\
\label{eq:h18}
\end{array}
\end{equation}
In the $\theta$-direction, if $\xi_{\half} = 0 (\theta = 0)$ and reflecting boundary conditions are imposed,
\begin{equation}
\begin{array}{lll} 
a_{R,0} = 0, & a_{L, 1} = 0, & \mbox{odd variables at $\theta = 0$} , \\
a_{R,0} = \left( 6 \, a_{1} + a_{L, 2} \right)/7, & a_{L, 1} = a_{R, 0} & \mbox{even variables at $\theta = 0$} , \\
\end{array}
\label{eq:h19}
\end{equation}
and if $\xi_{I + \half} = 0 (\theta = \pi)$,
\begin{equation}
\begin{array}{lll} 
a_{R,I} =  0, & a_{L, I+1} = 0, & \mbox{odd variables at $\theta = \pi$} , \\
a_{R, I} = \left( 6 \, a_{I} + a_{L, I-1} \right)/7, &a_{L, I+1} = a_{R, I} & \mbox{even variables at $\theta = \pi$} . \\
\end{array}
\label{eq:h20}
\end{equation}

In the presence of shocks, post-shock oscillations sometimes occur in some of the fluid variables -- e.g., entropy \citep[][Section 4]{CoWo84}. One method of suppressing these oscillations is to introduce some additional dissipation in the vicinity of a shock.
The method used here is to lower the order of the interpolation (i.e., flatten the interpolation profile) in the vicinity of a shock.
Thus, in the vicinity of a shock, $a_{L, I}$ and $a_{R, I}$ are modified as follows:
\begin{eqnarray}
 a_{L,I} &=& a_{i} f + a_{L, I} ( 1 - f ) , \nonumber \\
a_{R,I} &=& a_{i} f + a_{R, I} ( 1 - f ) ,
\label{eq:h21}
\end{eqnarray}
where the `flattening parameter,'  $0 \le f \le 1$, is zero away from shocks and approaches a user preset value $f_{\rm set}$ $\le 1$ in the limit of a strong shock with a steep profile.

The flattening parameter $f_{i}$ is computed similarly to that described in \citet{FrOlRi00} and begins with the computation of the pressure profile $p_{{\rm profile} \, i}$ given by
\begin{equation}
p_{{\rm profile} \, i} = \frac{ p_{i+1} - p_{i-1} }{ p_{i+2} - p_{i-2} } .
\label{eq:h21a}
\end{equation}
The presence of a shock is indicated if this quantity approaches 1, i.e., when $p_{i+2} - p_{i-2} \approx p_{i+1} - p_{i-1}$. For smooth flows $p_{i+2} - p_{i-2} \approx 2 \times ( p_{i+1} - p_{i-1} )$ and $p_{{\rm profile} \, i} \approx 1/2$. A pre-flattening parameter $f_{{\rm pre} \, i}$ is then computed from
\begin{equation}
f_{{\rm pre} \, i} = \max \left[ \left( p_{{\rm profile} \, i} - \omega_{1} \right) \times \omega_{2}, 0 \right] ,
\label{eq:h21b}
\end{equation}
where $\omega_{1}$ and $\omega_{1}$ are user selected parameters. 
We have found that 0.75 and 10 work well for $\omega_{1}$ and $\omega_{2}$ during the early stages of a simulation when the shock is nearly spherically symmetric, completely eliminating post-shock oscillations, and 0.6 and 10 work well thereafter better capturing oblique shocks. 
The pre-flattening parameter is set to zero if
\begin{equation}
\frac{ \left| p_{i+1} - p+{i-1} \right| }{ \min \left[ \max \left( p_{i+1} , 0.05 p_{i-1} \right),  \max \left( p_{i-1} , 0.05 p_{i+1} \right) \right] } < \frac{1}{3} ,
\label{eq:h21c}
\end{equation}
which indicates an insufficient pressure jump, and if
\begin{equation}
u_{i-1} < u_{i+1} ,
\label{eq:h21d}
\end{equation}
which indicates a velocity jump in the wrong direction. Finally, the flattening parameter is given by
\begin{equation}
f_{i} = \max \left\{ 0, \min \left[ f_{\rm set}, \max \left( f_{{\rm pre} \, i-1} , f_{{\rm pre} \, i}, f_{{\rm pre} \, i+1} \right) \right] \right\} .
\label{eq:h21e}
\end{equation} 
This limits the value of $f_{i}$ to $0 \le f_{\rm set}$ and sets the value $f_{i}$ based on the maximum value of $f_{\rm pre}$ in zone $i$ and the neighboring zones. Experimentation has indicated that $f_{\rm set} = 1$ works well when the shock is approximately spherically symmetric, and $f_{\rm set} = 0.5$ works well thereafter.

With the values of $a_{L, i}$ and $a_{R, i}$ for each zone $i$ determined, a piecewise parabolic interpolation function, $a(\xi)$, is constructed with $a(\xi)$ given by a parabolic profile in each zone:
\begin{equation}
a(\xi) = a_{L,i} + x ( \Delta a_{i} + a_{6,i}( 1 - x ) ) \qquad x = \frac{ \xi - \xi_{i - \half} }{ \xi_{i + \half} - \xi_{i - \half} }, \qquad \xi_{i - \half} \le \xi \le \xi_{i + \half} ,
\label{eq:h22}
\end{equation}
where 
\begin{equation}
\Delta a_{i} = a_{R,i} - a_{L,i} ,
\label{eq:h23}
\end{equation}
and where $a_{6,i}$. The parameter $a_{6,i}$ must now be determined so that Equation~(\ref{eq:h1}) is satisfied. The expressions for $a_{6,i}$ are given by
\begin{equation}
a_{6,i} = \left\{ \begin{array}{cc} 6 \left\{ a_{i} - \frac{1}{2} \left[ a_{L,i} \left( 1 - {\cal F}_{r}  \right) + a_{R,i} \left( 1 + {\cal F}_{r} \right) \right] \right\} {\cal G}_{r} & \mbox{$r$-direction} , \\ \left[ a_{i} \left( {\cal G}_{\theta} - {\cal F}_{\theta} \right) - a_{L,i} {\cal G}_{\theta} + a_{R,i} {\cal F}_{\theta} \right]/{\cal H}_{\theta} & \mbox{$\theta$-direction} , \\ 6 \left[ a_{i} - \frac{1}{2} \left( a_{L,i} + a_{R,i} \right) \right]  & \mbox{$\phi$-direction} , \end{array} \right.
\label{eq:h24}
\end{equation}
where 
\begin{equation}
{\cal F}_{r} = \left( y + 1/2 \right)/\left( 3 y^{2} + 3 y + 1 \right) ,
\label{eq:h25}
\end{equation}
\begin{equation}
{\cal G}_{r} = \left( 3 y^{2} + 3 y + 1 \right)/\left( 3 y^{2} + 3 y + 9/10 \right) ,
\label{eq:h26}
\end{equation}
\begin{equation}
y = \xi_{i-\half}/\Delta \xi_{i} ,
\label{eq:h27}
\end{equation}
and
\begin{equation}
{\cal F}_{\theta} = \cos(\xi_{i+\half}) - \left[ \sin(\xi_{i+\half}) - \sin(\xi_{i-\half}) \right]/\Delta x_{i} ,
\label{eq:h28}
\end{equation}
\begin{equation}
{\cal G}_{\theta} = \cos(\xi_{i-\half}) - \left[ \sin(\xi_{i+\half}) - \sin(\xi_{i-\half}) \right]/\Delta x_{i} ,
\label{eq:h29}
\end{equation}
\begin{equation}
{\cal H}_{\theta} = 2 \left[  \cos(\xi_{i-\half}) -  \cos(\xi_{i+\half} \right]/( \Delta x_{i} )^{2} - \left[ \sin(\xi_{i+\half}) - \sin(\xi_{i-\half}) \right]/\Delta x_{i} .
\label{eq:h30}
\end{equation}

In addition to the constraints imposed by Equations (\ref{eq:h11}) and (\ref{eq:h15})--(\ref{eq:h17}), the following monotonicity constraints are imposed to avoid the possibility of the interpolating function taking on values not between $a_{L, i}$ and $a_{R, i}$, which could otherwise lead to its developing spurious oscillations: 
\begin{equation}
a_{L,i, \rm mon} = \left\{ \begin{array}{cc} a_{R, i} \left( 1 + {\cal H}_{r}^{+} \right) - a_{i} {\cal H}_{r}^{+} & \mbox{$r$-direction} , \\
\left[ a_{R, i} \left( {\cal F}_{\theta} - {\cal H}_{\theta} \right) + a_{i} \left( {\cal G}_{\theta}  - {\cal F}_{\theta} \right) \right]/\left( {\cal G}_{\theta} - {\cal H}_{\theta} \right) & \mbox{$\theta$-direction} , \\
  3 a_{i} - 2 a_{R, i}  & \mbox{$\phi$-direction} , \end{array} \right.
\label{eq:h31}
\end{equation}
\begin{equation}
a_{R,i, \rm mon} = \left\{ \begin{array}{cc} a_{L, i} \left( 1 - {\cal H}_{r}^{-} \right) + a_{i} {\cal H}_{r}^{-} & \mbox{$r$-direction} , \\
\left[ a_{L, i} \left( {\cal G}_{\theta} - {\cal H}_{\theta} \right) + a_{i} \left( {\cal G}_{\theta}  - {\cal F}_{\theta} \right) \right]/\left( {\cal F}_{\theta} - {\cal H}_{\theta} \right)  & \mbox{$\theta$-direction} , \\
3 a_{i} - 2 a_{L, i}  & \mbox{$\phi$-direction} , \end{array} \right.
\label{eq:h32}
\end{equation}
where
\begin{equation}
{\cal H}_{r} ^{+} = \frac{6}{ 3 {\cal F}_{r} + 1/{\cal G}_{r} - 3 }, \qquad {\cal H}_{r} ^{-} = \frac{6}{ 3 {\cal F}_{r} - 1/{\cal G}_{r} + 3} .
\label{eq:h33}
\end{equation}
The need to use the monotonized expression for the left and right states arises if the parabola exceeds either of these states. With $\Delta a_{i}$ given by Equation~(\ref{eq:h23}), we thus have
\begin{eqnarray}
a_{L, i} &= a_{L,i, \rm mon} & \quad \mbox{if } \left( \Delta a_{i} \right)^{2} < \Delta a_{i} \, a_{6,i} , \\\label{eq:h34}
a_{R, i} &= a_{R,i, \rm mon} & \quad \mbox{if } \left( \Delta a_{i} \right)^{2} < - \Delta a_{i} \, a_{6,i} .
\label{eq:h35}
\end{eqnarray}
With these now monotonized values of $a_{L, i}$ and $a_{R, i}$, the quantities $\Delta a_{i}$, and $a_{6, i}$ are recomputed from Equations~(\ref{eq:h23}) and~(\ref{eq:h24}).
This completes the piecewise parabolic interpolation for the profile, $a(\xi)$, of zone-averaged variables $a_{i}$.

\subsection{Lagrangian Step}
\label{Lagrangian}

The equations describing the change in the hydrodynamic variables during the Lagrangian step are:
\begin{eqnarray}
\frac{d}{dt} \int_{V(t)} \rho \, dV &=& 0 , \label{eq:h51} \\
\frac{d}{dt} \int_{V(t)} (\rho \uvec) dV &=& - \int_{S(t)} p \nvec \, dS + \int_{V(t)} \rho \left[ - \nabla \edgrav + \fvec_{\nu} + \fvec_{\rm ccor} \right] dV , \label{eq:h52} \\
\frac{d}{dt} \int_{V(t)} \rho ( \edint + \edkin ) dV &=& - \int_{S(t)} p \uvec \cdot \nvec \, dS + \int_{V(t)} \rho \uvec \cdot \left[ - \nabla \edgrav + \fvec_{\nu} \right] dV , \label{eq:h53} \\
\frac{d}{dt} \int_{V(t)} (\rho \Ye) dV &=& \left[ \left. \pderiv{\Ye}{t} \right|_{ \nu-\rm interactions} \right] \mbox{, and} \label{eq:h54} \\
\frac{d}{dt} \int_{V(t)} (\rho X_{n}) dV &=& \left[ \left. \pderiv{X_{n}}{t} \right|_{ \nu-\rm interactions, \;nuclear\; reactions} \right] , \label{eq:h55}
\end{eqnarray}
where $V(t)$ and $S(t)$ are the volume and surface of a grid element or mass zone as it comoves with the fluid, $\nvec$ is a unit vector normal to $dS$ and pointing out of the mass zone, \edgrav\ is the gravitational potential, $\fvec_{\nu}$ is the specific neutrino stress, and $\fvec_{\rm ccor}$ denotes the centrifugal and Coriolis forces. The time derivatives are taken at constant mass (i.e., are Lagrangian time derivatives).
The expressions in the brackets on the right-hand sides of Equations~(\ref{eq:h54}) and (\ref{eq:h55}) denote the changes in the electron fraction, $\Ye$, due to neutrino transport and in the composition mass fractions, $X_{n}$, due to both neutrino transport and nuclear reactions, and are calculated elsewhere in the computational sweep, as described in Sections~\ref{dyedt} and \ref{app:nuc_network}.
During the Lagrangian hydrodynamics step these expressions are set to zero.

Equation~(\ref{eq:h53}) is a common formulation of energy conservation on which difference schemes are subsequently constructed \citep[e.g.,][]{StNo92a, BrNoSt95, FrOlRi00, Suth10}.
\chimera\ uses two alternative formulations of energy conservation, depending on the circumstances.
Using Equations~(\ref{eq:h51}) and (\ref{eq:h52}) in Equation~(\ref{eq:h53}), the following expression for $\edint$ can be derived:
\begin{equation}
\frac{d\edint}{dt} = \frac{p}{\rho} \left( \pderiv{\rho}{t} \right)_{S, \Ye} ,
\label{eq:h56}
\end{equation}
which is just the first law in the absence of changes in entropy and electron fraction. 
This equation is applicable for updating the specific internal energy during the Lagrangian step in regions well away from shocks.
It is more accurate numerically, in some cases, than updating the specific total energy as it does not ultimately involve the subtraction of potentially large values of the specific kinetic energy and the gravitational potential (if the latter is also included as a component in the specific total energy). 
In the vicinity of shocks, however, the specific total energy must be updated from the solution of the Riemann shock tube problem for the pressure and velocity at the zone interfaces. 
To do this, the term involving $\rho \uvec \cdot \nabla \edgrav$ on the right-hand side of the energy Equation~(\ref{eq:h53}) is transformed as follows:
\begin{eqnarray}
 \int_{V(t)} \rho \uvec \cdot \left[ - \nabla \edgrav \right] dV &=& \int_{V(t)} \left[ - \nabla \cdot \left( \rho \uvec \edgrav \right) + \edgrav \nabla \cdot \left( \rho \uvec \right) \right] dV
 \nonumber \\ 
&=& - \int_{V(t)} \left[ \nabla \cdot \left( \rho \uvec \edgrav \right) + \edgrav \pderiv{\rho}{t} \right] dV \nonumber \\ 
&=& - \int_{V(t)} \left[ \nabla \cdot \left( \rho \uvec \edgrav \right) + \pderiv{(\edgrav \rho)}{t} - \rho \pderiv{\edgrav }{t} \right] dV
 \nonumber \\ 
&=& - \int_{V(t)} \left[ \pderiv{(\edgrav \rho)}{t}  + \uvec \cdot \nabla \left( \rho \edgrav \right) + \rho \edgrav \nabla \cdot \uvec - \rho \pderiv{\edgrav }{t} \right] dV \nonumber \\ 
&=& - \frac{d}{dt} \int_{V(t)} \rho \edgrav dV +  \int_{V(t)} \rho \pderiv{\edgrav }{t} dV .
\label{eq:h57}
\end{eqnarray}
Substituting Equation~(\ref{eq:h57}) in Equation~(\ref{eq:h53}) results in the following equation for the total energy (including gravitational energy)
\begin{equation}
\frac{d}{dt} \int_{V(t)} \rho ( \edint + \edkin + \edgrav ) dV = - \int_{S(t)} p \uvec \cdot \nvec \, dS + \int_{V(t)} \left[ \rho \uvec \cdot \fvec_{\nu} + \rho \pderiv{ \edgrav }{t} \right]dV .
\label{eq:h58}
\end{equation}
Equation~(\ref{eq:h58}) is used in the radial-sweep to update the specific total energy in the vicinity of a shock. 
For the $\theta$- and $\phi$-sweeps, changes in the gravitational potential are very small and Equation~(\ref{eq:h53}) is used with the gradient of the gravitational potential on the right-hand side treated as a force.

To perform the Lagrangian update, the first step is to compute the displacement of the zone interfaces, after which Equations~(\ref{eq:h51}), (\ref{eq:h52}), (\ref{eq:h54}), and~(\ref{eq:h56}) or~(\ref{eq:h57}) are used to update $\rho$, the components of \uvec, \Ye, and \edint. 
The displacement of each zone interface during the Lagrangian step is determined by solving a Riemann problem for the velocity of the contact discontinuity at the zone interface.
This requires averages of the needed quantities over the domains of dependences of the left and right states. 
Rather than solving the exact Riemann problem, which is time consuming, as it is complicated and involves multiple calls to the EoS, \Chimera\ uses the approximate but very accurate method developed by \citet{CoGl85}.
This method parameterizes the EoS by the slowly varying quantity $\gamma$, given by 
\begin{equation}
\gamma = \frac{p}{\rho \edint } + 1 ,
\label{eq:h59}
\end{equation}
and the adiabatic exponent, $\Gamma$, defined by
\begin{equation}
\Gamma = \left( \pderiv{p}{\rho} \right)_{S, \Ye} .
\end{equation}
 
Solution of the approximate Riemann problem requires the values of the quantities $\rho$, $u$ (the component of velocity in the direction of the directional splitting), $p$, $\gamma$, and $\Gamma$ to the left and right of each zone interface. 
To maintain high-order accuracy, the values of each of these quantities are averaged over their domain of dependence of the zone interface as determined by the time step. 
This is accomplished for each zone interface by tracing the two characteristics from the interface at time $t + \Delta t$ to the $\xi$ axis at time $t$. 
Having speeds of $\pm c_{\rm s}$, where $c_{\rm s}$ is the local sound speed, the two characteristics intersect the $\xi$ axis on either side of the interface at the points $\xi_{i+\half} + c_{\rm s} \Delta t$ and $\xi_{i+\half} -  c_{\rm s} \Delta t$. 
Letting $a_{i}$ represent any one of the above quantities, the average, $\langle a \rangle_{L, \, i+\half}^{n+\half}$, of $a_{i}$ over the domain of dependence to the left of the zone interface, $\xi_{i+\half}$, is obtained by integrating the parabolic profile $a(\xi)$ of $a_{i}$ over the interval $\xi_{i+\half} -  c_{{\rm s},i} \Delta t$ to $\xi_{i+\half}$ and averaging, and is given by
\begin{equation}
\langle a \rangle_{L, \, i+\half} = a_{L, \, i} + \Delta a_{i} - \frac{ c_{{\rm s},i} \Delta t }{ 2 \Delta \xi_{i} } \left[ \Delta a_{i} - \left( 1 - \frac{4}{3} \frac{ c_{{\rm s},i} }{ 2 \Delta \xi_{i} } \right) a_{6 \, i} \right] ,
\label{eq:h60}
\end{equation}
where $\Delta a_{i}$ is given by Equation~(\ref{eq:h23}), $\Delta \xi_{i}$ is given by Equation~(\ref{eq:h8}), and $c_{{\rm s},i}^{2} = \Gamma_{i} p_{i}/\rho_{i}$. 
Likewise, the average, $\langle a \rangle_{R, \, i+\half}^{n+\half}$, of $a$ over the domain of dependence to the right of the zone interface, $\xi_{i+\half}$, is obtained by integrating the parabolic profile $a(\xi)$ of $a_{i}$ over the interval $\xi_{i+\half}$ to $\xi_{i+\half} + c_{{\rm s},i+1} \Delta t $ and averaging, and is given by
\begin{equation}
\langle a \rangle_{R, \, i+\half} = a_{L, \, i+1} + \frac{ c_{{\rm s},i+1} \Delta t }{ 2 \Delta \xi_{i+1} } \left[ \Delta a_{i+1} + \left( 1 + \frac{4}{3} \frac{ c_{{\rm s},i+1} \Delta t }{ 2 \Delta \xi_{i+1} } \right) a_{6,i+1} \right] .
\label{eq:h61}
\end{equation}

The time-averaged left and right states, $\langle p^{+} \rangle_{L, \, i+\half}$, $\langle p^{+} \rangle_{R, \, i+\half}$, $\langle u \rangle_{L, \, i+\half}$, $\langle u \rangle_{R, \, i+\half}$ of $p$ and $u$ are obtained from Equations~(\ref{eq:h60}) and (\ref{eq:h61}) above, and the time-averaged pressure is then corrected for the presence of gravitational, neutrino,  centrifugal, and Coriolis forces by
\begin{eqnarray}
\langle p \rangle_{L, \, i+\half} &=& \langle p^{+} \rangle_{L, \, i+\half} + \frac{ \Delta t \, \rho_{i} \, c_{{\rm s},i} }{ 2 } \left( - \nabla e_{{\rm grav}, \, i+\half}  + \fvec_{\nu, \, i+\half} + \fvec_{\rm ccor, \, i+\half} \right)_{r, \, \theta, \, \phi} , \\ \label{eq:h62}
\langle p \rangle_{R, \, i+\half} &=& \langle p^{+} \rangle_{R, \, i+\half} - \frac{ \Delta t \, \rho_{i+1} \, c_{{\rm s},i+1} }{ 2 } \left( - \nabla e_{{\rm grav}, \, i+\half}  + \fvec_{\nu, \, i+\half} + \fvec_{\rm ccor, \, i+\half} \right)_{r, \, \theta, \, \phi} . \label{eq:h63}
\end{eqnarray}

Given the time-averaged states of $p$ and $u$ to the left and right of each zone interface, $\xi_{i+\half}$, the time-averaged values, $p_{i+\half}^{n+\half}$ and $u_{i+\half}^{n+\half}$, of the pressure and velocity of the zone interface itself are computed by connecting $p_{i+\half}^{n+\half}$ and $u_{i+\half}^{n+\half}$ to the time-averaged left and right states by the Rankine-Hugoniot relations; that is:
\begin{equation}
\frac{ p_{i+\half}^{n+\half} - \langle p \rangle_{L, \, i+\half} }{\langle W \rangle_{L, \, i+\half}} + ( u_{i+\half}^{n+\half} - \langle u \rangle_{ L, \, i+\half} ) = 0,
\label{eq:h64}
\end{equation}
\begin{equation}
\langle W  \rangle_{L, \,i+\half}^{2} = \left[ \langle \gamma \rangle_{L}  \langle p \rangle_{L} \langle \rho \rangle_{L} \right]_{i+\half} \left[ 1 + \frac{ \langle \gamma \rangle_{L} + 1 }{2 \langle \gamma  \rangle_{L}}
\left(  \frac{p^{n+\half}}{\langle p \rangle_{L}} - 1 \right) \right]_{i+\half},
\label{eq:h65}
\end{equation}
\begin{equation}
\frac{ p_{i+\half}^{n+\half} - \langle p \rangle_{R, \, i+\half} }{\langle W \rangle_{R, \, i+\half}} + ( u_{i+\half}^{n+\half} - \langle u \rangle_{R, \, i+\half} ) = 0,
\label{eq:h66}
\end{equation}
\begin{equation}
\langle W \rangle_{R, \,i+\half}^{2} = \left[ \langle \gamma \rangle_{R} \langle p \rangle_{R} \langle \rho \rangle_{R} \right]_{i+\half} \left[ 1 + \frac{ \langle \gamma \rangle_{R} + 1 }{2 \langle \gamma \rangle_{R} }
\left( \frac{p^{n+\half}}{\langle p \rangle_{R}} - 1 \right) \right]_{i+\half}.
\label{eq:h67}
\end{equation}
Equations (\ref{eq:h64})--(\ref{eq:h67}) are iterated for $p_{i+\half}^{n+\half}$ and $u_{i+\half}^{n+\half}$ by the secant method. 

Having determined $p_{i+\half}^{n+\half}$ and $u_{i+\half}^{n+\half}$ for each of the zone interfaces, the Lagrangian update proceeds as follows.
With the values of $u_{i+\half}^{n+\half}$ determined, the zone interfaces are considered impenetrable and their positions are updated by
\begin{equation}
\xi_{i+\half}^{n+1'} = \xi_{i+\half}^{n} + \frac{ u_{i+\half}^{n+\half} \, \Delta t }{\cal R}; \qquad {\cal R} = \left\{ \begin{array}{cc} 1 & \mbox{$r$-direction} , \\
r_{i} & \mbox{$\theta$-direction} , \\
  r_{i} \sin \theta_{i} & \mbox{$\phi$-direction} , \end{array} \right.
\label{eq:h70}
\end{equation}
where the superscripts $n$ and $n+1'$ denote the value of a variable at time $t$ and at the end of the Lagrangian step at time $t + \Delta t$, respectively. A superscript $n+\half$ denotes a time-centered value of a variable.
We reserve the superscript $n+1$ for the value of a variable after both the Lagrangian and the remap step have been completed.
From Equation~(\ref{eq:h51}), the density is then updated by
\begin{equation}
\rho_{i}^{n+1'} = \rho_{i}^{n} \, \frac{ V_{i}^{n} }{ V_{i}^{n+1'} }; \qquad V_{i} = \left\{ \begin{array}{cc} \frac{1}{3} \left( r_{i+\half}^{3} - r_{i-\half}^{3} \right) & \mbox{$r$-direction} , \\
 r_{i} \left( \cos \theta_{i-\half} - \cos \theta_{i+\half} \right) & \mbox{$\theta$-direction} , \\
  r_{i} \sin \theta_{i} \left( \phi_{i+\half} - \phi_{i-\half} \right) & \mbox{$\phi$-direction} . \end{array} \right.
\label{eq:h71}
\end{equation}
Because of the conservation of mass in each zone during the Lagrangian step, as expressed by Equation~(\ref{eq:h51}), and in differenced form by Equation~(\ref{eq:h71}), Equation~(\ref{eq:h54}) for the change in \Ye\ and Equation~(\ref{eq:h55}) for the change in the composition mass fractions, $X_{i}$, with their right-hand sides set to zero, state the obvious: \Ye\ and $X_{i}$ are unchanged during the Lagrangian step.

Equation~(\ref{eq:h52}) with the help of Equation~(\ref{eq:h51}) can be rewritten in differential form as
\begin{equation}
\frac{d\uvec}{dt} = -\frac{1}{\rho} \nabla p - \nabla \edgrav + {\bf f}_{\nu} + {\bf f}_{\rm ccor} ,
\label{eq:h72}
\end{equation}
where, again, the time derivative is a Lagrangian time derivative.
In component form, Equation~(\ref{eq:h72}) becomes
\begin{eqnarray}
\frac{du_{r}}{dt} &=& - \frac{1}{\rho} \pderiv{p}{r} - \pderiv{\edgrav}{r} + f_{\nu,r} + \frac{ u_{\theta}^{2} + u_{\phi}^{2} }{r} , \nonumber \\
 \frac{du_{\theta}}{dt} &=& - \frac{1}{\rho} \frac{1}{r} \pderiv{p}{\theta} - \frac{1}{r} \pderiv{\edgrav}{\theta} + f_{\nu,\theta} + \frac{ u_{\phi}^{2} }{r \sin \theta } \cos \theta - \frac{ u_{r} u_{\theta} }{r}  , \label{eq:h73} \\
 \frac{du_{\phi}}{dt} &=& - \frac{1}{\rho} \frac{1}{r \sin \theta } \pderiv{p}{\phi} -  \frac{1}{r \sin \theta }\pderiv{\edgrav}{\phi} + f_{\nu,\phi} - \frac{ u_{\theta} u_{\phi} }{r \sin \theta } \cos \theta - \frac{ u_{r} u_{\phi} }{r} . \nonumber
\end{eqnarray}
For the radial sweep, the neutrino stress term, $f_{\nu,r}$, is computed as described by the right-most term on the right-hand side of Equation~(\ref{eq:ee39}). 
Because of the ray-by-ray approximation adopted by \chimera\ for neutrino transport, the values for $f_{\nu,\theta}$ and $f_{\nu,\phi}$ appearing in the $\theta$- and $\phi$-sweeps cannot be obtained directly as an outcome of the transport as can $f_{\nu,r}$ for the radial sweep. 
To include $f_{\nu,\theta}$ and $f_{\nu,\phi}$ in an approximate way, we regard the matter as being completely neutrino opaque at densities above $10^{12}$~\gcc, and completely transparent at lower densities.
The neutrino distribution in each zone is thus assumed to behave like an isotropic, completely relativistic gas at densities above $10^{12}$ \gcc, whose effect on the hydrodynamics is computed by means of their corresponding pressure, $p_{\nu}$, and specific energy, $e_{\nu}$. 
For densities below $10^{12}$~\gcc, $p_{\nu}$ and $e_{\nu}$ are set to zero. 
The neutrino stress for the $\theta$- and $\phi$-sweeps is either that of an isotropic gas entrained with the matter (above $10^{12}$~\gcc) or zero ($10^{12}$~\gcc). As a result, we set $f_{\nu,\theta}$ and $f_{\nu,\phi}$ to zero and include $p_{\nu}$ (nonzero above $10^{12}$~\gcc) in the material pressure, and the sum is PPM interpolated and incorporated into $p_{i+\half}^{n+\half}$ and is then used to update the specific internal energy, with  $e_{\nu,i}$ incorporated into and later extracted from $e_{{\rm int},i}$.

The finite difference approximations to Equations~(\ref{eq:h73}) are
\begin{eqnarray}
u_{r, \, i}^{n+1'} &=& u_{r, \, i}^{n} + \frac{ \frac{1}{2} \left( A_{i-\half}^{n+\half} + A_{i+\half}^{n+\half} \right) }{ V_{i}^{n+\half} } \frac{1}{\rho_{i}^{n+\half} } \left( p_{i-\half}^{n+\half} - p_{i+\half}^{n+\half} \right) \Delta t + \left( - \nabla_{r} \edgravi^{n+\half} + f_{r, \, \nu \, i}^{n} + f_{r,{\rm cenfugal}, i}^{n+\half} \right) \Delta t , \label{eq:h74} \\
 u_{\theta, \phi, \, i}^{n+1'} &=& u_{\theta, \phi, \, i}^{n} + \frac{ \frac{1}{2} \left( A_{i-\half}^{n+\half} + A_{i+\half}^{n+\half} \right) }{ V_{i}^{n+\half} } \frac{1}{\rho_{i}^{n+\half} } \left( (p + p_{\nu})_{i-\half}^{n+\half} - (p + p_{\nu})_{i+\half}^{n+\half} \right) \Delta t  \nonumber \\ 
 && + \left( - \nabla_{\theta, \phi} \, \edgravi^{n+\half} + f_{\theta, \phi,{\rm cenfugal}, i}^{n+\half} \right) \Delta t , \label{eq:h75}
\end{eqnarray}
where $A_{i-\half}^{n+\half}$ is defined by $\delta V_{i+\half} = A_{i-\half}^{n+\half} u_{i+\half}^{n+\half} \Delta t$. $\delta V_{i+\half}$ is the volume swept out by the change in the position of the $i^{th}+1$ interface in the time interval $\Delta t$ and is given by
\begin{equation}
A_{i+\half} = \left\{ \begin{array}{cc} {\ds \frac{1}{3} \left( ( r_{i+\half}^{n+1'} )^{2} + r_{i+\half}^{n+1' }r_{i+\half}^{n} + ( r_{i+\half}^{n} )^{2} \right)} & \mbox{$r$-direction} , \\
 {\ds \left( \cos \theta_{i+\half}^{n} -  \cos \theta_{i+\half}^{n+1'} \right)/\left( \theta_{i+\half}^{n+1''} - \theta_{i+\half}^{n} \right) } & \mbox{$\theta$-direction} , \\
  1 & \mbox{$\phi$-direction}. \end{array} \right.
\label{eq:h76}
\end{equation}
$f_{{\rm cenfugal}, i}^{n+\half}$ includes the last term in Equations~(\ref{eq:h73}) for the $r$-direction sweep and the second to last term in Equations~(\ref{eq:h73}) for the $\theta$-direction sweep.
The last term in Equation~(\ref{eq:h73}) for the $\theta$-direction sweep is an expression for the change in $u_{\theta}$ due to a change in the radial position by virtue of angular momentum conservation. Rather than including this term in Equation~(\ref{eq:h73}), its contribution to the change in $u_\theta$ is updated in the $r$-sweep by the equivalent expression
\begin{equation}
u_{\theta,i}^{n+1'} = u_{\theta,i}^{n} \, \frac{ r_{i}^{n} }{ r_{i}^{n+1'} } .
\label{eq:h77}
\end{equation}
Similarly, the last two terms in Equation~(\ref{eq:h73}) for the $\phi$-direction sweeps are expressions for angular momentum conservation about the z-axis due to a change in the $\theta$ and radial positions, respectively.
Rather than including these terms in Equation~(\ref{eq:h73}), their contributions to the change in $u_\phi$ are updated in the $\theta$- and $r$-direction sweeps by the equivalent expressions
\begin{equation}
u_{\phi,i}^{n+1'} = u_{\phi,i}^{n} \, \frac{ r_{i} \sin \theta_{i}^{n} }{ r_{i} \sin \theta_{i}^{n+1'} }, \qquad u_{\phi,i}^{n+1'} = u_{\phi,i}^{n} \, \frac{ r_{i}^{n} }{ r_{i}^{n+1'} } .
\label{eq:h78}
\end{equation}
The time-centered value of the gravitational potential, $\phi_{\rm g}$, is obtained by extrapolation, as described in Section~\ref{Gravity}.
The two centrifugal force terms are differenced as follows:
\begin{eqnarray}
f_{{\rm cenfugal}, i}^{n+\half} =& {\ds \frac{ \left( u_{\theta,i}^{n+\half} \right)^{2} + \left( u_{\phi,i}^{n+\half} \right)^{2} }{ r_{i}^{n+\half} } } & \mbox{$r$-direction} , \\
f_{{\rm cenfugal}, i}^{n+\half} =& {\ds \frac{ \left( u_{\phi,i}^{n+\half} \right)^{2} }{ r_{i}^{n+\half} \sin \theta_{i}^{n+\half} } \cos \theta_{i}^{n+\half} } & \mbox{$\theta$-direction} ,
\label{eq:centrifugal}
\end{eqnarray}
where the time-centering of $r_{i}$ in the $r$-direction and $\theta_{i}$ in the $\theta$-direction is performed explicitly by averaging their values at time $n$ and $n+1$, and the time-centering of the other variables is approximately accomplished by the symmetric way in which the directional splitting is performed, as described in Section~\ref{app:Dir_Splt}.
Finally, the neutrino stress term $f_{\nu,i}^{n}$ is not time centered. A second execution of the neutrino transport would be required to center it. 
This term is small and slowly varying in time, so we include it to first-order only and avoid the significant additional cost that would be paid were we to 
include it with second-order accuracy.

The specific internal energy is updated differently depending on whether the zone is in the vicinity of a shock or away from shocks. 
In the vicinity of a shock, the results of solving the Rankine-Hugoniot equations must be used, as the flow there is non-isentropic. 
In this case the specific total energy is updated as given, in general form, by Equation~(\ref{eq:h58}) for the radial sweep, and Equation~(\ref{eq:h53}) for the $\theta$- and $\phi$-sweeps. 
Specifically, for the radial-sweep, the specific total energy, $\edtot = \edint + \edkin + \edgrav$, is updated by
\begin{equation}
\edtoti^{n+1'} = \edtoti^{n} - \frac{ \left( A_{i+\half}^{n+\half} p_{i+\half}^{n+\half} u_{r, \,i+\half}^{n+\half} - A_{i-\half}^{n+\half} p_{i-\half}^{n+\half} u_{r, \, i-\half}^{n+\half} \right) }{ V_{i}^{n+\half} \rho_{i}^{n+\half} } \Delta t + \frac{1}{2} \left( u_{r, \, i}^{n} +  u_{r, \, i}^{n+1'} \right) f_{\nu,i}^{n} \Delta t \quad \mbox{$r$-direction} ,
\label{eq:h80}
\end{equation}
where $u_{i}^{n+1'}$ is given by Equation~(\ref{eq:h74}). 
The $\partial \edgrav/\partial t$ term in Equation~(\ref{eq:h58}) has been omitted at this stage but is included later in the radial sweep. 
For the $\theta$- and $\phi$-sweeps, the specific energy, $\edtotp = \edint + \edkin$, is updated by
\begin{eqnarray}
\left( \edtotp + e_{\nu} \right)_{i}^{n+1'} &=& ( \edtotp + e_{\nu} )_{i}^{n} \nonumber \\ 
 &&-\, \frac{ \left( A_{i+\half}^{n+\half} (p + p_{\nu})_{i+\half}^{n+\half} u_{i+\half}^{n+\half} - A_{i-\half}^{n+\half} (p + p_{\nu})_{i-\half}^{n+\half} \, u_{i-\half}^{n+\half} \right) }{ V_{i}^{n+\half} \rho_{i}^{n+\half} } \Delta t \nonumber \\
 &&+\, \frac{1}{2} \left( u_{i}^{n} +  u_{i}^{n+1'} \right) \left( - \nabla \edgravi^{n+\half} \right) \Delta t \quad \mbox{$\theta$-, $\phi$-directions} ,
\label{eq:h81}
\end{eqnarray}
followed by
\begin{equation}
\edtotp^{n+1'} = ( \edtotp + e_{\nu} )_{i}^{n+1'} - e_{\nu}^{n+1'} = ( \edtotp + e_{\nu} )_{i}^{n+1'} - e_{\nu}^{n} \left( \frac{ \rho^{n+1'} }{\rho^{n}} \right)^{4/3} .
\label{eq:h82}
\end{equation}

Away from shocks the specific energy could still be updated as above, but errors might then arise during the remap step when subtracting the specific kinetic energy from the specific total energy.
The problem arises from the use of $\langle \uvec \rangle^{2}$ in the expression for the specific kinetic energy rather than $\langle \uvec^{2} \rangle$.
The two expressions can differ importantly in supersonic flow and near reflection boundaries where the gradient of \uvec\ can be large \citep[see][for a discussion of this point]{BlLu93a}.
Computing $\langle \uvec^{2} \rangle$ would be costly in a multidimensional simulation, as it would involve a multidimensional integration over the components of \uvec, and may not be well defined.
Instead, \Chimera\ updates the specific internal energy using the first law of thermodynamics, assuming isentropic and constant-composition flow, as non-isentropic changes due to nuclear and neutrino sources are computed elsewhere by operator splitting.

The update of the specific internal energy thus takes the form
\begin{equation}
\edinti^{n+1'} = \edinti^{n} + \frac{ p_{i}^{n+\half} }{ \left( \rho_{i}^{n+\half} \right)^{2} } \left( \rho_{i}^{n+1'} - \rho_{i}^{n} \right) ,
\label{eq:h79}
\end{equation}
where the time-centering of the pressure $p$ is accomplished by a predictor-corrector loop, completing the Lagrangian step.

\subsection{Remap Step}
\label{Remap}

Following the Lagrangian step, in the case of the $\theta$- and $\phi$-sweeps the grid is remapped back to the configuration that prevailed before the Lagrangian step, thus making the combination of a Lagrangian step and a remap step effectively an Eulerian step.
In the case of the radial sweep, the grid is remapped back to a configuration specified by the regridder, which will be described in Section~\ref{sec:Regrid}.
Via the regridder options the user can specify that the grid following the Lagrangian step be left as is, remapped back to the configuration that prevailed before the Lagrangian step, or, by invoking one of the regridder algorithms, remapped to a configuration that tends to optimize the resolution of structures that develop during a simulation. 

\subsubsection{Remapping mass, momenta, and angular momenta}
\label{remap1}

For quantities like the mass, specific momenta (momenta per gram), and specific angular momenta, the remapping procedure is straightforward. 
Denoting, as before, the values of the grid and other variables after the Lagrangian step by the superscript $n+1'$, and after the remap step by the superscript $n+1$, the difference $\delta \xi$, between the values of the grid variables at $n+1'$ and $n+1$, given by
\begin{equation}
\delta \xi_{i+\half} = \xi_{i+\half}^{n+1'} - \xi_{i+\half}^{n+1}
\label{eq:h101}
\end{equation}
is computed, as is the volume $\delta V_{i+\half}$ contained within $\delta \xi_{i+\half}$.
The latter is given by
\begin{equation}
\delta V_{i+\half} = \left\{ \begin{array}{cc} {\ds \frac{1}{3} \left[ \left( \xi_{i-\half}^{n+1'} \right)^{3} - \left( \xi_{i+\half}^{n+1} \right)^{3} \right] } & \mbox{$r$-direction} , \\ 
{\ds r_{i} \left( \cos \xi_{i-\half}^{n+1} - \cos \xi_{i+\half}^{n+1'} \right) } & \mbox{$\theta$-direction} , \\
 {\ds r_{i} \sin \theta_{i} \left( \phi_{i+\half}^{n+1'} - \phi_{i+\half}^{n+1} \right) } & \mbox{$\phi$-direction} . \end{array} \right.
\label{eq:h102}
\end{equation}

If  $\delta \xi_{i+\half} > 0$, the grid interface $\xi_{i + \half}^{n+1}$ is placed by the remap step between $\xi_{i - \half}^{n+1'}$ and $\xi_{i + \half}^{n+1'}$.
The mass advected across the zone interface $\xi_{i+\half}$ is next computed by first constructing a PPM profile, $\rho(\xi)$, of the density $\rho_{i}^{n+1'}$.
The value, $\langle \rho \rangle_{L, \, i+\half}^{n+1'}$, of the interpolated density $\rho(\xi)$ averaged over the interval $\delta \xi_{i+\half}$ to the left of interface $\xi_{i+\half}^{n+1'}$ is then computed by Equation~(\ref{eq:h60}), with $\delta \xi_{i+\half}$ replacing $c_{{\rm s},i}\Delta t $.
The mass advected is then given by
\begin{equation}
\delta {\cal M}_{i+\half}^{n+1'} = \langle \rho \rangle_{L, \, i+\half}^{n+1'} \delta V_{i+\half} .
\label{eq:h103}
\end{equation}
The zone mass after the remap step, ${\cal M}_{i}^{n+1}$, is then computed from the the zone mass before the remap step, ${\cal M}_{i}^{n+1'}$, and the advected mass 
$\delta {\cal M}_{i+\half}^{n+1'}$ by
\begin{equation}
{\cal M}_{i}^{n+1} = {\cal M}_{i}^{n+1'} - \delta {\cal M}_{i+\half}^{n+1'} + \delta {\cal M}_{i-\half}^{n+1'} ,
\label{eq:h103q}
\end{equation}
and the density $\rho_{i}^{n+1}$ is then computed by
\begin{equation}
\rho_{i}^{n+1} = \frac{ {\cal M}_{i}^{n+1}}{ V_{i}^{n+1} } .
\label{eq:h103b}
\end{equation}
Having determined the advected mass $\delta {\cal M}_{i+\half}^{n+1'}$, the remapping of any specific (i.e., per gram) quantity $a_{i}^{n+1'}$ proceeds by determining the amount, $\delta {\cal A}_{i+\half}^{n+1'}$, of that quantity advected.
As in the case of the density, a piecewise parabolic interpolation profile, $a(\xi)$, of $a_{i}^{n+1'}$ is constructed and the average $\langle a \rangle_{L, \, i+\half}^{n+1'}$ of that quantity over the interval $\delta \xi_{i+\half}$ is computed using Equation~(\ref{eq:h60}). 
The mass of  $a(\xi)$ contained within $\delta \xi_{i+\half}$ is finally computed as the penultimate step by
\begin{equation}
\delta {\cal A}_{i+\half}^{n+1'} = \langle a \rangle_{L, \, i+\half}^{n+1'} \delta {\cal M}_{i+\half}^{n+1'} .
\label{eq:h104}
\end{equation}

If  $\delta \xi_{i+\half} < 0$, the grid interface $\xi_{i + \half}^{n+1}$ is placed by the remap step between $\xi_{i + \half}^{n+1'}$ and $\xi_{i + \frac{3}{2}}^{n+1'}$, and the quantity of $a(\xi)$ within $\delta \xi_{i+\half}$ is advected from the right to the left of interface $\xi_{i + \half}^{n+1}$. 
The procedure is the same as described above for the $\delta \xi_{i+\half} > 0$ case, except that $\langle \rho \rangle_{R, \, i+\half}^{n+1'}$ and $\langle a \rangle_{R,i+\half}^{n+1'}$ are computed from $\rho_{i+1}^{n+1'}$ and $a_{i+1}^{n+1'}$ by Equation~(\ref{eq:h61}), giving $\delta {\cal M}_{i+\half}^{n+1'}$ and $\delta {\cal A}_{i+\half}^{n+1'}$ to the right of interface $\xi_{i+\half}^{n+1'}$, with $\delta \xi_{i+\half}$ replacing $  c_{{\rm s},i} \Delta t$, as before, but now with a negative value of $\delta V_{i+\half}$.
Therefore, the quantity $\delta {\cal M}_{i+\half}^{n+1'}$ computed from $\langle \rho \rangle_{R,i+\half}^{n+1'}$ by an equation analogous to Equation~(\ref{eq:h103}), and the quantity $\delta {\cal A}_{i+\half}^{n+1'}$ computed from $\langle a \rangle_{R,i+\half}^{n+1'}$ by an equations analogous to Equation~(\ref{eq:h104}), are both negative.

The remap step is finally completed by performing the advection:
\begin{equation}
a_{i}^{n+1} = \frac{ a_{i}^{n+1'} {\cal M}_{i}^{n+1'}  - \delta {\cal A}_{i+\half}^{n+1'} + \delta {\cal A}_{i-\half}^{n+1'} }{ {\cal M}_{i}^{n+1} } .
\label{eq:h105}
\end{equation}
Negative values of $\delta {\cal A}_{i+\half}^{n+1'}$ or $\delta {\cal A}_{i-\half}^{n+1'}$ mean simply that the advection proceeds from right to left rather than in the other direction. 
The advection step is conservative since the same amount of a given quantity enters a zone as leaves the adjacent zone. 
Again, the sign of $\delta {\cal A}_{i+\half}^{n+1'}$ determines whether the advection is from zone $i$ to zone $i+1$, or vice versa.

\subsubsection{Remapping Composition and Electron Fraction}
\label{remap2}

The algorithm for the advection of the composition mass fractions, $X_{n}$, and net electron fraction, \Ye, depends on whether the matter on either side of the zone interface is in nuclear  statistical equilibrium (NSE) or not (non-NSE). 
If the the advection is between zones in NSE, the composition is given by the EoS as a function of the values of $\rho$, $\edint$, and \Ye\ of the material being advected.
In this case, no explicit advection of composition mass fractions needs to be performed. The advection of \Ye\ in this case proceeds as described in Section \ref{remap1} above. 
If the advection is between zones in non-NSE, then along with $\rho$ and $\edint$, the advection of the composition mass fractions, $X_{n}$, is carried out explicitly as described in Section \ref{remap1} and in accordance with the consistent multi-fluid advection method of \citet{PlMu99}. 
That is, the average value of each composition mass fraction $X_{n}$ to be advected is computed by Equation~(\ref{eq:h60}) or (\ref{eq:h61}) depending on whether $\delta \xi$ is positive or negative.
The resulting composition in the mass to be advected is then normalized to unity before performing the advection. 
In order that the net \Ye\ advected in this case be consistent with the net proton fraction of the advected composition, the \Ye\ advected is taken to be that of the net \Ye\ of the advected composition.

If the advection is from a zone in non-NSE to one in NSE, the composition of the material to be advected is computed as described above for the non-NSE to non-NSE case.
The material advected is then assumed to become part of the NSE material in the acceptor zone, and only the independent thermodynamic variables $\rho$, $\edint$, and \Ye\ of the material to be advected are advected. 
Finally, if material is advected from a zone in NSE to one in non-NSE, the material advected is first `deflashed,' that is, its NSE composition mass fractions are extracted from the EoS, stored in a temporary mass fraction array, and advected as described above for the case of two adjacent non-NSE zones.

\subsubsection{Multiple EoSs and the Energy Remap}
\label{remap3}

\chimera\ is designed to accommodate multiple EoSs that are applied in contiguous density ranges. As explained above in Section~\ref{StellarEoS}, for example, \chimera\ has used the LS EoS at densities above $10^{11}$ \gcc, and a different EoS below that density.
Since \chimera\ updates the specific energy directly rather than the temperature, and uses the updated specific energy and other needed thermodynamic variables to update the temperature, slight differences in the energy zeros at the boundary between two EoS's could result in unphysical temperature updates. 
Specific energy differences between two EoSs could also arise from peculiarities or approximations peculiar to each EoS. 
In either case, we will refer to the potential difference in specific energies given by two EoS's for the same thermodynamic state as a zero-energy offset.
Unphysical temperature updates could happen, for example, during the remap step if matter from a zone linked to one EoS is advected into an adjacent zone linked to a different EoS. 
The quantity of energy advected would contain the difference in the energy zeros as well as the physically relevant energy.
This problem could affect the radial sweep but not the $\theta$- or $\phi$-sweeps as the same EoS is always used along a $\theta$- or $\phi$-directed ray.

To avoid this problem \chimera\, overlaps by four radial zones in either direction the specific internal energy at the boundary between EoSs as shown in Figure~(\ref{fig:Energy_offset}), where radial zones $i-1$ and lower are tied to a particular EoS~A, while zones $i$ and higher are tied to a different EoS~B.
The two EoSs have different zero-energy offsets, as indicated by their vertical displacement and an overlap of four zones in either direction.
This enables PPM profiles of the specific internal energy for zones $i-1$ and below to be constructed using EoS~A up to the interface $i - \half$.
Likewise, PPM profiles of the specific internal energy for zones $i$ and above can be constructed using EoS~B.
The example in Figure~(\ref{fig:Energy_offset}) is one in which the zone interface $\xi_{i-\half}$ is remapped a distance $\delta \xi_{i-\half}$ to the left of its original position, its new position being indicated by the vertical red dashed line.
This entails that a quantity of specific internal energy contained within $\delta \xi_{i-\half}$ be advected from the left to the right of zone interface $\xi_{i-\half}$. 
\chimera\ performs this advection by advecting the energy within $\delta \xi_{i-\half}$ as given by EoS~A out of zone $i-1$, and advecting the energy within the same $\delta \xi_{i-\half}$ but as given by EoS~B into zone $i$.
Since the specific internal energy advected out of a zone and into the adjacent zone is consistent with the different EoSs tied to each of the two zones, the unphysical temperature jump that would occur if the zero-energy offset had not been accounted for in the advection is avoided.

\begin{figure}
\fig{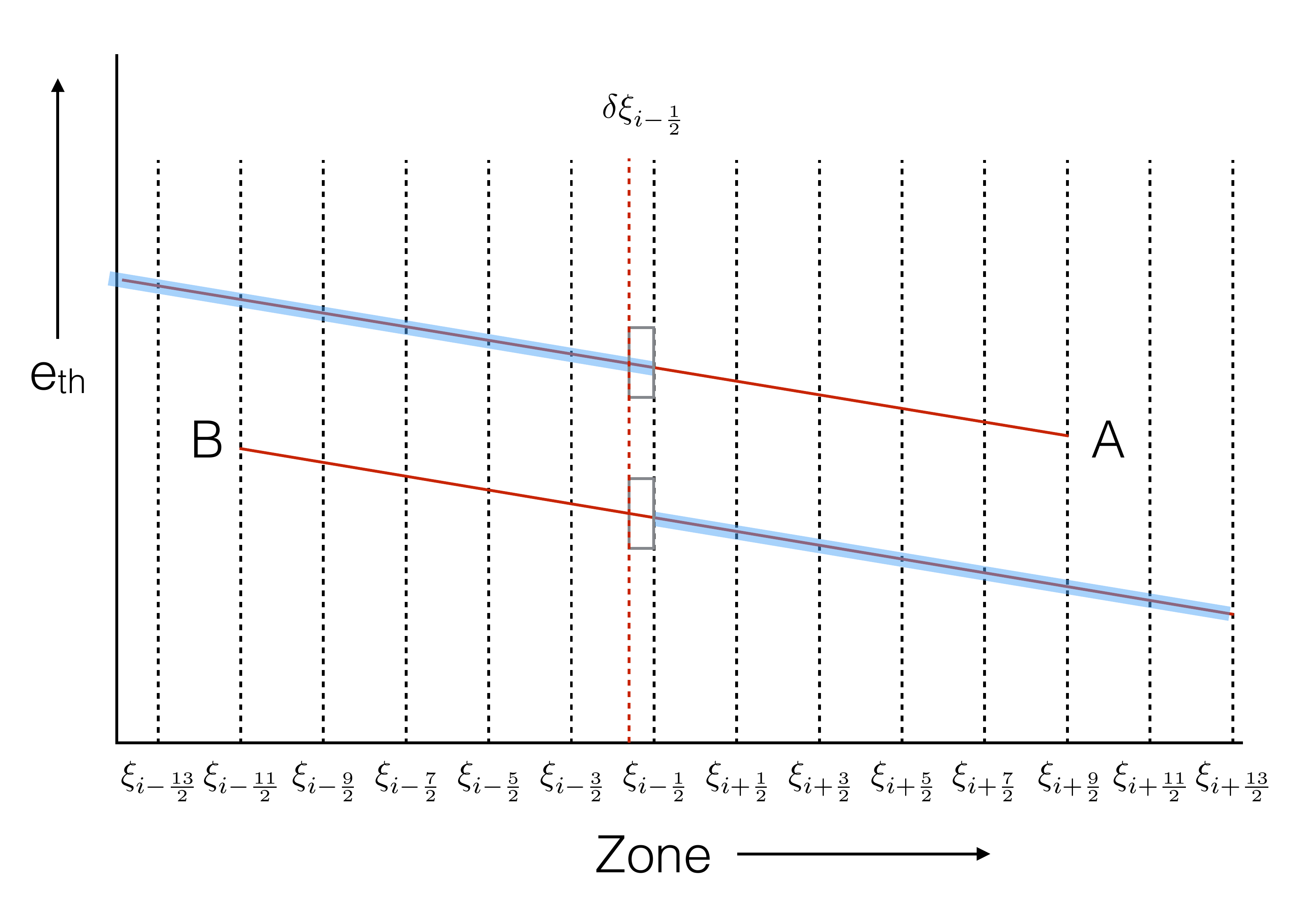}{0.6\textwidth}{}
\caption{\label{fig:Energy_offset}
Schematic representation of the advection of energy from the left to the right of zone interface $i - \frac{1}{2}$. Zones $i - 1$ and below are tied to EoS A while zones $i$ and above are tied to EoS B. 
EoS A and EoS B have a zero energy offset indicated by their vertical displacement from each other. 
In remapping zone interface $i - \half$ a distance $\delta \xi_{i-\half}$ from its initial location to the location indicated by the red vertical dashed line, the energy within $\delta \xi_{i-\half}$ given by EoS A must be advected from zone $i-1$ to zone $i$. 
This is accomplished by advecting the energy within $\delta \xi_{i-\half}$ given by EoS A out of zone $i-1$, and advecting the energy within the same interval $\delta \xi_{i-\half}$  but given by EoS B into zone $i$.
The four-zone overlap on each side exists so that a PPM profile of the energy in zone $i-1$ can be constructed from both EoSs.
}
\end{figure}

\subsubsection{Nuclear Binding Energy}
\label{remap4}

In advecting the specific energy between two adjacent zones during a remap, the specific internal energy is split into a nuclear binding energy component, \edbind, and the rest of the energy, and the two components are advected separately. For example, the specific internal energy, \edint, is split into \edbind\ and $\edth = \edint - \edbind$, and remapped as follows:
\begin{equation}
\edinti^{n+1} = \frac{ \left( \edthi^{n+1'} + \edbindi^{n+1'} \right) \delta {\cal M}_{i}^{n+1'} - \delta E_{{\rm th},i+\half}^{n+1'} - \delta E_{{\rm bind},i+\half}^{n+1'} + \delta E_{{\rm th},i-\half}^{n+1'} + \delta E_{{\rm bind},i-\half}^{n+1'} }{ \delta {\cal M}_{i}^{n+1} } ,
\label{eq:h106}
\end{equation}
where $e_{{\rm th},i}^{n+1'}$  and $e_{{\rm bind},i}^{n+1'}$ are the specific internal energy minus the specific binding energy, and the specific binding energy, respectively, of zone $i$, and $\delta E_{{\rm th},i \pm \half}^{n+1'}$ and $\delta E_{{\rm bind},i\pm\half}^{n+1'}$ are the internal energy minus the binding energy and the binding energy, respectively,  transferred through outer zone edge ($i+\half$) and inner zone edge ($i-\half$) of zone $i$. 
The masses $\delta {\cal M}_{i}^{n+1'}$ and $\delta {\cal M}_{i}^{n+1}$ are the masses of zone $i$ before and after the remap, respectively, as given by Equation~(\ref{eq:h103}).
This mode of energy advection is appropriate for advecting non-NSE material during a remap, as the non-NSE material being advected does not necessarily have the same composition, and therefore binding energy, as the original material in the zone from which it is being advected. 
The composition of the material being advected is obtained by integrating over the PPM profile constructed for each ionic mass fraction, and then normalizing the sum to unity.
Since the PPM profiles of different ionic mass fractions may be differently shaped, the composition that results after integrating over the portion of the profiles being advected  and  normalizing can result in a composition and binding energy different from the original. 
In the energy advection procedure described above, the binding energy of the advecting material is computed once its composition is ascertained, and the rest of the advected energy, $\delta e_{{\rm th}}$, is also computed by integrating the PPM profile of $\edth = \edint - \edbind$ over the advecting mass. 
The net energy advected should thus reflect both the correct nuclear binding energy of the advecting material, as well as its thermal component.

\subsubsection{Energy remap for the $\theta$- and $\phi$-sweeps and the preliminary remap for the radial sweep}
\label{remap5}

The energy remap described above is the ultimate step in the $\theta$- and $\phi$-sweep hydrodynamics, and the penultimate step in the radial sweep hydrodynamics. 
Away from shocks, the specific internal energy is remapped as given by Equation~(\ref{eq:h106}), as opposed to remapping the sum of the specific internal plus specific kinetic energy, $\edtotpi = \edint + \edkin$. 
This is permissible as the flow is isentropic apart from the contributions of nuclear and neutrino source terms, which have been included elsewhere in the radial sweep (see the text above Equation~(\ref{eq:h79}) for the motivation for this approach). 
In the vicinity of shocks, the specific energy, \edtotp, which has been evolved during the Lagrangian step using the Rankine-Hugoniot equations, must be remapped, and the specific internal energy is extracted afterwards.
We define $\edtotpp = \edtotp - \edbind = \edth + \edkin$ and remap \edtotpp\ and \edbind\ separately, as described above.
To use consistent values for the left and right states for calculating a PPM profile of \edtotpp (i.e., consistent with the velocity remap), we define
\begin{eqnarray}
e_{{\rm tot}'',L,i}^{n+1'} &=& e_{{\rm th},L,i}^{n+1'} + \frac{1}{2} \left[ \left( u_{r,L,i}^{n+1'} \right)^{2} + \left( u_{ \theta,L,i}^{n+1'} \right)^{2} + \left( u_{\phi,L,i}^{n+1'} \right)^{2} \right], \\ \nonumber
e_{{\rm tot}'',R,i}^{n+1'} &=& e_{{\rm th},R,i}^{n+1'} + \frac{1}{2} \left[ \left( u_{r,R,i}^{n+1'} \right)^{2} + \left( u_{ \theta,R,i}^{n+1'} \right)^{2} + \left( u_{\phi,R,i}^{n+1'} \right)^{2} \right] .
\label{eq:h107}
\end{eqnarray}
Having calculated the PPM profile of \edtotpp, the quantity of \edtotpp\ advected across a given zone interface is computed by Equation~(\ref{eq:h103}) or (\ref{eq:h104}), and the remapping of \edtotpp\ is performed by an equation analogous to Equation~(\ref{eq:h106}). 
After the remapping of \edtotpp\ and \edbind, the specific internal energy is extracted from their sum, \edtotp.

\subsubsection{Recomputation of the Gravitational Potential and the Computation of $\partial{\edgrav}/\partial{t}$}
\label{remap6}

Following the remap in the radial-sweep of the mass, momenta, angular momenta, the independent thermodynamic variables, and the preliminary remap of the specific energy, the specific gravitational potential, $\edgrav^{n+1}$, is computed. 
In the case in which the spherically symmetric component of the gravitational potential is computed by means of a general relativistic approximation, the remapped pressure, specific energy, and neutrino contributions are used as sources of gravity as well as the density, which necessitated the preliminary remap of the specific energy. 
It is at this stage of the radial-sweep that the quantity $\partial{\edgrav}/\partial{t}$ is computed and added, in accordance with Equation~(\ref{eq:h58}), to the specific total energy, $\edtot^{n+1'}$, that was computed by Equation~(\ref{eq:h80}) during the Lagrangian step. 
The most direct procedure for calculating $\partial{\edgrav}/\partial{t}$ would be to calculate the gravitational potential, $\edgrav^{n+1'}$, after the Lagrangian step, interpolate the initial gravitational potential, $\edgrav^{n}$, to the Lagrangian grid to get the quantity $e_{{\rm grav},I-L}^{n}$, and approximate $\partial{\edgrav}/\partial{t}$ by $\left( \edgrav^{n+1'} - e_{{\rm grav},I-L}^{n} \right)/\Delta t$. 
This would work well for one-dimensional simulations, but for multidimensional simulations a given radial grid edge, $\xi_{i+\half}$, after the Lagrangian step is a function of $\theta$ and/or $\phi$, making the gravitational potential difficult to compute at this point. 
Instead, the gravitational potential is computed after the remap of the radial grid and interpolated as a function of $\theta$ and/or $\phi$ to the Lagrangian grid, thereby obtaining $e_{{\rm grav},F-L}^{n+1}$. 
The time derivative of the gravitational potential added to $\edtot^{n+1'}$ is thus given, as a function of $\theta$ and/or $\phi$, by
\begin{equation}
\pderiv{ \edgrav} {t} = \frac{ e_{{\rm grav},F-L}^{n+1} - e_{{\rm grav},I-L}^{n} }{ \Delta t} .
\label{eq:h108}
\end{equation}

\subsubsection{Final Radial-Sweep Remap of the Total Energy}
\label{remap7}

The final remapping of the specific total energy (\edkin + \edint + \edgrav) in the radial sweep begins with the specific energy, $e_{{\rm tot}'',i}^{n+1'}$, given by
\begin{equation}
e_{{\rm tot}'',i}^{n+1'} = e_{{\rm tot},i}^{n+1'} - \edbindi^{n+1'} - \edkini^{n+1'} + \left( \pderiv{ \edgrav} {t} \right)_{i}^{n+\half} \Delta t ,
\label{eq:h109}
\end{equation}
where \edtoti\ has been updated during the Lagrangian step, as given by Equation~(\ref{eq:h80}). 
Consistent left and right states of $e_{{\rm tot}'',i}^{n+1'}$ are determined as specified by Equations~(\ref{eq:h107}), PPM profiles of $e_{{\rm tot}',i}^{n+1'} = e_{{\rm tot}'',i}^{n+1'} + \edkini^{n+1'}$ are then obtained, and the amount of $e_{{\rm tot}' \, i}^{n+1'}$ to be advected across the zone interface are given by Equation~(\ref{eq:h104}) and the procedure outlined in the discussion below it. 
Remapping then proceeds in accordance with an equation analogous to Equation~(\ref{eq:h106}), and the final specific internal energy is extracted from $\edtoti^{n+1} = e_{{\rm tot}',i}^{n+1} + \edbindi^{n+1}$ by
\begin{equation}
\edinti^{n+1} = \edtoti^{n+1} - \edkini^{n+1} - \edgravi^{n+1} .
\label{eq:h110}
\end{equation}
This completes the remap step of the radial-sweep hydrodynamics.

\subsubsection{Suppression of Carbuncles}

When shocks are aligned with one of the coordinate directions in multidimensional simulations they are susceptible to an ``odd-even decoupling'' or ``carbuncle'' instability \citep{Quir94,Liou00,SuBiBi03}. 
This can lead to a strong rippling of the shock front, which could, in turn, excite hydrodynamic instabilities in the post-shock region.
Our scheme for suppressing this instability is the use of a ``local oscillation filter'' similar in philosophy to that described by \citet{SuBiBi03}. This approach is local and does not affect the well-resolved features of the flow elsewhere.
To suppress the carbuncle instability in the radial direction, which is where this instability typically arises in a supernova simulation, the angular ($\theta$) and azimuthal ($\phi$) remap steps are each followed by an examination of the radial velocities along the angular and azimuthal rays to search for radial velocity extrema.
If there are at least three radial velocity extrema in any group of five adjacent zones, and if a shock is present, these zones are marked for ``smoothing.''
For the particular case in which zones $m$ and $m+1$ are marked for smoothing, the flux $\delta {\cal A}_{m+\half}$ of the quantity $a$, defined by
\begin{equation}
\delta{\cal A}_{m+\half} = c_{\rm smooth} \left( a_{m+1} - a_{m} \right) \min( {\cal M}_{m}, {\cal M}_{m+1} ) ,
\label{eq:h110a}
\end{equation}
is computed, where ${\cal M}_{m}$ is the mass of zone $m$, and $c_{\rm smooth}$ is an empirical parameter.
Experimentation has ahown that a value of 0.075 for $c_{\rm smooth}$ works well.
The final step in the procedure is to sweep across the angular and azimuthal rays and exchange the flux $\delta{\cal A}_{m+\half}$ between the zones marked for smoothing in a step analogous to that described by Equation~(\ref{eq:h105}). 
This reduces the difference in the values of the quantity $a$ between adjacent zones, thereby inhibiting the growth of this difference.
Applying this procedure to the quantities $u_{r}$ and $u_{\phi}$, and to $L_{\phi}$ (to smooth $u_{\theta}$), proved sufficiently robust to suppress the carbuncle instability.

\subsection{Radial Regridder}
\label{sec:Regrid}

The PPM Lagrangian-Remap format permits the grid after the Lagrangian step to be remapped to a grid other than the initial grid from which the Lagrangian step originated. While the $\theta$- and $\phi$-grids are remapped back to their initial grids after the Lagrangian step, making them effectively Eulerian, 
\Chimera\ uses the remapping freedom to provide the user with a number of remapping options for the radial grid to ensure that the grid continues to resolve important structures that arise during the course of a simulation. 
One option is for the radial grid to be remapped back to the initial grid after the Lagrangian step, making the grid effectively Eulerian as it is for the $\theta$ and $\phi$ grids. 
Another option is for the remapped grid to follow the mean motion of the fluid, referred to here and below as pseudo-Lagrangian, making the grid purely Lagrangian in the case of spherically symmetric fluid flow.

Currently a number of more sophisticated options are available specific to the pre-bounce or post-bounce phase of a \ccsn\ simulation.
For both the pre-bounce phase and the post-bounce phase, an inner-outer boundary dividing the radial grid into an inner and an outer section is determined based on a number of user selected criteria.
These criteria can differ between the pre-bounce and the post-bounce phases, and can differ at user selected time intervals during the post-bounce phase.
During both the pre-bounce and the post-bounce phases, the outer grid can be selected to be pseudo-Lagrangian, which is useful if there are sharp chemical discontinuities in non-NSE material that need to be preserved, or Eulerian if advection through the outer boundary of a prescribed distribution of material is important.
During the pre-bounce phase, the inner grid starts out as pseudo-Lagrangian, but blends into another grid between two user selected densities. 
This second grid is constructed so that adjacent zones satisfy $\Delta r_{i+1} = \mbox{ constant} \times \Delta r_{i}$, referred to here as a `zoomed grid,' with the properties that the width of the outer zone of this zoomed grid is equal to the zone width of the first zone of the outer grid, and the inner zone tends to a user selected width when that zone reaches $3 \times 10^{14}$ \gcc. 
The result is a smooth and smoothly evolving grid that can be tuned to provide the desired grid resolution at the proto-neutron star surface when it forms. 
During the post-bounce phase the inner grid remains a zoomed grid from the core center to a density of $10^{14}$ \gcc, with the central zone width such that it would attain a user selected zone width at a density of $3 \times 10^{14}$ \gcc.
A second zoomed grid covers the density range from $10^{14}$ to $10^{12}$ \gcc, and a third covers the density range from $10^{12}$ to $10^{10}$ \gcc.
Both of these latter two grids have the same number of zones, which are equal to a user selected value. 
This ensures there are a sufficient number of zones to resolve the neutrinosphere as the proto-neutron star shrinks and as the density cliff forms near its edge.
Finally, a fourth zoomed grid covers the region from the density of $10^{10}$ \gcc\ to the outer edge of the inner grid.
The result of this regridding is again a smooth and smoothly evolving user controlled grid designed to resolve the critical features that arise during the course of a \ccsn\ simulation.

\subsection{Gravity Solver}
\label{Gravity}

Self gravity can be chosen to be either one- or multi-dimensional, with a further choice of either a Newtonian or an approximate general relativistic monopole component. 
The approximate general relativistic gravitational potential used for the monopole component  in the latter case is a modified Tolman-Oppenheimer-Volkoff (TOV) potential suggested by \citet[][Case A]{MaDiJa06} and described briefly below in Section \ref{1Dgravity}. 
Multidimensional gravity is obtained by expanding the Newtonian gravitational potential in a multipole expansion as described by \citet{MuSt95} and below in Section~\ref{MDgravity_Axi}.
Approximate general relativistic multidimensional gravity is obtained by replacing the Newtonian monopole in the multipole expansion by the approximate general relativistic monopole.

\subsubsection{One-Dimensional Gravitational Potential}
\label{1Dgravity}

Newtonian monopole gravity is trivial.
The radial zone-edged and zone-centered gravitational accelerations, $g_{i+\frac{1}{2}}$ and $g_{i}$, respectively, are given by
\begin{equation}
g_{i+\frac{1}{2}} = - G M_{i+\frac{1}{2}} /R_{i+\frac{1}{2}}^{2}, \quad g_{i} = - G M_{i} /R_{i}^{2} ,
\label{eq:h111}
\end{equation}
where $M_{i+\frac{1}{2}}$ is the rest mass enclosed in a volume of radius $R_{i+\frac{1}{2}}$, $R_{i}$ the mass-averaged zone-centered radius, and $G$ the gravitational constant. 
The radial zone-edged and zone-centered gravitational potentials, \edenx{grav}{i+\half}\ and \edenx{grav}{i}, respectively, are given by
\begin{equation}
\edenx{grav}{i-\half} = \edenx{grav}{i+\half} - g_{i} \Delta R_{i}, \quad \edenx{grav}{i} = \edenx{grav}{i+1} - g_{i+\frac{1}{2}} \Delta R_{i+\frac{1}{2}} ,
\label{eq:h112}
\end{equation}
where at the outer edge of the radial grid
\begin{equation}
\edenx{grav}{I+\half} = - \frac{G M_{I+\frac{1}{2}} } {R_{I+\frac{1}{2}} }; \quad \edenx{grav}{I} = \edenx{grav}{I+\half} - g_{I} \left( R_{I+\frac{1}{2}} - R_{I} \right) .
\label{eq:h113}
\end{equation}

Approximate GR monopole gravity is computed by first iterating the following two equations \citep[][Case A]{MaDiJa06} for $M_{\rm TOV}$:
\begin{equation}
M_{{\rm TOV},I+\frac{1}{2}} = M_{{\rm TOV},I-\frac{1}{2}} + \Gamma_{{\rm TOV},i} \Delta V_{i} \left[ \rho_{i} + \frac{ \rho_{i} \left( \edenx{int}{i} + \edenx{\nu}{i} \right) + u_{r,i} {\cal F}_{\nu,{\rm flux} }/ (c^{2} \Gamma_{{\rm TOV},i} ) }{ c^{2} } \right] ,
\label{eq:h114}
\end{equation}
\begin{equation}
\Gamma_{{\rm TOV},i} = \sqrt{ 1 + \frac{ u_{r,i}^{2} - 2 G M_{{\rm TOV},I} /R_{i} }{ c^{2} } } ,
\label{eq:h115gam}
\end{equation}
where \edenx{int}{i}\ and \edenx{\nu}{i}\ are the specific energy densities of matter and neutrinos, respectively, and $u_{r,i}$ is the radial velocity. 
With $M_{{\rm TOV},I}$ computed as $\frac{1}{2} \left( M_{{\rm TOV},I-\frac{1}{2}} + M_{{\rm TOV},I+\frac{1}{2}} \right)$, the zone-centered gravitational force is computed by
\begin{equation}
g_{i} = - G \frac{ M_{{\rm TOV},I}  + 4\pi R_{i}^{3} \left(  p_{{\rm gas},i} + p_{\nu,i} \right)/c^{2} }{ R_{i}^{2} } \, \frac{ 1 + \left( \edenx{int}{i} + p_{{\rm gas},i}/\rho_{i} \right)/ c^{2} }{ \Gamma_{{\rm TOV},i}^{2} } ,
\label{eq:h115}
\end{equation}
where $p_{{\rm gas},i}$ and $p_{\nu,i}$ are the matter pressure and spherically averaged neutrino pressure, respectively.
Once $g_{i}$ is computed, $g_{i+\frac{1}{2}}$ is computed as $\left( g_{i} + g_{i+1} \right)/2$, with $g_{0+\frac{1}{2}} = 0$ and $g_{I+\frac{1}{2}}$ extrapolated from $g_{I}$ and $g_{I-\frac{1}{2}}$.
The zone-edged gravitational potential, \edenx{grav}{I+\half}, is then computed by Equation~(\ref{eq:h112}), and the zone-centered gravitational potential is given by $g_{i} = \left( g_{i-\frac{1}{2}} + g_{i+\frac{1}{2}} \right)/2$.

\subsubsection{Multipole Expansion of the Gravitational Potential - Axisymmetry}
\label{MDgravity_Axi}

To incorporate nonspherical gravity, \chimera\ uses a scheme based on the method described by \citet{MuSt95} of expanding the integral Newtonian Poisson equation in a multipole expansion. When implementing approximate general relativistic gravity, the Newtonian monopole is replaced with the approximate general relativistic monopole \citep[][Case~A]{MaDiJa06} described above. Multipole gravity is implemented in both axisymmetric and three-dimensional simulations.

For axisymmetric simulations, this scheme utilizes the identity
\begin{equation}
\frac{1}{ ({\bf r - r'}) } = \sum_{\ell = 0}^{\infty} \frac{ r_{<}^{\ell} }{ r_{>}^{\ell + 1} } P_{\ell} (\cos ( \theta' - \theta ) ,
\label{eq:h116}
\end{equation}
where 
\begin{equation}
\frac{ r_{<}^{\ell} }{ r_{>}^{\ell + 1} } = \Theta(r - r') \frac{ r'^{\ell} }{ r^{\ell+1} } + \Theta(r' - r) \frac{ r^{\ell} }{ r'^{\ell+1} } ,
\label{eq:h117}
\end{equation}
$\Theta(x)$ is the Heaviside function, and $P_{\ell}$ is the Legendre polynomial of order $\ell$, to expand the Poisson integral in a Legendre series:
\begin{eqnarray}
 \edgrav(r,\theta) &=& - G \int_{0}^{\infty} \frac{\rho({\bf r'}) dV' }{ ({\bf r - r'}) } \nonumber \\ 
&=& - G\int_{0}^{\infty} r'^{2} dr' \int_{-1}^{1} d\cos(\theta' - \theta) \int_{0}^{2\pi} d\phi'  \rho({\bf r'}) \sum_{\ell = 0}^{\infty} \frac{ r_{<}^{\ell} }{ r_{>}^{\ell + 1} }  \nonumber \\ 
&& \times \left( P_{\ell} (\cos\theta')P_{\ell} (\cos\theta) + 2 \sum_{m=1}^{\ell} \frac{ ( \ell - m)! }{ ( \ell + m)! } P_{\ell}^{m}( \cos\theta' ) P_{\ell}^{m}( \cos\theta ) \cos m(\phi' - \phi) \right) \nonumber \\ 
&=& - 2\pi G \sum_{\ell = 0}^{\infty} \, P_{\ell} (\cos\theta) \int_{0}^{\infty} r'^{2} dr' \int_{-1}^{1} d\cos(\theta') \rho({\bf r'}) \frac{ r_{<}^{\ell} }{ r_{>}^{\ell + 1} } P_{\ell} (\cos\theta').
\label{eq:h118}
\end{eqnarray}
Here, the $P_{\ell}^{m}$ are the associated Legendre functions and the integral over $\phi'$ has been performed utilizing the assumption of axisymmetry.
In differencing Equation~(\ref{eq:h118}), the spherical coordinate system utilized by \chimera\ enables the radius $r'$ to be integrated over each spherical shell and the potential to be computed at zone interfaces. Given the singular nature of the Poisson equation, this avoids the problem of the gravitational self-interaction, which can lead to a nonconvergence of the multipole expansion, as pointed out by \citet{Couch_gf13}.
Equation~(\ref{eq:h118}) in differenced form then becomes
\begin{eqnarray}
 e_{{\rm grav},i+\frac{1}{2}, j + \frac{1}{2}} &=&  - 2 \pi G P_{\ell \, j + \frac{1}{2} } \left[  \sum_{\ell = 0}^{N_{\ell}} \sum_{i' = 1}^{i} \frac{ r_{i'+ \frac{1}{2}}^{\ell+3} - r_{i'- \frac{1}{2}}^{\ell+3} }{ (\ell + 3) r_{i+ \frac{1}{2}}^{\ell+1} } {\cal P}_{\ell,i'} \right.  \nonumber \\ 
&&+ \left. \sum_{\ell = 1, (\ell \ne 2)}^{N_{\ell}} \sum_{i' = i}^{I-1} \frac{ r_{i'  + \frac{3}{2} }^{2-\ell} - r_{i'  + \frac{1}{2} }^{2-\ell} }{2-\ell} r_{i  + \frac{1}{2} }^{\ell} {\cal P}_{\ell,i'+1}
+ \sum_{i' = i}^{I-1} r_{i + \frac{1}{2} }^{2} \ln \frac{ r_{i' + \frac{3}{2} } }{ r_{i' + \frac{1}{2} } } {\cal P}_{2,i'+1}  \right] \quad i \ge 1,
 \label{eq:h119} \\
e_{{\rm grav},\frac{1}{2}, j + \frac{1}{2}} &=&  - 2 \pi G \left[ \sum_{i' = i}^{I-1} \frac{ r_{i'  + \frac{3}{2} }^{2} - r_{i'  + \frac{1}{2} }^{2} }{2} {\cal P}_{0 \, i'+1}
+ \frac{1}{2} r_{\frac{3}{2}}^{2} {\cal P}_{0,1}  \right] \quad i = 0,
\label{eq:h120}
\end{eqnarray}
where
\begin{equation}
{\cal P}_{\ell,i} = \sum_{j=1}^{J} P_{{\rm int},j+ \frac{1}{2} } \rho_{i,j}, \quad {\rm and} \quad P_{{\rm int},j+ \frac{1}{2} } =  \int_{\cos\theta_{j- \frac{1}{2}} }^{\cos\theta_{j+ \frac{1}{2}} } P_{\ell} (\cos\theta) d(\cos\theta) .
\label{eq:h121}
\end{equation}

The Legendre polynomials up to the specified order and the angular integrations of these polynomials over the angular zone widths are generated as an initialization step.
The Legendre polynomials are first computed at the angular zone edges using
\begin{equation}
P_{0,j+\frac{1}{2}} = 1, \quad P_{1,j+\frac{1}{2}} = \cos(\theta_{j+\frac{1}{2}}) ,
\label{eq:h122}
\end{equation}
and the recurrence relation for $\ell > 1$:
\begin{equation}
P_{\ell,j+\frac{1}{2}} = \frac{ 2 \ell - 1 }{\ell} \cos(\theta_{j+\frac{1}{2}}) P_{\ell - 1,j+\frac{1}{2}} - \frac{ \ell - 1 }{\ell} P_{\ell - 2,j+\frac{1}{2}}
\label{eq:h123} .
\end{equation}
The Legendre polynomials are then integrated over the zone widths, using
\begin{equation}
P_{{\rm int},0,j} \equiv \int_{\cos(\theta_{j-\frac{1}{2}}) }^{\cos(\theta_{j+\frac{1}{2}}) } P_{0}(y) dy = P_{1,j+\frac{1}{2}} - P_{1,j-\frac{1}{2}}
\label{eq:h124} ,
\end{equation}
and then
\begin{equation}
P_{{\rm int},\ell,j} \equiv \int_{\cos(\theta_{j-\frac{1}{2}}) }^{\cos(\theta_{j+\frac{1}{2}}) } P_{\ell}(y) dy = \frac{ P_{\ell-1,j+\frac{1}{2}} - P_{\ell-1,j-\frac{1}{2}} }{ 2 \ell + 1} - \frac{ P_{\ell+1,j+\frac{1}{2}} - P_{\ell+1,j-\frac{1}{2}} }{ 2 \ell + 1} .
\label{eq:h125}
\end{equation}
To assess the accuracy of the gravitational potential expansion as a function of the number of multipoles used, and for code verification, the gravitational potentials computed by Equations~(\ref{eq:h119}) and~(\ref{eq:h120}) for a Maclaurin spheroid are compared to those given by an exact analytic solution in Section~\ref{G_Potential_Test}.

\subsubsection{Multipole Expansion of the Gravitational Potential --- Non-Axisymmetry}
\label{MDgravity_nonAxi}

For non-axisymmetric simulations, this scheme utilizes the identity
\begin{equation}
\frac{1}{ ({\bf r - r'}) } = \sum_{\ell = 0}^{\infty} \sum_{m = - \ell}^{\ell} \frac{4 \pi}{2 \ell + 1 } \frac{ r_{<}^{\ell} }{ r_{>}^{\ell + 1} } Y_{\ell}^{m} (\theta, \phi ) Y_{\ell}^{m*} (\theta', \phi'),
\label{eq:h126}
\end{equation}
to expand the gravitational potential in spherical harmonics, where $Y_{\ell}^{m*}$ denotes the complex conjugate of the spherical harmonic $Y_{\ell}^{m}$, and $r_{<}^{\ell}/r_{>}^{\ell + 1}$ is defined by Equation~(\ref{eq:h117}). The expansion is given by
\begin{eqnarray}
\edgrav(r, \theta, \phi) &=& - G \int_{0}^{\infty} \frac{\rho({\bf r'}) dV' }{ ({\bf r - r'}) } , \nonumber \\ 
&=& - G \sum_{\ell = 0}^{\infty} \frac{4\pi}{2\ell+1} \sum_{m = -\ell}^{\ell} Y_{\ell}^{m} (\theta, \phi) \int_{0}^{\infty} (r')^{2} dr' \frac{ r_{<}^{\ell} }{ r_{>}^{\ell + 1} } \int_{4\pi} d\Omega' Y_{\ell}^{m*} (\theta', \phi') \rho(r', \theta', \phi') , \nonumber \\ 
&=& - G \sum_{\ell = 0}^{\infty} \frac{4\pi}{2\ell+1} \sum_{m = -\ell}^{\ell} Y_{\ell}^{m} (\theta, \phi) \left( \frac{1}{r^{\ell+1}} C_{(r)}^{\ell \, m} + r^{\ell} D_{(r)}^{\ell \, m} \right) ,
\label{eq:h127}
\end{eqnarray}
where
\begin{equation}
C_{(r)}^{\ell \, m} = \int_{4\pi} d\Omega' \, Y_{\ell}^{m \, *} (\theta', \phi') \int_{0}^{r} dr' (r')^{\ell+2} \rho(r', \theta', \phi')
= \frac{1}{\sqrt{4\pi}} \int_{0}^{r} dr' (r')^{\ell+2} A(r',\ell,m) ,
\label{eq:h128}
\end{equation}
\begin{equation}
D_{(r)}^{\ell \, m} = \int_{4\pi} d\Omega' \, Y_{\ell}^{m \, *} (\theta', \phi') \int_{r}^{\infty} dr' (r')^{1-\ell} \rho(r', \theta', \phi')
= \frac{1}{\sqrt{4\pi}} \int_{r}^{\infty} dr' (r')^{1-\ell} A(r',\ell,m) ,
\label{eq:h129}
\end{equation}
and where
\begin{equation}
A(r',\ell,m) = \int_{0}^{2\pi} d\phi' \int_{0}^{\pi} d \theta' \sin \theta' H_{(m)} \tilde{P}_{\ell}^{m}(\cos \theta')e^{-im\phi'} \rho(r', \theta'. \phi') ,
\label{eq:h130}
\end{equation}
and
 \begin{equation}
H_{(m)} = \left\{ \begin{array}{c@{\quad} l} 1 & m = 0 \\ \frac{1}{\sqrt{2}} & m \ne 0 \end{array} \right. .
\label{eq:h131}
\end{equation}
The occasional use of $i = \sqrt{-1}$ here should not cause confusion with the radial index $i$. 
The spherical harmonics $Y_{\ell}^{m} (\theta, \phi)$ have been written as functions of the normalized associated Legendre functions $\tilde{P}_{\ell}^{m}(\cos\theta)$:
 \begin{equation}
Y_{\ell}^{m} (\theta, \phi) = \frac{1}{\sqrt{4\pi}} \tilde{P}_{\ell}^{m} \cos\theta) e^{im\phi} H_{(m)},
\label{eq:h132}
\end{equation}
and the latter are given in terms of the unnormalized associated Legendre polynomials $P_{\ell}^{m} (\cos\theta)$ by
\begin{equation}
\tilde{P}_{\ell}^{m} (\cos\theta) = P_{\ell}^{m} (\cos\theta) \times  \left\{ \begin{array}{c@\quad r} \sqrt{2 \ell+1} & m = 0 \\ \sqrt{ \frac{ 2( 2\ell+1 )( \ell - m)!}{2 \ell+1}}  & m \ne 0 \end{array} \right. , 
\label{eq:h133}
\end{equation}
where $\tilde{P}_{\ell}^{m} (\cos\theta)$ satisfies the relation
\begin{equation}
\tilde{P}_{\ell}^{-m} (\cos\theta) = (-1)^{m} \tilde{P}_{\ell}^{m} (\cos\theta) .
\label{eq:h134}
\end{equation}
The normalized associated Legendre functions, $\tilde{P}_{\ell}^{m} (\cos\theta)$, and their integrals, $P_{\rm int}(\ell,m,j)$, defined below by Equation (\ref{eq:h144})
are calculated during an initial setup step by using the subroutines developed by NGA\footnote{(alf\_sr\_v121305) \url{http://earth-info.nga.mil/GandG/wgs84/gravitymod/new\_egm/new\_egm.html}} based on the algorithms of \citet{Paul_78} and \citet{gerstl_80}.

To transform $A(r',\ell,m)$ into a form suitable for calculation, we first change variables from $\theta'$ to $y' \equiv \cos\theta'$, to obtain
\begin{equation}
A(r',\ell,m) = H_{(m)} \int_{0}^{2\pi} d\phi' \int_{-1}^{1} d y' \tilde{P}_{\ell}^{m}(y')e^{-im\phi'} \rho(r', y'. \phi') ,
\label{eq:h135}
\end{equation}
which, separating real and imaginary parts, can be written
\begin{equation}
A(r',\ell,m) = A_{r}(r',\ell,m) + i A_{i}(r',\ell,m) ,
\label{eq:h136}
\end{equation}
where
\begin{equation}
A_{r}(r',\ell,m) = H_{(m)} \int_{0}^{2\pi} d\phi' \int_{-1}^{1} d y' \tilde{P}_{\ell}^{m}(y') \cos(m\phi') \rho(r', y', \phi') ,
\label{eq:h137}
\end{equation}
\begin{equation}
A_{i}(r',\ell,m) = - H_{(m)} \int_{0}^{2\pi} d\phi' \int_{-1}^{1} d y' \tilde{P}_{\ell}^{m}(y') \sin(m\phi') \rho(r', y', \phi') .
\label{eq:h138}
\end{equation}
Next define the variables
\begin{equation}
z_{1}' \equiv \sin(m\phi'), \quad z_{2}' \equiv \cos(m\phi') ,
\label{eq:h139}
\end{equation}
and substitute them into Equations (\ref{eq:h137}) and (\ref{eq:h138}) to obtain
\begin{equation}
A_{r}(r',\ell,m) = \frac{H_{(m)}}{m} \int dz_{1}' \int_{-1}^{1} d y' \tilde{P}_{\ell}^{m}(y') \rho(r', y',z_{1}') ,
\label{eq:h140}
\end{equation}
\begin{equation}
A_{i}(r',\ell,m) = \frac{H_{(m)}}{m} \int dz_{2}' \int_{-1}^{1} d y' \tilde{P}_{\ell}^{m}(y') \rho(r', y',z_{2}') .
\label{eq:h141}
\end{equation}
Note the symmetry conditions
\begin{equation}
A_{r}(r',\ell,-m) = (-1)^{m} A_{r}(r',\ell,m) , \qquad A_{i}(r',\ell,-m) = (-1)^{m+1} A_{i}(r',\ell,m) .
\label{eq:h142}
\end{equation}

To discretize the $A_{r}(r',\ell,m)$, introduce a $j$ index, $j = 1, \cdots, J$, and a $k$ index, $k = 1, \cdots, K$, to denote the zone centers of the angular variable, $\theta$, and the azimuthal variable, $\phi$, respectively, analogous to the index $i = 1, \cdots, I$, which is illustrated in Figure~\ref{g} and used in this section for the radial index. 
Half-integer values of $i$, $j$ and $k$ refer to zone edges. 
Primed and unprimed indices will refer to source and field quantities, respectively.
We have
\begin{equation}
A_{r}(r',\ell,m) = \sum_{k'=1}^{N_{k}} \sum_{j'=1}^{N_{j}} S_{\rm int}(m,k') P_{\rm int}(\ell,m,j') \rho(r',j',k'),
\label{eq:h143}
\end{equation}
where
\begin{equation}
P_{\rm int}(\ell,m,j') \equiv \int_{ y_{j'-\frac{1}{2}} }^{ y_{j'+\frac{1}{2}} } \tilde{P}_{\ell}^{m}(y') dy' ,
\label{eq:h144}
\end{equation}
and where
\begin{equation}
S_{\rm int}(m,k') \equiv \left\{ \begin{array}{c@\quad l} \frac{1}{m\sqrt{2}} \left[ \sin(m\phi_{k'+\frac{1}{2}} - \sin(m\phi_{k'-\frac{1}{2}}) \right] & m \ne 0 \\ \phi_{k'+\frac{1}{2}} - \phi_{k'-\frac{1}{2}} & m = 0 \end{array} \right. .
\label{eq:h145}
\end{equation}
Similarly,
\begin{equation}
A_{i}(r',\ell,m) = \sum_{k'=1}^{N_{k}} \sum_{j'=1}^{N_{j}} C_{\rm int}(m,k') P_{\rm int}(\ell,m,j') \rho(r',j',k'),
\label{eq:h146}
\end{equation}
where
\begin{equation}
C_{\rm int}(m,k') \equiv \left\{ \begin{array}{c@\quad l} \frac{1}{m\sqrt{2}} \left[ \cos(m\phi_{k'+\frac{1}{2}} - \cos(m\phi_{k'-\frac{1}{2}}) \right] & m \ne 0 \\ 0 & m = 0 \end{array} \right. .
\label{eq:h147}
\end{equation}

To construct the gravitational potential, we have from Equations~(\ref{eq:h127})--(\ref{eq:h129}) and (\ref{eq:h132})
\begin{eqnarray}
\edgrav(r, \theta, \phi) &=& - G \sum_{\ell = 0}^{\infty} \frac{4\pi}{2\ell+1} \sum_{m = -\ell}^{\ell} \frac{H_{(m)}}{\sqrt{4\pi}} \tilde{P}_{\ell}^{m}(\cos\theta) e^{im\phi} \left( \frac{1}{r^{\ell+1}} C_{(r)}^{\ell \, m} + r^{\ell} D_{(r)}^{\ell \, m} \right) , \nonumber \\ 
&=& - G \sum_{\ell = 0}^{\infty} \frac{4\pi}{2\ell+1} \sum_{m = -\ell}^{\ell} H_{(m)} \tilde{P}_{\ell}^{m}(\cos\theta) \nonumber \\ 
&& \times  \left[ \frac{ e^{im\phi} }{ r^{\ell+1} } \int_{0}^{r} dr' (r')^{\ell+2} A(r',\ell,m) + e^{im\phi} e^{\ell} \int_{r}^{\infty} dr' (r')^{1-\ell} A(r',\ell,m) \right] .
\label{eq:h148}
\end{eqnarray}
Now the expression $\tilde{P}_{\ell}^{m}(\cos\theta) e^{im\phi} A(r',\ell,m)$ in Equation (\ref{eq:h148}) can be written
\begin{eqnarray}
\tilde{P}_{\ell}^{m}(\cos\theta) e^{im\phi} A(r',\ell,m) &=& \tilde{P}_{\ell}^{m}(\cos\theta) \left[\cos(m\phi) \, A(r',\ell,m) + i \sin(m\phi) \, A(r',\ell,m) \right] \nonumber \\ 
&=& \tilde{P}_{\ell}^{m}(\cos\theta) \left[ B_{r}(r',\ell,m) + i B_{i}(r',\ell,m) \right],
\label{eq:h149}
\end{eqnarray}
where the real and imaginary parts of $B(r',\ell,m)$ are given by
\begin{eqnarray}
B_{r}(r',\ell,m,\phi) &=& \cos(m\phi) A_{r}(r',\ell,m) - \sin(m\phi) A_{i}(r',\ell,m) , \\ 
\label{eq:h150}
B_{i}(r',\ell,m,\phi) &=& \sin(m\phi) A_{r}(r',\ell,m) + \cos(m\phi) A_{i}(r',\ell,m) .
\label{eq:h151}
\end{eqnarray}
Using the symmetry conditions expressed by Equations (\ref{eq:h134}) and (\ref{eq:h142}), we have
\begin{equation}
\tilde{P}_{\ell}^{-m}(\cos\theta) B_{r}(r',\ell,-m,\phi) = \tilde{P}_{\ell}^{m}(\cos\theta) B_{r}(r',\ell,m,\phi) ,
\label{eq:h152}
\end{equation}
and
\begin{equation}
\tilde{P}_{\ell}^{-m}(\cos\theta) B_{i}(r',\ell,-m,\phi) = - \tilde{P}_{\ell}^{m}(\cos\theta) B_{i}(r',\ell,m,\phi) ,
\label{eq:h153}
\end{equation}
which gives
\begin{equation}
\tilde{P}_{\ell}^{-m}(\cos\theta) B_{r}(r',\ell,-m,\phi) + \tilde{P}_{\ell}^{m}(\cos\theta) B_{r}(r',\ell,m,\phi) = 2 \tilde{P}_{\ell}^{m}(\cos\theta) B_{r}(r',\ell,m,\phi),
\label{eq:h154}
\end{equation}
and
\begin{equation}
\tilde{P}_{\ell}^{-m}(\cos\theta) B_{i}(r',\ell,-m,\phi) + \tilde{P}_{\ell}^{m}(\cos\theta) B_{i}(r',\ell,m,\phi) = 0 .
\label{eq:h155}
\end{equation}
Thus, the imaginary part of Equation (\ref{eq:h149}) cancels out when summed over negative and positive $m$, and Equation~(\ref{eq:h154}) for the real part of $B(r',\ell,-m,\phi)$ can be used to replace the summation of $m$ from $-\ell$ to $\ell$ to a summation from $0$ to $\ell$.

The construction of the gravitational potential in \Chimera\ begins with the computation of $A(r',\ell,m)$, defined in Equation~(\ref{eq:h130}) and discretized as described by Equations~(\ref{eq:h143})--(\ref{eq:h147}). 
With $P_{\rm int}$, $S_{\rm int}$, and $C_{\rm int}$ given by Equations~(\ref{eq:h144}), (\ref{eq:h145}), and (\ref{eq:h147}), respectively, and computed in the initial setup,  $A(r',\ell,m)$ is computed from
\begin{equation}
A(i',\ell,m) = \sum_{k'=1}^{K} \sum_{j'=1}^{J} \left[ S_{\rm int}(m,k') P_{\rm int}(\ell,m,j') \rho_{i',j',k'} + i C_{\rm int}(m,k') P_{\rm int}(\ell,m,j') \rho_{i',j',k'} \right] .
\label{eq:h156}
\end{equation}
Using Equations (\ref{eq:h154}) and (\ref{eq:h155}), the quantities $\phi_{\rm in}(r)$ and $\phi_{\rm out}(r)$, given respectively by $e^{im\phi} C_{(r)}^{\ell \, m}$ and $e^{im\phi} D_{(r)}^{\ell \, m}$ (Equations~(\ref{eq:h128}) and (\ref{eq:h129})), are then computed recursively by
\begin{eqnarray}
\phi_{{\rm in} \, \frac{1}{2}, k+\frac{1}{2}}(\ell,m) &=& 0 , \nonumber \\ 
\phi_{{\rm in} \, i+\frac{1}{2} , k+\frac{1}{2}}(\ell,m) &=& \phi_{{\rm in} \, i-\frac{1}{2}, k+\frac{1}{2}}(\ell,m) \left( \frac{r_{i+\frac{1}{2}} }{r_{i+\frac{3}{2}} } \right)^{\ell+1} 
+ B_{r}(r',\ell,m,\phi_{k+\frac{1}{2}}) \frac{ r_{i+\frac{1}{2}}^{2} - r_{i-\frac{1}{2}}^{2} \left( \frac{r_{i+\frac{1}{2}} }{r_{i+\frac{3}{2}} } \right)^{\ell+1} }{\ell + 3} ,
\label{eq:h157}
\end{eqnarray}
\begin{eqnarray}
\phi_{{\rm out} \, I+\frac{1}{2}, k+\frac{1}{2}}(\ell,m) &=& 0 , \nonumber \\ 
\phi_{{\rm out} \, i+\frac{1}{2}, k+\frac{1}{2}}(\ell,m) &=& \phi_{{\rm out} \, i+\frac{3}{2}, k+\frac{1}{2}}(\ell,m) \left( \frac{r_{i+\frac{3}{2}} }{r_{i+\frac{5}{2}} } \right)^{\ell} 
+ B_{r}(r',\ell,m,\phi_{k+\frac{1}{2}}) \frac{ r_{i+\frac{5}{2}}^{2} \left( \frac{r_{i+\frac{3}{2}} }{r_{i+\frac{5}{2}} } \right)^{\ell} - r_{i-\frac{3}{2}}^{2} }{2 - \ell} ,
\label{eq:h158}
\end{eqnarray}
if $\ell \ne 2$, and
\begin{eqnarray}
\phi_{{\rm out} \, i+\frac{1}{2}, k+\frac{1}{2}}(\ell,m) &=& \phi_{{\rm out} \, i+\frac{3}{2}, k+\frac{1}{2}}(\ell,m) \left( \frac{r_{i+\frac{3}{2}} }{r_{i+\frac{5}{2}} } \right)^{\ell} 
+ B_{r}(r',\ell,m,\phi_{k+\frac{1}{2}}) r_{i+\frac{3}{2}}^{\ell} \ln \left( \frac{ r_{i+\frac{5}{2}} }{ r_{i+\frac{3}{2}} } \right) ,
\label{eq:h159}
\end{eqnarray}
if $\ell = 2$, and
\begin{equation}
\phi_{{\rm out} \, \frac{1}{2}, k+\frac{1}{2}}(0.0) = \phi_{{\rm out} \, \frac{3}{2}, k+\frac{1}{2}}(0.0) + \frac{1}{2} \Re A(1,0,0) r_{i+\frac{5}{2}}^{2} 
\label{eq:h160}
\end{equation}
if $\ell = 0$. The gravitational potential is finally calculated as
\begin{equation}
e_{{\rm grav},i+\frac{1}{2}, j + \frac{1}{2}, k + \frac{1}{2}} = \sum_{\ell = 0}^{N_{\ell}} \sum_{m = 0}^{\ell} \frac{- G}{2 \ell + 1} \tilde{P}_{\ell}^{m}(\cos\theta_{j}) \left( \phi_{{\rm in} \, i-\frac{1}{2}}(\ell,m,k) + \phi_{{\rm out} \, i-\frac{1}{2}}(\ell,m,k) \right) \left\{ \begin{array}{c@\quad c} 1 & \ell = 0 \nonumber \\ \sqrt{2} & \ell \ne 0 \end{array} \right.
\label{eq:h161}
\end{equation}
To assess the accuracy of the gravitational potential expansion as a function of multipole number and to verify the code, the gravitational potentials computed by Equation~(\ref{eq:h161}) for a Maclaurin spheroid are compared to those given by an exact analytic solution in Section~\ref{G_Potential_Test}.

\section{Hydrodynamics Test Problems}
\label{sec:Hydro_Test}

We have subjected the \chimera\ hydrodynamics code modules to a number of test problems. The key problems are described here.

\subsection{Point Blast Explosion}
\label{point_blast}
 
The ability of a supernova code to simulate a spherical outgoing shock is an important first test. 
An analytic solution for a spherical outgoing shock is available for the `point-blast explosion' problem, which consists of the instantaneous deposition of an amount of energy, $E_{0}$, at a point in a zero-gravity, stationary, uniform medium of constant density, $\rho_{0}$. 
The energy $E_{0}$ is required to be very large in comparison with the initial energy of the medium. 
The analytic solution for this problem was found by \citet{Tayl50} and \citet{Sedo59}, and various parts of the solution can be found in a number of text books, such as \citet{MiMi84} or \citet{ZeRa66}, with the most complete account being given in \citet{LaLi59}. 
Because of typos in the publications cited above, we have rederived the solution and have found that Equation~(99.8) of \citet{LaLi59} should be multiplied by the square of the normalized radius (see below) and that the expression for $\nu_{5}$ in Equation~(99.10) of \citet{LaLi59} should be replaced by 
\begin{equation}
\nu_{5} = \frac{2}{\gamma -1} .
\label{eq:ht1}
\end{equation}

To set up this problem, a uniform, $\gamma = 5/3$ gas maintained in hydrostatic equilibrium within a spherical volume by an external pressure boundary condition and having a constant density $\rho_{0}  = 0.1$ \gcc\ was divided into 200 zones of equal, 1-cm width. 
The gas was  given a constant ambient temperature of $10^{-8}$ MeV. A point explosion at the center of the spherical mass was simulated by instantaneously depositing an energy, $E_{0} \simeq 6.06 \times 10^{17}$ ergs, by increasing the temperature of the first zone to 1~MeV. 
Rather than using a simple gamma-law EoS, we used the \chimera\ non-NSE EoS, consisting of half neutrons and half protons. 
The electron and photon contributions were set to zero.
In this way, the EoS used was equivalent to a gamma-law EoS with $\gamma = 5/3$, but the \Chimera\ EoS machinery (e.g., composition remapping, EoS interpolation) described in Section~\ref{app:Eos} was tested as well. 

\begin{figure}
\gridline{
\fig{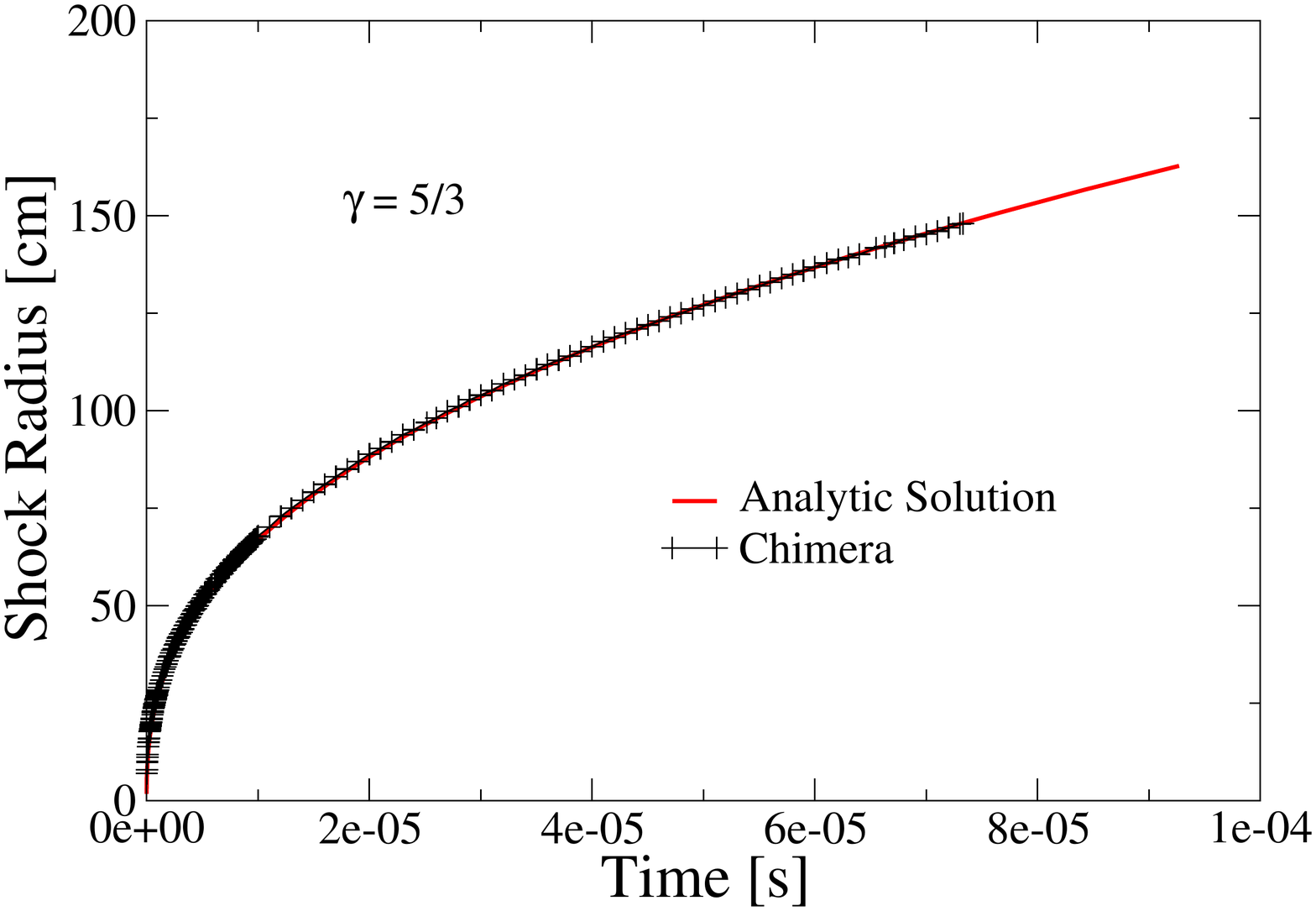}{0.5\textwidth}{(a)}
\fig{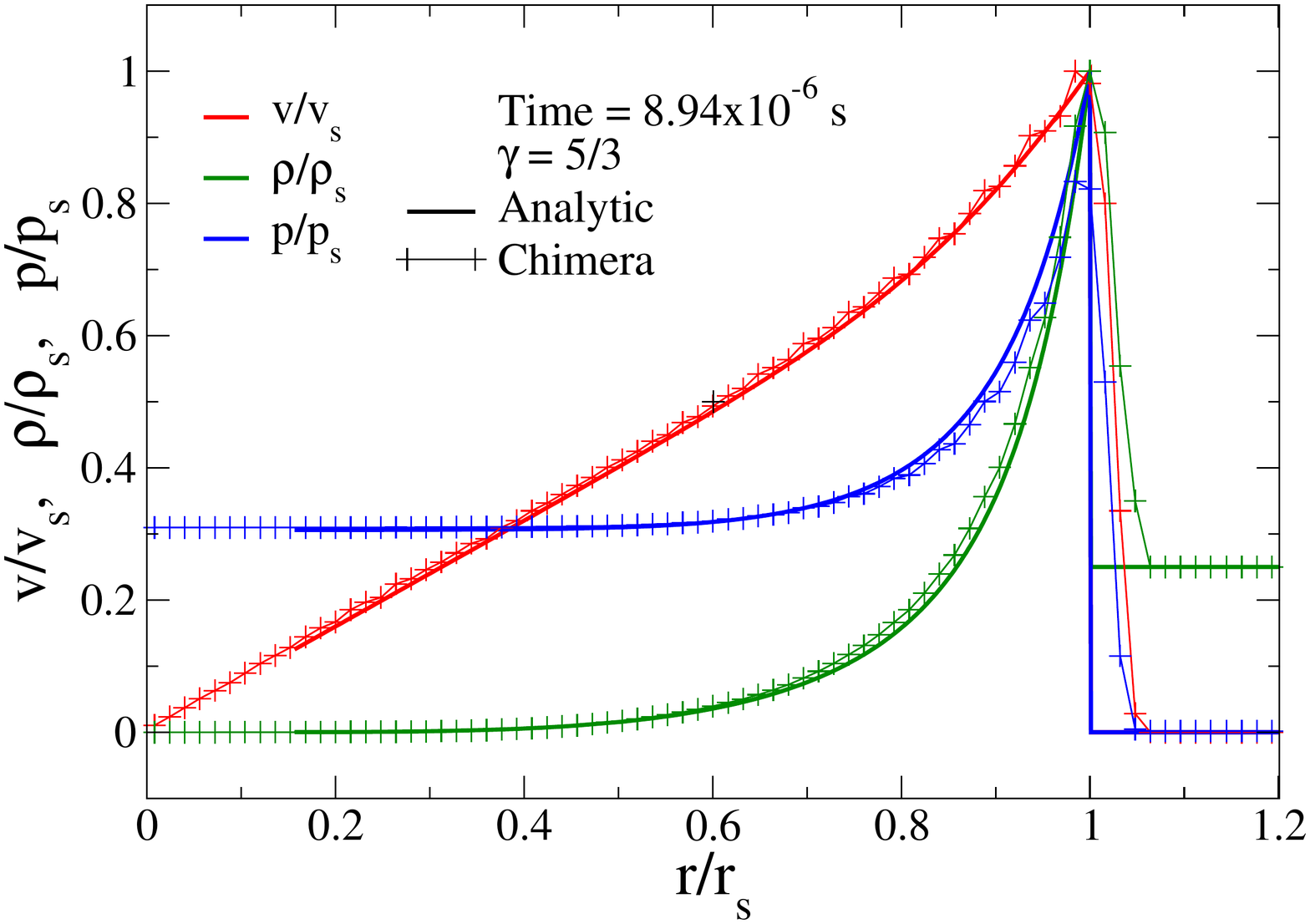}{0.5\textwidth}{(b)}
}
\caption{\label{fig:point_blast}
Panel (a): Shock radii versus time for the point-blast explosion problem for \chimera\ (black plus signs) and the analytic expression (red line) given by Equations~(\ref{eq:ht2}) and (\ref{eq:ht3}).
Panel (b): Velocity (red), density (green), and pressure (blue) behind the shock as a function of radius, normalized by their immediate post-shock values for \chimera\ (pluses) relative to the analytic solution given by Equations~(99.8) and (99.10) of \citet{LaLi59} with the corrections noted in the text.
}
\end{figure}

Denoting by $r_{s}$ the distance of the shock from the origin, the time-dependence of $r_{s}$ is given by
\begin{equation}
r_{s} = \xi_{s} \left( \frac{ E_{0}t^{2} }{ \rho_{0} } \right)^{1/5} ,
\label{eq:ht2}
\end{equation}
where $\xi_{s}$ is determined by
\begin{equation}
\xi_{s}^{5} \frac{32 \pi}{ 25 \left( \gamma^{2} - 1 \right) } \int_{0}^{1} \left( \xi^{4} \frac{\rho}{\rho_{s}} \left( \frac{u}{u_{s}} \right)^{2} + \xi^{9} \frac{p}{p_{s}} \right) d\xi = 1,
\label{eq:ht3}
\end{equation}
where $u$, $\rho$, and $p$ are the fluid velocity, density, and pressure, respectively, between the origin and the shock. The subscript `$s$' denotes their immediate post-shock values.
Numerically integrating Equation~(\ref{eq:ht3}), using Equations~(99.10) of \citet{LaLi59}, gives $\xi_{s} = 1.17$.

A comparison of $r_{s}$ as a function of time, computed by \chimera\ versus the analytic result given by Equation~(\ref{eq:ht2}) is shown in Figure~\ref{fig:point_blast}(a) and demonstrates agreement to  within a few percent.
Figure~\ref{fig:point_blast}(b), shows that the velocity, density, and pressure of the fluid behind the shock  normalized by their immediate post-shock values also agree well with the analytic solutions.

\subsection{Sod Shock Tube Problem}
\label{Shock_Tube}

A standard hydrodynamics problem admitting an analytic solution is the Sod shock tube problem \citep{Sod78}. 
As Sod's formulation was done in the context of a plane-parallel geometry, we approximate such a geometry with \Chimera's spherical grid by working with a variable $x = R - r$, where $R = 10^{5}$ cm and $-2 \le r \le 2$.
We cover the $-2 \le r \le 2$ range of $r$ with a grid consisting of 200 zones, initially equally spaced. 
Following Sod's original formulation \citep{Sod78}, we set up a Riemann problem with pressure, $p = 1$ ergs cm$^{-3}$, density, $\rho = 1$~\gcc, velocity, $u = 0$~\cmps\ , for $x < 0$, and $p = 0.1$~ergs cm$^{-3}$, $\rho = 0.125$~\gcc, and $u = 0$~\cmps\ , for $x > 0$. 
The same EoS was used as described above for the point blast problem, resulting in a constant $\gamma$ of $5/3$. 
The results of the test at time $t = 0.5$~s are shown in Figure~\ref{fig:Shock_tube}.

\begin{figure}
\gridline{
\fig{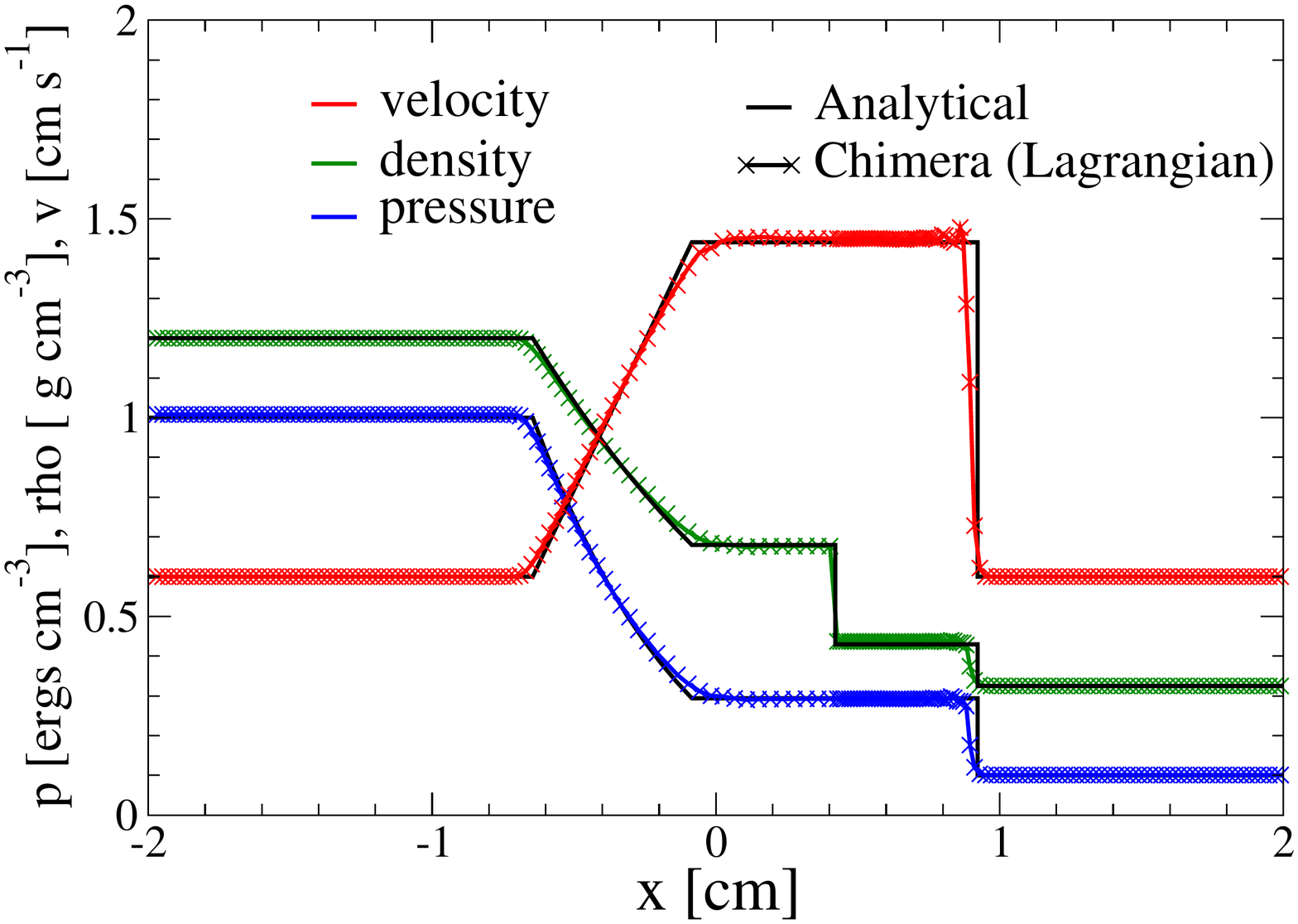}{0.5\textwidth}{(a)}
\fig{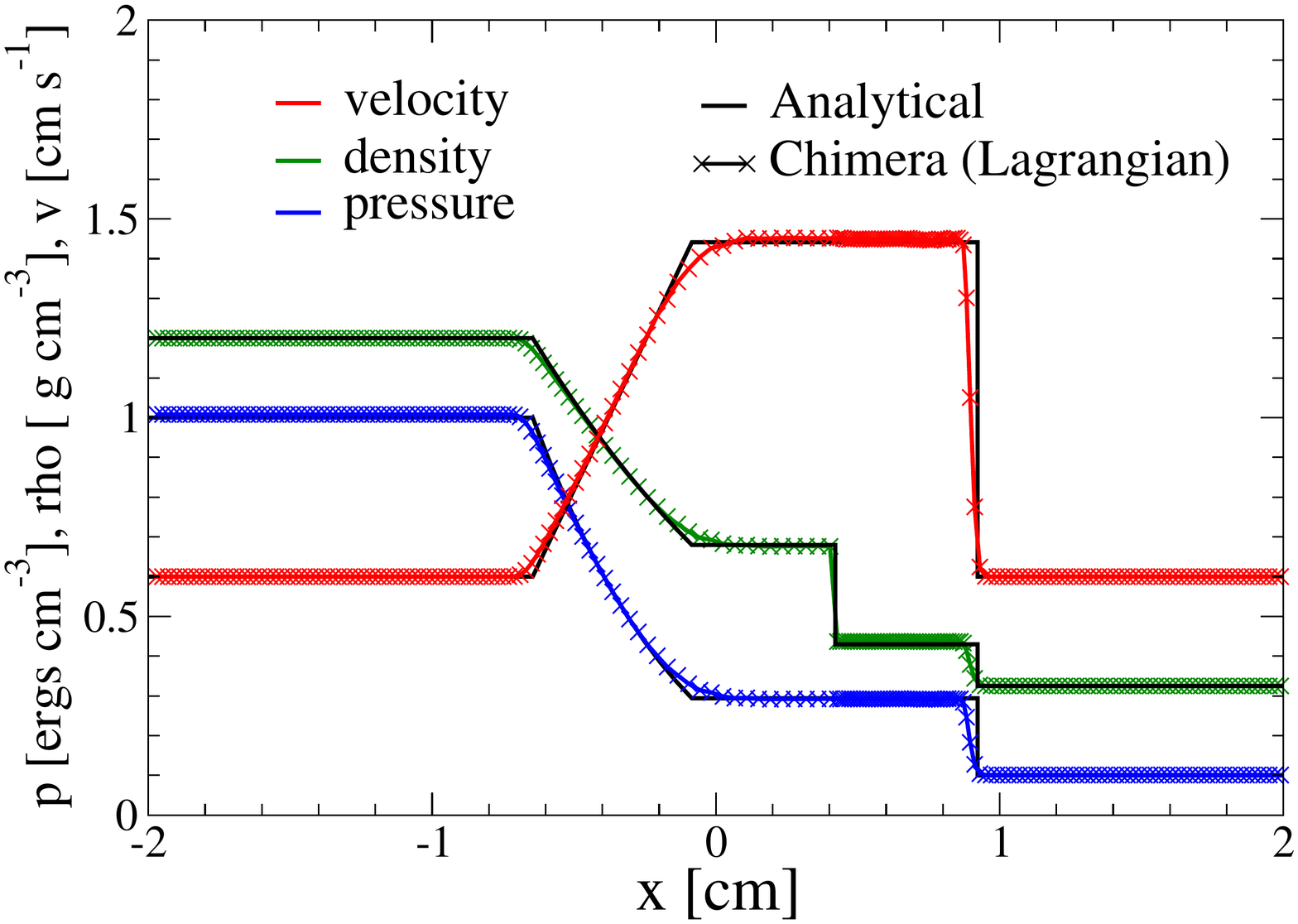}{0.5\textwidth}{(b)}
}
\gridline{
\fig{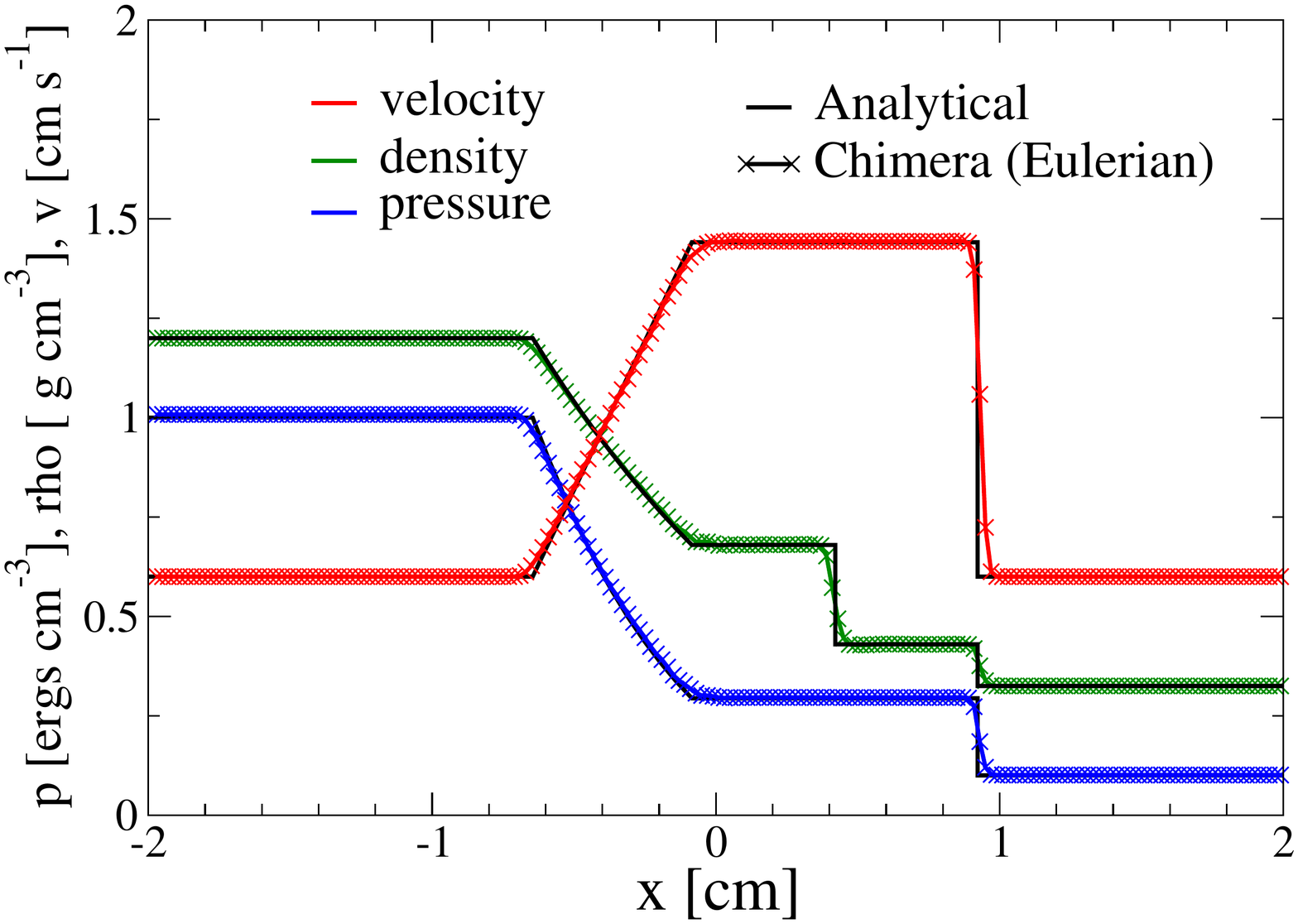}{0.5\textwidth}{(c)}
}
\caption{\label{fig:Shock_tube}
Comparison of the analytic solution for $\gamma = 5/3$ (solid lines) and  \Chimera\ (crosses) solutions to the Sod shock tube problem for velocity (red), density (green), and pressure (blue), as a function of distance from the initial discontinuity at 0.5~s.
\Chimera\ simulations were performed (a) in Lagrangian mode with the values of the flatten parameters suggested by \citet{CoWo84}, (b) in Lagrangian mode with a different set of flatten parameters that suppress the low-amplitude post-shock oscillations, and (c) in Eulerian mode with the values of the flatten parameters suggested by \citet{CoWo84}.
}
\end{figure}

Numerical hydrodynamics schemes employing Riemann solvers, such as the scheme employed in \Chimera, can introduce low-amplitude, post-shock oscillations in flows involving shocks unless extra dissipation is added. 
To suppress these, \Chimera\ introduces a small amount of dissipation by reducing locally the order of the interpolation scheme in the neighborhood of sufficiently strong shocks \citep[see][Section 4 \& Appendix]{CoWo84}. 
In particular, the left- and right-hand states are modified by Equations~(4.1) of \citet{CoWo84}, namely
\begin{eqnarray}
a_{L,i}^{\rm flat} &=& a_{i}^{n} f_{i} + a_{L,i} ( 1 - f_{i} ) , \nonumber \\
a_{R,i}^{\rm flat} &=& a_{i}^{n} f_{i} + a_{R,i} ( 1 - f_{i} ) ,\label{eq:ht4}
\end{eqnarray}
where the `flattening parameter' $(0 \le f_{i} \le 1)$ determines the mixture of first-order and the higher-order PPM interpolations in constructing the left and right states.
For the test results shown in Figure~\ref{fig:Shock_tube}(a), \Chimera\ was run in Lagrangian mode with the \citet{CoWo84} suggested value of 0.75 for $\omega^{(1)}$ in the equation following their Equation~(A.2), and a maximum value of the flattening parameter, $f_{i} = 0.5$. 
As can be seen, the agreement between the analytical and the numerical results is very good, but very-low-amplitude, post-shock oscillations are evident, particularly in the velocity.
In Figure~\ref{fig:Shock_tube}(b), we show the results of the same test except that the parameter $\omega^{(1)}$  was set to 0.6, and the maximum value of the flattening parameter was set to 1. 
This introduces more dissipation, and there is now no evidence of any post-shock oscillations in the numerical solutions.
Finally, Figure~\ref{fig:Shock_tube}(c) shows the results of the test with \Chimera\ now run in Eulerian mode with the \citet{CoWo84} suggested value of $\omega^{(1)}$ and the maximum value of $f_{i}$ set to 0.5, as in the test shown in Figure~\ref{fig:Shock_tube}(a).
The agreement between the numerical and analytical results is again very good, and there is no sign of post-shock oscillations.

\subsection{Shu-Osher Shock Tube Problem}
\label{Shu_Shock_Tube}

A test suggested by \citet{ShOs89} involves structure, testing the resolution of the numerical hydrodynamics scheme: A moving shock interacts with sine waves in density. Initially, $\rho = 3.85713$~\gcc, $v = 2.639369$~\cmps, and $P = 10.33333$ erg cm$^{-3}$, for $x < -0.8$, and $\rho = 1 + 0.2 \sin 5x$~\gcc, $v = 0$~\cmps, and $P = 1$ erg cm$^{-3}$, for $x > -0.8$.
The results are plotted in Figure~\ref{fig:Shu_Shk_Tube_d}.
The black line shows the solution obtained with a grid of 3200 evenly spaced zones, which is taken as the reference solution.
The dashed green line and the red Xs show the solution obtained with a grid of 800 and 200 zones, respectively. Clearly, the solution has converged with a grid of 800 zones.
With 200 zones, the solution still shows the detailed structure, albeit with somewhat reduced amplitude.

\begin{figure}
\fig{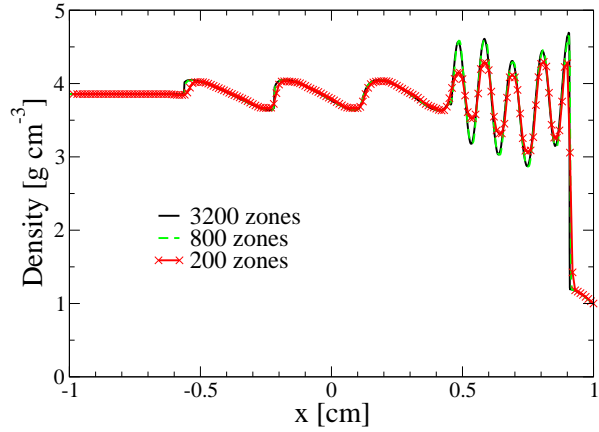}{0.5\textwidth}{}
\caption{\label{fig:Shu_Shk_Tube_d}
Evolved density for the shock tube test suggested by \citet{ShOs89}, for a grid of 3200 (black solid), 800 (dashed green), and 200 (red with X) zones.
}
\end{figure}

\subsection{Radial Advection Test}
\label{R_Wind}

A test of the radial advection and, in particular, the geometry corrections for a spherical geometry, given the choice of such a geometry in \Chimera, consists of solving the Euler equations with an adiabatic $\Gamma$-law EoS for an initial self-similar radial outflow problem \citep{Mign14} described by
\begin{equation}
\rho(\xi,0) = \rho_{0}(\xi) , \quad u(\xi,0) = \alpha_{0} \xi , \quad p(\xi,0) = \frac{1}{\Gamma},
\label{eq:ht5}
\end{equation}
where $\rho(\xi,0)$ is an arbitrary function and $\alpha_{0}$ is a constant. An exact analytic solution of this problem is given, for spherical symmetry, by
\begin{equation}
\rho(\xi,t) = \left( \frac{\alpha(t)}{\alpha_{0}} \right)^{3} , \quad u(\xi,t) = \alpha_{t} \xi , \quad p(\xi,0) = \frac{1}{\Gamma} \left( \frac{\alpha(t)}{\alpha_{0}} \right)^{3\Gamma}, \quad \mbox{with} \quad \alpha(t) = \alpha_{0}/( 1 + \alpha_{0}t ).
\label{eq:ht6}
\end{equation}
A test that focuses on the remapping procedure is that of a self-similar outflow at constant density with the imposition of constant pressure, as described by \citet{BlLu93} and \citet{Mign14}.
 Following their example, we set $\alpha_{0} = 100$ and utilize a grid of 100 evenly spaced zones. 
The solution is characterized by constant velocity/radius and constant density.
The solution of this problem for the first 10 zones, where the geometry-dependent corrections in the remap are most important, is shown in Figure \ref{fig:Radial_Wind_1}. The numerical solution exhibits a constant velocity/radius that matches the analytic solution.

\begin{figure}
\fig{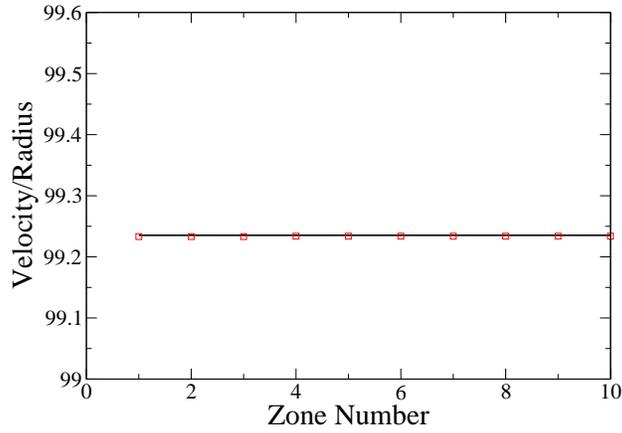}{0.5\textwidth}{}
\caption{\label{fig:Radial_Wind_1}
Numerical solution for velocity/radius for the constant density radial outflow problem. The solid line is the analytic solution with velocity/radius = 99.24 and the numerical solution is shown by the square symbols for the first 10 zones, where the geometry-dependent interpolations are most important.
}
\end{figure}

\subsection{Angular Advection Test}
\label{Advection}

\begin{figure}
\gridline{
\fig{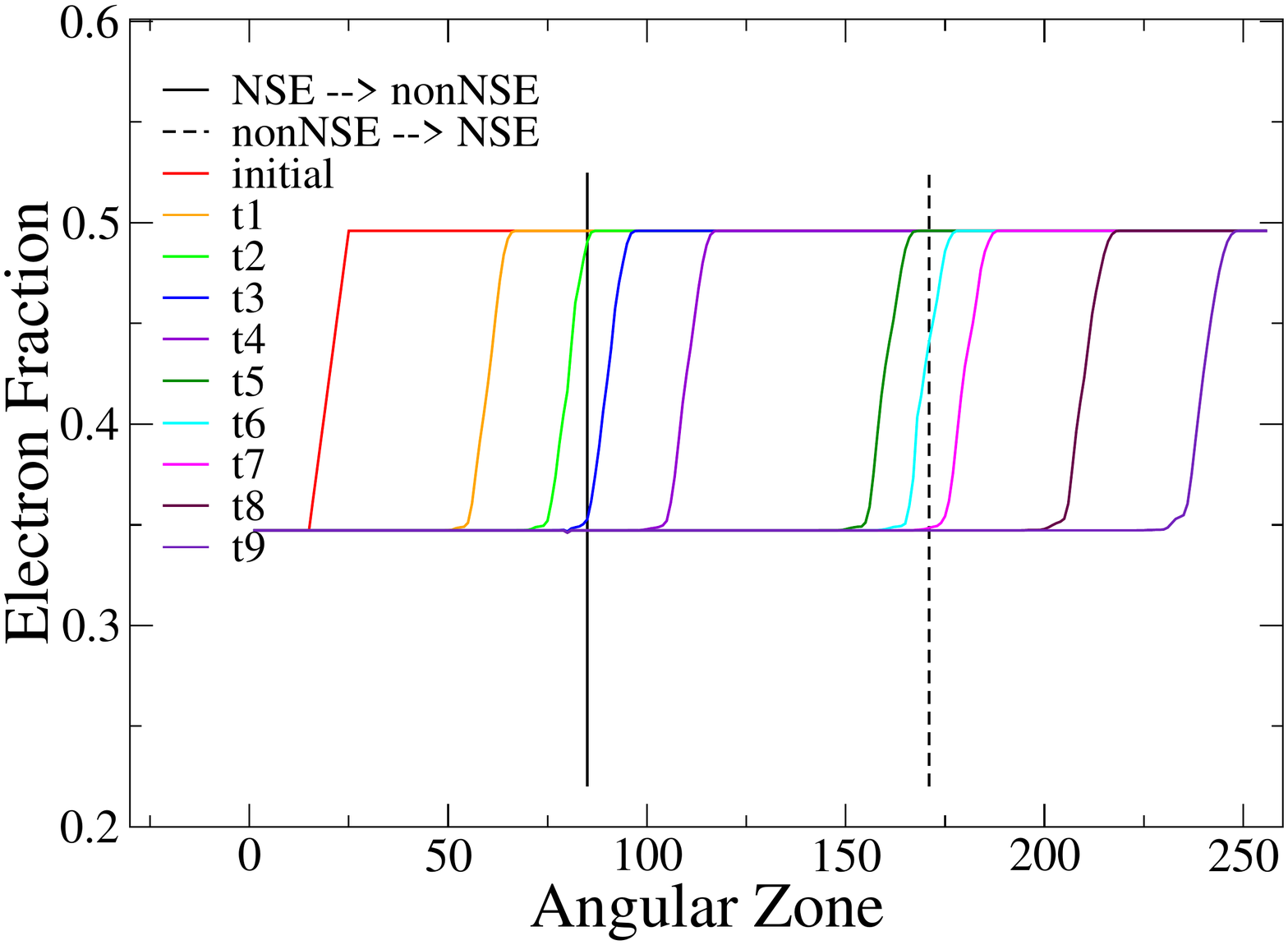}{0.5\textwidth}{(a)}
\fig{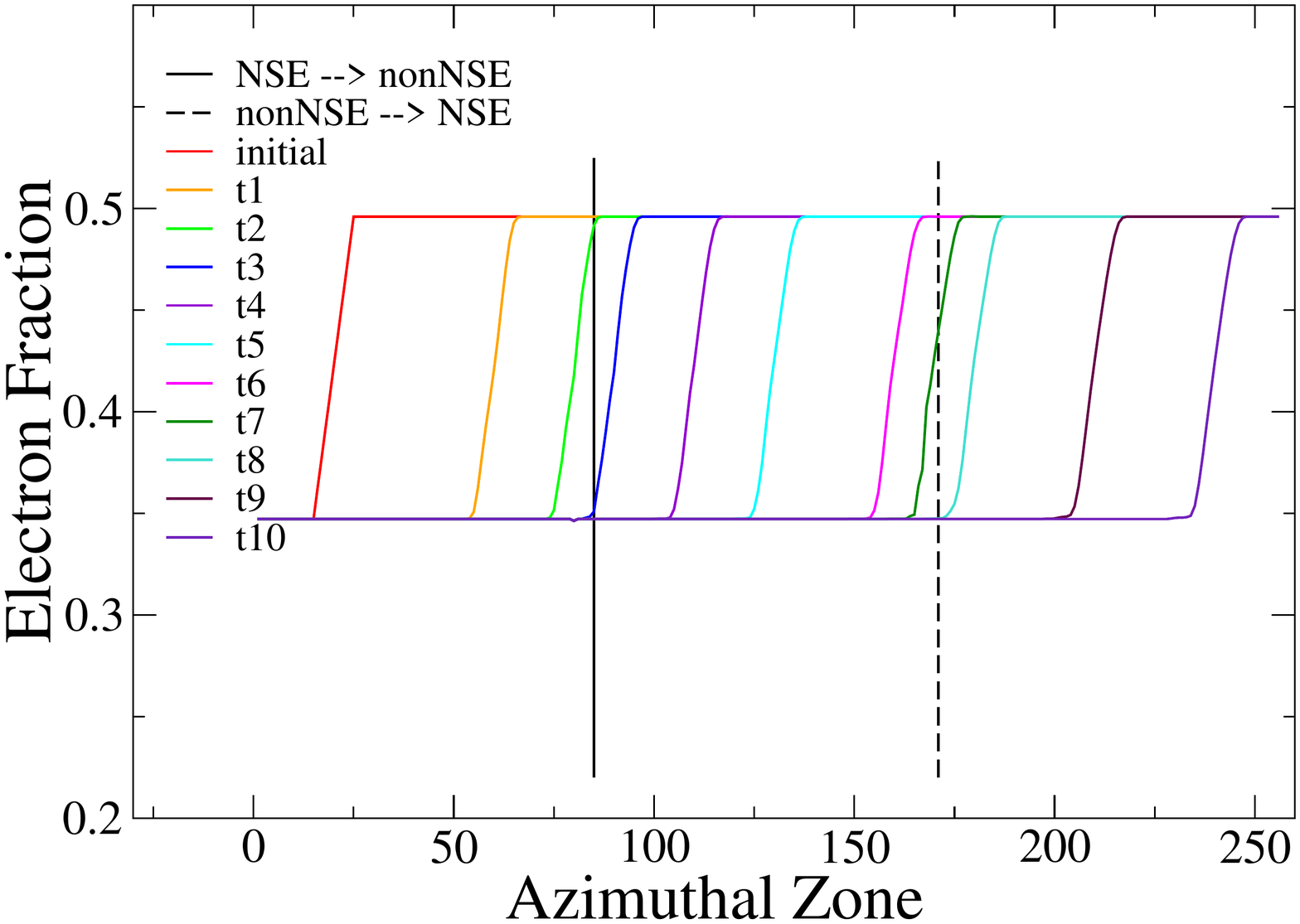}{0.5\textwidth}{(b)}
}
\caption{\label{fig:Ye_adv}
Advection of an electron fraction pattern across an NSE to non-NSE transition (solid vertical line) followed by a non-NSE to NSE transition (dashed vertical line) at several times (colored curves) in the (a) $\theta$-direction and (b) $\phi$-direction.
}
\end{figure}

As a test of the angular advection algorithms and the logic involved in deflashing a zone (transitioning it from NSE to non-NSE) and flashing a zone (transitioning it from non-NSE to NSE), an electron fraction \Ye\ pattern was advected in angle ($\theta$) across most of 256 angular zones, beginning at zone 18 ($\theta = 0.22$) and continuing across the grid. 
The \Ye\ pattern consisted of a linear rise over 10 zones, from a value of 0.3472 to a value of 0.4960. A density of \den{2.47}{7}\ and a temperature of 0.45~MeV were chosen, which are representative of conditions where material flows in or out of NSE.
Between zones 80 and 81 ($\theta = 0.98$) a transition from NSE to non-NSE was imposed, and between zones 168 and 169 ($\theta = 2.1$) a transition from non-NSE to NSE was imposed. 
As can be seen in Figure~\ref{fig:Ye_adv}(a), the pattern in \Ye\ is nicely preserved as it is advected across the angular grid.
A similar test with similar results performed for the azimuthal advection algorithms is shown in Figure~\ref{fig:Ye_adv}(b).

\subsection{Energy Conservation Test}
\label{E_Conserve}

\begin{figure}
\gridline{
\fig{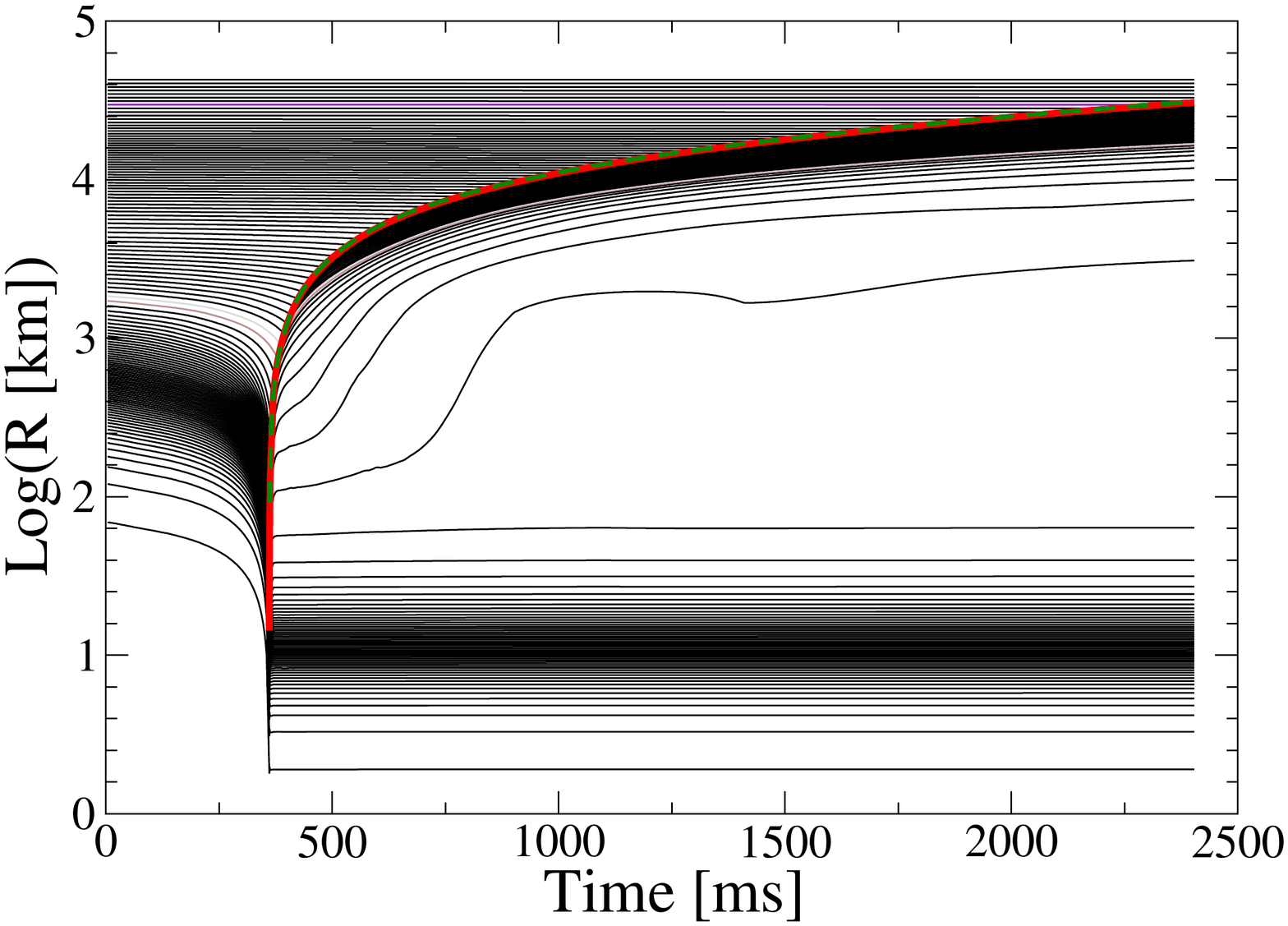}{0.5\textwidth}{(a)}
\fig{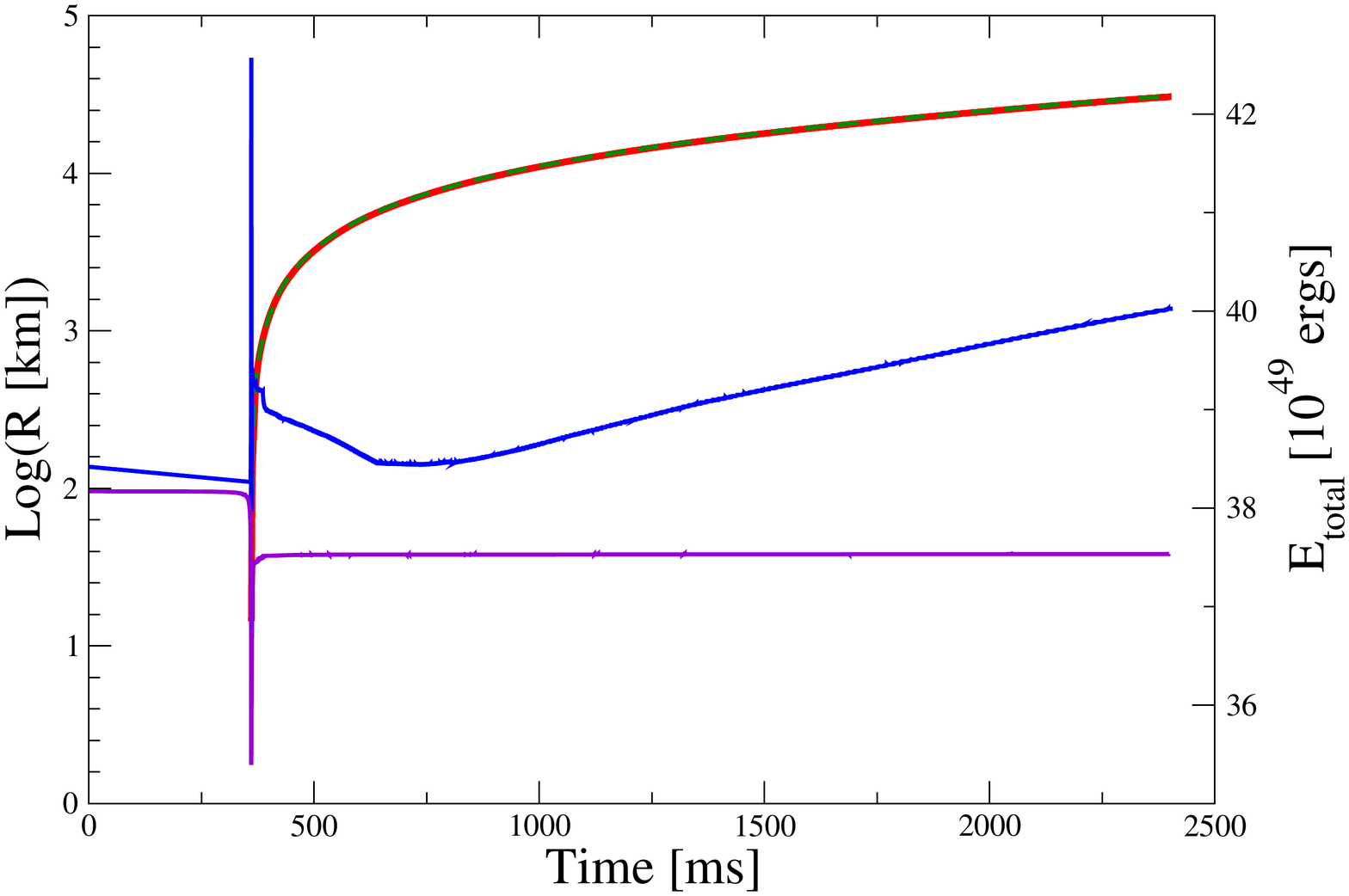}{0.5\textwidth}{(b)}
}
\caption{\label{fig:N_Hydro12}
Panel (a): Newtonian hydrodynamics core collapse simulations, NH$_{\rm par}$ and NH$_{\rm tot}$, as described in the text. 
Black lines show the Lagrangian trajectories of zones enclosing increments of 0.025 \msun\ for the NH$_{\rm par}$ simulation. 
Shock trajectories for simulations NH$_{\rm par}$ and NH$_{\rm tot}$ are shown by the (solid red) and (dashed green) line, respectively. 
Panel (b): Shock trajectories for the two simulations are plotted as in Panel (a), and the total energy with arbitrary offset (but the same offset for the two simulations) (right scale) for simulations NH$_{\rm par}$ (blue) and NH$_{\rm tot}$ (violet).
}
\end{figure}
}

\begin{figure}
\gridline{
\fig{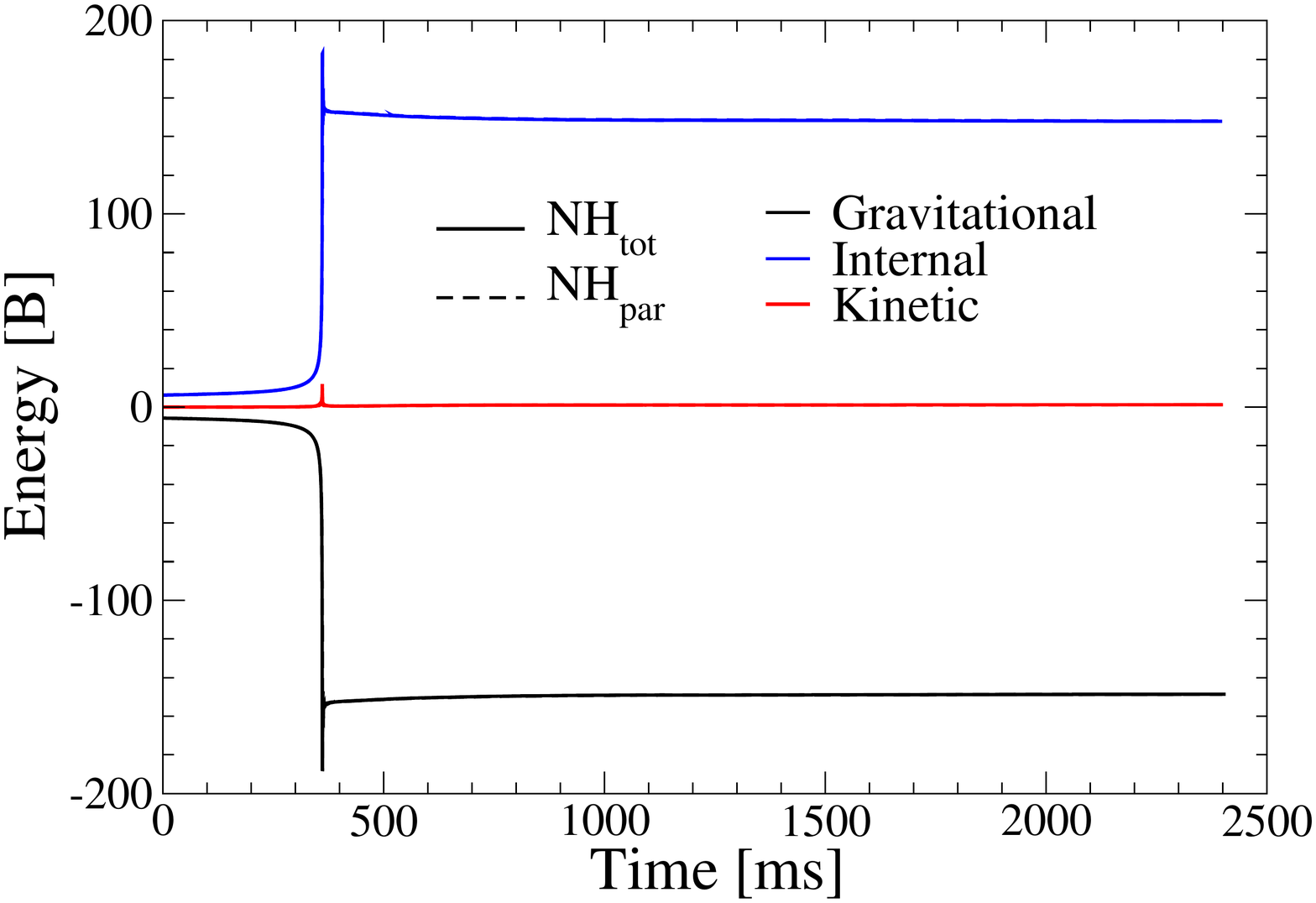}{0.5\textwidth}{(a)}
\fig{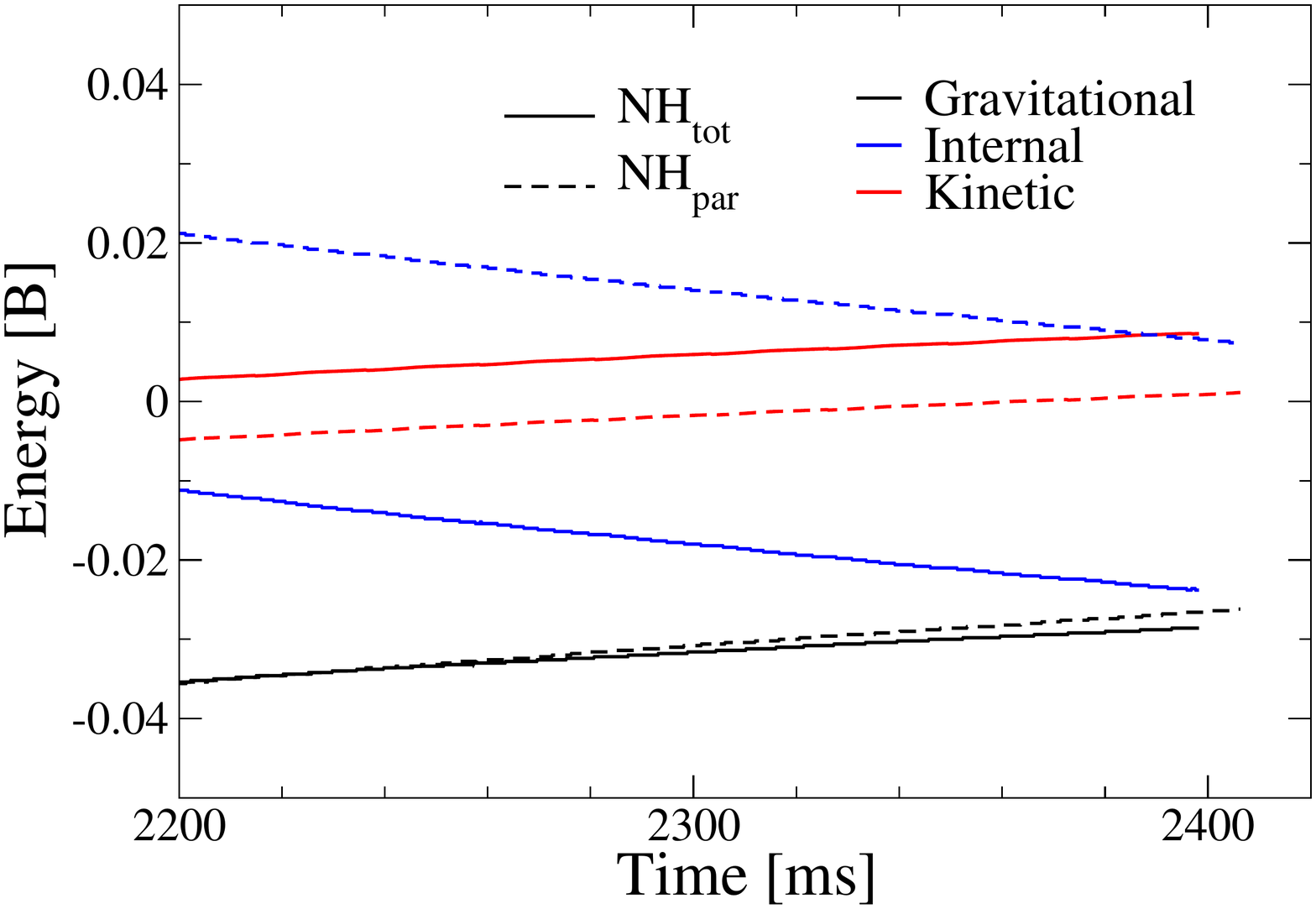}{0.5\textwidth}{(b)}
}
\caption{\label{fig:N_Hydro34}
Panel (a): Gravitational, internal and kinetic energy components of the total energy for NH$_{\rm tot}$ (solid lines) and NH$_{\rm par}$ (dashed lines). The difference between the two simulations is too small to discern at the scale of a Bethe. Panel (b): Gravitational, internal, and kinetic energy components of the final 200 ms for NH$_{\rm tot}$ (solid lines) and NH$_{\rm par}$ (dashed lines), with an arbitrary offset (but the same for corresponding pairs of energy components), shown at a scale for which differences are apparent.
}
\end{figure}

When run in normal mode, the \chimera\ hydrodynamics does not impose total energy conservation, where the total energy is defined by the integral on the left-hand side of Equation (\ref{eq:a66}); therefore, its ability to conserve total energy is a rigorous test of the hydrodynamics algorithms. 
As a test of \chimera's ability to conserve energy with the realistic EoS described in Section~\ref{StellarEoS} and its numerical implementation, we performed two Newtonian hydrodynamics simulations initiated from a 15-\msun \citet{WoHe07} progenitor and carried out for 2 seconds post-bounce, at which point the bounce shock had traversed 38,000~km of our 43,000~km extent of the radial grid.

In the first simulation, referred to as NH$_{\rm par}$, \chimera\ was run in its normal mode. 
Away from shocks, the specific internal energy was updated during the Lagrangian step by the first law, Equation~(\ref{eq:h79}), and remapped as described by Equation~(\ref{eq:h106}).
In the vicinity of a shock, the specific total energy, defined immediately above Equation (\ref{eq:h80}), was evolved during the Lagrangian step by Equation~(\ref{eq:h80}), Equation~(\ref{eq:h108}), and Equation~(\ref{eq:h109}), and remapped by an equation analogous to Equation~(\ref{eq:h106}), with the specific internal energy then extracted by Equation~(\ref{eq:h110}).
In the second simulation, referred to as NH$_{\rm tot}$, the specific total energy was updated during the Lagrangian step and remapped for all zones, whether they were in the vicinity of a shock or not.
Run in this mode \chimera\ automatically conserves the total energy apart from the source term given by Equation~(\ref{eq:h108}, which is added to the total energy after the Lagrangian step. 
Total energy conservation during the Lagrangian step can be ascertained by multiplying Equation~(\ref{eq:h80}) by the mass $V_{i}^{n+\half} \rho_{i}^{n+\half}$ and noting that the terms involving the pressure cancel in pairs when summed over the zones, leaving only the surface terms, and noting also that the term $f_{\nu \, i}^{n}$ is zero for these hydrodynamics-only runs. 
During the remap step, the terms on the right-hand side of  Equation~(\ref{eq:h106}) cancel in pairs after multiplying by the zone mass $\delta {\cal M}_{i}^{n+1}$.
Careful bookkeeping kept track of all nonphysical changes in the specific energy having no dynamical effect, such as occurs most importantly when material is advected between adjacent spatial zones characterized by different EoSs with slightly different energy zeros, as described in Section \ref{remap3}.

The results of this test are shown in Figure~\ref{fig:N_Hydro12}. 
The Lagrangian trajectories, at 0.025~\msun\ intervals, for the NH$_{\rm tot}$ simulation shown in Figure~\ref{fig:N_Hydro12}(a) were almost identical to those of the NH$_{\rm par}$ simulation and thus are not shown.
The shock trajectories (red and dashed green lines) for the two simulations, \added{shown in both panels of Figure~\ref{fig:N_Hydro12}} are essentially on top of each other.
The  total energy (blue and violet lines) are shown in Figure~\ref{fig:N_Hydro12}(b).
The same arbitrary constant has been added to the energy of each simulation to bring the energies within the range of the energy ordinate.
They are gratifyingly flat, with both showing a small blip at bounce, and the NH$_{\rm par}$ simulation (blue line) showing a small rise of about $10^{49}$ ergs from 700~ms to the end of the simulation.
In that simulation there is also a small decline in total energy from bounce to about 700~ms, which we traced to the remapping of the velocity, thus conserving momentum by design, rather than remapping the square of the velocity, which would conserve kinetic energy by design.
This slight decline in kinetic energy does not appear in the evolved total energy of the NH$_{\rm tot}$ simulation (violet line). In this case, the specific total energy is the remapped quantity and therefore automatically conserved, modulo the small source term on the right-hand side of the specific total energy equation, Equation~(\ref{eq:h108}), as noted above.
Figures~\ref{fig:N_Hydro34}(a)~and~\ref{fig:N_Hydro12}(b) plot the evolution of the gravitational, internal, and kinetic energy components of the total energy for the simulations NH$_{\rm par}$ (dashed lines) and NH$_{\rm tot}$ (solid lines).
Figure~\ref{fig:N_Hydro34}(a) plots the evolution of the energy components of the two simulations, but at a scale for which differences are not readily discernible. 
Figure~\ref{fig:N_Hydro34}(b) plots the last 200 ms of the two simulations at a much finer scale.
The same arbitrary constant has been added to each pair of energy components of the two simulations so they fit within the energy range of the ordinate. 
It is seen that the energy differences are small, with most of the energy differences arising from the $\sim 3 \times 10^{49}$ ergs difference in the internal energy and the $\sim 1 \times 10^{49}$ ergs difference in the kinetic energy.

Finally, an important quantity relating to the explosion energy obtained in \ccsne\ simulations and frequently used in comparing simulation outcomes between groups is the ``diagnostic energy" $E_{\rm diag}$, which is the sum of the gravitational, thermal, and kinetic energy in each zone in turn summed over all zones for which the sum in that zone is positive. 
At 2500 ms from the initiation of core collapse, the diagnostic energy for both of the above models had essentially become constant and was 0.477 B for the NH$_{\rm tot}$ simulation and 0.475 B for the NH$_{\rm par}$ simulation.

\begin{figure}
\gridline{
\fig{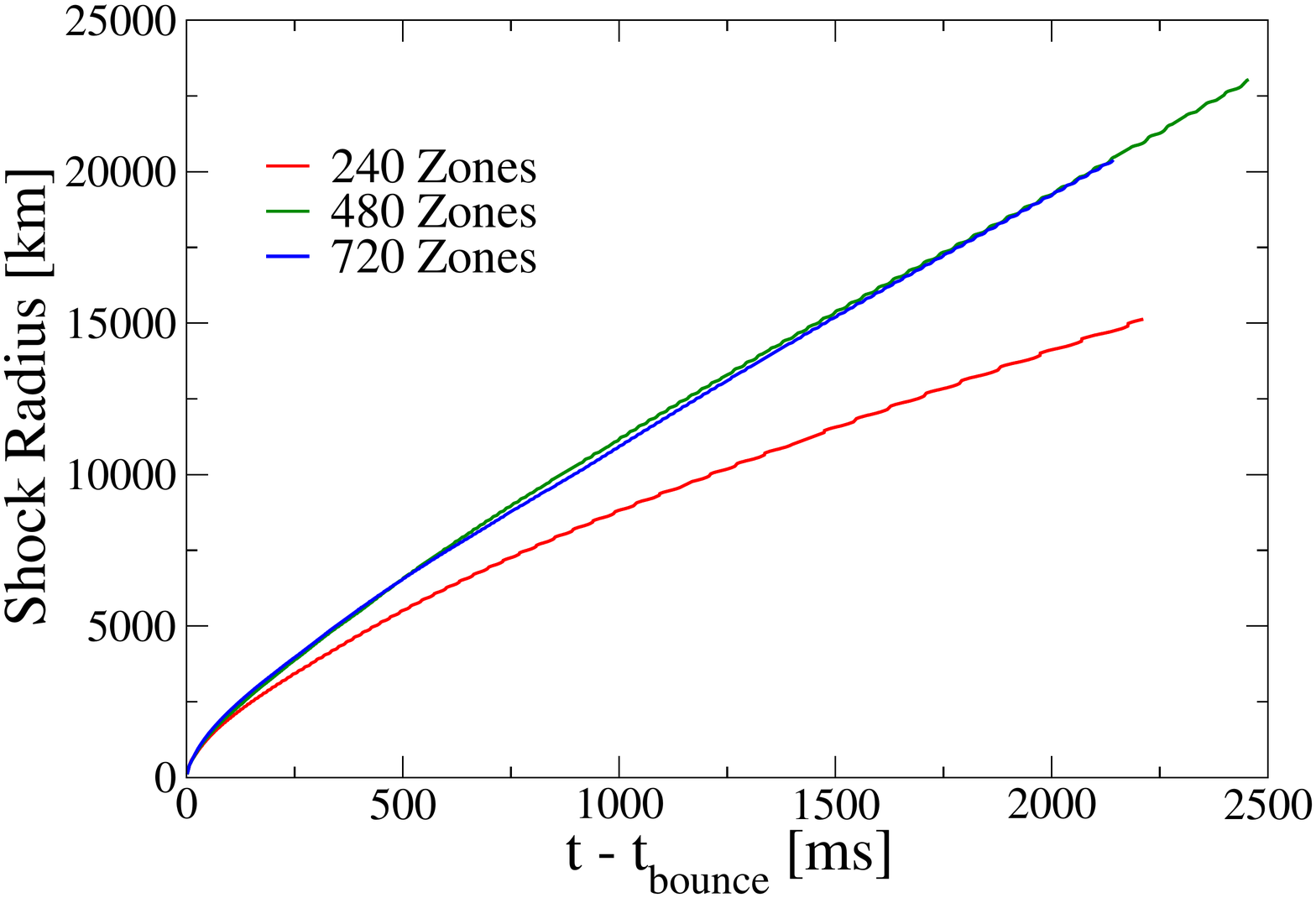}{0.5\textwidth}{(a)}
\fig{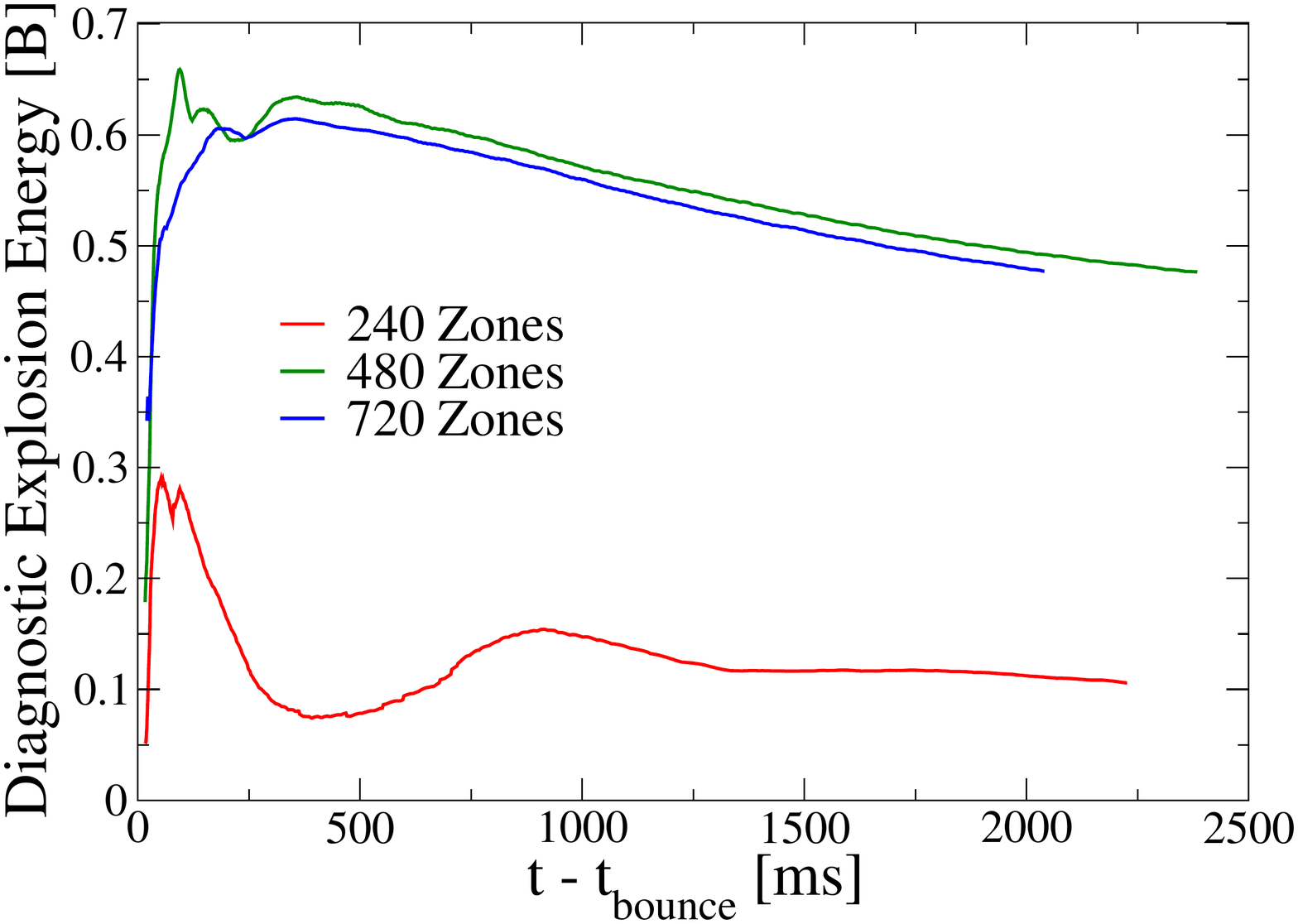}{0.5\textwidth}{(b)}
}
\caption{\label{fig:N_Hydro56}
Panel (a): Shock trajectory as a function of post-bounce time for the simulation of model NH$_{\rm par}$ with 240, 480, and 720 radial zones. Panel (b): Diagnostic energy (sum of the total kinetic, gravitational, and internal minus rest-mass energies) as a function of post-bounce time.
}
\end{figure}

Both simulations, NH$_{\rm par}$ and NH$_{\rm tot}$, were performed on an adaptive radial grid of 720 zones, as described in Section \ref{sec:Regrid}.
To ascertain whether this grid resolution results in a radially converged numerical solution, we have performed the NH$_{\rm par}$ simulation with 240 and 480 zones. The results are shown in Figure~\ref{fig:N_Hydro56}. It is clear that the numerical solutions have essentially converged at a radial grid resolution of 480 zones but not with a radial grid resolution of 240 zones.

\subsection{Gravitational Potential Tests}
\label{G_Potential_Test}

To verify the accuracy of the gravitational potential expansions given in Sections~\ref{MDgravity_Axi} and \ref{MDgravity_nonAxi}, and to ascertain an appropriate maximum number of multipoles to use in simulations, we have compared with an analytic solution the results of the axisymmetric, Equations~(\ref{eq:h119}) and~(\ref{eq:h120}), and non-axisymmetric, Equation~(\ref{eq:h161}), expansions for the gravitational potential of a Maclaurin spheroid.
We consider the spheroid described by \citet{CoGrFl13}, viz., a spheroid of uniform density of $\rho = 1$ \gcc\ embedded in a background of vanishing density. 
The spheroid has a semi-major axis of 1~m, an eccentricity of 0.9, and is located in a spherical volume of radius 2~m.
So that Equations~(\ref{eq:h119}), (\ref{eq:h120}), and~(\ref{eq:h161}) can be used to compute the gravitational potential, we set up a spherical computational grid with the spheroid located at the center. 
A $720 \times 240$ grid is used for the axisymmetric multipole expansion, reflecting the grid resolution in some of our recent 2D simulations, and a grid of $540 \times 180 \times 180$ is used for the non-axisymmetric multipole expansion, reflecting the angular resolution used in some of our 3D simulations.
Where the boundary of the spheroid cuts through a given zone, the density of that zone is adjusted in proportion to the percentage of the zone interior to the boundary.
The analytic solution that we use for a point interior to the spheroid is given by Equation~(21) of \citet{CoGrFl13}, and for a point exterior to the spheroid we use the solution given by Equation~(1a) of \citet{HoCrCr18}. 

\begin{figure}
\gridline{
\fig{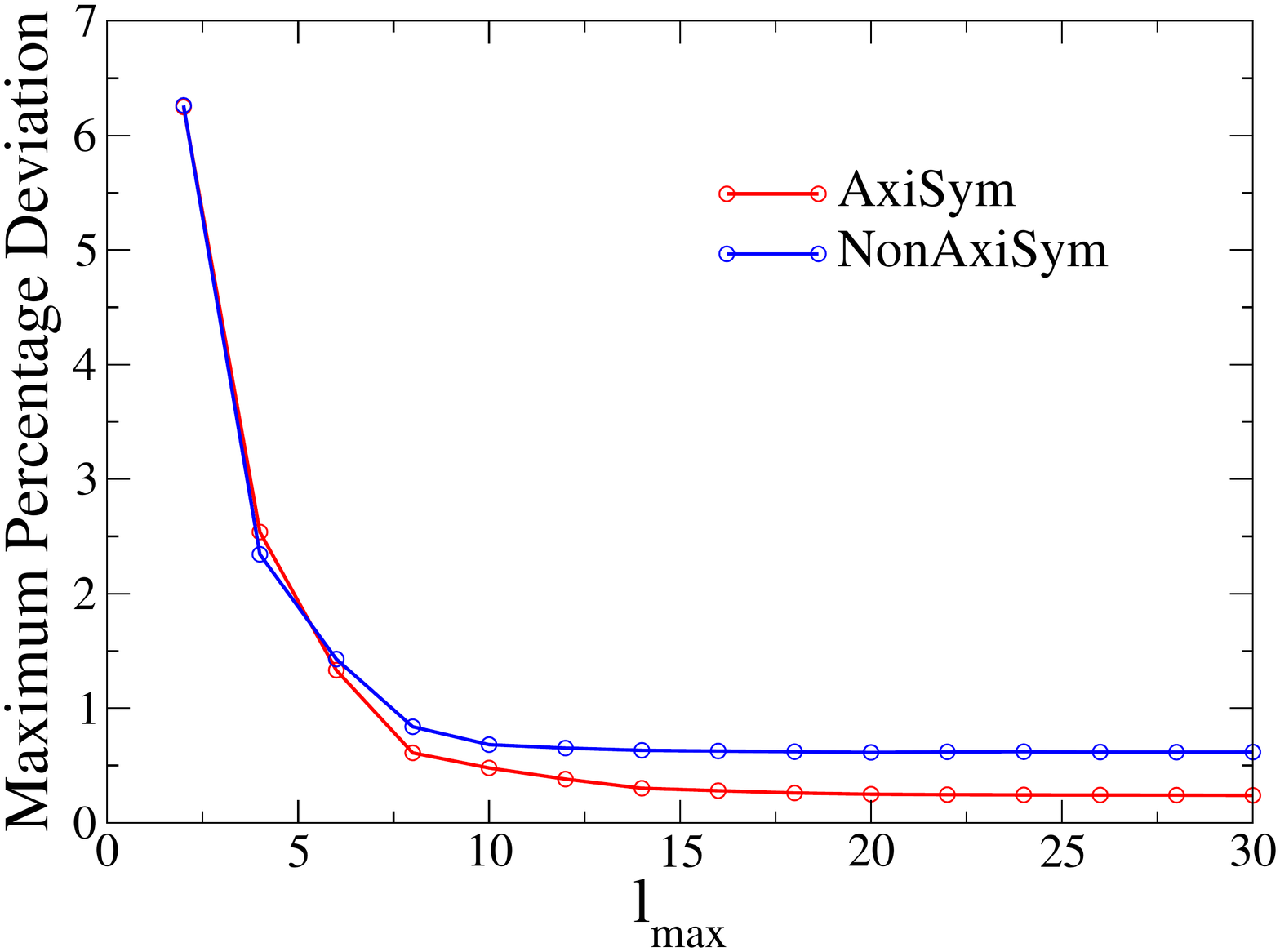}{0.5\textwidth}{(a)}
\fig{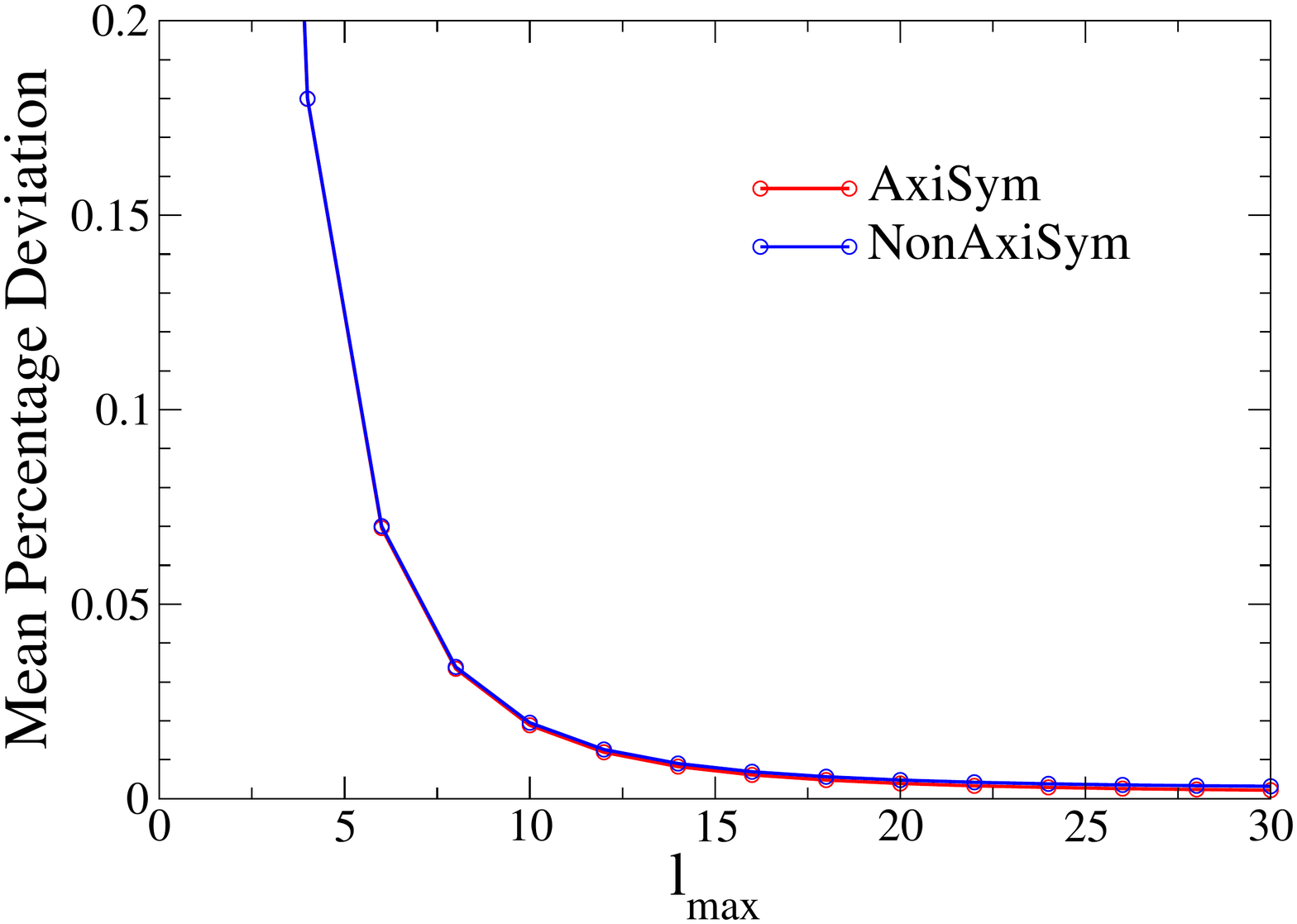}{0.5\textwidth}{(b)}
}
\caption{\label{fig:g_potential_error}
Panel (a): The maximum percentage deviation of the gravitational potentials computed by the axisymmetric multipole expansion (red line) and the non-axisymmetric multipole expansion (blue line) from the analytic solution, as a function of $\ell_{\rm max}$, the maximum multipole used.
Panel (b): Similar to Panel (a) but for the zone-weighted mean deviation of the multipole gravitational potentials from the analytic solution, as a function of $\ell_{\rm max}$.
}
\end{figure}

The results of the comparisons are shown in Figure~\ref{fig:g_potential_error}. 
The zone-weighted mean of the deviation from the analytic solutions of our multipole expansions decreases with $\ell_{\rm max}$, the maximum multipole used, and is below 0.02 percent for $\ell_{\rm max} \ge 10$. 
The maximum deviation also decreases with $\ell_{\rm max}$ and is below 1\% for $\ell_{\rm max} \ge 10$.
The choice of $\ell_{\rm max}$ in a simulation is obviously a compromise between accuracy and computational time.
\Chimera\ simulations typically use a value of $\ell_{\rm max} = 10$.

\section{Neutrino Transport}
\label{app:transport}

Neutrino transport is a key process that must be modeled accurately in the simulation of \ccsne\. The neutrino-driven explosion mechanism depends sensitively on the coupling to matter of a small fraction of the enormous neutrino luminosity that ensues upon the collapse of the stellar core. 
Additionally, accurate neutrino transport modeling is important for computing the neutrino emission expected from a given nearby \ccsne\, to best enable us to work backward from the sequence of neutrino detections accompanying such an event to establishing details of the explosion mechanism in the deep interior of the stellar core. 
In this section we describe our algorithms for modeling neutrino transport. 
Sections \ref{BoltzEq}--\ref{dyedt} provide (i) the derivation of the neutrino Boltzmann equation, which establishes the metric and independent variables we use and forms the basis of our transport scheme, (ii) the angular moment equations obtained from the Boltzmann equation, (iii) our method of flux limiting, (iv) of operating splitting the resultant transport equations into a transport piece and an energy advection piece, and (v) the derivation of the terms required for coupling neutrinos to the matter hydrodynamics. 
Finally, in Section \ref{sec:LTS} by means of Equations (\ref{eq:a58}), (\ref{eq:a59}), (\ref{eq:a62a}) followed by Equations (\ref{eq:a62}), (\ref{eq:a62h}), and (\ref{eq:a62i}), we present the full differencing scheme used to advance the neutrino transport through a Lagrangian step. 
A number of velocity-dependent terms are dropped in closing the angular moment equations by replacing the first angular moment equation by a diffusion like equation with a flux-limiter.
This resulted in transport solutions near shocks that were less than satisfactory due to the neutrino distribution moments being defined relative to the fluid frame and therefore subject to the effects of the large velocity discontinuities there. 
Section \ref{trans_shock} describes a modification of our transport scheme that more accurately models these discontinuities in the neutrino distribution moments in the presence of shocks. 
This modification replaces Equations~(\ref{eq:a58}) and (\ref{eq:a59}) for the Lagrangian step by Equations (\ref{eq:a99}) and (\ref{eq:a93}), respectively.

Neutrino energy advection during transport is operator split from spatial advection and is described in Section \ref{neu_e_adv}. Our scheme for updating the neutrino zeroth angular moment during a Lagrangian step is detailed by Equations (\ref{eq:a78})--(\ref{eq:a81}). 
Because our energy grid is tied to a lapse function (Equation~(\ref{eq:a8a}), Section \ref{app:e_advct_lapse} describes the use of the neutrino energy advection machinery developed in Section \ref{neu_e_adv} to update the neutrino distribution due to changes in the lapse function resulting from changes in the configuration of the core during a hydrodynamics step. 
As our neutrino transport scheme is based on a Lagrangian-Remap formulation of numerical hydrodynamics, a transport remap step must follow the Lagrangian update.
Section \ref{spatial_adv} describes our scheme for spatially remapping the neutrino distributions that is associated with the remapping of the grid following a Lagrangian hydrodynamics step. 
The scheme is summarized by Equations (\ref{eq:a84}) and (\ref{eq:a85}). 
Section \ref{eddington} specifies the scalar Eddington factors used to represent higher angular moments of the neutrino distributions in terms of lower moments, the former of which appear in the equations for the lowest moments.

The derivation of the neutrino transport equations and the energy--momentum transfer between neutrinos and the background matter have been carried out, with an eye toward future developments of \chimera, in full spherically symmetric general relativity. Several approximations have been made, however, in the current versions of \Chimera. Some velocity terms have been dropped in deriving the diffusion-like equation and the flux-limiter relating the first moment of the neutrino distribution function to the zeroth moment [Equations (\ref{eq:a14}) and (\ref{eq:a15})]. 
Additionally, the proper length parameter $\Gamma$, Equation~(\ref{eq:a38}), has been set equal to unity. 
It was found that retaining $\Gamma$ as computed by Equation~(\ref{eq:a38}) does not significantly affect the neutrino transport during a \ccsn\ simulation \citep{BrDeMe01}.
What we do retain is the redshift parameter $a$ given by Equations~(\ref{eq:a46}) and (\ref{eq:a48}), as this parameter does significantly affect the luminosity and mean energy of neutrinos emerging from the neutrinosphere \citep{BrDeMe01}.

\chimera\ employs multi- neutrino energy transport, and to delineate the neutrino energy grid structure we use indices $k+\frac{1}{2}, \; k = 0, 1, 2, \cdots, N_{\epsilon}$ -- i.e., half-integer values -- to denote energy-grid zone edges. 
Energy-grid zone centers are denoted by indices $k, \; k = 1, 2, \cdots, N_{\epsilon}$ -- i.e., with integer values.
With $\epsilon_{k+\frac{1}{2}}$ denoting the value of the neutrino energy at zone edges, the value of the neutrino energy at zone centers is defined by
\begin{equation}
\epsilon_{k} = \frac{1}{3} \left( \epsilon_{k+\frac{1}{2}}^{2} + \epsilon_{k+\frac{1}{2}} \epsilon_{k-\frac{1}{2}} + \epsilon_{k-\frac{1}{2}}^{2} \right),
\label{eq:a01}
\end{equation}
so that
\begin{equation}
\epsilon_{k}^{2} \Delta \epsilon_{k} =  \frac{1}{3} \left( \epsilon_{k+\frac{1}{2}}^{3} - \epsilon_{k-\frac{1}{2}}^{3} \right) ,
\label{eq:a01b}
\end{equation}
where $\Delta \epsilon_{k} = \epsilon_{k+\frac{1}{2}} - \epsilon_{k-\frac{1}{2}}$.
With this definition of  $\epsilon_{k}$, $4\pi \epsilon_{k+\frac{1}{2}}^{2} \Delta \epsilon_{k}$ is the energy-space volume between zone edges $\epsilon_{k-\frac{1}{2}}$ and $\epsilon_{k+\frac{1}{2}}$.

\subsection{Boltzmann Equation}
\label{BoltzEq}

The time evolution of the invariant occupation function $f = f(x,p)$ [number of neutrinos per state at the phase point $(x,p)$] is given by the coordinate invariant Boltzmann equation \citep[e.g.,][]{Lind66}
\begin{equation}
\frac{df}{d\ell} = \frac{dx^{\alpha}}{d\ell} \pderiv{f}{x^{\alpha}} +  \frac{dp^{\alpha}}{d\ell} \pderiv{f}{p^{\alpha}} = \left( \frac{df}{d\ell} \right)_{S} ,
\label{eq:a1}
\end{equation}
where $\ell$ is the affine path-length, which we choose to be defined such that
\begin{equation}
p^{\alpha} = \frac{dx^{\alpha}}{d\ell} ,
\label{eq:a2}
\end{equation}
and the right-hand side of Equation (\ref{eq:a1}) denotes the change in $f$ due to sources -- i.e., emission, absorption, and scattering. 
In the spirit of the ray-by-ray approximation in which transport along each radial ray is assumed to be spherically symmetric, we will consider the spherically symmetric form of the transport equation. 
In addition, we will express the Boltzmann equation in the comoving or fluid frame with the intention of executing our neutrino transport along with the Lagrangian radial  hydrodynamics step, to be followed by a remap of both the matter and the neutrino quantities to the original grid or to the grid displaced according to a regridding algorithm. We will hereafter denote all quantities defined with respect to the fluid frame by a subscript `0.'

The derivation, leading to Equation (\ref{eq:a8}), of the Boltzmann transport equation for spherically symmetric spacetimes, beginning with Equation~(\ref{eq:a1}), can be found with varying detail in a number of references \citep[e.g.,][]{Lind66, Cast72, MiMi84, MeMa89, BaMyCo89}, and we include enough detail  so that important quantities used subsequent to Equation~(\ref{eq:a8}) are clearly defined. 
We assume that neutrinos follow null geodesics between localized interactions, so that their paths between interactions are given by
\begin{equation}
0 = \frac{d \pcm^{\alpha}}{d \ell} 
 + \left\{ \begin{array}{c}
     \alpha    \\
 \!\! \beta \;\; \gamma \!\!
\end{array} \right\} \pcm^{\beta} \frac{d x^{\gamma}}{d \ell}  \quad \Rightarrow \quad \frac{d \pcm^{\alpha}}{d \ell} = - \left\{ \begin{array}{c}
     \alpha    \\
 \!\! \beta \;\; \gamma \!\!
\end{array} \right\} \pcm^{\beta} \pcm^{\gamma} ,
\label{eq:a2a}
\end{equation}
where $\left\{ \begin{array}{c}     \alpha    \\ \!\! \beta \;\; \gamma \!\! \end{array} \right\}$ is a Christoffel symbol of the second kind (connection coefficients in the coordinate basis), usually denoted by $\Gamma_{\beta \gamma}^{\alpha}$, but we reserve the latter symbol to denote the Ricci rotation coefficients (connection coefficients in the orthonormal basis). 

To evaluate the source functions we choose a local, comoving, orthonormal set of basis vectors (${\bf e_{t}}, {\bf e_{m}}, {\bf e_{\theta}}, {\bf e_{\phi}}$) parallel to a spherical polar coordinate basis to resolve the components of the neutrino four-momentum. 
In terms of this orthonormal set of basis vectors, the components of four-vectors, such as the four-momentum, $\pcm^{\hat {\rm a}}$, are denoted by characters with hatted latin indices. 
In terms of the original components, $\pcm^{\alpha}$, the components with respect to the orthonormal basis are given by
\begin{equation}
\pcm^{\hat {\rm a}} = \epsilon_{\alpha}^{\hat {\rm a}} \pcm^{\alpha} \mbox{      with inverse      } \pcm^{\alpha} = \epsilon_{\hat {\rm a}}^{\alpha} \pcm^{\hat {\rm a}} = \left( \epsilon_{\alpha}^{\hat {\rm a}} \right)^{-1} \pcm^{\hat {\rm a}} .
\label{eq:a3d}
\end{equation}
Substituting the second part of Equation~(\ref{eq:a3d}) into Equation~(\ref{eq:a2a}) and rearranging the indices, ${d p^{\hat {\rm a}}}/{d\ell}$ is given by
\begin{equation}
\frac{d \pcm^{\hat {\rm a}}}{d\ell} 
= - \epsilon^{\hat {\rm a}}_{\alpha} \epsilon_{\hat {\rm c}}^{\gamma} 
\left[ \left( \frac{\partial (\epsilon^{\alpha}_{\hat {\rm b}})}
{\partial x^{\gamma}} \right)_{x^{\gamma \ne \alpha}}
 + \left\{ \begin{array}{c}
     \alpha    \\
 \!\! \beta \;\; \gamma \!\!
\end{array} \right\} 
\epsilon^{\beta}_{\hat {\rm b}} \right] 
\pcm^{\hat {\rm b}} \pcm^{\hat {\rm c}} = - \Gamma_{\hat {\rm b} \hat {\rm c}}^{\hat {\rm a}} 
\pcm^{\hat {\rm b}} \pcm^{\hat {\rm c}} ,
\label{eq:a3e}
\end{equation}
where the Ricci rotation coefficients, $\Gamma_{\hat {\rm a} \hat {\rm c}}^{\hat {\rm b}}$, are defined by Equation~(\ref{eq:a3e}).
Using Equations~(\ref{eq:a2}), (\ref{eq:a3d}), and~(\ref{eq:a3e}), Equation~(\ref{eq:a1}), written in terms of the $p^{\hat {\rm a}}$, becomes
\begin{equation}
\pcm^{\hat {\rm b}} \left[ \epsilon^{\alpha}_{\hat {\rm b}} \left( \frac{\partial \fcm} 
{\partial x^{\alpha}} \right)_{x^{\beta \ne \alpha},\pcm^{\hat {\rm c}}} 
  - \Gamma_{\hat {\rm b} \hat {\rm c}}^{\hat {\rm a}} 
 \pcm^{\hat {\rm c}} \left( \frac{\partial \fcm}{\partial \pcm^{\hat {\rm a}}} 
\right)_{x^{\gamma},\pcm^{{\hat {\rm d}}\ne \hat {\rm a}}} 
 \right] = \left( \frac{\delta \fcm}{\delta \ell} \right)_{S} . 
\label{eq:a3f}
\end{equation}

The components of $\pcm^{\hat{\rm a}}$ with respect to the above orthonormal basis are given by
\begin{equation}
(\pcm^{\hat{\rm 0}},\pcm^{\hat{\rm 1}},\pcm^{\hat{\rm 2}},\pcm^{\hat{\rm 3}}) \equiv (\pcm^{\hat{\rm 0}},\bar{p}_{_{\! 0}}) = \frac{1}{c} \left[ \epscm, \epscm \mucm, \epscm \sqrt{ 1 - \mucm^{2} } \cos\phi_{p}, \epscm \sqrt{ 1 - \mucm^{2} } \sin\phi_{p} \right] ,
\label{eq:a4}
\end{equation}
where $\epscm$ and $\mucm$ are the neutrino energy and the direction cosine of the neutrino three-momentum with respect to the radial direction, respectively, both measured in the comoving frame. 
Because of the mass shell condition
\begin{equation}
\pcm^{\hat 0} =  \sqrt{ \left( \pcm^{\hat 1} \right)^{2} + \left( \pcm^{\hat 2} \right)^{2} + \left( \pcm^{\hat 3} \right)^{2} } ,
\end{equation}
only three of the four $\pcm^{\hat a}$'s are independent. 
We choose these independent components to be $\pcm^{\hat {\imath}}, \hat {\imath} = 1, 2, 3$.
In terms of these independent four-momentum components, the derivatives, $\partial/\partial \pcm^{\hat \imath}$ are given by
\begin{eqnarray}
\left( \frac{\partial \fcm}{\partial \pcm^{\hat 1}} \right)_{\pcm^{2'}, \pcm^{3'}}& =& c\mucm \left( \frac{\partial \fcm}{\partial \epscm} \right)_{\mucm} 
 + \frac{c(1 - \mucm^{2})}{\epscm} \left( \frac{\partial \fcm}{\partial \mucm} \right)_{\epscm} , \label{eq:a4c} \\ 
\left( \frac{\partial \fcm}{\partial \pcm^{\hat 2}} \right)_{\pcm^{1'}, \pcm^{3'}}& =& c \left( 1 - \mucm^{2} \right)^{1/2} \cos (\phi_{p}) \left( \frac{\partial \fcm}{\partial \epscm} \right)_{\mucm} 
 + c \left( 1 - \mucm^{2} \right)^{1/2} \cos (\phi_{p}) \left( \frac{\partial \fcm}{\partial \mucm} \right)_{\epscm} , \mbox{and} \nonumber  \\
\left( \frac{\partial \fcm}{\partial \pcm^{\hat 3}} \right)_{\pcm^{1'}, \pcm^{2'}} &=& c \left( 1 - \mucm^{2} \right)^{1/2} \sin (\phi_{p}) \left( \frac{\partial \fcm}{\partial \epscm} \right)_{\mucm} 
 + c \left( 1 - \mucm^{2} \right)^{1/2} \sin (\phi_{p}) \left( \frac{\partial \fcm}{\partial \mucm} \right)_{\epscm}. \nonumber
\end{eqnarray}

We choose a synchronous gauge with a spherically symmetric, orthogonal metric given by
\begin{equation}
ds^{2} = g_{\alpha \beta} dx^{\alpha} dx^{\beta} = - a^{2}(t,m) c^{2} dt^{2} + b^{2}(t,m) dm^{2} + R^{2}(t,m) \left( d\theta^{2} + \sin^{2}\theta \, d\phi^{2} \right) ,  
\label{eq:a3}
\end{equation}
where $x^{0} = ct$, $x^{1} = m$, $x^{2} = \theta$, and $x^{3} = \phi$.
With this metric the transformation functions relating coordinate and orthonormal bases are given by
\begin{equation}
\epsilon_{\hat {0}}^{0} = \frac{1}{a}, \quad \epsilon_{\hat {1}}^{1} = \frac{1}{b},  \quad \epsilon_{\hat 2}^{2} = \frac{1}{R}, \quad \epsilon_{\hat 3}^{3} = \frac{1}{R\sin \theta}, \quad \epsilon_{\hat {a}}^{\beta} = 0 \; {\rm if} \; \hat {a} \ne \beta.
\label{eq:a3basis}
\end{equation}

We choose this form of the metric so that the relativistic equations will closely parallel the Newtonian fluid equations, and various levels of Newtonian approximations can be easily made.
The metric function $a = a(t,m)$ is the lapse function and relates the interval of proper time of an observer attached to the motion of a fluid element to an interval of coordinate time, and defined so that coordinate and proper time are equal at infinity. 
The metric function $b = b(t,m)$ will be chosen so that the coordinate $m$ can be identified with the enclosed rest mass. 
The metric function $R$ is the areal radius (i.e., the 2-sphere area $= 4 \pi R^{2}$). 
The condition that ties the coordinate system to the comoving frame is that the four-velocity of the fluid, $u^{\nu}$, be given by $u^{\nu} \equiv { dx^{\nu} }/{ d\tau } \equiv c\, { dx^{\nu} }/{ ds } = [u^{0}, 0, 0, 0 ]$, which with the metric given by Equation~(\ref{eq:a3}) requires that $u^{0} = {c}/{a}$. 
This definition of $u^{\nu}$ implies that
\begin{equation}
u = \frac{1}{a} \frac{dR}{dt} ,  
\label{eq:a3vel}
\end{equation}
where $u$ is the first component of the four-velocity as observed from a frame of constant areal radius $R$ \citep{MaWh67}. 
To specify $b$ so that $m$ can identified with the enclosed rest mass, we note that the rest mass density $\rho$ satisfies the local conservation condition
\begin{equation}
0 = (\rho u^{\alpha})_{;\alpha} = \frac{1}{\sqrt{g}} \pderiv{}{x^{\nu}} \left( \sqrt{g} \rho u^{\nu} \right)  \quad \Rightarrow  \quad 0 = \pderiv{}{t}\left( b R^{2} \rho \right) \quad \Rightarrow \quad b R^{2} \rho = \mbox{constant} ,
\label{eq:a3b}
\end{equation}
where the semicolon denotes covariant differentiation and where we have used the expression for $u^{\nu}$ given immediately above Equation~(\ref{eq:a3vel}).

With the metric (Equation \ref{eq:a3}) the proper volume, $dV$, is given by $dV = 4\pi R^{2} b\, dm$, or, in terms of the rest mass $dM$ contained in $dV$, by $dV = {dM}/{\rho}$. 
It follows from these two expressions for $dV$ that the requisite choice of $b$ is 
\begin{equation}
b= \frac{1}{ 4 \pi R^{2} \rho }.
\label{eq:a3c}	
\end{equation}

With the above choices for a spherical, comoving coordinate system and a comoving, orthonormal four-vector basis, the coordinate invariant volume elements become
\begin{equation}
dV = \sqrt{-g}\, \epsilon_{\delta \alpha \beta \gamma} u^{\delta} d_{1}x^{\alpha} d_{2}x^{\beta} d_{3}x^{\gamma} \rightarrow b\, dm\, R^{2} d\Omega ,
\label{eq:a5}
\end{equation}
\begin{equation}
dP = \sqrt{-g}\, \epsilon_{ijk} u^{\delta} \frac{ d_{1}\pcm^{i} \, d_{2}\pcm^{j} \, d_{3}\pcm^{k} }{ \pcm } \rightarrow \frac{1}{c^{2}} \epscm \, d\epscm \, d\Omega_{p} ,
\label{eq:a6} 
\end{equation}
where $g$ is the determinant of the metric, and $\epsilon_{\delta \alpha \beta \gamma}$ and $\epsilon_{ijk}$ are the Levi-Civita alternating symbols.
The invariant distribution function, \fcm, introduced at the beginning of this section, is defined so that the number of world lines crossing the volume element, $dV$, with four-momenta in the range $dP$ about ${\bf \pcm}$ is given by \citep{Lind66}
\begin{equation}
dN = \frac{1}{h^{3}} \fcm(x,{\bf p}_{_{\! 0}}) ( - {\bf p}_{_{\! 0}} \cdot {\bf u} ) dV dP \rightarrow \frac{1}{(hc)^{3}} \fcm b\, dm R^{2} d\Omega \, \epscm^{2} \, d \epscm \, d\Omega_{p} ,
\label{eq:a7}
\end{equation}
where the right-hand sides of Equations~(\ref{eq:a5}) -- (\ref{eq:a7}) are the expressions for $dV$, $dP$, and $dN$ in our choice of coordinate system and four-vector basis. 
Finally, evaluating the transformation coefficients from the metric (Equation~\ref{eq:a3}), and using them for the Ricci coefficients in Equation~(\ref{eq:a3f}), and using Equations (\ref{eq:a4}) and (\ref{eq:a4c}), the Boltzmann equation becomes
\begin{eqnarray}
 \frac{\epscm}{c} \left\{ \frac{1}{ac} \left( \pderiv{\fcm}{t} \right)_{m,\epscm,\mucm} 
+ \frac{\mucm }{b} \left( \pderiv{\fcm}{m} \right)_{t,\epscm,\mucm} \right. \nonumber \\ 
 - \left[  \frac{1}{b} \pderiv{\ln a}{m} \mucm 
+ \frac{1}{ac} \pderiv{\ln b}{t} \mucm^{2} 
+ \frac{1}{ac} \pderiv{\ln R}{t} ( 1 - \mucm^{2} ) \right] 
\epscm \left( \pderiv{\fcm}{\epscm} \right)_{t,m,\mucm}  \nonumber \\ 
 + \left.  \left[ \frac{ \mucm }{ac}  \left( \pderiv{\ln R}{t} - \pderiv{\ln b}{t} \right) 
- \frac{1}{b} \left( \pderiv{\ln a}{m} - \pderiv{\ln R}{m} \right) \right] 
( 1 - \mucm^{2} ) \left( \pderiv{\fcm}{\mucm} \right)_{t,m,\epscm} \right\} \nonumber \\ 
 = \left( \frac{ d\fcm  }{ d\ell }\right)_{ \! S } ,
\label{eq:a8}
\end{eqnarray}
Note that with the substitutions $a = e^{\Phi}$, $b = e^{\Lambda}$, and $\Gamma = (1/b) ( \partial R / \partial m )_{t}$, Equation~(\ref{eq:a8}) reduces to Equation~(3.7) of \citet{Lind66}.
For economy of notation and for clarity of presentation we have suppressed the dependency of \fcm\ on $t$, $m$, $\epscm$, and $\mucm$, and the dependencies of the metric variables a, b, $\Gamma$, and R on $t$ and $m$, and will do so with new dependent variables as they are introduced as long as this brings no ambiguity.

We now modify Equation~(\ref{eq:a8}) by transforming from independent variables $(t, m, \epscm, \mucm)$ to $(t, m, \Ecm, \mucm)$ where
\begin{equation}
\Ecm = a \epscm .
\label{eq:a8a}
\end{equation}
Therefore, with this variable transformation
\begin{equation}
\fcm(t,m,\epscm,\mucm) = \fcm\left(t,m, \frac{\Ecm}{a}, \mucm\right) = \fcm'(t,m,\Ecm,\mucm)
\label{eq:a8b}
\end{equation}
Though mathematically imprecise, we will use the same symbol to describe the neutrino distribution functions, originally functions of $(t,m,\epscm,\mucm)$, as functions of $(t,m,\Ecm,\mucm)$.
We also define $\Gamma$ by $\Gamma = (1/b) ( \partial{R}/ \partial{m} )_{t}$ so that derivatives with respect to $m$ at constant time can be replaced by derivatives with respect to $R$, which is directly related to our grid, by the identity
 \begin{equation}
 \left( \pderiv{X}{m} \right)_{t} = \left( \pderiv{X}{R} \right)_{t} \left( \pderiv{R}{m} \right)_{t} = b \Gamma \left( \pderiv{X}{R} \right)_{t} \equiv b \Gamma X_{,R} .
 \end{equation}
  Applying the above definition of $\Gamma$ and variable transformation to Equation~(\ref{eq:a8}) gives 
\begin{eqnarray}
 \frac{\Ecm}{ac} \left\{\frac{1}{ac} \left( \pderiv{\fcm}{t} \right)_{m, \mucm, \Ecm} + \mucm \Gamma \left( \pderiv{\fcm}{R} \right)_{t, \mucm, \Ecm} \right. \nonumber \\ 
 + \frac{1}{ac} \left[ \pderiv{\ln a}{t} -  \mucm^{2} \pderiv{\ln b}{t} - ( 1 - \mucm^{2} ) \pderiv{\ln R}{t} \right] \Ecm \left( \pderiv{\fcm}{\Ecm} \right)_{t, m, \mucm} \nonumber \\ 
 + \left.  \left[ \frac{ \mucm }{ac} \left( \pderiv{\ln R}{t} - \pderiv{\ln b}{t} \right)  - \Gamma \left( \pderiv{\ln a}{R} - \frac{1}{R} \right) \right] 
( 1 - \mucm^{2} ) \left( \pderiv{\fcm}{\mucm} \right)_{t, m, \Ecm} \right\} \nonumber \\ 
 = \left( \frac{ d\fcm }{ d\ell } \right)_{ \! {\rm S} } .
\label{eq:a9}
\end{eqnarray}

The transformation given by Equation (\ref{eq:a8a}) affords several advantages. 
First, with constant values of $\Ecmi{k+\frac{1}{2}}$ anchoring the energy grid, the energy grid $\epscmi{ k+\frac{1}{2}}$ will be scaled to higher values as the lapse dips below unity at high densities -- i.e., for constant $\Ecmi{k+\frac{1}{2}}$the neutrino energy will scale as $\epscmi{ k+\frac{1}{2}} \propto 1/a$. 
This will permit a smaller upper bound to the neutrino energy grid farther out from the center where $a \simeq 1$, resulting in the grid energies being more closely spaced there for a given number of grid points, while still permitting sufficient energy headroom at high densities to accommodate the high-energy neutrinos that are produced at the high matter densities prevailing near the core center. 
Second, the radial derivative of $a$ in the factor multiplying the energy derivative of \fcm\ in Equation~(\ref{eq:a8}) has been replaced by a time derivative in Equation~(\ref{eq:a9}). 
This factor now contains terms involving only time derivative, and therefore vanishes for a static space-time. 
Thus, apart from energy-changing interactions, for a static space-time there will be no flow of neutrinos through the neutrino energy grid as they propagate outward. 
The gravitational redshifting will consequently be accomplished automatically.
 In non-static space-times the advection of neutrinos through the energy grid can be performed algebraically (Sections~\ref{neu_e_adv} and \ref{app:e_advct_lapse}), 
 another advantage of this scheme. 
 Using this choice of energy gridding, we will assume that the space-time is constant over a time step and make a small correction to the neutrino distribution at the end of the time step to correct for the change in $a$ during a time step and other processes that shift the neutrinos in energy, such as their advection across spatial zones with differing lapses.

\subsection{Moment Equations}

Let the $n$-th angular moment of \fcm\ be denoted by \psimomcm{n}, that is

\begin{equation}
\psimomcm{n}(R, t, \Ecm) = \frac{1}{4\pi} \int_{\Omega} d\Omega \mucm^{n} \fcm(R, t, \Ecm, \mucm) = \frac{1}{2} \int_{-1}^{1} d\mucm \mucm^{n} \fcm(R, t, \Ecm, \mucm) ,
\label{eq:a10}
\end{equation}
and let 
\begin{equation}
\left( \frac{ d\fcm }{ d\ell } \right)_{ \! S }^{(n)} \equiv \left( \frac{ d }{ d\ell } \fcm(m, t, \Ecm, \mucm) \right)_{ \! S }^{(n)} \equiv  \frac{1}{2} \int_{-1}^{1} \mucm^{n} d\mucm\, \left( \frac{ d }{ d\ell } \fcm(m, t, \Ecm, \mucm) \right)_{ \! S } .
\label{eq:a11}
\end{equation}
Then the first two angular moments of Equation (\ref{eq:a9}) are given by
\begin{eqnarray}
\frac{\Ecm}{ac} \left\{ \frac{1}{ac} \left( \pderiv{\psimomcm{0}}{t} \right)_{m,\Ecm} + \Gamma  \left( \pderiv{ \psimomcm{1} }{R} \right)_{t,\Ecm} \right.  \nonumber \\ 
 + \frac{1}{ac} \Ecm \left[ \pderiv{\ln a}{t} \left( \pderiv{\psimomcm{0}}{\Ecm} \right)_{t,m}  - \pderiv{\ln b}{t} \left( \pderiv{\psimomcm{2}}{\Ecm} \right)_{t,m}  - \pderiv{\ln R}{t} \left( \pderiv{}{\Ecm} \right)_{t,m} \left( \psimomcm{0} - \psimomcm{2} \right) \right]  \nonumber \\ 
 \left. - \frac{1}{ac} \left( \pderiv{\ln R}{t} - \pderiv{\ln b}{t} \right) \left( \psimomcm{0} - 3 \psimomcm{2} \right) + \Gamma \left( - \pderiv{\ln a}{R} + \frac{1}{R} \right) 2 \psimomcm{1} \right\} \nonumber \\ 
 = \left( \frac{ d\fcm }{ d\ell } \right)_{ \! S }^{(0)}
\label{eq:a12}
\end{eqnarray}
and
\begin{eqnarray}
 \frac{\Ecm}{ac} \left\{ \frac{1}{ac} \left( \pderiv{\psimomcm{1}}{t} \right)_{m,\Ecm} + \Gamma \left( \pderiv{ \psimomcm{2} }{R} \right)_{t,\Ecm} \right.  \nonumber \\ 
 + \frac{1}{ac} \Ecm \left[ \pderiv{\ln a}{t} \left( \pderiv{\psimomcm{1}}{\Ecm } \right)_{t,m}  - \pderiv{\ln b}{t} \left( \pderiv{\psimomcm{3}}{ \Ecm } \right)_{t,m}  - \pderiv{\ln R}{t} \left( \pderiv{}{ \Ecm } \right)_{t,m} \left( \psimomcm{1} - \psimomcm{3} \right) \right]  \nonumber \\ 
 \left. - \frac{1}{ac} \left( \pderiv{\ln R}{t} - \pderiv{\ln b}{t} \right) \left( 2 \psimomcm{1} - 4 \psimomcm{3} \right) - \Gamma \left( - \pderiv{\ln a}{R} + \frac{1}{R} \right) ( \psimomcm{0} -  3 \psimomcm{2} ) \right\}  \nonumber \\ 
 = \left( \frac{ d\fcm }{ d\ell } \right)_{ \! S }^{(1)} .
\label{eq:a13}
\end{eqnarray}

\subsection{Flux Limiting}
\label{flux_limiting}

To close this system of moment equations we employ flux limiting to derive a relation between \psimomcm{1}\ and \psimomcm{0}.
In analogy with the procedure described in \citet{LePo81}, let the scalar Eddington factors $\eta^{(n)}$ be defined as the ratio of \psimomcm{n}\ to \psimomcm{0}\ so that $\psimomcm{n} = \eta^{(n)} \psimomcm{0}$. 
Substituting $\psimomcm{n} = \eta^{(n)} \psimomcm{0}$ in Equations~(\ref{eq:a12}) and (\ref{eq:a13}), solving Equation~(\ref{eq:a12}) for $( \partial\psimomcm{0}/\partial t )_{m, \Ecm}$ and substituting the result into Equation~(\ref{eq:a13}) gives
\begin{eqnarray}
\left\{ \frac{1}{ac} \left( \pderiv{\eta^{(1)}}{t} \right)_{m,\Ecm} - \Gamma  \eta^{(1)} \left( \pderiv{\eta^{(1)}}{R} \right)_{t,\Ecm} + \frac{1}{ac} \left( \pderiv{\ln b}{t} - \pderiv{\ln R}{t} \right) \Ecm  \left( \pderiv{\eta^{(2)}}{\Ecm} \right)_{t,m} \eta^{(1)} \right. \nonumber \\ 
 + \frac{1}{ac} \left( \pderiv{\ln R}{t} - \pderiv{\ln b}{t} \right) \left( - \eta^{(1)} - 3 \eta^{(1)} \eta^{(2)} + 4 \eta^{(3)} \right) + \Gamma \left( \pderiv{\ln a}{R} - \frac{1}{R} \right) \left( 1 + 2 \left( \eta^{(1)} \right)^{2} -  3 \eta^{(2)} \right) \nonumber \\ 
 \left. + \Gamma \left( \pderiv{ \eta^{(2)} }{R} \right)_{t,\Ecm} + \frac{1}{ac} \left( \pderiv{\ln a}{t} - \pderiv{\ln b}{t} \right) \Ecm \left( \pderiv{\eta^{(1)}}{\Ecm} \right)_{t,m} - \frac{1}{ac} \pderiv{\ln R}{t} \Ecm \left( \pderiv{ \left( \eta^{(1)} - \eta^{(3)} \right) }{\Ecm} \right)_{t,m} \right\} \psimomcm{0} \nonumber \\ 
+ \left\{ \frac{1}{ac} \left( \pderiv{\ln b}{t} - \pderiv{\ln R}{t} \right) \left( \eta^{(1)} \eta^{(2)} - \eta^{(3)} \right) \right\} \Ecm \left( \pderiv{\psimomcm{0}}{\Ecm} \right)_{t,m}  \nonumber \\ 
  - \Gamma \left( \pderiv{\psimomcm{0}}{R} \right)_{t,\Ecm} \left( \eta^{(1)} \right)^{2} + \frac{ac}{\Ecm} \eta^{(1)} \left( \frac{ d\fcm }{ d\ell } \right)_{ \! S }^{(0)} + \Gamma \left( \pderiv{ \psimomcm{0} }{R} \right)_{t,\Ecm} \eta^{(2)} = \frac{ac}{\Ecm} \left( \frac{ d\fcm }{ d\ell } \right)_{ \! S }^{(1)} .
\label{eq:a14}
\end{eqnarray}
The factors multiplying $\psi_{0}^{(0)}$ and $( \partial \psi_{0}^{(0)}/\partial{E_{0}} )_{t,m}$ in Equation~(\ref{eq:a14}) in both the diffusion limit ($\eta^{(1)} \rightarrow 0$; $\eta^{(2)} \rightarrow \rfrac{1}{3}$; $\eta^{(3)} \rightarrow 0$) and the free streaming limit ($\eta^{(n)} \rightarrow 1$) are zero.
We therefore make the approximation that these two factors are zero everywhere.
Equation~(\ref{eq:a14}) then becomes
\begin{equation}
\frac{ac}{\Ecm} \eta^{(1)} \left( \frac{ d\fcm }{ d\ell } \right)_{ \! S }^{(0)} + \Gamma \left( \pderiv{ \psimomcm{0} }{R} \right)_{t,\Ecm} \left( \eta^{(2)} - \left( \eta^{(1)} \right)^{2} \right) = \frac{ac}{\Ecm} \left( \frac{ d\fcm }{ d\ell } \right)_{ \! S }^{(1)} .
\label{eq:a15}
\end{equation}
Multiplying the right-hand side of Equation (\ref{eq:a15}) by \psimomcm{1}/\psimomcm{1}, i.e., unity, and solving for \psimomcm{1}, we get
\begin{equation}
\psimomcm{1} 
 = \frac{ \ds - \Gamma \left( \eta^{(2)} - (\eta^{(1)})^{2} \right) \left( \frac{ \partial \psimomcm{0} }{ \partial R } \right)_{\! t,\Ecm} - \eta^{(1)} \frac{ac}{\Ecm} \left( \frac{ d\fcm }{ d\ell } \right)_{ \! S }^{(0)} }{ \ds - \frac{ac}{\Ecm} \frac{1}{ \psimomcm{1} } \left( \frac{ d\fcm }{ d\ell } \right)_{ \! S }^{(1)} }.
\label{eq:a16}
\end{equation}
In the diffusion limit, $\eta^{(1)} \rightarrow 0$ and $\eta^{(2)} \rightarrow \rfrac{1}{3}$, and Equation~(\ref{eq:a16}) reduces to the standard diffusion equation if we neglect the second term in the numerator and identify the transport mean free path, $\lambda^{(t)}$, as 
\begin{equation}
\frac{1}{ \lambda^{(t)} } =  - \frac{ac}{\Ecm} \frac{1}{ \psimomcm{1} } \left( \frac{ d\fcm }{ d\ell } \right)_{ \! S }^{(1)}   .
\label{eq:a17}
\end{equation}
In regimes other than the diffusion regime we regard $[ \eta^{(2)} - (\eta^{(1)})^{2} ]$ as a free parameter, which we write as 
\begin{equation}
{\cal F} = 3 \left[ \eta^{(2)} - (\eta^{(1)})^{2} \right] .
\label{eq:a18}
\end{equation}
Using Equations (\ref{eq:a17}) and (\ref{eq:a18}) in Equation (\ref{eq:a16}), we get a diffusion-like equation for \psimomcm{1}:
\begin{equation}
\psimomcm{1} = - \frac{ \lambda^{(t)} }{3} {\cal F} \, \Gamma \left( \frac{ \partial \psimomcm{0} }{ \partial R } \right)_{\! t,\Ecm}  . 
\label{eq:a19}
\end{equation}

Equations (\ref{eq:a12}) and (\ref{eq:a19}) for each energy zone and for each neutrino species (\nue, \nuebar, \numt, \numtbar) with a prescription for ${\cal F}$ are the MGFLD equations that are solved in \Chimera\ for the B-series simulations.
The modification of this scheme in the vicinity of shocks, used for our C-series and later simulations, is described in Section~\ref{trans_shock}.
The parameter ${\cal F}$ is referred to as the ``flux-limiter," and should be unity in the diffusion limit and tend to zero in such a way that \psimomcm{1}~= \psimomcm{0}\ in the limit of free streaming.

The derivation of Equations~(\ref{eq:a19}) and (\ref{eq:a17}) differ from that of the corresponding equations (Equations (A25) and (A26)) in \citet{Brue85}. 
To derive a diffusion-like equation, there the derivative $\partial \psimomcm{1}/\partial t$ in the nonrelativistic version of Equation~(\ref{eq:a13}) and all velocity dependent terms were set to zero. 
However, the results are similar.
Apart from the relativistic time dilation factor $a$, the expressions for $1/ \lambda_{i}^{(t)}$, given here by Equation~(\ref{eq:a17}) and given in \citet{Brue85} by Equation~(A26), are the same.
Equation~(\ref{eq:a16}) for \psimomcm{1} here is similar to Equation~(A25)  \psimomcm{1} in \citet{Brue85}, with the exception here of the relativistic proper distance correction $\Gamma$ and the factor of [$ \eta^{(2)} - (\eta^{(1)})^{2} ]$, which we take as our flux limiter. 
The flux limiter in \citet{Brue85} is introduced as a modification of $\lambda_{i}^{(t)}$.
Finally, the terms in the numerator of Equations~(\ref{eq:a16}) here and in Equation~(A25) of \citet{Brue85} are neglected, other than the term involving $\partial \psimomcm{0}/\partial R$.
They involve the redistribution of neutrinos in angle due to anisotropies in the source terms, but are small given they depend on the product of \psimomcm{0} and \psimomcm{1}.
Equation~(\ref{eq:a19}) is given in differenced form by Equation~(\ref{eq:a59}).

\subsection{Flux-Limiter}

The flux limiter, ${\cal F}$, constructed to satisfy the diffusion and free-streaming limits, consists of two parts.
The first part is a specific implementation of the usual scheme for interpolating between these two limits, namely
\begin{equation}
{\cal F}_{\rm intrp} \equiv {\cal F}_{\rm intrp}(R, t, \epscm) = \left( 1 + \frac{1}{3} \lambda_{i+\frac{1}{2}}^{(t)}(\epscm) \left[ \frac{ \ds \left| \Gamma \pderiv{ \psimomcm{0}(\Ecm) }{ R } \right| }{ \psimomcm{0}(\Ecm) } \right] \right)^{-1}.
\label{eq:a20}
\end{equation}
From Equations (\ref{eq:a19}) and (\ref{eq:a20}) we see that in the optically-thick, diffusion regime, for which $\lambda^{(t)}(\epscm) \rightarrow 0$, ${\cal F}_{\rm intrp} \rightarrow 1$ and Equation~(\ref{eq:a19}) limits to the diffusion equation
\begin{equation}
\psimomcm{1}(\Ecm) \rightarrow - \frac{ \lambda_{i+\frac{1}{2}}^{(t)}(\epscm) }{3} \Gamma \pderiv{ \psimomcm{0}(\Ecm) }{ R } ,
\label{eq:a21}
\end{equation}
while in the optically-thin, free-streaming regime, where $\lambda^{(t)}(\epscm) \rightarrow \infty$, Equation~(\ref{eq:a19}) limits to the free-streaming condition
\begin{equation}
\psimomcm{1}(\Ecm) = \psimomcm{0}(\Ecm).
\label{eq:a22}
\end{equation}

As it stands, this scheme suffers from the generic problem of an overly rapid transition to the free-streaming limit (i.e., the angular distribution becomes too forwardly peaked) when matter goes from optically thick to optically thin abruptly,  such as when the `density cliff' forms in the post-bounce core of a supernova progenitor.
To avoid this problem, a second piece of the flux limiter is constructed.
It prevents the neutrino angular distribution from becoming more forwardly peaked than the geometrical limit.
This geometrical limit can be expressed for $R > R_{\nu}$, where $R_{\nu} \equiv R_{\nu}(\epscm)$ is the radius of the neutrinosphere for a particular flavor and energy, by $\psimomcm{1} = \frac{1}{2} [1 + \mucmi{\nu}(\epscm)]\psimomcm{0}$.
This expression assumes that the neutrino distribution function \fcm\ is constant for rays satisfying $\mucm < \mucmi{\nu}$, and zero otherwise, where $\mucmi{\nu}$ is the cosine of the angle subtended between a line extending from a point at $R$ to the neutrinosphere limb and a line extending from the point at $R$ to the core center, as observed by a comoving observer.
This geometrical piece of the flux-limiter is then given by
\begin{equation}
{\cal F}_{\rm geom} \equiv {\cal F}_{\rm geom}(R, t, \epscm) = \frac{ \frac{1}{2} ( 1 + \mucmi{ \nu} ) \psimomcm{0} }{ \frac{1}{3} \lambda_{i}^{(t)} \left| \Gamma \pderiv{ \psimomcm{0} }{ R }  \right| }  \mbox{ if } R > R_{\nu} ,
\label{eq:a23}
\end{equation}
and ${\cal F}_{\rm geom}(R, t, \epscm) = 1$ interior to the neutrinosphere, $R \le R_{\nu}(\epscm)$.
The comoving angle \mucmi{ \nu} is given in terms of the fixed frame angle $\mu_{\nu}$ by
\begin{equation}
 \mucmi{\nu}(\epscm) = \frac{ \mu_{\nu} + \beta }{ 1 - \mu_{\nu} \beta },
\label{eq:a24}
\end{equation}
where $\beta = v/c$, and the fixed frame angle $\mu_{\nu}$ is given by
\begin{equation}
 \mu_{\nu} = \sqrt{ 1 - \left( \frac{ R_{\nu} }{ R } \right)^{2} {\cal G} },
\label{eq:a25}
\end{equation}
where ${\cal G}$ is the correction for gravitational ray bending given by
\begin{equation}
{\cal G}  = \sqrt{ \frac{ 1 - 2 G M_{g} /R c^{2} }{ 1 - 2 G M_{g} /R_{\nu} c^{2} } },
\label{eq:a26}
\end{equation}
where $M_{g}$ is the gravitational mass. 
The geometrical flux limiter, ${\cal F}_{\rm geom}$, by itself would relate \psimomcm{1} to \psimomcm{0} outside the neutrinosphere by
\begin{equation}
\psimomcm{1} = \frac{1}{2} \left( 1 + \mucmi{ \nu} \right) \psimomcm{0}
\label{eq:a26b}
\end{equation}
The final flux limiter is given by
\begin{equation}
{\cal F} = \min [ {\cal F}_{\rm intrp}, {\cal F}_{\rm geom} ]
\label{eq:a26a}
\end{equation}

\subsection{Operator Splitting}

To solve Equation (\ref{eq:a12}) along with Equation (\ref{eq:a19}), we operator split Equation~(\ref{eq:a12}) into a transport equation and an energy advection equation
\begin{equation}
\left( \pderiv{\psimomcm{0}}{t} \right)_{m,\Ecm}  = \left( \pderiv{\psimomcm{0}}{t} \right)_{m,\Ecm}^{T} + \left( \pderiv{\psimomcm{0}}{t} \right)^{E}_{m,\Ecm} ,
\label{eq:a29}
\end{equation}
where $( \partial \psimomcm{0} / \partial t )_{m,\Ecm}^{T}$, the rate of change of \psimomcm{0}\ due to sources and transport, is given by
\begin{equation}
\frac{1}{ac} \left( \pderiv{\psimomcm{0}}{t} \right)_{m, \Ecm}^{T} + \Gamma \left( \pderiv{\psimomcm{1}}{R} \right)_{t,\Ecm} - \Gamma \left( \pderiv{ \ln a}{R} - \frac{1}{R} \right) 2 \psimomcm{1} = \frac{ac}{\Ecm} \left( \frac{ d }{ d\ell } \fcm(x,\pcm) \right)_{ \! S }^{(0)} ,
\label{eq:a30}
\end{equation}
and $( \partial \psimomcm{0} / \partial t )^{E}_{m,\Ecm}$, the rate of change of \psimomcm{0}\ due to energy advection, is given such that
\begin{eqnarray}
\left( \pderiv{\psimomcm{0}}{t} \right)_{m,\Ecm}^{E} + \Ecm \left[  \pderiv{\ln a}{t} \left( \pderiv{\psimomcm{0}}{\Ecm} \right)_{t,m} - \pderiv{\ln b}{t} \left( \pderiv{\psimomcm{2}}{\Ecm} \right)_{t,m} 
 - \pderiv{\ln R}{t} \left( \pderiv{}{\Ecm} \right)_{t,m} \left( \psimomcm{0} - \psimomcm{2} \right) \right] \nonumber \\ 
 - \left[ \pderiv{\ln R}{t} - \pderiv{\ln b}{t} \right] \left( \psimomcm{0} - 3 \psimomcm{2} \right) = 0 .
\label{eq:a31}
\end{eqnarray}
In advancing \psimomcm{0}\ and \psimomcm{1}\ over a time step, the pair of Equations~(\ref{eq:a30}) and (\ref{eq:a19}) are solved in one step along with the associated energy and lepton conservation equations introduced below, then Equation (\ref{eq:a31}) is solved in a second step.
We will refer to these two separate steps as a transport step and an energy advection step, respectively, and describe in more detail below how each is performed.

\subsection{Einstein's Equations}

In order to derive the Einstein equations, which are needed to obtain expressions for the gravitational mass, $M_{g}$, and the metric parameters $\Gamma$ and $a$, we need the stress-energy tensor, ${\cal T} = $$^{(m)}{\cal T} + $$^{(\nu)}{\cal T}$, where $^{(m)}{\cal T}$ and $^{(\nu)}{\cal T}$ are the matter and neutrino contributions, respectively. 
Following \citet{MiMi84} we begin with the definition of the radiation (i.e., neutrino) stress-energy tensor 
\begin{equation}
^{(\nu)}{\cal T}^{\alpha \beta} = \frac{ c^{2} }{ h^{3} } \int \sum_{q} \fcmi{q} \, p^{\alpha} p^{\beta} \frac{ d^{3} p }{ cp^{0} } ,
\end{equation}
where the sum $q$ is over all neutrino species.
In the local comoving orthonormal frame, the components of the stress-energy tensor are
\begin{equation}
^{(\nu)}{\cal T}^{\hat{\rm a}\hat{\rm b}} = \frac{ c^{2} }{ (hc)^{3} } \int \sum_{q} \fcmi{q} \, \pcm^{\hat{\rm a}} \pcm^{\hat{\rm b}} \epscm d\epscm \, d\Omegacm . 
\label{eq:a32}
\end{equation}
where as before we use latin indices with hats here to distinguish components with respect to the local comoving orthonormal frame from those with respect to the coordinate basis (distinguished using Greek letters).

Using Equations~(\ref{eq:a4}) for $p^{\hat{\rm a}}$ and $p^{\hat{\rm b}}$ in Equation~(\ref{eq:a32}) for $^{(\nu)}{\cal T}^{\hat{\rm a}\hat{\rm b}}$, the radiation stress-energy tensor $^{(\nu)}{\cal T}^{\hat{\rm a}\hat{\rm b}}$ can be written in terms of the local neutrino energy density, $E_{\nu}$, flux, $F_{\nu}$, and pressure, $P_{\nu}$, as
\begin{equation}
\left( E_{\nu}, F_{\nu}/c, P_{\nu} \right) = \frac{ 1 }{ (hc)^{3} } \int \sum_{q} \fcmi{q} \epscm^{3} d\epscm \mucm^{(0, 1, 2)} d\Omegacm 
= \frac{ 4\pi }{ (hc)^{3} } \int \sum_{q} \psimomcm{0, 1, 2} \epscm^{3} d\epscm .
\label{eq:a33}
\end{equation} 
Transforming $^{(\nu)}{\cal T}^{\hat{\rm a}\hat{\rm b}}$ written in terms of $E_{\nu}, F_{\nu}/c, P_{\nu}$ back to the coordinate basis using Equations~(\ref{eq:a3basis}), and adding the stress-energy components of a perfect fluid, given by
\begin{equation}
^{(m)}T^{\alpha \beta} = ( \rho c^{2} + E_{\rm m} + P_{\rm m} ) \frac{ u^{\alpha} u^{\beta} }{c^{2} } + P_{\rm m}  g^{\alpha \beta} ,
\label{eq:a34}
\end{equation}
with  $u^{\alpha}$ defined immediately above Equation~(\ref{eq:a3vel}) and  $g_{\alpha \beta}$ given by Equation~(\ref{eq:a3}), the combined matter--neutrino stress energy tensor is given by
\begin{equation}
{\cal T}^{\alpha \beta}
= \left(
\begin{array}{cccc}
\ds \frac{1}{ a^{2} } \left( \rho c^{2} + E_{m} + E_{\nu} \right) & \ds \frac{1}{bac} F_{\nu} & 0 & 0 \\
\ds \frac{1}{bac} F_{\nu} & \ds \frac{1}{b^{2}} \left( P_{m} + P_{\nu} \right) & 0 & 0 \\
0 & 0 & \ds \frac{1}{ R^{2} } \left( P_{m} + \frac{1}{ 2} \left( E_{\nu} - P_{\nu} \right) \right) & 0 \\
0 & 0 & 0 & \ds \frac{1}{ R^{2} \sin^{2} \theta } \left( P_{m} + \frac{1}{2} \left( E_{\nu} - P_{\nu} \right) \right)
\end{array}
\right),
\label{eq:a35}
\end{equation}
where $\rho$ is the proper rest mass density, and $E_{m}$ and $P_{m}$ are the matter internal energy density and pressure, respectively.  

The Einstein field equations are given by 
\begin{equation}
- \frac{8\pi G}{c^{4}} \left( T_{\nu}^{\;\; \mu} \right) = {\cal R}_{\nu}^{\;\; \mu} - \frac{1}{2} g_{\nu}^{\;\; \mu} {\cal R}_{\lambda}^{\;\; \lambda} ,
\label{eq:a36}
\end{equation}
where ${\cal R}_{\nu}^{\;\; \mu}$ is the Ricci tensor.
After some algebra \citep[e.g.,][]{MaWh66, MaWh67} we find from the Einstein equations that the gravitational mass is given by
\begin{equation}
  M_{\rm g} = M_{\rm g}(t,R) = \int_{0}^{R} \left[ ( \rho + \frac{E_{\rm m}}{c^{2}} + \frac{E_{\nu}}{c^{2}} ) + \frac{1 }{c^{4}} \frac{ u F_{\nu} }{ \Gamma } \right] 4 \pi R^{2} dR ,
\label{eq:a37}
\end{equation}
and the metric parameter $\Gamma$ is given by
\begin{equation}
\Gamma = \sqrt{ 1 + \frac{u^{2}}{c^{2}} - \frac{ 2 M_{\rm g} G }{Rc^{2}} } ,
\label{eq:a38}
\end{equation}
where $u$ is the velocity, $u = {a}^{-1} ( { dR / dt} )_{m}$.
Einstein equations can also be used to derive the radial equation of motion, which is given by
\begin{equation}
\frac{1}{a} \frac{du}{dt} = - G \frac{M_{\rm g}}{R^{2}} - \frac{\Gamma^{2}}{ \rho w } \pderiv{ P_{{\rm m}} }{R} - \frac{4\pi G}{c^{2}} R ( P_{\rm m} + P_{\nu} ) - \frac{\Gamma}{ \rho w } \frac{4\pi c }{(hc)^{3}} \int \epscm^{2} d\epscm \; \sum_{q} \left( \frac{ d }{ d\ell } f_{q}(x,p) \right)_{ \! S }^{(1)} ,
\label{eq:ee39}
\end{equation}
where $w$, the specific enthalpy, is given by
\begin{equation}
w = 1 + \frac{ E_{m} + P_{m} }{\rho c^{2} }.
\label{eq:a47}
\end{equation}
and where the sum over $\nu$ is a sum over all neutrino and antineutrino species.
Equation (\ref{eq:ee39}), with $w = \Gamma = a = 1$ (Newtonian approximation), and with the centrifugal term added, is the radial Equation~(\ref{eq:h73}) and, in differenced form, Equation~(\ref{eq:h74}).
Here we consider the neutrino component of these equations; i.e., the last term of Equation~(\ref{eq:ee39}), which corresponds to the term $f_{\nu \, r}$ in the radial Equation~(\ref{eq:h73}).
The neutrino contributions to the $\theta$- and $\phi$-components of the velocity and energy hydrodynamics equations are described in the text following Equations~(\ref{eq:h73}) and in Equations (\ref{eq:h75}), (\ref{eq:h81}), and (\ref{eq:h82}).

Using Equation (\ref{eq:a2}) and the first of Equation~(\ref{eq:a3basis}), the last term of Equation~(\ref{eq:ee39}) in the Newtonian approximation can be written as
\begin{equation}
f_{\nu,r} = - \frac{\Gamma}{ \rho w } \frac{4\pi c }{(hc)^{3}} \int \epscm^{2} d\epscm \; \sum_{q} \left( \frac{ d }{ d\ell } f_{q}(x,p) \right)_{ \! S }^{(1)} 
\rightarrow - \frac{1}{ \rho } \frac{4\pi}{(hc)^{3}} \int \epscm^{3} d\epscm \; \sum_{q} \left( \frac{1}{c} \frac{ d }{ dt } f_{q}(x,p) \right)_{ \! S }^{(1)}
\label{eq:aa47}
\end{equation}
The differenced form of the expression given by the last term in Equation~(\ref{eq:aa47}) for $f_{\nu,r}$ is given in Section~\ref{sec:LTS}.

\subsection{Matter--Neutrino Energy--Momentum Exchange}

To determine the energy-momentum exchange between the matter and neutrinos we begin with the hydrodynamics equation
\begin{equation}
^{(m)}T_{\alpha \; ;\beta}^{\beta} = G_{\alpha} ,
\label{eq:a40}
\end{equation}
where as before the semicolon denotes covariant differentiation. 
The left-hand side is the divergence of the stress-energy tensor of the matter, and $G^{\alpha}$ is the four-force density -- i.e., the negative of the matter to neutrino energy-momentum transfer rate per unit volume. 
To determine the latter we operate on Equation~(\ref{eq:a8}), or Equation~(\ref{eq:a9}), by the negative of the four-momentum density operator (the integral of the product of the invariant momentum volume element and the neutrino four-momentum), which in the orthonormal basis is given by 
\begin{equation} 
- \frac{1}{(hc)^{3}} \int \sum_{q} \frac{ \epscm^{2} d\epscm d\Omega_{p} } { \epscm/c} \frac{\epscm}{c} \left( 1, \mucm, (1 - \mucm^{2})^{1/2} \cos\phi_{p}, (1 - \mucm^{2})^{1/2} \sin\phi_{p} \right) ,
\end{equation}
and use Equations (\ref{eq:a33}) and (\ref{eq:a5}) to get
\begin{eqnarray}
\lefteqn{- \frac{1}{(hc)^{3}} \int \epscm^{2} d\epscm d\Omega_{p} \left( 1, \mucm, (1 - \mucm^{2})^{1/2} \cos\phi_{p}, (1 - \mucm^{2})^{1/2} \sin\phi_{p} \right) \epscm \left\{ \frac{1}{a} \left( \pderiv{\fcm}{t} \right)_{m,\epscm,\mucm} \cdots \right\} }  \nonumber \\ 
 && = \left[ - \frac{1}{a} \left( \frac{\partial E_{\nu} }{\partial t} \right)_{m,\Ecm,\mucm} - \cdots,  - \frac{1}{ca}  \left( \frac{\partial F_{\nu} }{\partial t } \right)_{m,\Ecm,\mucm} - \cdots, 0 - \cdots, 0 - \cdots \right] \nonumber \\ 
&& = - \frac{4\pi c}{(hc)^{3}} \int \epscm^{2} d\epscm \; \left( \sum_{q} \left( \frac{ d }{ d\ell } \fcmi{q} \right)_{ \! S }^{(0)}, \sum_{q} \left( \frac{ d }{ d\ell } \fcmi{q} \right)_{ \! S }^{(1)}, 0, 0 \right)  \nonumber \\
&& = \left( G^{\hat{1}}, G^{\hat{2}}, G^{\hat{3}}, G^{\hat{4}} \right) .
\label{eq:a41}
\end{eqnarray}
Transforming Equation (\ref{eq:a41}) to the coordinate basis we have for $G^{\alpha}$,
\begin{equation}
G^{\alpha} = - \frac{4\pi c}{(hc)^{3}} \int \epscm^{2} d\epscm \; \left( \frac{1}{a} \sum_{q} \left( \frac{ d }{ d\ell } \fcmi{ q}  \right)_{ \! S }^{(0)}, \frac{1}{b} \sum_{q} \left( \frac{ d }{ d\ell } \fcmi{ q} \right)_{ \! S }^{(1)}, 0, 0 \right) .
\label{eq:a42}
\end{equation}

The energy equation is obtained by projecting Equation (\ref{eq:a40}) along the fluid four-velocity: 
\begin{equation}
u^{\alpha \, (m)}T_{\alpha \; ;\beta}^{\beta} = u^{\alpha} G_{\alpha} = - \frac{4\pi c^{2}}{(hc)^{3}} \int \epscm^{2} d\epscm \; \sum_{q} \left( \frac{ d }{ d\ell } \fcmi{ q} \right)_{ \! S }^{(0)} ,
\label{eq:a43}
\end{equation}
where we have used Equation~(\ref{eq:a42}) for $G^{\alpha}$.
Using Equation~(\ref{eq:a34}) for $^{(m)}T^{\alpha \beta}$ on the left-hand side of Equation~(\ref{eq:a40}), invoking conservation of rest mass [$u^{\beta}_{\; ;\beta} = - u^{\beta} \rho_{,\beta}/\rho$], and denoting by $e_{\rm m}$ the internal energy per unit rest mass ($e_{\rm m} = E_{\rm m}/\rho$), $u^{\alpha \, (m)}T_{\alpha \; ;\beta}^{\beta}$ is given by
\begin{equation}
u^{\alpha \, (m)}T_{\alpha \; ;\beta}^{\beta} = \rho \left[ \frac{1}{a} \pderiv{e_{m}}{t} + P_{m} \frac{1}{a} \pderiv{(1/\rho )}{t} \right] ,
\label{eq:a43a}
\end{equation}
and Equation~(\ref{eq:a43}) becomes
\begin{equation}
\frac{1}{a} \left( \pderiv{e_{\rm m}}{t} \right)_{m} = - p_{\rm m} \left( \frac{1}{a} \pderiv{ \left( 1 / \rho  \right) }{t} \right)_{m} - \frac{4\pi c^{2}}{(hc)^{3}} \frac{1}{\rho}\int \epscm^{2} d\epscm \; \sum_{q} \left( \frac{ d }{ d\ell } \fcmi{ q} \right)_{ \! S }^{(0)} ,
\label{eq:a44}
\end{equation}
which states that changes in the matter internal energy arise from work by local compression or expansion and from energy exchange with neutrinos.

The other nontrivial equation in (\ref{eq:a40}) is the radial equation
\begin{equation}
^{(m)}T_{\; \; \; \; ;\beta}^{1\beta} = G^{1} .
\label{eq:a45}
\end{equation}
With $^{(m)}T_{\; \; \; \; ;\beta}^{1\beta}$ given by
\begin{equation}
^{(m)}T_{\; \; \; \; ;\beta}^{1\beta} = \frac{1}{b^{2}} \left[ \pderiv{\ln a}{m} \left( \rho c^{2} + E_{m} + P_{m} \right) + \pderiv{P_{m}}{m} \right] 
= \frac{\Gamma}{b} \left[ \pderiv{\ln a}{R} \left( \rho c^{2} + E_{m} + P_{m} \right) + \pderiv{P_{m}}{R} \right] ,
\label{eq:a45a}
\end{equation}
Equation (\ref{eq:a45}) gives
\begin{equation}
\frac{ d \ln a }{dR} = - \left( \frac{ dP_{m} }{dR}  + \frac{4\pi c }{ \Gamma (hc)^{3}} \int \epscm^{2} d\epscm \; \sum_{q} \left( \frac{ d }{ d\ell } \fcmi{ q} \right)_{ \! S }^{(1)}  \right) \frac{1}{\rho w c^{2} } ,
\label{eq:a46}
\end{equation}
where $w$ is given by Equation (\ref{eq:a47}).

Equation~(\ref{eq:a46}) relates the spatial gradient of the matter pressure, $P_{\rm m}$, and the rate of neutrino--matter momentum exchange to the spatial gradient of the lapse function.
Upon integrating $a$ from the surface using Equation~(\ref{eq:a46}), we start with the surface (subscript s) boundary condition, which we take to be the Schwarzschild solution on the exterior, and neglect the small deviations from this induced by the radiation; namely,
\begin{equation}
a_{s} = \left( 1 - \frac{ 2 G M_{s} }{ c^{2} R_{s} } \right) \frac{1}{\Gamma_{s}} ,
\label{eq:a48}
\end{equation}
in order that the coordinate time is that of a distant clock.
In practice our models are extended enough that $a_{s}$ is extremely close to unity.
The differencing of the lapse is given below by Equations~(\ref{eq:a62k}) and (\ref{eq:a62l}).

\subsection{Matter--Neutrino Lepton Exchange}
\label{dyedt}

The local lepton number density of the matter, $^{(m)}n_{\ell}$, is given from charge conservation by 
\begin{equation}
^{(m)}n_{\ell} = n_{\rm p} = n_{\rm e^{-}} - n_{\rm e^{+}} = \frac{\rho}{m_{\rm B}} \Ye ,
\label{eq:a49}
\end{equation}
where $n_{\rm p}$, $n_{\rm B}$, $n_{\rm e^{-}}$, and $n_{\rm e^{+}}$ are the proper proton, baryon, electron, and positron number densities, respectively, $\Ye$ is the proton fraction, and $m_{\rm B}$ is the mean baryon mass.
Referring to Equation (\ref{eq:a7}), the local lepton number density in neutrinos, $^{(\nu)}n_{\ell}$, is given by
\begin{equation}
^{(\nu)}n_{\ell} = \frac{1}{(hc)^{3}} \int \left( \fcmi{ \nue} - \fcmi{ \nuebar} \right)  \epscm^{2} d \epscm d\Omega_{p} .
\label{eq:a50}
\end{equation}
Transport produces global changes in $^{(\nu)}n_{\ell}$, which by itself does not change $^{(m)}n_{\ell}$ (or \Ye).
Ignoring the effects of transport, the total lepton number, $^{\rm total}n_{\ell} = ^{(m)}n_{\ell} + ^{(\nu)}n_{\ell}$, is a locally conserved quantity, and this allows us to relate the change in \Ye\ to the change in $^{(\nu)}n_{\ell}$.
In particular,
\begin{equation}
0 = \left( ^{\rm total}n_{\ell} u^{\alpha} \right)_{;\alpha} = \left( \frac{\rho}{m_{\rm B}} \Ye u^{\alpha} \right)_{;\alpha} + \left( ^{(\nu)}n_{\ell} u^{\alpha} \right)_{;\alpha} .
\label{eq:a51}
\end{equation}
Expanding the covariant derivative of \Ye\ with the use of Equation (\ref{eq:a3b}), which expresses rest mass conservation, Equation (\ref{eq:a51}) can be written
\begin{equation}
\left( \frac{\rho}{m_{\rm B}} \Ye u^{\alpha} \right)_{;\alpha} = \frac{\rho}{m_{\rm B}} \frac{1}{a} \left( \pderiv{ \Ye }{t} \right)_{m} = - \left( ^{(\nu)}n_{\ell} u^{\alpha} \right)_{;\alpha} .
\label{eq:a52}
\end{equation}
Now expanding the covariant derivative of $^{(\nu)}n_{\ell}$, Equation (\ref{eq:a52}) becomes
\begin{equation}
\frac{\rho}{m_{\rm B}} \frac{1}{a} \left( \pderiv{ \Ye }{t} \right)_{m} 
 = - \frac{\rho}{a} \left( \pderiv{ \left( ^{(\nu)}n_{\ell}/\rho \right) }{t} \right)_{m}.
\label{eq:a53}
\end{equation}
With the use of Equations (\ref{eq:a8}) (without the transport terms), (\ref{eq:a2}), (\ref{eq:a3c}), and (\ref{eq:a7}), the term on the right-hand side of Equation (\ref{eq:a53}) becomes
\begin{eqnarray}
 - \frac{\rho}{a} \left( \pderiv{ \left( ^{(\nu)}n_{\ell}/\rho \right) }{t} \right)_{m} = && - \frac{1}{(hc)^{3}} \int \epscm^{2} d \epscm d\Omega_{p} \left( \frac{1}{a} \left( \pderiv{ \fcmi{ \nue} }{t} \right)_{m} - \frac{1}{a} \left( \pderiv{ \fcmi{ \nuebar} }{t} \right)_{m} \right)  \nonumber \\ 
&& = - \frac{1}{(hc)^{3}} \int \epscm^{2} d \epscm d\Omega_{p} \frac{ c^{2} }{ \epscm } \left[ \left( \frac{ d  \fcmi{ \nue} }{ d\ell } \right)_{ \! S } - \left( \frac{ d  \fcmi{ \nuebar} }{ d\ell } \right)_{ \! S } \right]  \nonumber \\
&& = \frac{c}{(hc)^{3}} \int \epscm^{2} d \epscm  \left[ \left( j_{\nue}(\epscm) \left( 1 - \psimomcmi{0}{\nue}(\epscm) \right) - \frac{ \psimomcmi{0}{\nue}(\epscm) }{ \lambda^{a}_{\nu_{\rm e}}(\epscm)} \right) \right . \nonumber \\
&& \, \, \left. - \left( j_{\bar{\nu}_{\rm e}}(\epscm) \left( 1 - \psimomcmi{0}{\nuebar}(\epscm) \right) - \frac{ \psimomcmi{0}{\nuebar}(\epscm) }{ \lambda^{a}_{\nuebar}(\epscm)} \right) \right] ,
\label{eq:a54}
\end{eqnarray}
where the last expression in Equation (\ref{eq:a54}) arises from the fact that the only terms that do not vanish upon integrating over solid angle and neutrino number are emission and absorption, and the latter are isotropic.
Here $j_{\nu_{\rm e}}(\epscm)$ and $1/\lambda^{a}_{\nue}(\epscm)$ are the electron-neutrino emission and absorption per state per unit length, and  $j_{\nuebar}(\epscm)$ and $1/\lambda^{a}_{\nuebar}(\epscm)$ are the electron-antineutrino counterparts.
Equations (\ref{eq:a51})--(\ref{eq:a54}) give the following equation for the evolution of the electron fraction due to sources:
\begin{equation}
\left( \pderiv{ \Ye }{t} \right)_{m} = - \frac{4\pi ac}{(hc)^{3}} \, \frac{ m_{\rm B}}{\rho} \int \epscm^{2} d\epscm  \left[ \left( j_{\nu_{\rm e}}(\epscm) \left( 1 - \psimomcmi{0}{\nue}(\epscm) \right) - \frac{ \psimomcmi{0}{\nue}(\epscm) }{ \lambda_{\nue}(\epscm)} \right) - \left( j_{\nuebar}(\epscm) \left( 1 - \psimomcmi{0}{\nuebar}(\epscm) \right) - \frac{ \psimomcmi{0}{\nuebar}(\epscm) }{ \lambda_{\bar{\nu}_{\rm e}}(\epscm)} \right) \right] .
\label{eq:a55}
\end{equation}

\subsection{Lagrangian Transport Step}
\label{sec:LTS}

In the Lagrangian transport step, the pair of transport Equations (\ref{eq:a30}) and (\ref{eq:a19}) (or, rather, Equation~(\ref{eq:a57}) in place of Equation~(\ref{eq:a30}), where Equation~(\ref{eq:a57}) is Equation~(\ref{eq:a30}) written in conservative form) are solved along with the neutrino-specific term of the energy Equation~(\ref{eq:a44}), namely,
\begin{equation}
\frac{1}{a} \left( \pderiv{e_{\rm m}}{t} \right)_{m} =  - \frac{4\pi c^{2}}{(hc)^{3}} \int \epscm^{2} d\epscm \; \sum_{q} \left( \frac{ d }{ d\ell } \fcmi{ q} \right)_{ \! S }^{(0)} ,
\label{eq:a56}
\end{equation}
and Equation~(\ref{eq:a55}) for the change in the matter proton number. 
The pair of Equations~(\ref{eq:a30}) and (\ref{eq:a19}) for each neutrino species and Equation~(\ref{eq:a55}) are solved together fully implicitly, while Equation~(\ref{eq:a56}) is solved immediately afterwards, as described in more detail below. 
It was found that executing Equation~(\ref{eq:a56}) subsequent to the others leads to a substantially better conditioned set of equations and no loss of stability and that the internal energy update could then be easily omitted for those radial zones for which the internal energy update is derived from a total energy update. 
The role of neutrinos in the latter case will be described later. 
In either case the updated temperature is obtained from the updated internal energy and electron fraction by numerically inverting the equation of state. 

Equation (\ref{eq:a30}), written in conservative form, is given by
\begin{equation}
\frac{1}{a c} \left( \pderiv{ \psimomcmi{0}{q} }{t} \right)_{m,a\epsilon}^{(m)} + \Gamma \frac{ a^{2} }{ R^{2} } \left( \pderiv{}{R} \left(\frac{ R^{2} }{ a^{2} }\psimomcmi{1}{q}\right) \right)_{t,e} = \frac{c}{\epscm} \left( \frac{ d }{ d\ell } \fcmi{ q}(x,p) \right)_{ \! S }^{(0)} ,
\label{eq:a57}
\end{equation}
where we have omitted the superscript $T$. 
Consider a radial mesh where, as before, the zone edges are labeled by $i+\rfrac{1}{2}$ and the zone centers by $i$.
Consider also an energy mesh labeled similarly by $k$. Let the subscript $q$ denote, as before, the neutrino species, and let the superscript $n$ denote the $n$-th time slice. 
Then for each neutrino species, Equation~(\ref{eq:a57}) is differenced conservatively as follows:
\be
& {\ds \frac{1}{a_{i}c} \frac{ \psimomcmin{0}{i,k,q}{n+1} - \psimomcmin{0}{i,k,q}{n} }{ \Delta t^{n+\frac{1}{2}} } + 
} & \nonumber \\ 
& {\ds + \frac{a_{i}^{2}}{ \Vol{i} } \left[ \frac{1}{a_{i+\frac{1}{2}}^{2}}\Area{i+\frac{1}{2}} \psimomcmin{1}{i+\frac{1}{2},k,q}{n+1} - \frac{1}{a_{i-\frac{1}{2}}^{2}}\Area{i-\frac{1}{2}}  \psimomcmin{1}{i-\frac{1}{2},k,q}{n+1} \right] = \frac{c}{\epscm} \left( \frac{ d }{ d\ell } \fcm(x,p) \right)_{ \! S \; i,k,q }^{(0) \, n+1} ,
} &
\label{eq:a58} 
\ee
with \psimomcmin{1}{i+\frac{1}{2},k,q}{n+1}\ given by
\begin{equation}
 \psimomcmin{1}{i+\frac{1}{2},k,q}{n+1} = - \frac{ \lambda^{(t) \, n+1}_{i+\frac{1}{2},k,q} }{3} {\cal F}^{n+1}_{i+\frac{1}{2},k,q} \Gamma^{n+1}_{i+\frac{1}{2}} \frac{ \psimomcmin{0}{i+1,k,q}{n+1} -  \psimomcmin{0}{i,k,q}{n+1} }{ R_{i+1} - R_{i} }  , 
\label{eq:a59}
\end{equation}
(or by its modification given by Equation (\ref{eq:a93}) below), where $\Area{i+\frac{1}{2}} = 4\pi R_{i+\frac{1}{2}}^{2}$ is the proper area of the outer boundary of zone $i$, $R_{i}  = ( R_{i-\frac{1}{2}} + R_{i+\frac{1}{2}} )/2$, $\Vol{i} = 4\pi \left( R_{i+\frac{1}{2}}^{3} -  R_{i-\frac{1}{2}}^{3} \right)/3 \Gamma$ is the proper volume enclosed between zones edges $i-\rfrac{1}{2}$ and $i+\rfrac{1}{2}$, $\Delta t^{n+\frac{1}{2}} = t^{n+1} - t^{n}$, and $\lambda^{(t) \, n+1}_{i+\frac{1}{2},k,m} = \sqrt{ \lambda^{(t) \, n+1}_{i,k,m} \lambda^{(t) \, n+1}_{i+1,k,m} }$.
To demonstrate that  Equation~(\ref{eq:a58}) is conservative in neutrino number, first multiply it by $c \, \Vol{i} \, \Delta t^{n+\frac{1}{2}}/a_{i}^{2}$ to  rewrite it as
\begin{eqnarray}
 \frac{1}{a^{3}_{i}} \left( \psimomcmin{0}{i,k,q}{n+1} - \psimomcmin{0}{i,k,q}{n} \right) \Vol{i} = && - \frac{1}{a_{i+\frac{1}{2}}^{3}}\Area{i+\frac{1}{2}} \psimomcmin{1}{i+\frac{1}{2},k,q}{n+1} a_{i+\frac{1}{2}} \Delta t^{n+\frac{1}{2}} c  + \frac{1}{a_{i-\frac{1}{2}}^{3}}\Area{i-\frac{1}{2}} \psimomcmin{1}{i-\frac{1}{2},k,q}{n+1} a_{i-\frac{1}{2}} \Delta t^{n+\frac{1}{2}} c   \nonumber \\ 
 &&+ a_{i} \Delta t^{n+\frac{1}{2}} \frac{c}{a^{3}_{i}} \Vol{i} \times \frac{c}{\epscm} \left( \frac{ d }{ d\ell } \fcm(x,p) \right)_{ \! S \; i,k }^{(0)\, n+1}.
\label{eq:a60} 
\end{eqnarray}
According to the discussion following Equation~(\ref{eq:a9}) and our choice of the lapse function limit, $a(R, t) \rightarrow 1$ as $R \rightarrow \infty$, a neutrino propagating along a geodesic passing through $R_{i-\frac{1}{2}}$, $R_{i}$, and $R_{i+\frac{1}{2}}$ in a static space-time will have locally measured energies at these points related by 
\begin{equation}
\epscmi{ i-\frac{1}{2}} a_{i-\frac{1}{2}} = \epscmi{ i} a_{i} = \epscmi{ i+\frac{1}{2}} a_{i+\frac{1}{2}} = \epscmi{\infty} .
\label{eq:a61}
\end{equation}
Thus, multiplying Equation~(\ref{eq:a60}) by $(hc)^{-3}(\epscmi{k \; \infty})^{2} \Delta \epscmi{k \; \infty}$ and using Equation~(\ref{eq:a61}), we get 
\begin{eqnarray}
 \frac{ (\epscmi{i, k})^{2} \Delta \epscmi{i, k} }{(hc)^{3}} \left( \psimomcmin{0}{i,k,q}{n+1} - \psimomcmin{0}{i,k,q}{n} \right) \Vol{i} &=& - \Area{i+\frac{1}{2}} \frac{ c \, (\epscmi{ i+\frac{1}{2}, k})^{2} \Delta \epscmi{ i+\frac{1}{2}, k}}{(hc)^{3}} \psimomcmin{1}{i+\frac{1}{2},k,q}{n+1} a_{i+\frac{1}{2}} \Delta t^{n+\frac{1}{2}} \nonumber \\
 &&+ \Area{i-\frac{1}{2}} \frac{ c \, (\epscmi{ i-\frac{1}{2}, k})^{2} \Delta \epscmi{i-\frac{1}{2}, k}}{(hc)^{3}} \psimomcmin{1}{i-\frac{1}{2},k,q}{n+1} a_{i-\frac{1}{2}} \Delta t^{n+\frac{1}{2}}\nonumber \\  
 &&+ a_{i} \Delta t^{n+\frac{1}{2}} \frac{ (\epscmi{ i, k})^{2} \Delta \epscmi{ i, k} } {(hc)^{3}} \Vol{i} \times \left( \frac{ d }{ dt } \fcm \right)_{ \! S \; i,k }^{(0)\, n+1} .
\label{eq:a62} 
\end{eqnarray}	
The left-hand side of Equation~(\ref{eq:a62}) is the change in the number of neutrinos in $\Vol{i}$ between energies $\epscmi{ i,k}$ and $\epscmi{ i,k} + \Delta \epscmi{ i,k}$ during the proper time interval $a_{i} \Delta t^{n+\frac{1}{2}}$. The first term on the right-hand side subtracts the number of neutrinos that leave through the outer surface of  $\Vol{i}$ between energies $\epscmi{ i+\frac{1}{2},k}$ and $\epscmi{ i+\frac{1}{2},k} + \Delta \epscmi{ i+\frac{1}{2},k}$ in proper time $a_{i+\frac{1}{2}} \Delta t^{n+\frac{1}{2}}$. The second term on the right-hand side adds the number of neutrinos that enter through the inner surface of  $\Vol{i}$ between energies $\epscmi{ i-\frac{1}{2},k}$ and $\epscmi{ i-\frac{1}{2},k} + \Delta \epscmi{ i-\frac{1}{2},k}$ in proper time $a_{i-\frac{1}{2}} \Delta t^{n+\frac{1}{2}}$. The last term is the net number of neutrinos emitted, absorbed, or scattered in/out of $\Vol{i}$ between energies $\epscmi{ i,k}$ and $\epscmi{ i,k} + \Delta \epscmi{ i,k}$ in proper time $a_{i} \Delta t^{n+\frac{1}{2}}$.

Equation (\ref{eq:a55}) is differenced straightforwardly as
\begin{eqnarray}
 \frac{1}{a_{i}c} \frac{  \Yein{i}{n+1} - \Yein{i }{n} }{\Delta t} = && - \frac{4\pi }{(hc)^{3}} \, \frac{ m_{\rm B}}{\rho_{i}} \sum_{k=1}^{N_{k}} \epscmi{ i,k}^{2} \Delta \epscmi{ i,k}  \nonumber \\ 
&& \times \left[ \left( j_{\nue \, i,k}^{n+1} \left( 1 - \psimomcmin{0}{ i,k,\nue}{n+1} \right) - \frac{ \psimomcmin{0}{ i,k,\nue}{n+1} }{ \lambda_{\nue \, i,k}^{n+1}} \right) - \left( j_{\nuebar \, i,k}^{n+1} \left( 1 - \psibarmomcmin{0}{ i,k,\nue}{n+1} \right) - \frac{ \psibarmomcmin{0}{ i,k,\nue}{n+1} }{ \lambda_{\nuebar \, i,k}^{n+1}} \right) \right] .
\label{eq:a62a}
\end{eqnarray}
During the Lagrangian transport step, the set of Equations~(\ref{eq:a58}) and (\ref{eq:a59}), modified in the presence of shocks as given by Equations~(\ref{eq:a93}) and (\ref{eq:a99}), and (\ref{eq:a62a}) are solved implicitly for \Yein{i}{n+1}\ and \psimomcmin{0}{ i,k,q}{n+1}\ for each neutrino species $q$.

The temperature change accompanying the Lagrangian transport step is computed following the solutions of Equations~(\ref{eq:a58}), (\ref{eq:a59}), and (\ref{eq:a62a}), from the change in the internal energy as given by the following differenced version of Equation~(\ref{eq:a56}):
\begin{equation}
\frac{1}{a_{i}c} \frac{ e_{ {\rm int},i }^{n+1} - e_{ {\rm int},i }^{n} }{ \Delta t } = - \frac{4\pi }{(hc)^{3}} \, \frac{ m_{\rm B}}{\rho_{i}}  \sum_{q=1}^{N_{\nu}} {\cal N}_{{\rm stwt},q } \sum_{k=1}^{N_{k}} \epscmi{ i,k}^{3} \Delta \epscmi{ i,k} \frac{c}{\epscm} \left( \frac{ d }{ d\ell } \fcm(x,p) \right)_{ \! S \, i, k, q }^{(0) \, n+1} ,
\label{eq:a62b}
\end{equation}
and the change in \Ye\ given by Equation~(\ref{eq:a62a}), where $N_{\nu}$ is the number of neutrino species (typically four), and ${\cal N}_{{\rm stwt},q }$ is the statistical weight of each neutrino species, $q$, (typically 1 for \nue\ and \nuebar, and 2 for \numt\ and \numtbar).
Specifically,
\begin{equation}
T_{i}^{n+1} = T_{i}^{n} + \frac{  e_{ {\rm int},i }^{n+1} - e_{ {\rm int}, i }^{n} + \left( \Yein{i}{n+1} - \Yein{i }{n} \right) \left( \ds \frac{ d e_{ {\rm int}}}{ d\Ye } \right)_{ \! \rho, \Ye \; i }^{ n+\frac{1}{2}} }{ \ds \left( \frac{ d e_{ {\rm int} }}{ dT } \right)_{ \! \rho, \Ye \; i }^{ n+\frac{1}{2}} } .
\label{eq:a62d}
\end{equation}

It was found that the temperature change during a time step always had very little effect on the terms in the transport equation, and therefore on the stability of the difference scheme.
On the other hand, including the temperature change as part of the implicit solution of Equations~(\ref{eq:a58}), (\ref{eq:a59}), and (\ref{eq:a62a}) caused the system of equations to be rather ill-conditioned.
It was therefore deemed numerically expedient to solve for the temperature change after, rather than simultaneously with, the solution of the transport equations.
The solution of Equations~(\ref{eq:a58}), (\ref{eq:a59}), and (\ref{eq:a62a}) followed by Equation (\ref{eq:a62b}) and (\ref{eq:a62d}) completes the Lagrangian transport step.

\subsection{Solution of the Transport Equations}
\label{trans_solution}

The numerical solution of the set of transport equations from time step $n$ to time step $n+1$ proceeds by an outer iteration for the corrections $\delta \psimomcmin{0}{ i,k,q}{n,\,i+1}$ and $\delta \Yein{i }{n, \, i+1}$ to the quantities $\psimomcmin{0}{ i,k,q}{n, \, i}$ and $\delta \Yein{i }{n, \, i}$; i.e.,
\begin{equation}
\psimomcmin{0}{ i,k,q}{n,\,i+1} = \psimomcmin{0}{ i,k,q}{n, \, i} + \delta \psimomcmin{0}{ i,k,q}{n,\,i+1}, \qquad \Yein{i }{n, \, i+1} = \Yein{i }{n, \, i} + \delta \Yein{i }{n, \, i+1}
\label{eq:a62aa}
\end{equation}
where the superscript `$i$' denotes the quantities after the ith iteration. 
When the corrections have become sufficiently small, the quantities $\psimomcmin{0}{ i,k,q}{n, \, i+1}$ and $\Yein{i }{n, \, i+1}$ are regarded as having converged to $\psimomcmin{0}{ i,k,q}{n+1}$ and $\Yein{i }{n+1}$, respectively. The criterion for convergence is that 
\begin{equation}
\delta \psimomcmin{0}{ i,k,q}{n,\,i+1} < \frac{ {\rm tol}_{\psi} }{ \max( \psimomcmin{0}{ i,k,q}{n, \, i+1}, \, {\rm tolmin}_{\psi} ) } \qquad
\delta \Yein{i }{n, \, i+1} < \frac{ {\rm tol}_{Y_{\rm e} } }{ \max(\Yein{i }{n, \, i+1}, \, {\rm tolmin}_{Y_{\rm e} } ) } 
\label{eq:a62ab}
\end{equation}
for all $\psimomcmin{0}{ i,k,q}{n, \, i+1}$ and $\Yein{i }{n, \, i+1}$, where typically ${\rm tol}_{\psi} = 10^{-6}$, ${\rm tol}_{Y_{\rm e} } = 10^{-6}$, ${\rm tolmin}_{\psi} = 1$, and ${\rm tolmin}_{Y_{\rm e} } = 0.1$.

The transport equations, Equations~(\ref{eq:a58}) and (\ref{eq:a59}), when linearized comprise a set $N_{\nu} \times N_{k}$ coupled linear equations of the form:
\be
& {\ds \mbox{\small LHS0}^{i}_{i,n_{q}+k} + 
 \mbox{\small LHS0}^{i}_{{\rm p0p} \, i,n_{q}+k} \delta \psimomcmin{0}{ i,k,q}{n,\,i+1}
+ \mbox{\small LHS0}^{i}_{{\rm p0m} \,  \, i,n_{q}+k } \delta \psimomcmin{0}{ i-1,k,q}{n,\,i+1} 
+ \mbox{\small LHS0}^{i}_{{\rm p0} \,  \, i,n_{q}+k } \delta \psimomcmin{0}{ i+1,k,q}{n,\,i+1}
} & \nonumber \\
& {\ds = \mbox{\small RHS0}^{i}_{i,n_{q}+k} +  \mbox{\small RHS0}^{i}_{{\rm y} \, i,n_{q}+k} \delta Y_{{\rm e},i}^{n, \; i+1} 
+ \sum_{q'=1}^{N_{\nu}} \sum_{k'=1}^{N_{k}} \mbox{\small RHS0}^{i}_{{\rm p0} \, i,n_{q}+k, n'_{q'} + k'} \delta \psimomcmin{0}{ i,k',q'}{n,\,i+1} ,
} &  
\label{eq:a62ac}
\ee
where the coefficients `{\small LHS0}' and `{\small RHS0}' are independent of the corrections, and where $n_{q} = (N_{\nu} - 1) \times q$, and the notation `{\small LHS0}' and `{\small RHS0}' denote the left-hand side (transport) and right-hand side (sources) of the transport equations, respectively. Equation (\ref{eq:a62a}) provides an additional equation, which is coupled to the preceding set of equations, and when linearized is of the form
\be
& {\ds \mbox{\small LHS0}^{{\rm Ye} \, i}_{i} + \mbox{\small LHS0}^{{\rm Ye} \, i}_{{\rm y} \, i} \delta Y_{{\rm e},i}^{n, \, i+1}
 = \mbox{\small RHS0}^{{\rm Ye} \, i}_{i} +  \mbox{\small RHS0}^{{\rm Ye} \, i}_{{\rm y} \, i} \delta Y_{{\rm e}, i}^{n, \; i+1} 
} & \nonumber \\
& {\ds + \sum_{k=1}^{N_{\rm e}} \mbox{\small RHS0}^{{\rm Ye} \, i}_{{\rm p0} \, i, k,q=1} \delta \psimomcmin{0}{i, k, q=1}{n,\, i+1}
+  \sum_{k=1}^{N_{\rm e}} \mbox{\small RHS0}^{{\rm Ye} \, i}_{{\rm p0} \, i,k,q=2} \delta \psimomcmin{0}{i, k, q=2}{n,\, i+1}
}, &  
\label{eq:a62ad}
\ee
where $q=1$ and $q=2$ refer to \nue\ and \nuebar, respectively.
Equations (\ref{eq:a62ac}) and (\ref{eq:a62ad}) are combined to form a matrix equation for the corrections, of the form
\begin{equation}
\sum_{\alpha' = 1}^{N_{\nu} \times N_{k}+1} A_{i,\alpha, \alpha'} \delta u_{i,\alpha'} + Bm_{i,\alpha} \delta u_{i-1,\alpha} + Bp_{i,\alpha} \delta u_{i+1,\alpha} + C_{i,\alpha}, \quad i = 1, \cdots, I, \; \alpha = 1, \cdots, N_{\nu} \times N_{k}+1,
\label{eq:a62ae}
\end{equation}
where the corrections $\delta \psimomcmin{0}{ i,k,q}{n,\,i+1}$ and $\delta Y_{{\rm e},i}^{n, \; i+1}$ are incorporated in the array $\delta u_{i,\alpha}, \alpha = 1, \cdots, N_{\nu} \times N_{k}+1$. 

     The boundary condition at the center is simply that the neutrino flux is zero; i.e., $\psimomcmin{1}{i=1, k, q}{} = 0$. 
Constant luminosity is assumed at the surface, which requires that $\psimomcmin{1}{I+1, k, q}{} R_{I+1}^{2}  = \psimomcmin{1}{I, k, q}{} R_{I}^{2}$.
Following the discussion immediately above Equation (\ref{eq:a23}), a geometric relation between \psimomcmin{1}{I, k, q}{} and \psimomcmin{0}{I, k, q}{} of the form $\psimomcmin{1}{I, k, q}{} = (1/2) ( 1 + \mucmi{ I, k, q} ) \psimomcmin{0}{I, k, q}{}$ is assumed. Then, without velocity or GR corrections to \mucmi{ I, k, q}, the surface boundary condition for \psimomcmin{0}{I, k, q}{} becomes
\begin{equation}
\psimomcmin{0}{I+1, k, q}{} = \psimomcmin{0}{I, k, q}{}  \frac{ R_{I}^{2} - \frac{1}{4} R_{\nu \, k,q} }{ R_{I+1}^{2} - \frac{1}{4} R_{\nu \, k,q} },
\label{eq:a62af}
\end{equation}
where $R_{\nu \, k,q}$ is the neutrinosphere radius for neutrinos of energy $k$ and type $q$.

\subsection{Neutrino Stress and the Lapse}
\label{stress}

Once \psimomcmin{0}{i, k}{n+1}\ has been obtained, the neutrino--matter stress $f_{\nu,r}$ given by Equation~(\ref{eq:aa47}) and used in Equations~(\ref{eq:h74}) and (\ref{eq:h80}) for the radial hydrodynamics solve is computed as follows.
Using Equation~(\ref{eq:a17}) and then (\ref{eq:a19}) in Equation~(\ref{eq:aa47}), we have
\begin{eqnarray}
 f_{\nu,r} &=& - \frac{\Gamma}{ \rho w } \frac{4\pi c }{(hc)^{3}} \in	t \epscm^{2} d\epscm \; \sum_{q} \left( \frac{ d }{ d\ell } f_{q}(x,p) \right)_{ \! S }^{(1)} \nonumber \\ 
&= & \frac{\Gamma}{ \rho w } \frac{4\pi }{(hc)^{3}} \int \epscm^{3} d\epscm \; \sum_{q} \frac{ \psimomcmi{1}{q}( \epscm )}{ \lambda_{q}^{t}(\epscm) } \nonumber \\
&= & - \frac{\Gamma}{ \rho w } \frac{4\pi }{(hc)^{3}} \int \epscm^{3} d\epscm \; \sum_{q} \frac{ 1 }{3} {\cal F}_{q}(\epscm) \, \Gamma \left( \frac{ \partial \psimomcmi{0}{q}(\Ecm) }{ \partial R } \right)_{\! t,\Ecm} \nonumber  \\
&\rightarrow & - \frac{1}{ \rho } \frac{4\pi }{(hc)^{3}} \int \epscm^{3} d\epscm \; \sum_{q} \frac{ 1 }{3} {\cal F}_{q}(\epscm) \,  \left( \frac{ \partial \psimomcmi{0}{q}(\Ecm) }{ \partial R } \right)_{\! t,\Ecm}
\label{eq:a62c}
\end{eqnarray}
where in the last equation we have set $\Gamma$ and $w$ to their nonrelativistic values of unity. Equation (\ref{eq:a62c}) is differenced as
\begin{equation}
f_{\nu,r \, i+\frac{1}{2}} = - \frac{1}{ \rho_{i+\frac{1}{2}} } \frac{4\pi }{(hc)^{3}} \frac{ 1 }{3} \sum_{k=1}^{N_{k}} \sum_{q=1}^{N_{\nu}} {\cal N}_{{\rm stwt},q } {\cal F}^{n+1}_{i+\frac{1}{2},k,q} A_{i+\frac{1}{2}} \rho_{i+\frac{1}{2}}
\frac{  \epscmi{ i+1,k}^{3} \Delta \epscmi{ i+1,k} \psimomcmi{0}{i+1,k,q} - \epscmi{i,k}^{3} \Delta \epscmi{ i,k} \psimomcmi{0}{i,k,q} }{ \frac{1}{2} \left( \Delta M_{i} + \Delta M_{i+1} \right) } .
\label{eq:a62e}
\end{equation}
The replacement
\begin{equation}
\epscmi{i+\frac{1}{2},k}^{3} \Delta \epscmi{ i+\frac{1}{2},k} \frac{ \psimomcmi{0}{i+1,k,q} - \psimomcmi{0}{i,k,q} }{ \left( R_{i+1} - R_{i} \right) }
\rightarrow A_{i+\frac{1}{2}} \rho_{i+\frac{1}{2}}
\frac{  \epscmi{ i+1,k}^{3} \Delta \epscmi{ i+1,k} \psimomcmi{0}{i+1,k,q} - \epscmi{i,k}^{3} \Delta \epscmi{ i,k} \psimomcmi{0}{i,k,q} }{ \frac{1}{2} \left( \Delta M_{i} + \Delta M_{i+1} \right) }
\label{eq:a62f}
\end{equation}
has been made so the final expression for the neutrino stress will have the correct limiting behavior in the short neutrino mean free path limit (${\cal F}^{n+1}_{i+\frac{1}{2},k,q} \rightarrow 1$)
\begin{equation}
f_{\nu,r \, i+\frac{1}{2}} \rightarrow \sum_{k=1}^{N_{k}} \sum_{q=1}^{N_{\nu}} {\cal N}_{{\rm stwt},q } \frac{ \left( p_{\nu \, i+1,k,q} - p_{\nu \, i,k,q} \right) A_{i+\frac{1}{2} } }{ \frac{1}{2} \left( \Delta M_{i} + \Delta M_{i+1} \right) } ,
\label{eq:a62g}
\end{equation}
which is the force per unit mass exerted by an isotropic relativistic gas, and in the long neutrino mean free path limit which, with the help of Equation (\ref{eq:a19}), is given by
\begin{equation}
f_{\nu,r \, i+\frac{1}{2}} \rightarrow \frac{1}{ \rho_{i+\frac{1}{2} } } \frac{4\pi }{(hc)^{3}} \sum_{k=1}^{N_{k}} \epscmi{i,k}^{3} \Delta \epscmi{ i,k}  \sum_{q=1}^{N_{\nu}} {\cal N}_{{\rm stwt},q } \frac{ \psimomcmi{1}{i+\frac{1}{2},k,q}}{ \lambda_{i+\frac{1}{2},k,q}^{t} } .
\label{eq:a62h}
\end{equation}
where we assume that at large $R$, where the neutrino mean free paths are large, $\epscmi{i+1,k} \simeq \epscmi{i,k}$. As before, $\Delta M_{i}$ denotes the zone mass.
Finally, the zone-centered neutrino stress is computed as the average of the edge values, viz.,
\begin{equation}
f_{\nu,r \, i} = \frac{1}{2} \left( f_{\nu,r \, i+\frac{1}{2}} + f_{\nu,r \, i-\frac{1}{2}} \right) .
\label{eq:a62i}
\end{equation}

To update the lapse given by Equation (\ref{eq:a46}), note first that a comparison of Equation (\ref{eq:a46}) with Equation (\ref{eq:aa47}) shows that
\begin{equation}
\left(  \frac{4\pi c }{ \Gamma (hc)^{3}} \int \epscm^{2} d\epscm \; \sum_{q} \left( \frac{ d }{ d\ell } \fcmi{ q} \right)_{ \! S }^{(1)}  \right) \frac{1}{\rho w c^{2} } 
= \frac{ f_{\nu,r} }{ \Gamma^{2} c^{2} } .
\label{eq:a62j}
\end{equation}
Setting $\Gamma = 1$, $M_{\rm g} \; [\mbox{given by Equation (\ref{eq:a37})}] = M \equiv M_{\rm rest \, mass}$, and $w = 1$, the  update of the lapse begins at the outer edge with 
\begin{equation}
a_{I+1} = a_{I+\frac{1}{2}} = \left( 1 - \frac{ 2 G M_{I+\frac{1}{2}} }{ c^{2} R_{I+\frac{1}{2}} } \right) ,
\label{eq:a62k}
\end{equation}
then inward to the center with
\begin{equation}
a_{i} = a_{i+1} \exp \left[ \left( p_{i+1} - p_{i} - \frac{\Delta M_{i+\frac{1}{2}} }{ 4 \pi R_{i+\frac{1}{2}}^{2} \rho_{i+\frac{1}{2}} } f_{\nu,r \, i+\frac{1}{2}}      \right)/\rho_{i+\frac{1}{2}} c^{2} \right] ,
\label{eq:a62l}
\end{equation}
where the zone-centered quantities defined at zone edges are arithmetic averages of their zone-centered values on either side of the zone edge, and $\Delta M_{i+\frac{1}{2}}/4 \pi R_{i+\frac{1}{2}}^{2} \rho_{i+\frac{1}{2}}$ has been used for $R_{i+1} - R_{i}$. Zone-edged values of $a$ are defined by $a_{i+\frac{1}{2}} = ( a_{i} + a_{1+1})/2$.

\subsection{Transport through a Shock}
\label{trans_shock}

Comparisons of transport tests with a Boltzmann solver revealed a shortcoming of our flux-limited diffusion scheme when encountering a discontinuity in the fluid velocity -- namely, a shock.
The problem arises from our neglect of the velocity-dependent terms in going from Equation~(\ref{eq:a14}) to Equation~(\ref{eq:a15}), which should be important for computing the transport at the shock.
The reason was traced ostensibly to the fact that both $\psi^{(0)}$ and $\psi^{(1)}$, being defined with respect to the fluid frame, are therefore physically both discontinuous at a discontinuity in the fluid velocity.
However,  as a result of the fact $\psi^{(1)}$ is given by the gradient of $\psi^{(0)}$ (e.g., Equation~(\ref{eq:a19}) and its differenced version, Equation (\ref{eq:a59})) $\psi^{(0)}$ is constrained to be continuous across the shock. 
In the case in which the fluid velocity ahead of the shock has a large negative radial velocity that discontinuously transitions to a much smaller negative post-shock radial velocity, which characterizes the change in the fluid velocity across the bounce shock as it stagnates and then revives, both $\psi^{(0)}$ and $\psi^{(1)}$ should exhibit, outwardly, a positive discontinuous change through the shock. 
However, $\psi^{(0)}$, rather than exhibiting this discontinuous behavior, ramps up over many zones to its final pre-shock value ahead of the shock.
This behavior of $\psi^{(0)}$ is required in order that the required flux through the shock be computed via Equation~(\ref{eq:a59}).
This rise of $\psi^{(0)}$ through the post-shock region, occurring in the neutrino heating region in the case of the bounce shock, has the unfortunate effect of causing a rise in the mean neutrino energy
\begin{equation}
\epscmi{  {\rm rms}} = \sqrt{ \int_{0}^{\infty} \psi^{(0)}_0(\epscm) \epscm^{5} d \epsilon / \int_{0}^{\infty} \psi^{(0)}_{0}(\epscm) \epscm^{3} d \epsilon } .
\label{eq:a88}
\end{equation}
The result is an unphysical additional neutrino heating in the region behind the bounce shock, potentially causing the shock to revive too soon. 

\begin{figure}
\gridline{\fig{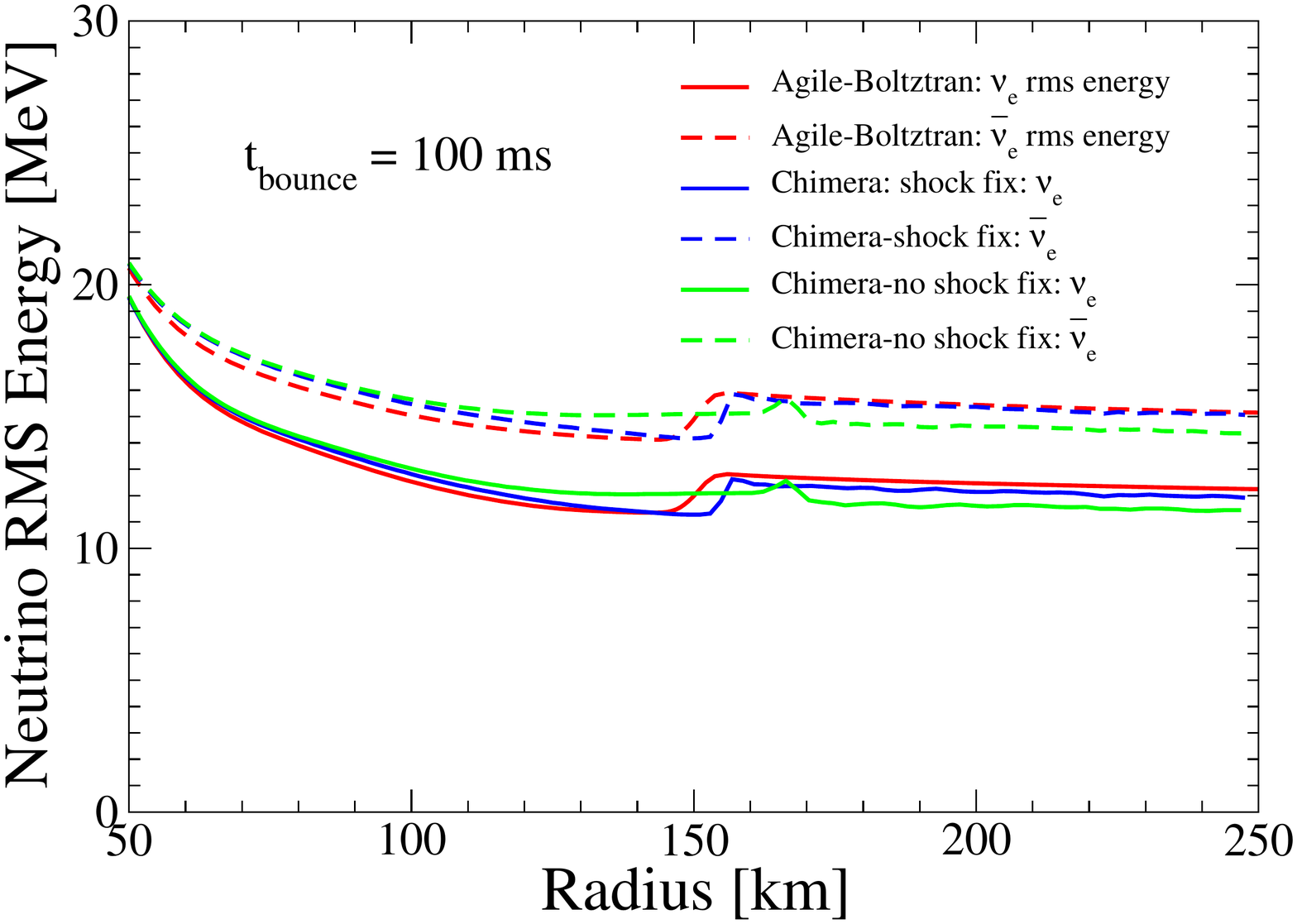}{0.5\textwidth}{(a)}
              \fig{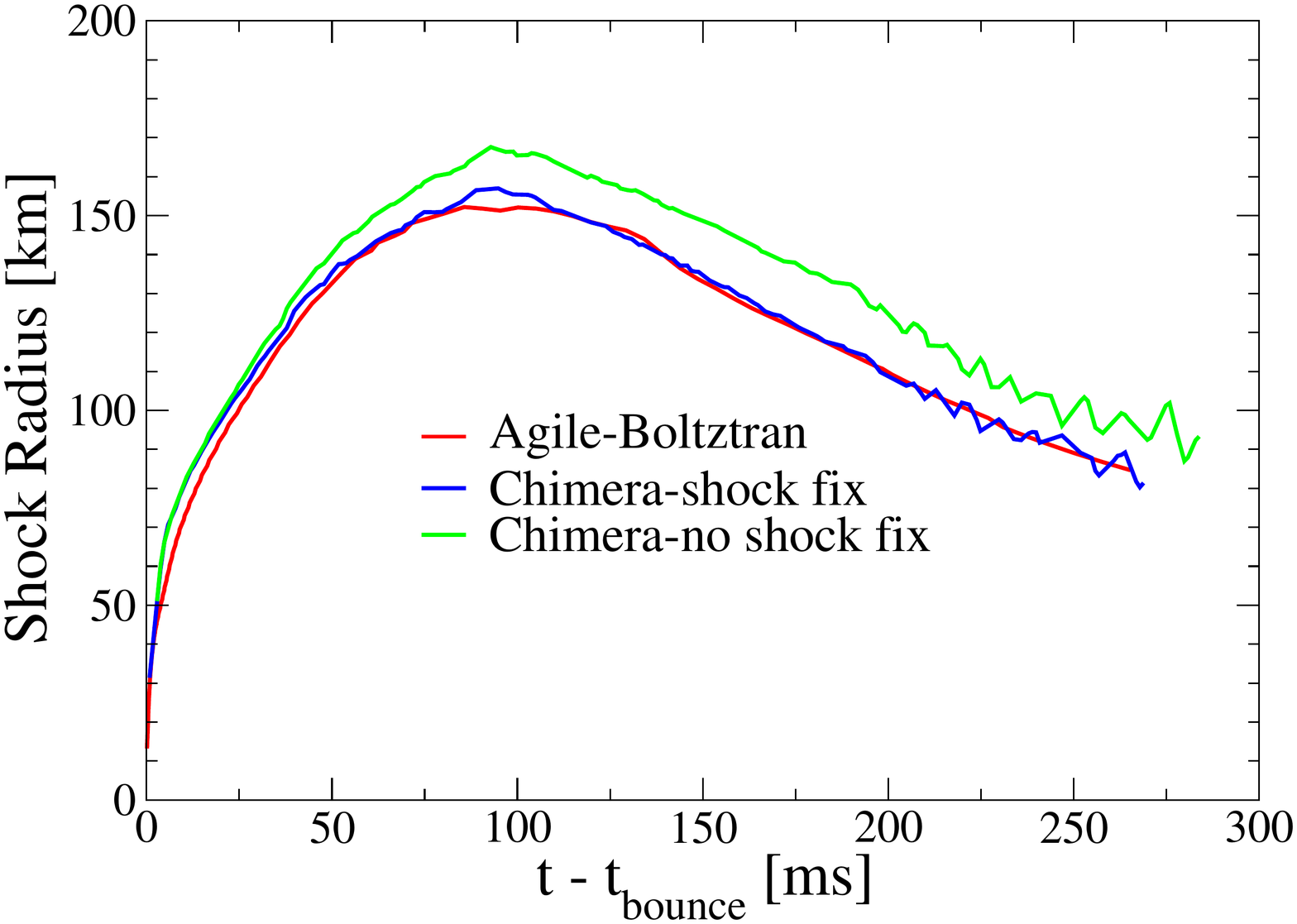}{0.5\textwidth}{(b)}
               }
\gridline{
              \fig{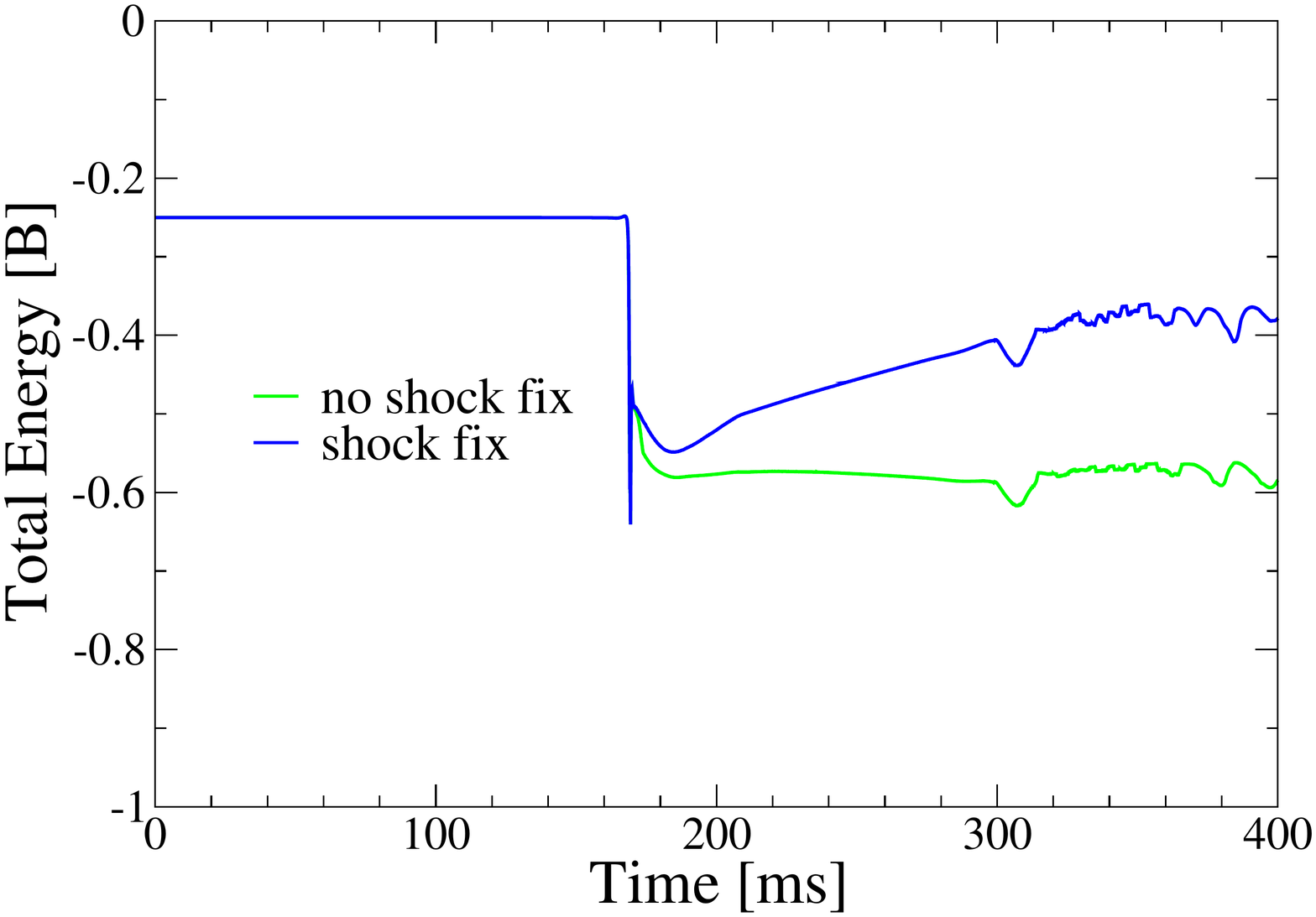}{0.5\textwidth}{(c)}
              \fig{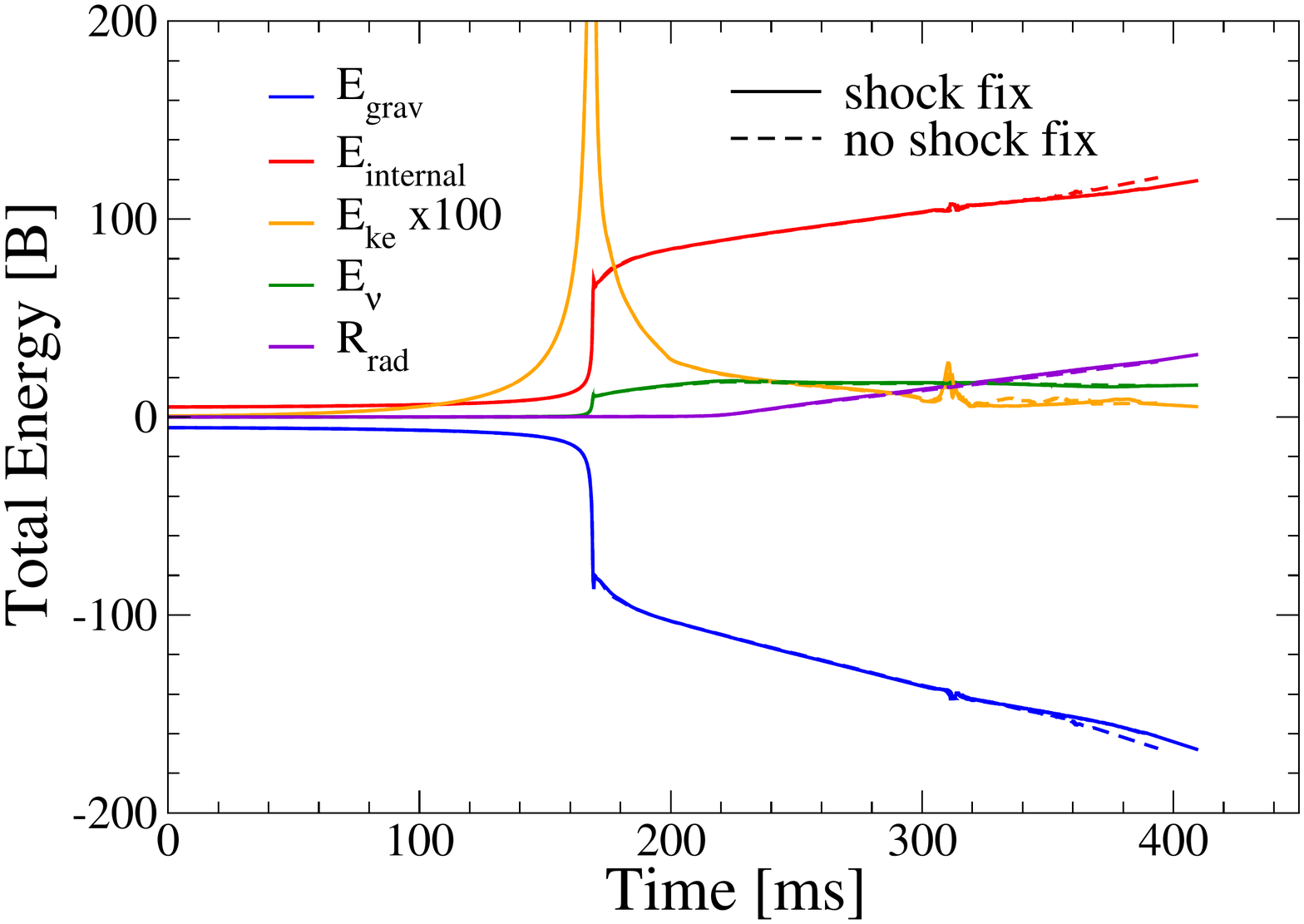}{0.5\textwidth}{(d)}
               }
\caption{\label{fig:AB_vs_CH}
Panel (a): \nue-rms and \nuebar-rms energies at $t_{\rm bounce} = 100$~ms; Panel (b):  shock radius as a function of post-bounce time using \chimera\ with (blue) and without (green) the shock transport algorithm, compared to a calculation using \agileboltztran\ (red); 
Panel (c): total energy as a function of time for the two \chimera\ simulations redone using Newtonian gravity; and Panel (d): $E_{\rm grav}$,  total gravitational potential energy, $E_{\rm internal}$,  total internal energy, $E_{\rm ke}$,  total kinetic energy, $E_{\nu}$,  total on-grid neutrino energy, and $E_{\rm rad}$,  neutrino energy lost by neutrinos exiting the grid for the \Chimera\ runs plotted in Panel (c) .
The sum of these energies is the total energy plotted in Panel (c).
}
\end{figure}

We illustrate this problem by comparing 1D simulations performed using \chimera\ and the relativistic Boltzmann code \agileboltztran, which is described in Section~\ref{comparisons}.
The progenitor used is the 15-\msun\ model evolved to the point of core collapse by \citet{WoHe07}, and the neutrino physics is that described in \citet{Brue85}. 
The rather straightforward collection of neutrino physics used in this comparison is chosen so that its implementation in both codes can be easily made identical. 
The unphysical rise in the \nue-rms and \nuebar-rms energies as computed by \chimera\ is illustrated by the green lines in Figure~\ref{fig:AB_vs_CH}(a).
Compared with the rms energies computed by \agileboltztran\ (plotted by the red lines) the rms energies computed by \chimera\ are 1--2 MeV higher by the time the neutrinos reach the shock. 
A consequence of this neutrino rms energy rise is illustrated in Figure~\ref{fig:AB_vs_CH}(b) by the increase in the shock radius due to increased neutrino heating in the \chimera\ simulation (shown in green) as compared with that given by the \agileboltztran\ simulation (shown in red).
The shock position in the \chimera\ simulations has been pushed about 15~km farther out in radius compared with its position as given by \agileboltztran. 

To describe the causes of this problem in more detail and our shock transport algorithm designed to eliminate this problem, consider Equation~(\ref{eq:a59}) at a velocity discontinuity situated at zone interface $i+\rfrac{1}{2}$. 
In this circumstance  \psimomcmin{0}{i+1,k}{n+1}\ and \psimomcmin{0}{i,k}{n+1}\  are defined in two different reference frames.
In the case of the outwardly directed bounce shock the former would be defined in the pre-shock frame, the latter in the post-shock frame. 
To avoid the unphysical ramping up of $\psi^{(0)}$ toward the shock in the post-shock frame as we approach the shock from below, \psimomcmin{0}{i,k,q}{n+1}\ is replaced by \psimomcmin{0}{i,k,q}{n+1, \, \uparrow}\  in Equation~(\ref{eq:a59}), where \psimomcmin{0}{i,k,q}{n+1, \, \uparrow}\  is \psimomcmin{0}{i,k,q}{n+1}\  transformed to the same reference frame as $\psimomcmin{0}{i+1,k,q}{n+1}\  (\epscmi{ i})$; i.e., the reference frame of radial zone $i+1$. 
Note that the energy argument $\epscmi{ i}$, referenced by the subscript $k$, is the same in the transformed and untransformed function $\psi^{(0)}$, reflecting the velocity independence of the \chimera\ energy grid.

To obtain an expression for \psimomcmin{0}{i,k,q}{n+1, \, \uparrow}, let the quantities with subscripts $i$ and $i+1$ denote their values defined with respect to the frame corresponding to radial coordinate $i$ and $i+1$, respectively.  
The transformed neutrino functions will be denoted by \psimomcmin{0}{i,k,q}{n+1, \, \uparrow}, as above, to emphasize that they are transformed from frame $i$ to frame $i+1$ rather than originally residing in radial zone $i+1$.
Suppressing the indices $k$, $n$, and $q$, we use the invariance of $f_{0}( \epscm, \mucm )$ -- e.g., $\fcmi{i+f}(\epscmi{i+1}, \mucmi{i+1} ) = \fcmi{i}(\epscmi{i}, \mucmi{i} )$ -- and the transformation properties of the independent variables \epscmi{} and \mucmi{},
\begin{equation}
\beta \equiv \beta_{i+\frac{1}{2}} = \frac{ u_{i+1} - u_{i} }{c}, \quad \gamma \equiv \gamma_{i+\frac{1}{2}} = \frac{1}{\sqrt{ 1 - \beta^{2} } }, \quad \epscmi{ i+1} = \gamma \epscmi{ i} ( 1 - \beta \mucmi{i} ), \quad  \mucmi{  i+1} = \frac{ \mucmi{  i} - \beta }{ 1 - \beta \mucmi{  i} } ,
\label{eq:a90}
\end{equation}
to derive Equation (\ref{eq:a91}) below, where we have kept terms to order $\mathcal{O}(v/c)$.
The use of \psimomcmin{0}{i,k,q}{n+1, \, \uparrow}\ in place of \psimomcmin{0}{i,k,q}{n+1}\ in Equation~(\ref{eq:a59}) enables the correct flux to be computed through zone interface $i+\rfrac{1}{2}$, while the difference between \psimomcmin{0}{i,k,q}{n+1}\ and \psimomcmin{0}{i,k,q}{n+1, \, \uparrow}\ constitutes the discontinuity in \psimomcm{0}\ at the shock. 
It must be noted that the replacement of \psimomcmin{0}{i,k,q}{n+1}\ in Equation~(\ref{eq:a59}) by \psimomcmin{0}{i,k,q}{n+1, \, \uparrow}\ is only performed when the flux is being computed at interface $i+\rfrac{1}{2}$.
When computed at interface $i-\rfrac{1}{2}$, \psimomcmin{0}{i,k,q}{n+1}\ is not modified.

To implement this part of the shock transport algorithm, \chimera\ computes $d\psimomcmin{0}{i,k,q}{{\rm shock}}$, the difference between \psimomcmin{0}{i,k,q}{n+1, \, \uparrow}\ and \psimomcmin{0}{i,k,q}{n+1}, 
\begin{equation}
d\psimomcmin{0}{i,k,q}{{\rm shock}} = \psimomcmin{0}{i,k,q}{n+1, \, \uparrow} - \psimomcmin{0}{i,k,q}{n+1} = \beta_{i+\frac{1}{2}} \left\{ - \psimomcmi{1}{i,k,q} + \frac{ \epscmi{ i, k+\frac{1}{2}}^{3} \psimomcmi{1}{i,k+\frac{1}{2},q} - \epscmi{ i,  k - \frac{1}{2}}^{3} \psimomcmi{1}{i,k-\frac{1}{2},q} } { \frac{1}{3} \left( \epscmi{ i, k+\frac{1}{2}}^{3} - \epscmi{ i, k - \frac{1}{2}}^{3} \right) } \right\}
\label{eq:a91}
\end{equation}
in the presence of a shock, and $d\psimomcmin{0}{i,k,q}{{\rm shock}} = 0$ otherwise, where
\begin{equation}
 \psimomcmi{1}{i,k+\frac{1}{2},q} = \frac{1}{2} \left( \psimomcmi{1}{i,k,q} + \psimomcmi{1}{i,k+1,q} \right)
\label{eq:a92}
\end{equation}
and the zone-centered value $\psimomcmi{1}{i,k,q} = \eta_{i,k,q}^{(1)} \psimomcmi{0}{i,k,q}$, where  $\eta$ is defined in Section~\ref{eddington}.
With $d\psimomcmin{0}{i,k,q}{{\rm shock}}$ computed, Equation~(\ref{eq:a59}) is modified to read
\begin{equation}
\psimomcmin{1}{i+\frac{1}{2},k,q}{n+1} = - \frac{ \lambda^{(t) \, n+1}_{i+\frac{1}{2},k,q} }{3} {\cal F}^{n+1}_{i+\frac{1}{2},k,q} \Gamma^{n+1}_{i+\frac{1}{2}} \frac{ \psimomcmin{0}{i+1,k,q}{n+1} - \psimomcmin{0}{i,k,q}{n+1} - d\psimomcmin{0}{i,k,q}{n+1, {\rm shock}}  }{ R_{i+1} - R_{i} }  .
\label{eq:a93}
\end{equation}

A modification is needed in Equation (\ref{eq:a58}) to complete this scheme. 
In Equation~(\ref{eq:a58}) the change in \psimomcmin{0}{i,k,q}{n+1}\ due to transport is determined by the fluxes through the outer zone edge, $i+\rfrac{1}{2}$, proportional to \psimomcmin{1}{i+\frac{1}{2},k,q}{n+1}, and by the fluxes through the inner zone edge, $i-\rfrac{1}{2}$, proportional to \psimomcmin{1}{i-\frac{1}{2},k,q}{n+1}.
However, \psimomcmin{1}{i+\frac{1}{2},k,q}{n+1}\ at the outer zone edge is computed in the $i+1$ frame by Equation~(\ref{eq:a93}) with $\psimomcmin{0}{i,k,q}{n+1 \, \uparrow} = \psimomcmin{0}{i,k,q}{n+1} + d\psimomcmin{0}{i,k,q}{n+1, {\rm shock}}$ replacing \psimomcmin{0}{i,k,q}{n+1}, while \psimomcmin{1}{i-\frac{1}{2},k,q}{n+1}\ at the inner zone edge is computed in the $i$-th frame.
Consistency is restored by replacing \psimomcmin{1}{i+\frac{1}{2},k,q}{n+1}\ in Equation~(\ref{eq:a58}) by \psimomcmin{1}{i+\frac{1}{2},k,q}{n+1, \, \downarrow}, where the latter is the former transformed from the frame at $i+1$ to the frame at $i$.
This will ensure that the fluxes computed through both the inner and outer edges of zone $i$ are with respect to the same frame.

To derive an expression for \psimomcmin{1}{i+\frac{1}{2},k,q}{n+1, \, \downarrow}, we will be transforming quantities from the frame at radial zone $i+1$ to the frame at radial zone $i$.
To keep the notation manageable, we will again suppress the indices $k$, $n$, and $q$, and denote quantities defined at frame $i$ and $i+1$, as before, by subscript $i$ and $i+1$, respectively, and neutrino functions transformed from frame $i+1$ to frame $i$ by a down-arrow to distinguish them from neutrino functions actually residing at radial zone $i$.
In particular, the quantity \psimomcmin{1}{i+\frac{1}{2},k,q}{n+1} is computed in the $i + 1$ frame by Equation~(\ref{eq:a93}) above, and we will continue to denote its value when transformed to the $i$ frame by \psimomcmin{1}{i+\frac{1}{2},k,q}{n+1, \, \downarrow}. 
The transformation is obtained by equating the number of neutrinos passing through an area $dS$ perpendicular to the radial direction in a time $dt$ as seen from the two frames \citep[][p.413]{MiMi84}; namely,
\begin{eqnarray}
d\Ncmi{i} &=& \int_{-1}^{1} d\mucmi{ i} \fcmi{ \, i}( \epscmi{ i}, \mucmi{ i} ) \,  \epscmi{ i}^{2} \, d\epscmi{ i} \, c \, \mucmi{ i} \, dS \dtcmi{ i} , {\rm and } \nonumber \\ 
d\Ncmi{i+1} &=& \int_{-1}^{1} d\mucmi{ i+1} \fcmi{ \, i+1}( \epscmi{ i+1}, \mucmi{ i+1} ) \,  \epscmi{ i+1}^{2} \,  d\epscmi{ i+1} \, c \, \mucmi{ i+1} \, dS \dtcmi{ i+1} .
\label{eq:a94}
\end{eqnarray}
Using $\epscmi{ i+1} \, \mucmi{ i+1} \, d\mucmi{ i+1} = \epscmi{ i} \, \mucmi{ i} \, d\mucmi{ i}$ in the expression for $d\Ncmi{i+1}$, and transforming the remaining variables to the $i$-frame, we get
\begin{eqnarray}
d\Ncmi{i+1} =&& \int_{-1}^{1} d\mucmi{ i} \fcmi{ \, i+1} \left( \gamma \epscmi{ i} ( 1 - \beta \mucmi{ i} ), \frac{ \mucmi{ i} - \beta }{ 1 - \beta \mucmi{ i} } \right) \nonumber \\ 
 && \times \gamma \epscmi{ i} ( 1 - \beta \mucmi{ i} ) \epscmi{ i} \,  d\epscmi{ i} \, c \, \frac{ \mucmi{ i} - \beta}{ 1 - \beta \mucmi{ i} } \, dS \, \gamma \, \dtcmi{ i} .
\label{eq:a95}
\end{eqnarray}
Dividing by $c \, \epscmi{ i}^{2} \, d\epscmi{ i} \, dS \, \dtcmi{ i}$ and considering terms to order $\mathcal{O}(v/c)$,
\begin{eqnarray}
\psimomcmin{1}{i+1}{\downarrow}(\epscmi{ i})  
&=& \int_{-1}^{1} d\mucmi{ i} \fcmi{ f}( \epscmi{ i}, \mucmi{ i} ) \, \mucmi{ i} 
= \frac{ d\Ncmi{f} }{ c \, \epscmi{ i}^{2} \, d\epscmi{ i} \, dS \, \dtcmi{ i} }
=  \frac{ d\Ncmi{i+1} }{ c \, \epscmi{ i}^{2} \, d\epscmi{ i} \, dS \, \dtcmi{ i} } \nonumber \\ 
&=& \int_{-1}^{1} d\mucmi{ i} \fcmi{ \, i+1} \left( \epscmi{ i} ( 1 - \beta \mucmi{ i} ), ( \mucmi{ i} - \beta )( 1 + \beta \mucmi{ i} ) \right) \, ( 1 - \beta \mucmi{ i} ) \, ( \mucmi{ i} - \beta )( 1 + \beta \mucmi{ i} ) \nonumber \\ 
&\simeq& \int_{-1}^{1} d\mucmi{ i} \fcmi{ \, i+1} \left( \epscmi{ i} - \beta \epscmi{ i} \mucmi{ i} ), ( \mucmi{ i} - \beta ( 1 - \mucmi{ i}^{2} ) \right) \, ( \mucmi{ i} - \beta  ) \nonumber \\ 
&\simeq& \int_{-1}^{1} d\mucmi{ i} \left[ \mucmi{ i} \fcmi{ \, i+1} \left( \epscmi{ i}, \mucmi{ i} \right) - \mucmi{ i} \pderivoo{\fcm}{\mucm}{i} \beta ( 1 - \mucmi{ i}^{2} ) - \mucmi{ i} \pderivoo{\fcm}{\epscm}{i} \beta \epscmi{ i} \mucmi{ i} - \beta \fcmi{ \, i+1} \left( \mucmi{ i}, \epscmi{ i} \right) \right]  \nonumber \\ 
&=& \psimomcmi{1}{i+1} ( \epscmi{ i} ) - \frac{ \beta }{ \epscmi{ i}^{2} } \pderivo{ \epscmi{ i}^{3} \psimomcmi{2}{i+1}( \epscmi{ i} ) }{ \epscmi{ i} } .
\label{eq:a96}
\end{eqnarray} 
To implement this part of the shock transport algorithm, \chimera\ computes $d\psimomcmin{1}{i+\frac{1}{2},q}{ {\rm shock}}(\epscmi{ i})$, the difference between $\psimomcmin{1} {i+1,q}{\downarrow} \, (\epscmi{ i})$ and $\psimomcmi{1}{i+1,q} ( \epscmi{ i} )$, both defined at zone edge $i + \rfrac{1}{2}$, by
\begin{equation}
d\psimomcmin{1}{i+\frac{1}{2},k,q}{{\rm shock}}
= - \beta_{i+\frac{1}{2}} \frac{ \epscmi{ i, k+\frac{1}{2}}^{3} \psimomcmi{2}{i+\frac{1}{2},k+\frac{1}{2},q} - \epscmi{ i,  k - \frac{1}{2}}^{3} \psimomcmi{2}{i+\frac{1}{2},k-\frac{1}{2},q} } { \frac{1}{3} \left( \epscmi{ i, k+\frac{1}{2}}^{3} - \epscmi{ i,  k - \frac{1}{2}}^{3} \right) } ,
\label{eq:a97}
\end{equation}
in the presence of a shock, and by $d\psimomcmin{1}{i+\frac{1}{2},k,q}{{\rm shock}} = 0$, otherwise, where
\begin{equation}
 \psimomcmi{2}{i,k+\frac{1}{2},q} = \frac{1}{2} \left( \psimomcmi{2}{i,k,q} + \psimomcmi{2}{i,k+1,q} \right)
\label{eq:a98}
\end{equation}
and where $\psimomcmi{2}{i,k,q} = \eta_{i,k,q}^{(2)}\psimomcmi{0}{i,k,q}$. 
With $d\psimomcmin{1}{i+\frac{1}{2},k,q}{ {\rm shock}}$ computed, Equation~(\ref{eq:a58}) is modified to read
\be
& {\ds \frac{1}{a_{i}c} \frac{ \psimomcmin{0}{i,k,q}{n+1} - \psimomcmin{0}{i,k,q}{n} }{ \Delta t^{n+\frac{1}{2}} } + 
} & \nonumber \\ 
& {\ds + \frac{a_{i}^{2}}{ \Vol{i} } \left[ \frac{1}{a_{i+\frac{1}{2}}^{2}}\Area{i+\frac{1}{2}} \left( \psimomcmin{1}{i+\frac{1}{2},k,q}{n+1}  + d\psimomcmin{1}{i+\frac{1}{2},k,q}{ {\rm shock}} \right) - \frac{1}{a_{i-\frac{1}{2}}^{2}}\Area{i-\frac{1}{2}} \psimomcmin{1}{i-\frac{1}{2},k,q}{n+1} \right] = \frac{c}{\epscm} \left( \frac{ d }{ d\ell } \fcm(x,p) \right)_{ \! S \; i,k,q }^{(0) \, n+1} .
} &
\label{eq:a99} 
\ee

Numerical experiments have shown that rolling in this algorithm from $| v_{i+1} - v_{i} | =$ 0.01--0.02 c with a corresponding rollout of the energy advection algorithm -- i.e., at the shock, this algorithm replaces the energy-advection algorithm -- gave excellent results. 
In the above comparisons of the neutrino rms energies and shock trajectories computed with the \chimera\ and \agileboltztran codes, the \chimera\ results with the use of the above shock transport algorithm, plotted in Figure~\ref{fig:AB_vs_CH} by the lines in blue, show much better agreement with the results from \agileboltztran. 
The discontinuities in \psimomcm{0}\ are clearly evident, as is the absence of the unphysical rise in \psimomcm{0}\ as the shock is approached from below.
Further examples of \chimera\ test results without the use of this scheme (Series B) and with the use of this scheme (Series C) are given in Sections~\ref{trans_tests} and \ref{comparisons} and demonstrate the scheme's ability to give accurate transport solutions across a shock.

As a test of the ability of \chimera\ to conserve total energy, the simulations described above were redone with the Newtonian gravitational potential substituted for the general relativistic monopole potential. Substitution of the general relativistic monopole in place of the Newtonian monopole in the multipole expansion of the self-gravitational potential is not conservative and, therefore, not suitable for a test of total energy conservation. We also set the lapse function to unity.
The results are shown in Figure~\ref{fig:AB_vs_CH}, panels (c) and (d). 
In both simulations, energy is conserved with high accuracy up to bounce, where a discontinuous change in energy of about -0.35 B occurs in both simulations.
Thereafter, the total energy of the simulation with the shock transport algorithm turned off is essentially flat for the next $\sim$140 ms, after which it decreases by about -0.1 B during the subsequent $\sim$100 ms.
The total energy of the simulation with the shock transport algorithm turned on increases by about 0.1 B for the $\sim$140 ms following bounce, and then remains essentially flat for the next $\sim$100 ms. 
The components of the total energy of the two \chimera\ simulations are plotted in panel (d). 
There are small differences discernible in the components of the total energy in the two simulations, particularly between 300 and 400 ms.

\subsection{Neutrino Energy Advection Step}
\label{neu_e_adv}

Following the Lagrangian transport step, \chimera\  executes the Lagrangian energy advection step, which consists of solving Equation~(\ref{eq:a31}), which has been operator split along with the ``transport'' Equation~(\ref{eq:a30}) from Equation~(\ref{eq:a12}). 
Omitting the superscript $E$ from $\left( \partial \psimomcm{0}/\partial t \right)_{m,\Ecm}$ with the understanding that this time derivative will hereafter refer to the change in \psimomcm{0} due to the energy advection step, and considering for the moment only the terms in Equation~(\ref{eq:a31}) involving \psimomcm{0}\ (terms involving \psimomcm{2}\ will be considered shortly), we rearrange these terms as follows:
\begin{eqnarray}
\left( \pderiv{\psimomcm{0}}{t} \right)_{m,\Ecm} + \Ecm \left( \pderiv{\ln a}{t} - \pderiv{\ln R}{t} \right) \left( \pderiv{\psimomcm{0}}{\Ecm} \right)_{m,t} - \left[ \pderiv{\ln R}{t} -  \pderiv{\ln b}{t} \right] \psimomcm{0} \nonumber \\ 
 = \frac{a^{3}}{bR^{2}} \left( \pderiv{ }{t} \left( \frac{bR^{2}}{a^{3}} \psimomcm{0} \right) \right)_{m,\Ecm} + \frac{1}{\Ecm^{2}} \pderiv{}{\Ecm} \left[ \Ecm \left( \pderiv{\ln a}{t} - \pderiv{\ln R}{t} \right) \Ecm^{2} \psimomcm{0} \right]_{m,t} ,
\label{eq:a63}
\end{eqnarray}
or, multiplying by $\Ecm^{2}(bR^{2}/a^{3})$,
\begin{equation}
\left( \pderiv{ }{t} \left( \frac{bR^{2}}{a^{3}} \Ecm^{2} \psimomcm{0} \right) \right)_{m,\Ecm} + \pderiv{}{\Ecm} \left[ \Ecm \left( \pderiv{\ln a}{t} - \pderiv{\ln R}{t}  \right) \left( \frac{bR^{2}}{a^{3}} \Ecm^{2} \psimomcm{0} \right) \right]_{t, R} .
\label{eq:a64}
\end{equation}
Note that we could have multiplied Equation~(\ref{eq:a63}) by any power of $\Ecm$ in deriving the final form of Equation~(\ref{eq:a64}), but the second term would always have been left with one more power of $\Ecm$ than the first term, and this feature will be important below.
The motivation for our choice of the quantity $\Ecm^{2} \psimomcm{0}$ is two fold: (1) this choice enables us to absorb the inhomogeneous terms involving \psimomcm{0}\ into the energy derivative operator, and (2) it facilitates a physical interpretation, described below, which provides a useful guide for differencing, and is based on the fact that $\Ecm^{2} \psimomcm{0}$ is proportional to the neutrino number density per unit energy.

Consider now the terms in Equation~(\ref{eq:a31}) involving \psimomcm{2}.
To derive an expression which, after multiplying by $\Ecm^{2}$, is similar to the second term in  Equation~(\ref{eq:a64}), we rearrange the terms involving \psimomcm{2}\ as
\begin{equation}
  \left( \pderiv{\ln b}{t} - \pderiv{\ln R}{t} \right) \left( \pderiv{\psimomcm{2}}{\Ecm} \right)_{m,t} + 3 \left[ \pderiv{\ln R}{t} -  \pderiv{\ln b}{t} \right] \psimomcm{2}
 = - \frac{1}{\Ecm^{2}} \pderiv{}{\Ecm} \left[ \Ecm \left( \pderiv{\ln b}{t} - \pderiv{\ln R}{t} \right) \Ecm^{2} \psimomcm{2} \right]_{m,t} .
\label{eq:a65}
\end{equation}
Substituting Equation (\ref{eq:a64}) and (\ref{eq:a65}) (the latter multiplied by $\Ecm^{2}$) in Equation~(\ref{eq:a31}), the energy advection equation becomes
\begin{eqnarray}
 0 &=& \left( \pderiv{ }{t} \left( \frac{bR^{2}}{a^{3}} \Ecm^{2} \psimomcm{0} \right) \right)_{m,\Ecm} + \pderiv{}{\Ecm} \left[ \Ecm \left( \pderiv{\ln a}{t} - \pderiv{\ln R}{t} \right) \Ecm^{2} \frac{bR^{2}}{a^{3}} \psimomcm{0}
- \Ecm \left( \pderiv{\ln b}{t} - \pderiv{\ln R}{t} \right) \Ecm^{2} \frac{bR^{2}}{a^{3}}\psimomcm{2} \right]_{m,t} \nonumber \\ 
 &=& \left( \pderiv{ }{t} \left( \frac{bR^{2}}{a^{3}} \Ecm^{2} \psimomcm{0} \right) \right)_{m,\Ecm}
+ \pderiv{}{\Ecm} \left[ \Ecm \left\{ \left( \pderiv{\ln a}{t} - \pderiv{\ln R}{t} \right) - \left( \pderiv{\ln b}{t} - \pderiv{\ln R}{t} \right) \eta^{(2)} \right\} \Ecm^{2} \frac{bR^{2}}{a^{3}} \psimomcm{0} \right]_{m,t} ,
\label{eq:a66}
\end{eqnarray}
where we have again introduced the scalar Eddington factor, $\eta^{(2)} = \eta^{(2)}(R, t, \epscm)$, defined by
\begin{equation}
\psimomcm{2} = \eta^{(2)} \psimomcm{0} ,
\label{eq:a67}
\end{equation}
as originally introduced at the beginning of Section~\ref{flux_limiting} and defined mathematically in Section~\ref{eddington}. Finally, defining the quantities \Psimomcm{0}\ and ${\cal V}$ as
\begin{equation}
\Psimomcm{0} \equiv \Psimomcm{0}(m, t, \Ecm) = \frac{ bR^{2} }{a^{3}} \Ecm^{2} \psimomcm{0},
\label{eq:a68}
\end{equation}
and
\begin{equation}
{\cal V} \equiv {\cal V}(m, t, \Ecm) = \Ecm \left\{ \left( \pderiv{\ln a}{t} - \pderiv{\ln R}{t} \right) - \left( \pderiv{\ln b}{t} - \pderiv{\ln R}{t} \right) \eta^{(2)} \right\},
\label{eq:a69}
\end{equation}
Equation (\ref{eq:a66}) can be written in the compact form:
\begin{equation}
\left( \pderiv{ \Psimomcm{0} }{t} \right)_{m,\Ecm} + \pderiv{}{\Ecm} \left( {\cal V} \Psimomcm{0} \right)_{m,t} = 0 .
\label{eq:a70}
\end{equation}
If we take into account the factor of $\Ecm$ in ${\cal V}$ and neglect the possible dependence of $\eta^{(2)}$ on $\Ecm$ over a time step, then writing ${\cal V} = \Ecm \Vcm$ we use the fact that $\Vcm$ is now independent of $\Ecm$ to write Equation~(\ref{eq:a70}) as 
\begin{equation}
\left( \pderiv{ \Psimomcm{0} }{t} \right)_{m,\Ecm} + \Ecm  \Vcm \left( \pderiv{ \Psimomcm{0} }{\Ecm} \right)_{m,t} + \Vcm \Psimomcm{0} = 0 ,
\label{eq:a71}
\end{equation}
which has the solution
\begin{equation}
\Psi_{_{\! 0}}^{(0) \, n+1}(m, t^{n+1}, \Ecm^{n+1}) = \exp \left( - \int_{t^{n}}^{t^{n+1}}\Vcm dt \right) \Psimomcm{0} \left[ m, t^{n}, \Ecm^{n+1} \, \exp \left( - \int_{t^{n}}^{t^{n+1}} \Vcm dt  \right) \right] ,
\label{eq:a72}
\end{equation}
where the general solution has been particularized by requiring that $\Psimomcm{0}(m, t^{n+1}, \Ecm^{n+1}) \rightarrow \Psimomcm{0}(m, t_{i}^{n}, \Ecm^{n})$ as $t^{n+1} \rightarrow t^{n}$, where $\Ecm^{n+1}$ and $\Ecm^{n}$ are related by
\begin{equation}
\Ecm^{n} = \Ecm^{n+1} \exp \left( - \int_{t^{n}}^{t^{n+1}} \Vcm dt \right) ,
\label{eq:a73}
\end{equation}
and where Equation (\ref{eq:a71}) has been integrated over the time step $t^{n} \rightarrow t^{n+1}$.

A convenient expression from which the finite differencing of Equation (\ref{eq:a72}) can be developed, which additionally lends itself to a physical interpretation, is obtained by (i) using Equation~(\ref{eq:a68}) to transform $\Psimomcm{0}$ back to \psimomcm{0}\ in Equation~(\ref{eq:a72}), (ii) multiplying the result by ${(4\pi)^{2}}{( hc)^{-3} } m \Delta \Ecm^{n+1}$, where $m$ is a given comoving mass, and (iii) recalling that $b= 1 / (4 \pi R^{2} \rho )$. 
The result is
\begin{equation}
\frac{4\pi}{( hc)^{3} } V^{n+1} \frac{  ( \Ecm^{n+1})^{2} \Delta \Ecm^{n+1} }{ (a^{n+1})^{3}} \psimomcm{0}(m, t^{n+1}, \Ecm^{n+1}) = \frac{4\pi}{(hc)^{3} } V^{n} \frac{ (\Ecm^{n})^{2} \Delta \Ecm^{n} }{ (a^{n})^{3}} \psimomcm{0}(m, t^{n}, \Ecm^{n}) ,
\label{eq:a74}
\end{equation}
where $V = m/\rho$, and Equation (\ref{eq:a73}) (first factor on the right-hand side of Equation~(\ref{eq:a72})) has been used to replace $\Delta \Ecm^{n+1}$ by $\Delta \Ecm^{n}$ on the right-hand side of Equation~(\ref{eq:a74}).
The left-hand side of Equation~(\ref{eq:a74}) is $\Delta N_{\nu}(m, t^{n+1}, \Ecm^{n+1})$, the number of neutrinos of energy $\epscm^{n+1} = \Ecm^{n+1}/a^{n+1}$ within the energy width $\Delta \epscm^{n+1} = \Delta \Ecm^{n+1}/a^{n+1}$ in a comoving spatial volume $V$ of mass $m$ after the energy advection step, and the right-hand side is $\Delta N_{\nu}(m, t^{n}, \Ecm^{n})$, the number of the same set of neutrinos before the energy advection step. 
Equation~(\ref{eq:a74}) therefore states that the neutrinos, numbering $\Delta N_{\nu}(m, t^{n}, \Ecm^{n})$, in a comoving volume $V^{n}$ with energies between $\epscm^{n} = \Ecm^{n}/a^{n}$ and  $\epscm^{n} + \Delta \epscm^{n} = ( \Ecm^{n} + \Delta \Ecm^{n} )/a^{n}$ are shifted in energy, while conserving number, to energies between $\epscm^{n+1} = \Ecm^{n+1}/a^{n+1} $ and  $\epscm^{n+1} + \Delta \epscm^{n+1} = ( \Ecm^{n+1} + \Delta \Ecm^{n+1} )/a^{n+1}$, where $\Ecm^{n}$ and  $\Ecm^{n+1}$ are related by Equation~(\ref{eq:a73}).

As a particular example of this scheme, consider the neutrino diffusion limit in which $\eta^{(2)} = \rfrac{1}{3}$. Recalling that $b = 1 / (4 \pi R^{2} \rho )$, Equation~(\ref{eq:a69}) with $\Ecm$ factored out becomes
\begin{equation}
\Vcm = \left[ \left( \pderiv{\ln a}{t} - \pderiv{\ln R}{t} \right) - \frac{1}{3} \left( \pderiv{\ln b}{t} - \pderiv{\ln R}{t} \right) \right] 
= \left( \pderiv{\ln a}{t} + \frac{1}{3} \pderiv{\ln \rho}{t} \right) .
\label{eq:a75}
\end{equation}
It follows that
\begin{equation}
\exp \left( \int_{t^{n}}^{t^{n+1}} \Vcm dt \right) = \exp \left( \int_{t^{n}}^{t^{n+1}} \left( \pderiv{\ln a}{t} + \frac{1}{3} \pderiv{\ln \rho}{t} \right) dt \right) = \frac{ a^{n+1} (\rho^{n+1})^{1/3} }{ a^{n} (\rho^{n})^{1/3} } ,
\label{eq:a76}
\end{equation}
and using Equation (\ref{eq:a76}) in Equation~(\ref{eq:a73}) we finally get
\begin{equation}
\Ecm^{n+1} = \Ecm^{n} \frac{ a^{n+1} (\rho^{n+1})^{1/3} }{ a^{n} (\rho^{n})^{1/3} } \quad \Rightarrow \quad \epscm^{n+1} = \epscm^{n} \left( \frac{ \rho^{n+1} }{ \rho^{n} } \right)^{1/3} ,
\label{eq:a77}
\end{equation}
which states that in the diffusion limit the neutrino energy scales as $\rho^{1/3}$ under compression or expansion, which is the expected property of a relativistic gas.

Given the solutions, Equations~(\ref{eq:a73}) and (\ref{eq:a74}), of the neutrino energy advection Equation~(\ref{eq:a31}), the \chimera\ numerical scheme for updating the neutrinos in energy proceeds in three steps, and is implemented during the radial sweep (Figure~\ref{fig:sweep}) after the Lagrangian hydrodynamics step in which the density $\rho$ and the areal radius $R$ are updated.
The steps are as follows:

\begin{enumerate}
\item We begin with an energy Lagrangian step arising from the changes in $\rho$, $R$, or $a$.
This shifts the neutrino energies to their new values and modifies $\psimomcm{0}$ through Equations~(\ref{eq:a73}) and (\ref{eq:a74}) in a way that conserves the total number of neutrinos in a given comoving fluid volume.
Labeling the values of quantities after this Lagrangian step with the superscript $n'+1$, Equations~(\ref{eq:a73}) and (\ref{eq:a74}), in differenced form, give for this step
\begin{equation}
\psimomcmin{0}{i, k, q}{n} \rightarrow \psimomcmin{0}{i, k, q}{n'+1}
=  \frac{ ( \rho_{i} a_{i}^{3} )^{n'+1} }{ ( \rho_{i} a_{i}^{3} )^{n} } \frac{ ( \Ecmi{ k}^{2} \Delta \Ecmi{ k} )^{n} }{ ( \Ecmi{ k}^{2} \Delta \Ecmi{ k} )^{n'+1} } \psimomcmin{0}{i, k, q}{n} ,
\label{eq:a78}
\end{equation}
where the energy grid is displaced by
\begin{equation}
\Ecm^{n}  \rightarrow \Ecmi{ k}^{n'+1} =  \Ecmi{ k}^{n} \exp \left( \int_{t^{n}}^{t^{n+1}} \Vcmi{i} dt \right) \simeq \left[ 1 + \left( \frac{da_{i}^{n+\frac{1}{2}} }{a_{i}^{n}} - \frac{dR_{i}^{n+\frac{1}{2}} }{R_{i}^{n}} \right) - \left( \frac{db_{i}^{n+\frac{1}{2}} }{b_{i}^{n}} - \frac{dR_{i}^{n+\frac{1}{2}} }{R_{i}^{n}} \right) \eta_{i, k, q}^{(2) \, n} \right] \Ecmi{ k}^{n} ,
\label{eq:a79}
\end{equation}
and
\begin{equation}
\epscmi{ i, k}^{n'+1} = \frac{ a_{i}^{n} }{ a_{i}^{n'+1} } \frac{ \Ecmi{ k}^{n'+1}}{\Ecmi{ k}^{n}} \epscmi{ i, k}^{n} ,
\label{eq:a80}
\end{equation}
where $da$, $dR$, and $db$ are the changes in $a$, $R$, and $b$ result from the Lagrangian hydrodynamics step or the update of the gravitational potential.
\item Next, we perform a remap of the energy grid, which has been displaced during step 1 from $\Ecmi{ k}^{n}$ to $\Ecmi{ k}^{n'+1}$, back to the initial set of values $\Ecmi{ k}^{n}$.
That is, letting quantities with the superscript $n+1$ denote their final values at the completion of the remap, this step entails
\begin{equation}
\Ecmi{ k}^{n'+1} \rightarrow \Ecmi{ k}^{n+1} = \Ecmi{ k}^{n},  \quad \psimomcmin{0}{i, k, q}{n'+1} \rightarrow \psimomcmin{0}{i, k, q}{n+1} .
\label{eq:a81}
\end{equation}
This step uses the PPM advection technology and is described in more detail in Section \ref{app:e_advct_lapse}. 
Briefly, for a given radial zone $i$ the quantities $\epscmi{i,k}^{2} \psimomcmi{0}{i,k,q}$, which are proportional to the neutrino number density per unit energy, are given piecewise parabolic profiles as a function of energy, which are then averaged over the displacement of the energy grid occasioned by the remap. 
These averages provide a high-order representation of the neutrino fluxes at the energy grid edges, which are used to remap the energy grid to its final values. 
While the neutrino number densities are automatically conserved in this step, the fluxes are scaled by a constant overall factor of the order unity to ensure that the total neutrino energy in a given spatial volume before and after this remap is conserved as well.
(This remap step just shifts the energy grid and should therefore conserve both neutrino number and energy.)
For more details of this latter procedure, refer to Equations~(\ref{eq:p6}) -- (\ref{eq:p8}) and the accompanying discussion.
\item In the third step the neutrino occupation probabilities \psimomcm{0} are checked to ensure that none exceed unity. If $\psimomcmi{0}{i,k,q} > 1$ for one or more values of $k$ for a given $i$,  each such $\psimomcmi{0}{i,k,q}$ is then subject to the following variant of the algorithm described in \citet[][eq. B8]{Brue85}, which conserves number and energy but limits  $\psimomcm{0} \le 1$. 

To describe the algorithm used here, let $\Delta^{(-)} \psimomcmi{0}{i,k,q}$ and $\Delta^{(-)} \psimomcmi{0}{i,k+2,q}$ correspond to the number of neutrinos removed from zone $(i,k)$ and zone $(i,k+2)$, respectively, and let $\Delta^{(+)} \psimomcmi{0}{i,k+1,q}$ correspond to the number of neutrinos added to zone $(i,k+1)$. Then specifying $\Delta^{(-)} \psimomcmi{0}{i,k,q}$ and using the conservation of number and energy is sufficient to fix the values of $\Delta^{(+)} \psimomcmi{0}{i,k+1,q}$ and $\Delta^{(-)} \psimomcmi{0}{i,k+2,q}$ so that number and energy is conserved. In particular, neutrino number conservation requires that
\begin{equation}
0 = - \epscmi{ i,k}^{2} \Delta \epscmi{ i,k} \Delta^{(-)} \psimomcmi{0}{i,k,q}
+ \epscmi{ i,k+1}^{2} \Delta \epscmi{ i,k+1} \Delta^{(+)} \psimomcmi{0}{i,k+1,q}
- \epscmi{ i,k+2}^{2} \Delta \epscmi{ i,k+2} \Delta^{(-)} \psimomcmi{0}{i,k+2,q} ,
\label{eq:a81a}
\end{equation}
while energy conservation requires that
\begin{equation}
0 = - \epscmi{ i,k}^{3} \Delta \epscmi{ i,k} \Delta^{(-)} \psimomcmi{0}{i,k,q} 
+ \epscmi{ i,k+1}^{3} \Delta \epscmi{ i,k+1} \Delta^{(+)} \psimomcmi{0}{i,k+1,q}
- \epscmi{ i,k+2}^{3} \Delta \epscmi{ i,k+2} \Delta^{(-)} \psimomcmi{0}{i,k+2,q} .
\label{eq:a81b}
\end{equation}
For a given $\Delta^{(-)} \psimomcmi{0}{i,k,q}$, Equations (\ref{eq:a81a}) and (\ref{eq:a81b}) give
\begin{equation}
\Delta^{(+)} \psimomcmi{0}{i,k+1,q} = \frac{ \left(  \epscmi{ i,k+2} -  \epscmi{ i,k} \right) \epscmi{ i,k}^{2} \Delta \epscmi{ i,k} }
{ \left(  \epscmi{ i,k+2} -  \epscmi{ i,k+1} \right) \epscmi{ i,k+1}^{2} \Delta \epscmi{ i,k+1} } \Delta^{(-)} \psimomcmi{0}{i,k,q} ,
\label{eq:a81c}
\end{equation}
and
\begin{equation}
\Delta^{(-)} \psimomcmi{0}{i,k+2,q} = \frac{ \left(  \epscmi{ i,k+1} -  \epscmi{ i,k} \right) \epscmi{ i,k}^{2} \Delta \epscmi{ i,k} }
{ \left(  \epscmi{ i,k+2} -  \epscmi{ i,k+1} \right) \epscmi{ i,k+2}^{2} \Delta \epscmi{ i,k+2} } \Delta^{(-)} \psimomcmi{0}{i,k,q} .
\label{eq:a81d}
\end{equation}
Proceeding from $k = 0$ to $N_{k}-2$, $\psimomcmi{0}{i,k,q}$ is checked and if found to exceed unity, then $\Delta^{(-)} \psimomcmi{0}{i,k,q} = \psimomcmi{0}{i,k,q} - 1$ is applied to Equations (\ref{eq:a81c}) and (\ref{eq:a81d}) and the computed values of $\Delta^{(+)} \psimomcmi{0}{i,k+1,q}$ and $\Delta^{(-)} \psimomcmi{0}{i,k+2,q}$ are added and removed from zones $(i,k+1)$ and $(i,k+2)$, respectively. The neutrino energies $\epscmi{ i,N_{k}-1}$ and $\epscmi{ i,N_{k}}$ are always chosen large enough that they are never overfilled. Finally, if $\Delta^{(-)} \psimomcmi{0}{i,k+2,q}$ is found to exceed $\psimomcmi{0}{i,k+2,q}$, $\Delta^{(-)} \psimomcmi{0}{i,k,q}$ is adjusted to prevent $\psimomcmi{0}{i,k+2,q}$ from becoming negative. In this case the algorithm only works approximately. In practice, conservation of leptons is always found to be conserved to machine accuracy, so if the above algorithm is ever employed, it never encounters a value of $\Delta^{(-)} \psimomcmi{0}{i,k,q}$ that would cause $\psimomcmi{0}{i,k+2,q}$ to become negative.
\end{enumerate}

\subsection{Neutrino Energy Advection due to Changes in the Lapse}
\label{app:e_advct_lapse}

The lapse function, $a$, is given by Equation~(\ref{eq:a46}) and in differenced form by Equations~(\ref{eq:a62k}) and (\ref{eq:a62l}).
During the evolution of the core, the lapse function $a_{i}$ of mass shell $i$ may change as a result of a hydrodynamics step due to changes in the configuration of the core. 
New values of the lapse are computed in \chimera\ after the gravitational potential update, which follows the Lagrangian hydrodynamics step and remap, and after the gravitational potential update, which immediately precedes the transport solve. 
A change in lapse also affects neutrinos radially advected between adjacent zones, as described in Section \ref{spatial_adv}, as these zones may have different lapses.
Because our energy grid is tied to the lapse, these changes in the lapse change the energy grid, viz.,  $\epscmi{ i,k+\frac{1}{2}} = \Ecmi{ k+\frac{1}{2}}/a_{i}$. 
As before, the $\Ecmi{ k+\frac{1}{2}} = \epscmi{ k+\frac{1}{2},\infty}$ are the energy grid edges at radius infinity. 
Let $a_{i}^{n} \rightarrow a_{i}^{n+1}$ be the change in the lapse as a result of the evolution preceding a gravitational potential update. 
Let the superscript $n$ and $n+1$ now denote the values of quantities given after the prior energy advection update and the current energy advection update, respectively. 
Then the change in the energy grid from $n$ to $n+1$ is given by
\begin{equation}
\epscmi{ i,k+\frac{1}{2}}^{n} = \Ecmi{k+\frac{1}{2}}/a_{i}^{n} \rightarrow \epscmi{ i,k+\frac{1}{2}}^{n+1} = \Ecmi{k+\frac{1}{2}}/a_{i}^{n+1} ,
\label{eq:p1}
\end{equation}
and $\psimomcmi{0}{i,k}$ must change in such a way that the total number and energy of neutrinos in each radial zone $i$ remain the same.
Ignoring factors common to both sides, we must have
\begin{equation}
N_{\nu \, i,q}^{n} = \sum_{k=1}^{N_{\epsilon}} \psimomcmin{0}{i,k,q}{n} \, \Delta ( \epscmi{ i,k}^{n})^{3} = N_{\nu \, i,q}^{n+1} = \sum_{k=1}^{N_{\epsilon}} \psimomcmin{0}{i,k,q}{n+1} \, \Delta ( \epscmi{ i,k}^{n+1})^{3} ,
\label{eq:p2}
\end{equation}
which is just Equation (\ref{eq:a74}) with $V^{n+1} = V^{n}$ and $\Ecm^{n+1} = \Ecm^{n}$.
As in the energy remap step following the energy advection step described above, we also must have
\begin{equation}
E_{\nu \, i,q}^{n} = \sum_{k=1}^{N_{\epsilon}} \psimomcmin{0}{i,k,q}{n} \, \Delta ( \epscmi{i,k}^{n})^{3} \, \epscmi{i,k}^{n} = E_{\nu \, i,q}^{n+1} = \sum_{k=1}^{N_{\epsilon}} \psimomcmin{0}{i,k,q}{n+1} \,  \Delta ( \epscmi{i,k}^{n+1})^{3} \, \epscmi{ i,k}^{n+1} ,
\label{eq:p3}
\end{equation}
where
\begin{equation}
\Delta ( \epscmi{ i,k}^{n})^{3} = \frac{4\pi}{3} \left[ \left( \epscmi{ i,k+\frac{1}{2}}^{n} \right)^{3} - \left( \epscmi{  i,k- \frac{1}{2}}^{n} \right)^{3} \right] \equiv 4\pi \left( \epscmi{ i,k}^{n} \right)^{2} \Delta \epscmi{ i,k}^{n} .
\label{eq:p4}
\end{equation}
The quantity $\Delta ( \epscmi{ i,k}^{n+1})^{3}$ is defined similarly, and the equivalence sign is a consequence of our definition of $\epscmi{ i,k}$, Equation~(\ref{eq:a01}). 

To compute \psimomcmin{0}{i,k,q}{n+1}\ from \psimomcmin{0}{i,k,q}{n}\ due to a change in the lapse, the quantities $( \epscmi{ i,k}^{n} )^{2}\psimomcmin{0}{i,k,q}{n}$, related to the neutrino number density per unit energy, are PPM interpolated to the grid edges $\epscmi{ i,k+\frac{1}{2}}^{n}$. 
From the interpolated $( \epscmi{ i,k}^{n} )^{2}\psimomcmin{0}{i,k+\frac{1}{2}, q}{n}$, the flux \Flx{i+\frac{1}{2},k+\frac{1}{2}, q}\ across the grid boundary -- i.e., the number of neutrinos in the overlapping shells between the old and new grid -- is computed from the grid displacement given by Equation~(\ref{eq:p1}). 
With \Flx{i,k+\frac{1}{2}, q}\ in hand,  \psimomcmin{0}{i,k,q}{n+1}\  is then computed from \psimomcmin{0}{i,k,q}{n}\ by performing the advection step
\begin{equation}
\psimomcmin{0}{i,k,q}{n+1} \, \Delta ( \epscmi{ i,k}^{n+1})^{3} = \psimomcmin{0}{i,k,q}{n} \, \Delta ( \epscmi{ i,k}^{n})^{3} - \Flx{i,k+\frac{1}{2},q} + \Flx{i,k-\frac{1}{2},q} .
\label{eq:p5}
\end{equation}
Equation (\ref{eq:p5}), when summed from 1 to $N_{\epsilon}$, automatically conserves neutrino number if the fluxes at the boundaries are zero; i.e., $\Flx{i,1-\frac{1}{2},q} = \Flx{i,N_{\epsilon}+\frac{1}{2},q} = 0$, as the two terms on the right-hand side of this equation will then sum to zero, leaving us with Equation~(\ref{eq:p2}). 
However, neutrino energy is not necessarily conserved. 
To ensure energy conservation we modify Equation~(\ref{eq:p5}) by multiplying the two flux terms by a constant factor $\xi_{i,q}$; i.e.,
\begin{equation}
\psimomcmin{0}{i,k,q}{n+1} \, \Delta ( \epscmi{ i,k}^{n+1})^{3} = \psimomcmin{0}{i,k,q}{n} \, \Delta ( \epscmi{ i,k}^{n})^{3} + \xi_{i,q} \left( - \Flx{i,k+\frac{1}{2},q} + \Flx{i,k-\frac{1}{2},q} \right) .
\label{eq:p6}
\end{equation}
Equation (\ref{eq:p6}) with any constant value for the parameter $\xi_{i,q}$ automatically conserves total neutrino number, as does Equation~(\ref{eq:p5}), but the parameter is now adjusted so that Equation~(\ref{eq:p6}) conserves energy as well. 
To determine $\xi_{i,q}$, we equate the initial neutrino energy, $E_{\nu \, i,q}$, to the final energy, $E_{\nu,f,q}$, the latter being given by the right-hand side of Equation~(\ref{eq:p6}) summed over $k$.
We get, substituting Equation~(\ref{eq:p6}) into Equation~(\ref{eq:p3}),
\begin{eqnarray}
 E_{\nu \, i,q} &=& \sum_{k=1}^{N_{\epsilon}} \psimomcmin{0}{i,k,q}{n+1} \, \Delta ( \epscmi{ i,k}^{n+1})^{3} \, \epscmi{ i,k}^{n+1} \nonumber \\ 
 &=& \sum_{k=1}^{N_{\epsilon}} \left[  \psimomcmin{0}{i,k,q}{n} \, \Delta ( \epscmi{ i,k}^{n})^{3} + \xi_{i,q} \left( - \Flx{i,k+\frac{1}{2},q} + \Flx{i,k-\frac{1}{2},q} \right) \right] \epscmi{ i,k}^{n+1} .
\label{eq:p7}
\end{eqnarray}
Solving Equation (\ref{eq:p6}) for $\xi_{i,q}$ gives
\begin{equation}
\xi_{i,q} = \frac{ E_{\nu \, i,q} - \sum_{k=1}^{N_{\epsilon}}  \psimomcmin{0}{i,k,q}{n} \, \Delta ( \epscmi{ i,k}^{n})^{3} \, \epscmi{ i,k}^{n+1} }{ \left( - \Flx{i,k+\frac{1}{2},q} + \Flx{i,k-\frac{1}{2},q} \right) \, \epscmi{ i+\frac{1}{2},k}^{n+1} } .
\label{eq:p8}
\end{equation}
Equation (\ref{eq:p6}), with $\xi_{i,q}$ given by Equation~(\ref{eq:p8}), conserves total neutrino number and energy and is the method by which \chimera\ advances  \psimomcmin{0}{i,k,q}{n}\ due to a change in the lapse.

\subsection{Neutrino Spatial Advection Step}
\label{spatial_adv}

As described in Section \ref{sec:Regrid}, following the Lagrangian hydrodynamics step in the case of the $\theta$- and $\phi$-sweeps, the grid is remapped back to the configuration that prevailed before the Lagrangian step.
In the case of the radial sweep, the grid is remapped back to a configuration specified by the regridder.
In either case the zero moments of the neutrino distribution functions must be advected through the grid in the remapping step.
The procedure is similar to that described in Section~\ref{remap1} for the mass specific quantities like the mass specific momenta, and mass specific angular momenta, but simpler, since the quantities to be remapped are volume specific rather than mass specific, thus omitting the necessity of first computing the advected mass.

If $\delta \xi_{i+\half} < 0$, where $\delta \xi_{i+\half}$is the displacement of the grid element $\xi_{i+\half}$ during the remap step, the advection proceeds from the left of $\xi_{i+\half}$ to the right; i.e., from zone $i$ to $i+1$.
If $a_{i,k,q}^{n+1'}$ represents  \psimomcmin{0}{i,k,q}{n+1'}, the zeroth angular moment of neutrinos of species $q$, radial/angular/azimuthal index $i$, and energy zone $k$, a PPM interpolation profile, $a_{i,k,q}(\xi)$, of $a_{i,k,q}^{n+1'}$ is constructed, as described in Section \ref{Interpolation}. The average, $\langle a \rangle_{L, \, i+\half, k,q}^{n+1'}$, of $a_{i,k,q}^{n+1'}$ (``L" denoting ``left") over the zone interface displacement $\delta \xi_{i+\half}$ is then computed, and the volume displacement $\delta V_{i+\half}^{n+1'}$ is computed from Equation~(\ref{eq:h102}). The quantity of $a_{i,k,q}^{n+1'}$ advected is then finally computed by
\begin{equation}
\delta {\cal A}_{i+\half, k, q}^{n+1'} = \langle a \rangle_{L, \, i+\half, k, q}^{n+1'} \delta V_{i+\half}^{n+1'} .
\label{eq:a84}
\end{equation}
If $\delta \xi_{i+\half} > 0$, during the remap the advection proceeds from the right to the left of interface $\xi_{i+\half}^{n+1'}$; i.e., from $i+1$ to $i$.
In this case the quantity advected, $\delta {\cal A}_{i+\half, k, q}^{n+1'}$, is computed from an equation similar to Equation~(\ref{eq:a84}) but using $\langle a \rangle_{R, \, i+\half, k, q}^{n+1'}$ (``R" denoting ``right") computed from $a_{i+1,k,q}^{n+1'}$.

In either case the advection then proceeds as
\begin{equation}
a_{i,k,q}^{n+1} = \frac{ a_{i,k,q}^{n+1'} \, V_{i}^{n+1'}  - \delta {\cal A}_{i+\half, k, q}^{n+1'} + \delta {\cal A}_{i-\half, k, q}^{n+1'} }{ V_{i}^{n+1} } ,
\label{eq:a85}
\end{equation}
where again we denote the variables after the Lagrangian step but before the remap step by the superscript $n+1'$, and after the remap step by the superscript $n+1$.

When \chimera\ is run in relativistic mode, the lapse function becomes a function of the location in the core, and the neutrino energy zones also become a function of location through the lapse, as given by Equation~(\ref{eq:a8a}).
In this case \chimera\ adds a couple of steps to the procedure described above for remapping \psimomcm{0}. 
If $\delta \xi_{i+\half} < 0$, neutrinos are advected from zone $i$ to zone $i+1$, and account must be taken of the different lapse functions in these two zones.
This is accomplished by appropriately advecting in energy  $\langle a \rangle_{L, \, i+\half, k, q}^{n+1'}$ before computing $\delta {\cal A}_{i+\half, k, q}^{n+1'}$ by Equation~(\ref{eq:a84}). 
This energy advection is performed by the algorithm described by Equations~(\ref{eq:a78}) -- (\ref{eq:a81}), with $a_{i}$ replacing $a^{n}$, $a_{i+1}$ replacing $a^{n+1'}$, and with $\rho^{n} = \rho^{n+1'}$, $dR = 0$, and $db = 0$. 
If $\delta \xi_{i+\half} > 0$, the neutrinos are advected from zone $i+1$ to $i$ and a similar modification of $\langle a \rangle_{R, \, i+\half, k, q}^{n+1'}$ is performed by the neutrino energy advection algorithms just described but with $a_{i+1}$ replacing $a^{n}$, and $a_{i}$ replacing $a^{n+1'}$.

\subsection{Scalar Eddington Factors}
\label{eddington}

The scalar Eddington factor, $\eta^{(2)} = \psimomcm{2}/ \psimomcm{0}$,  appears in the neutrino energy advection Equations~(\ref{eq:a66}), (\ref{eq:a69}), and (\ref{eq:a79}).
Consider first the case for  $R > R_{\nu}$, where $R_{\nu}$ is the radius of the neutrinosphere, and is a function of both neutrino energy and flavor. 
Analogous to the derivation of the geometric piece of the flux limiter, Equation~(\ref{eq:a23}), we use Equation~(\ref{eq:a10}) for \psimomcm{2}\ to write
\begin{equation}
\psimomcm{2} = \frac{1}{3} \left( 1 + \mucm + \mucm^{2} \right) \psimomcm{0} ,
\label{eq:a82}
\end{equation}
where this assumes that the neutrino distribution function \fcm\ is constant for rays satisfying $\mucm < \mucmi{\nu}$, and zero otherwise, where \mucmi{\nu}\ is defined in the text above Equation~(\ref{eq:a23}). 
The angle \mucm\ is given by Equations~(\ref{eq:a24})--(\ref{eq:a26}).

Where $R < R_{\nu}$, the diffusion limit applies, \psimomcm{1}\ is well defined, and we solve the equation $\psimomcm{1} = \frac{1}{2} (1 + \mucmi{ \nu})\psimomcm{0}$ for $\mucm$ to get
\begin{equation}
\mucm = 2 \eta^{(1)} - 1 ,
\label{eq:a83}
\end{equation}
where the flux factor $\eta^{(1)}$ is the ratio of \psimomcm{1}\ to \psimomcm{0}. 
Using Equation~(\ref{eq:a83}) in Equation~(\ref{eq:a82}), we get the correct behavior in the strong diffusion limit, where $\psimomcm{1} \rightarrow 0$, as this implies that $\eta^{(1)} \rightarrow 0$ and $\eta^{(2)} \rightarrow \rfrac{1}{3}$.

Versions of \chimera\ \added{subsequent to version C)} have used the Minerbo closure \citep{Mine78}, constructed on the maximum entropy principle. 
This closure is the classical limit of the of closure constructed by \citet{CeBl94} on the same principle but using the Fermi-Dirac distribution, and is much easier to implement and gives very similar results \citep{muAbUr17}. 
For this closure, the quantity $p$ is defined by
\begin{equation}
p = \frac{1}{3} + \frac{ 2 \left( \eta^{(1)} \right)^{2} }{ 15 } \left[ 3 - \eta^{(1)} + 3 \left( \eta^{(1)} \right)^{2} \right],
\label{eq:a86}
\end{equation}
and $\eta^{(2)}$ is given by
\begin{equation}
\eta^{(2)} = \frac{3 ( 1 - p )}{2} \eta_{\rm thick}^{(2)} + \frac{ 3 p - 1 }{2} \eta_{\rm thin}^{(2)} ,
\label{eq:a87}
\end{equation}
where $\eta_{\rm thick}^{(2)} = 1/3$, $\eta_{\rm thin}^{(2)} = 1$, and as before, $\eta^{(1)} = \psi^{(1)}/\psi^{(0)}$. The closure given by Equations~(\ref{eq:a86}) and (\ref{eq:a87}) gives results slightly closer to those given by a Boltzmann solver than those given by Equations~(\ref{eq:a82}) and (\ref{eq:a83}).

\section{Stationary State Transport Tests}
\label{trans_tests}

In the spirit of the tests suggested by \citet{MuJaDi10}, involving the stationary solution of the transport equations, we set up several analogous test problems.
As in \citet{MuJaDi10}, we consider a central source consisting of a homogeneous isothermal sphere of radius R of energy-independent absorption opacity and zero scattering opacity radiating into a medium of negligible absorption and scattering opacities.

\subsection{Gravitational Redshift}
\label{grav_redshift}

\begin{figure}
\gridline{\fig{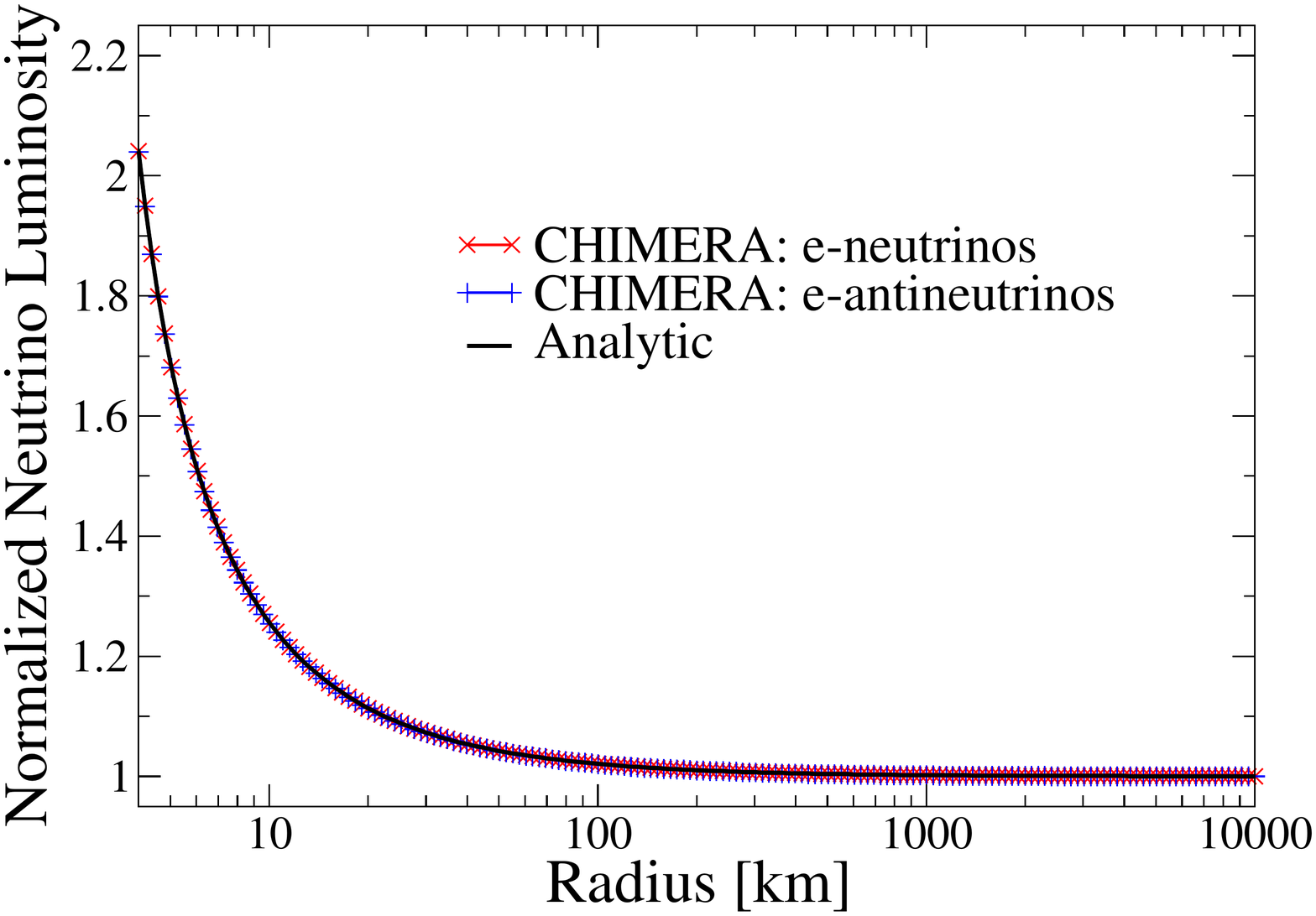}{0.5\textwidth}{(a)}
              \fig{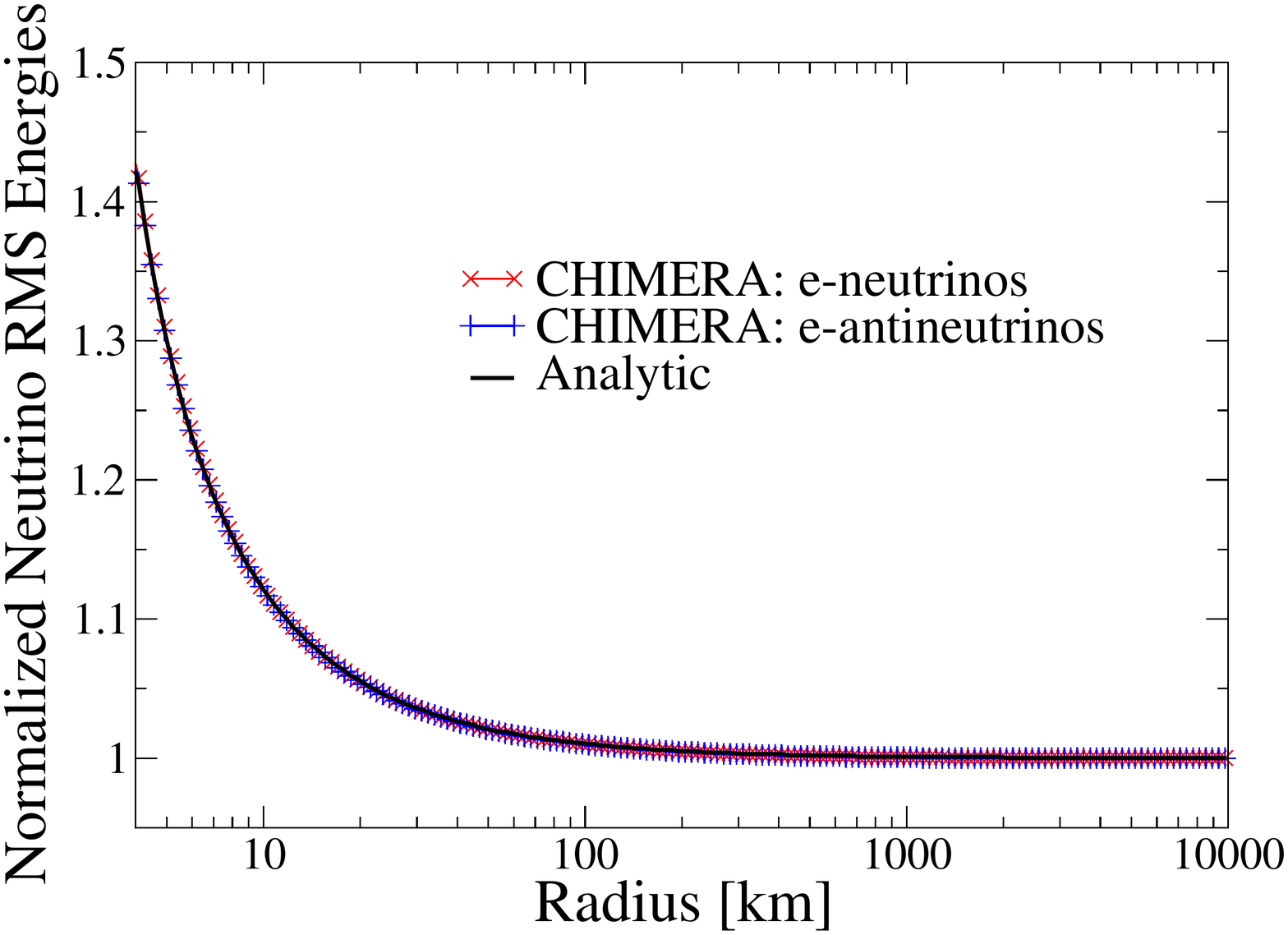}{0.5\textwidth}{(b)}
               }
\caption{\label{Lapse_comparison}
Panel (a): Comparison of \nue\ (red) and \nuebar\ (blue) luminosities with the analytic solution of Equation~(\ref{eq:t4}). 
Panel (b): Same as (a), but for the mean energies, compared to the analytic solution given by Equation~(\ref{eq:t7}).
The solutions have been normalized to unity at $R = 1000$~km.
}
\end{figure}

In this test we assume that the medium external to the central source is static, and impose a lapse profile as a function of R given by
\begin{equation}
a(R) = \sqrt{ 1 - R_{0}/R }, \quad R_{0} = 2.04 \mbox{ km}, \; R > 4 \mbox{ km} ,
\label{eq:t1}
\end{equation}
where $R_{0}$ is chosen so that $a(4\mbox{ km}) = 0.70$.
Equation~(\ref{eq:t1}) for $a(R)$ is suggested by the behavior of $a(R)$ outside a spherically symmetric source of gravity in the post-Newtonian approximation and numerically approximates the behavior of $a(R)$ in realistic core collapse models $\sim$50--100~ms after bounce.
An analytic expression for $\psimomcm{1}(R, \Ecm)$ can be derived for this problem, from which follows an analytic expression for the neutrino luminosity $L(R)$. With the assumption of a static medium, the derivatives with respect to $t$ all vanish, and Equation~(\ref{eq:a12}) becomes
\begin{equation}
\left( \pderiv{ \psimomcm{1} }{R} \right)_{t,\Ecm} + \left( - \pderiv{\ln a}{t} + \frac{1}{R} \right) 2 \psimomcm{1}
 = \frac{a^{2}}{R^{2}}  \left( \pderiv{ }{R} \left( \frac{R^{2}}{a^{2}} \psimomcm{1}\right) \right)_{t,\Ecm}
 = 0 ,
\label{eq:t2}
\end{equation}
which gives
\begin{equation}
\psimomcm{1}(R, \Ecm) = \frac{( a/a_{p} )^{2}}{( R/R_{p} )^{2}} \psimomcm{1}(R_{p}, \Ecm) ,
\label{eq:t3}
\end{equation}
where $R_{p}$ is some fiducial radius and $a_{p} = a(R_{p})$.
Using Equation~(\ref{eq:t3}), the neutrino luminosity, $L(R)$, is then given by
\begin{equation}
L(R) = \frac{4 \pi R^{2}}{ c^{2} h^{3} } \int_{0}^{\infty} \epsilon^{3} d\epsilon \psimomcm{1}(R, \Ecm)
= \frac{4 \pi R^{2}}{ c^{2} h^{3} } \int_{0}^{\infty} \frac{ \Ecm^{3} d\Ecm }{a^{4}} \frac{( a/a_{p} )^{2}}{( R/R_{p} )^{2}} \psimomcm{1}(R_{p}, \Ecm) \propto \frac{1}{a^{2}(R)}.
\label{eq:t4}
\end{equation}
The luminosity from the central source thus goes as $a^{-2}(R)$ in the surrounding static medium for this problem.
Figure~\ref{Lapse_comparison}(a) compares the analytic expression (Equation~\ref{eq:t4}) with the solutions for the \nue\ and \nuebar\ luminosities given by \Chimera.
The numerical solutions agree extremely well with the analytic solution, which is not surprising, given that the variable transformation (Equation~\ref{eq:a8a}) and the subsequent differencing of the transport equation essentially guarantee this agreement.

An analytic expression for the mean neutrino energy can also be derived. With the assumption of a static medium, Equation~(\ref{eq:a13}) becomes
\begin{equation}
\psimomcm{0} \left( \pderiv{\eta^{(2)}}{R} \right)_{t,\Ecm} + \eta^{(2)} \left( \pderiv{\psimomcm{0}}{R} \right)_{t,\Ecm}  - \left( - \pderiv{\ln }{t} + \pderiv{\ln R}{t} \right) \left( 1 - 3 \eta^{(2)} \right) \psimomcm{0},
\label{eq:t5}
\end{equation}
where the scalar Eddington factor $\eta^{2}$ is defined below Equation~(\ref{eq:a13}).
In the limit of a small angular diameter source, $\eta^{(2)} \rightarrow 1$ and Equation~(\ref{eq:t5}) reduces to an equation for \psimomcm{0}\ similar to Equation~(\ref{eq:t2}) for \psimomcm{1}\ , with a solution similar to Equation~(\ref{eq:t3}).
More generally, we note that the condition of the energy independence of the absorption and emission opacities, in addition to the static condition, implies that $\eta^{(2)}(R, \Ecm) = \eta^{(2)}(R)$. Therefore, Equation~(\ref{eq:t5}) for \psimomcm{0}\ depends only on $R$, and the solution can be written as
\begin{equation}
\psimomcm{0}(R, \Ecm) = \chi(R) \psimomcm{0}(R_{p}, \Ecm) ,
\label{eq:t6}
\end{equation}
where again $\psimomcm{0}(R_{p}, \Ecm)$ is the value of $\psimomcm{0}(R, \Ecm)$ at some fiducial radius $R_{p}$ and $\chi(R)$ is the solution of Equation~(\ref{eq:t5}) for $\chi(R)$ with $\psimomcm{0}(R_{p}, \Ecm)$ substituted for $\psimomcm{0}(R, \Ecm)$,  under the condition that $\chi(R_{p}) = 1$.
The mean energy is therefore given by
\begin{equation}
\langle \Ecm \rangle \frac{ \int_{0}^{\infty} \epscm^{3} d\epscm \psimomcm{0}(R, \Ecm) }{ \int_{0}^{\infty} \epscm^{2} d\epscm \psimomcm{0}(R, \Ecm) } = \frac{ {a^{-4}(R)} \int_{0}^{\infty} \Ecm^{3} d\Ecm \psimomcm{0}(R_{p}, \Ecm) }{ {a^{-3}(R)} \int_{0}^{\infty} \Ecm^{2} d\Ecm \psimomcm{0}(R_{p}, \Ecm) } \propto \frac{1}{a(R)} .
\label{eq:t7}
\end{equation}

For this problem, the mean energy from the central source thus goes as $a^{-1}(R)$ in the surrounding static medium.
Figure~\ref{Lapse_comparison}(b) shows that the analytic expression (Equation~\ref{eq:t7}) agree extremely well with the solutions for the \nue\ and \nuebar\ mean energies given by \Chimera.

\subsection{Imposed Shock Velocity Profile}
\label{shock_v}

\begin{figure}
\gridline{\fig{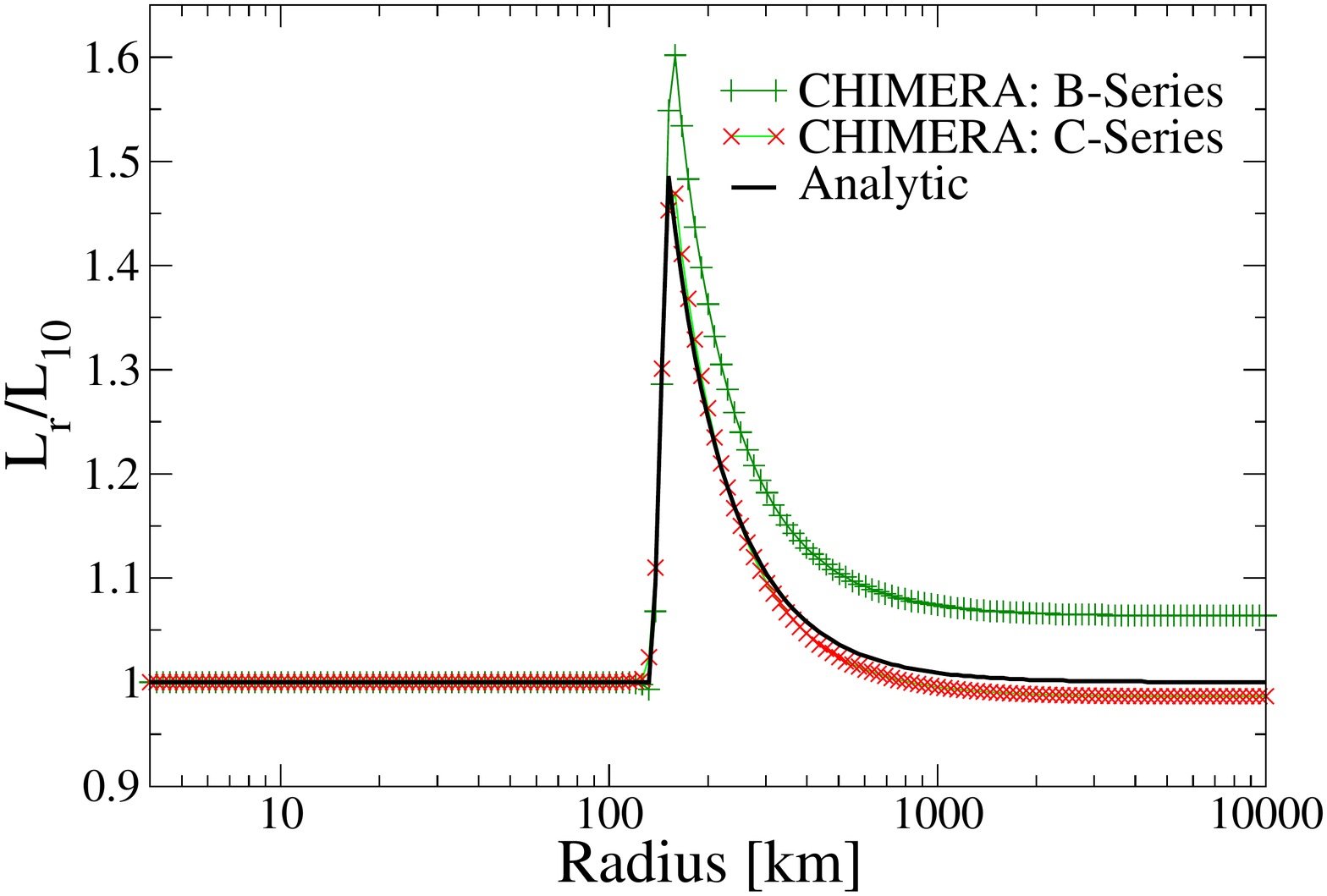}{0.5\textwidth}{(a)}
              \fig{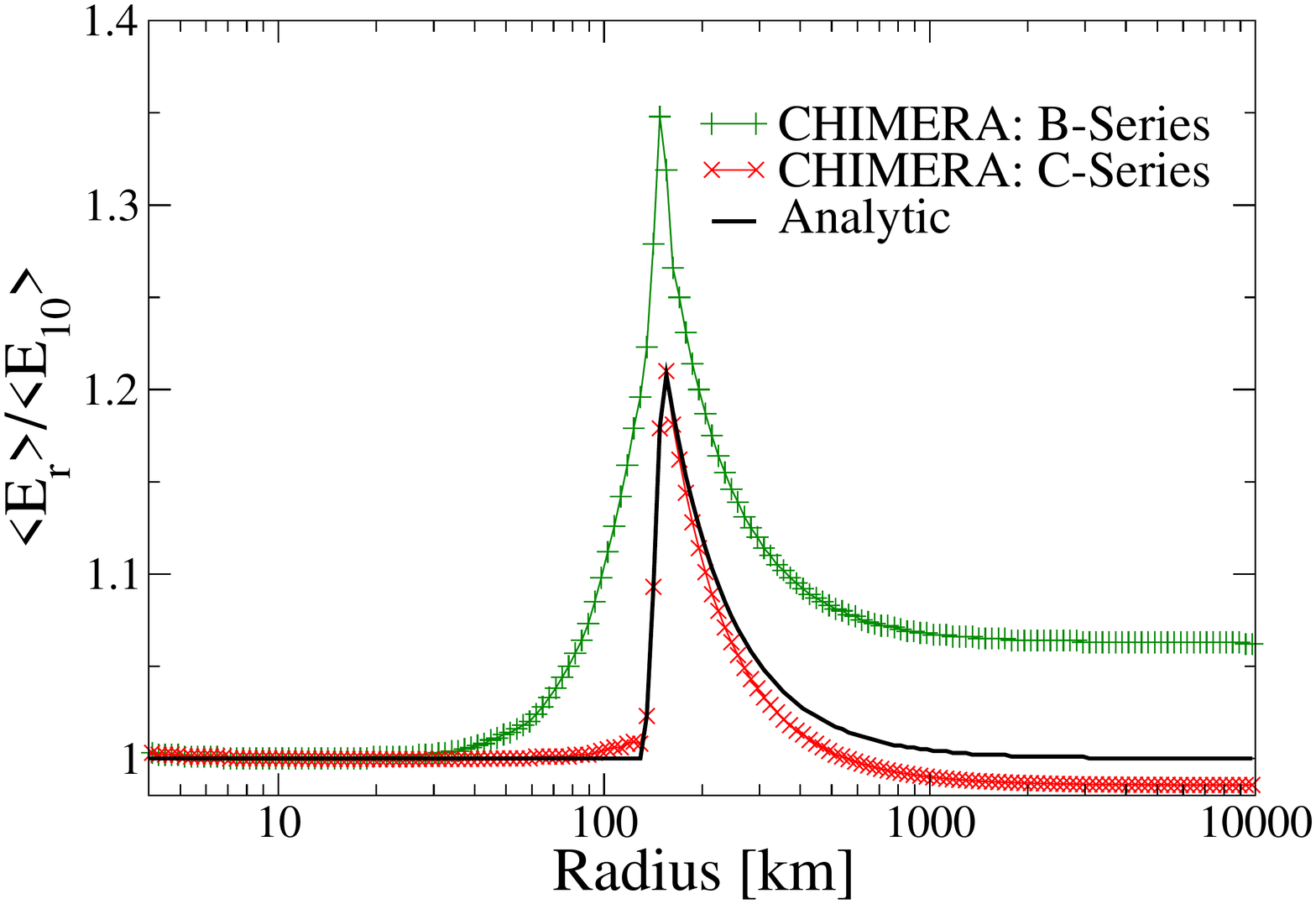}{0.5\textwidth}{(b)}
               }
\caption{\label{fig:ImposedShock}
Comoving \nue\ (a) luminosities and (b) mean energies computed with B-series \chimera\ (green) and C-series \chimera\ (red; including the shock transport treatment in Section~\ref{trans_shock}) compared with the analytic solution for luminosity (Equation~(\ref{eq:t11}), Panel (a)) and mean energy (Equation~(\ref{eq:t12}), Panel (b)) for the velocity profile specified in Equation~(\ref{eq:t8}). Solutions have been normalized to unity at $R = 10$ km. Normalized results for \nuebar\ are similar but the actual \nuebar\ luminosities are considerably reduced in magnitude and the actual mean energies are slightly reduced in magnitude.
}
\end{figure}

In this test a radial velocity field is imposed that mimics the velocity profile encountered during the accretion phase of a \ccsn. The velocity profile suggested by \citet{MuJaDi10}  is
\begin{equation}
v_{r} = \left\{ \begin{array}{cl} 0, & r < 135 \mbox{ km}; \\
-0.2c \ds \frac{ r - 135 \mbox{ km} }{ 15 \mbox{ km}}, & 135 \mbox{ km} \le r < 150 \mbox{ km}; \\
-0.2c \left( \ds \frac{ 150 \mbox{ km} }{ r } \right)^{2}, & 150 \mbox{ km} \le r .
\end{array} \right.
\label{eq:t8}
\end{equation}
Inside the homogeneous, isotropic central spherical source of 4 km radius we again turn off all scattering opacities and, rather than employ a frequency independent absorption opacity as in the preceding test, employ instead the \citet{Brue85} free nucleon absorption opacity corresponding to the state ($\rho, T, \Ye$) = ($ 10^{11}~\gcc, 4 \mbox{ MeV}, 0.5$).
Outside the central source the absorption and scattering opacities vanish, so the luminosity $L$ and the mean neutrino energy $\langle E \rangle$ are constant with $R$ in the lab frame. The nonzero velocity regime is at a large enough radius, compared with the source radius, that the neutrino flow can be well approximated as radially free streaming. In this case both $f(\epsilon,\mu)$ and $\fcm(\epscm,\mucm)$, the invariant neutrino occupation probabilities, vanish except at $\mu = \pi$, $\mucm = \pi$, respectively. The comoving neutrino energy density $\Ecmi{ \nu}$ is then related to the lab frame energy density $E_{\nu}$ by 
\begin{eqnarray}
\Ecmi{ \nu} &=& \frac{2\pi}{(hc)^{3}} \int \fcm( \epscm, \mucm ) \, \epscm^{3} \, d\mucm \, d\epscm
= \frac{2\pi}{(hc)^{3}} \int \fcm( \epscm, \pi ) \, \epscm^{3} \, d\epscm \nonumber \\ 
& = & \frac{2\pi}{(hc)^{3}} \int f( \epsilon, \mu ) \, \epscm^{2} \, \epsilon \, d\mu \, d\epsilon
= \gamma^{2} ( 1 - \beta )^{2} \frac{2\pi}{(hc)^{3}} \int f( \epsilon, \pi ) \, \epsilon^{3} \, d\epsilon
= \gamma^{2} ( 1 - \beta )^{2}  E_{\nu} . \label{eq:t9}
\end{eqnarray}
Likewise, the comoving neutrino number density, $\Ncmi{ \nu}$, is related to the lab frame number density, $N_{\nu}$, by
\begin{equation}
\Ncmi{\nu} = \frac{2\pi}{(hc)^{3}} \int \fcm( \epscm, \mucm ) \, \epscm^{2} \, d\mucm \, d\epscm
= \frac{2\pi}{(hc)^{3}} \int f( \epsilon, \mu ) \, \epscm \, \epsilon \, d\mu \, d\epsilon
= \gamma ( 1 - \beta ) N_{\nu} .
\label{eq:t10}
\end{equation}
The comoving neutrino luminosity is related to the (constant) lab frame luminosity by
\begin{equation}
\Lcmi{\nu} = 4\pi R^{2} c \Ecmi{ \nu} = \gamma^{2} ( 1 - \beta^{2} ) 4\pi R^{2} c  E_{\nu} = \gamma^{2} ( 1 - \beta )^{2} L_{\nu} = \frac{ 1 - \beta }{ 1 + \beta } \times \mbox{constant,}
\label{eq:t11}
\end{equation}
and the comoving mean neutrino energy, $\langle \Ecmi{ \nu} \rangle$, is related to the (constant) lab frame mean neutrino energy, $\langle E_{\nu} \rangle$,  by
\begin{equation}
\langle \Ecmi{ \nu} \rangle = \frac{ \Ecmi{ \nu} }{ \Ncmi{ \nu} } = \gamma ( 1 - \beta ) \frac{ E_{\nu} }{ N_{\nu} } = \gamma ( 1 - \beta ) \times \mbox{constant} .
\label{eq:t12}
\end{equation}

Figure~\ref{fig:ImposedShock}  shows the results of the B-series and the C-series \chimera\ transport, versus the analytical expressions given by Equations~(\ref{eq:t11}) and (\ref{eq:t12}) for the neutrino luminosity and mean energies, respectively.
Measured from zero, the luminosity and the mean energy as given by the B-series transport overshoot the correct luminosity and mean energy at the shock by $\sim 7$\% and $\sim 13$\%, respectively, and these solutions remain above the analytic solution at large $R$ by $\sim$6\%.
Moreover, the B-series solution for the comoving mean energy fails to resolve the steep rise of the mean energy at the shock, as given by the analytic solution.
The C-series solutions for both the luminosity and the mean energy, however, agree quite well with the analytic solutions, only deviating below the analytic solution at large $R$ by $\sim 1.5$\%, demonstrating the impact of the added special treatment of transport across the shock, discussed in Section~\ref{trans_shock}.

\section{Transport Sources}
\label{app:source}

Solving the neutrino radiation hydrodynamics also requires detailed neutrino--matter interactions to couple the radiation field to the fluid, which drives heating, cooling, and changes in \Ye.
Table~\ref{tab:opacities} lists the scattering, absorption-emission, and pair-production opacities currently incorporated in \Chimera.
For each zone, the logarithms of the opacities are stored at each of the eight corners of a cell in $\rho$--$T$--\Ye\ lattice space that surrounds the $(\rho, T, \Ye)$ value of each zone. 
The values at $(\rho, T, \Ye)$ and the \Ye-derivatives needed by the Jacobian for the solution of the  transport equation are then obtained by a three-dimensional interpolation from the eight corner values in the same manner as for the thermodynamic variables of the EoS. 
The spacing of the $\rho$--$T$--\Ye\ grid matches the user-selected spacing of the EoS grid, but is typically 20 points per decade in $\log \rho$ and $\log T$, with 0.01 intervals in $\Ye$. 
This resolution ensures that interpolated opacities match those that are computed directly, to within 1\%. 

\begin{deluxetable*}{lll}
\tabletypesize{\scriptsize}
\tablecaption{Summary of Neutrino Opacities\label{tab:opacities}}
\tablecolumns{3}
\tablewidth{0pt}
\tablehead{
\colhead{Process} & \colhead{Description} & \colhead{References}
}
\startdata
$\nu + e^{\pm} \leftrightharpoons \nu + e^{\pm}$ & Section \ref{scatnu_em}, \ref{scatnu_ep} & \citet{Brue85, MeBr93c} \\
$\nu + A \leftrightharpoons \nu + A$ & Section \ref{scatnu_A} & \citet{BrMe97, Horo97} \\
$\nu + n, p \leftrightharpoons \nu + n, p$ & Section \ref{scatnu_np} & \citet{RePrLa98, BuRaJa06} \\
$\nue + n \leftrightharpoons e^{-} + p$ & Section \ref{abem_np} &  \citet{Brue85, MeBr93c} \\
$\nuebar + p \leftrightharpoons e^{+} + n$ & Section \ref{abem_np} &  \citet{Brue85, MeBr93c} \\
$\nue + A' \leftrightharpoons e^{-} + A$ & Section \ref{abem_nuc} &  \citet{LaMaSa03, HiMeMe03} \\
$\nue + \nuebar \leftrightharpoons e^{-} + e^{+}$ & Section \ref{pairgen},  \ref{eepairgen} &  \citet{Brue85, MeBr93c} \\
$n, p + n, p \leftrightharpoons n, p + n, p + \nue + \nuebar$ & Section \ref{pairgen},  \ref{brem} & \citet{HaRa98} \\
\enddata
\end{deluxetable*}

\newpage
\subsection{Scattering: General}
\label{scatgen}

In spherical symmetry, the rate of change of the neutrino occupation probability, $f(\mucm,\epscm)$, (we still suppress the dependence of $f$ on $t$ and $R$), due to scattering process ``XX'' is given by
\begin{eqnarray}
\left( \frac{ df(\mucm, \epscm) }{ dt } \right)_{ \! {\rm S, \, scat \, XX}}&= &[ 1 - f(\mucm, \epscm) ] \frac{1}{(hc)^{3}} \int_{0}^{\infty} \epscm'^{2} d\epscm' \int_{-1}^{1} d\mucm' f(\mucm', \epscm') \int_{0}^{2\pi} d\beta' R_{\rm XX}^{\rm in}(\epscm, \epscm', \cos\theta ) \nonumber \\ 
&& - f(\mucm, \epscm) \frac{1}{(hc)^{3}} \int_{0}^{\infty} \epscm'^{2} d\epscm' \int_{-1}^{1} d\mucm' [1 - f(\mucm', \epscm') ] \int_{0}^{2\pi} d\beta' R_{\rm XX}^{\rm out}(\epscm, \epscm', \cos\theta ) , 
\label{eq:d1}
\end{eqnarray}
where $R_{\rm XX}^{\rm out}(\epscm, \epscm', \cos\theta )$ is the ``out-scattering kernel,'' (i.e., the neutrino-unblocked rate per final and initial neutrino state for scattering from energy $\epscm$ to energy $\epscm'$ through angle $\theta$), $R_{\rm XX}^{\rm in}(\epscm, \epscm', \cos\theta )$ is the ``in-scattering kernel,'' (i.e., the neutrino-unblocked rate per final and initial neutrino state for scattering from energy $\epscm'$ to energy $\epscm$ through angle $\theta$), and $\theta$ is given in terms of  the individual neutrino propagation directions, $(\mucm, \varphi)$ and $(\mucm', \varphi')$, by 
\begin{equation}
\cos \theta = \mucm \mucm' + \sqrt{ (1 - \mucm^{2} )(1 - \mucm'^{2} ) }\, \cos \delta \varphi ,
\label{eq:d2}
\end{equation}
where $\delta \varphi =\varphi' - \varphi$ is the azimuthal direction between the incident and scattered neutrino.
The Fermi blocking of neutrino states is incorporated explicitly in Equation~(\ref{eq:d1}) by the $( 1 - f )$ factors rather than in the scattering kernels. 
The scattering kernels $R_{\rm XX}^{\rm out/in}$ have the symmetry
\begin{equation}
R_{\rm XX}^{\rm in}(\epscm, \epscm',\cos\theta) = R_{\rm XX}^{\rm out}(\epscm', \epscm,\cos\theta),
\label{eq:d3}
\end{equation}
which follows simply from the fact that an in-scattering from $\epscm'$ to $\epscm$ is the same as an out-scattering from $\epscm'$ to $\epscm$.
Additionally, the kernels are related to each other by detailed balance at $\beta = 1/(k T)$:
\begin{equation}
R_{\rm XX}^{\rm in}(\epscm, \epscm',\cos\theta)  = e^{\beta (\epscm' - \epscm ) } R_{\rm XX}^{\rm out}(\epscm, \epscm',\cos\theta),
\label{eq:d4}
\end{equation}
which follows by substituting equilibrium Fermi-Dirac distributions for $f$ in Equation~(\ref{eq:d1}) and setting the left-hand side of that equation to zero.
Both Equations~(\ref{eq:d3}) and (\ref{eq:d4}) should be respected in any approximation scheme.

For neutrino scattering, the complication that the occupation functions $f$ and $f'$ are expressed in terms of $\mucm$ and $\mucm'$, respectively, while the scattering kernels are expressed in terms of $\theta$, where  $\mucm$, $\mucm'$, and $\theta$ are related by Equation~(\ref{eq:d2}), is overcome by Legendre expanding the scattering kernels and keeping only the terms to first order; i.e.,
\begin{eqnarray}
 R_{\rm XX}^{\rm in/out}(\epscm,\epscm',\cos\theta) &=& \frac{1}{2} \sum_{\ell=0}^{\infty} (2\ell + 1) \Phi_{\ell ,{\rm XX}}^{\rm in/out}(\epscm,\epscm') P_{\ell}( \cos\theta ) \nonumber \\ 
& \simeq& \frac{1}{2} \Phi_{0,{\rm XX}}^{\rm in/out}(\epscm,\epscm') + \frac{3}{2} \Phi_{1,{\rm XX}}^{\rm in/out}(\epscm,\epscm') \cos\theta , \label{eq:d5}
\end{eqnarray}
where the Legendre coefficients are given by
\begin{equation}
  \Phi_{\ell,{\rm XX}}^{\rm in/out}(\epscm,\epscm') = \int_{-1}^{1} d(\cos\theta) P_{\ell}( \cos\theta ) R_{\rm XX}^{\rm in/out}(\epscm,\epscm',\cos\theta).
\label{eq:d6}
\end{equation}
Note that Equations~(\ref{eq:d3}) and (\ref{eq:d4}) imply
\begin{eqnarray}
\Phi_{\ell,{\rm XX}}^{\rm in}(\epscm,\epscm') &=& \Phi_{\ell,{\rm XX}}^{\rm out}(\epscm',\epscm)  \label{eq:d7down} , \\
\Phi_{\ell,{\rm XX}}^{\rm in}(\epscm,\epscm') &=& e^{\beta (\epscm' - \epscm ) } \Phi_{\ell , {\rm XX}}^{\rm out}(\epscm,\epscm') \label{eq:d7up} .
\end{eqnarray}

Applying the moment operators $(4\pi)^{-1} \int d\Omega$ and $(4\pi)^{-1} \int \mucm d\Omega$ to Equation~(\ref{eq:d1}) and using the definitions in Equations~(\ref{eq:d2}), (\ref{eq:d5}), and (\ref{eq:a10}),  we get the moments of the scattering terms of the collision integral:
\begin{eqnarray}
\left( \frac{ df(\mucm, \epscm) }{ dt } \right)_{ \! {\rm S, \, scat} }^{(0,1)} &=& \frac{2\pi}{(hc)^{3}} \left[ 1 - \psimomcm{0}(\epscm), - \psimomcm{1}(\epscm) \right] \int_{0}^{\infty} \epscm'^{2} d\epscm' \Phi_{0, {\rm XX}}^{\rm in}(\epscm,\epscm') \psimomcm{0}(\epscm')  \nonumber \\ 
&&+ \frac{2\pi}{(hc)^{3}} \left[ -3 \psimomcm{1}(\epscm), 1 -3 \psimomcm{2}(\epscm) \right] \int_{0}^{\infty} \epscm'^{2} d\epscm' \Phi_{1,{\rm XX}}^{\rm in}(\epscm,\epscm') \psimomcm{1}(\epscm')  \nonumber \\ 
&&- \frac{2\pi}{(hc)^{3}} \left[ \psimomcm{0}(\epscm), \psimomcm{1}(\epscm) \right] \int_{0}^{\infty} \epscm'^{2} d\epscm' \Phi_{0, {\rm XX}}^{\rm out}(\epscm,\epscm') \left[ 1 - \psimomcm{0}(\epscm') \right] \nonumber \\ 
&& + \frac{2\pi}{(hc)^{3}} \left[ 3 \psimomcm{1}(\epscm), 3 \psimomcm{2}(\epscm) \right] \int_{0}^{\infty} \epscm'^{2} d\epscm' \Phi_{1,{\rm XX}}^{\rm out}(\epscm,\epscm') \psimomcm{1}(\epscm') .  \label{eq:d8}
\end{eqnarray}

\begin{figure}
\gridline{\fig{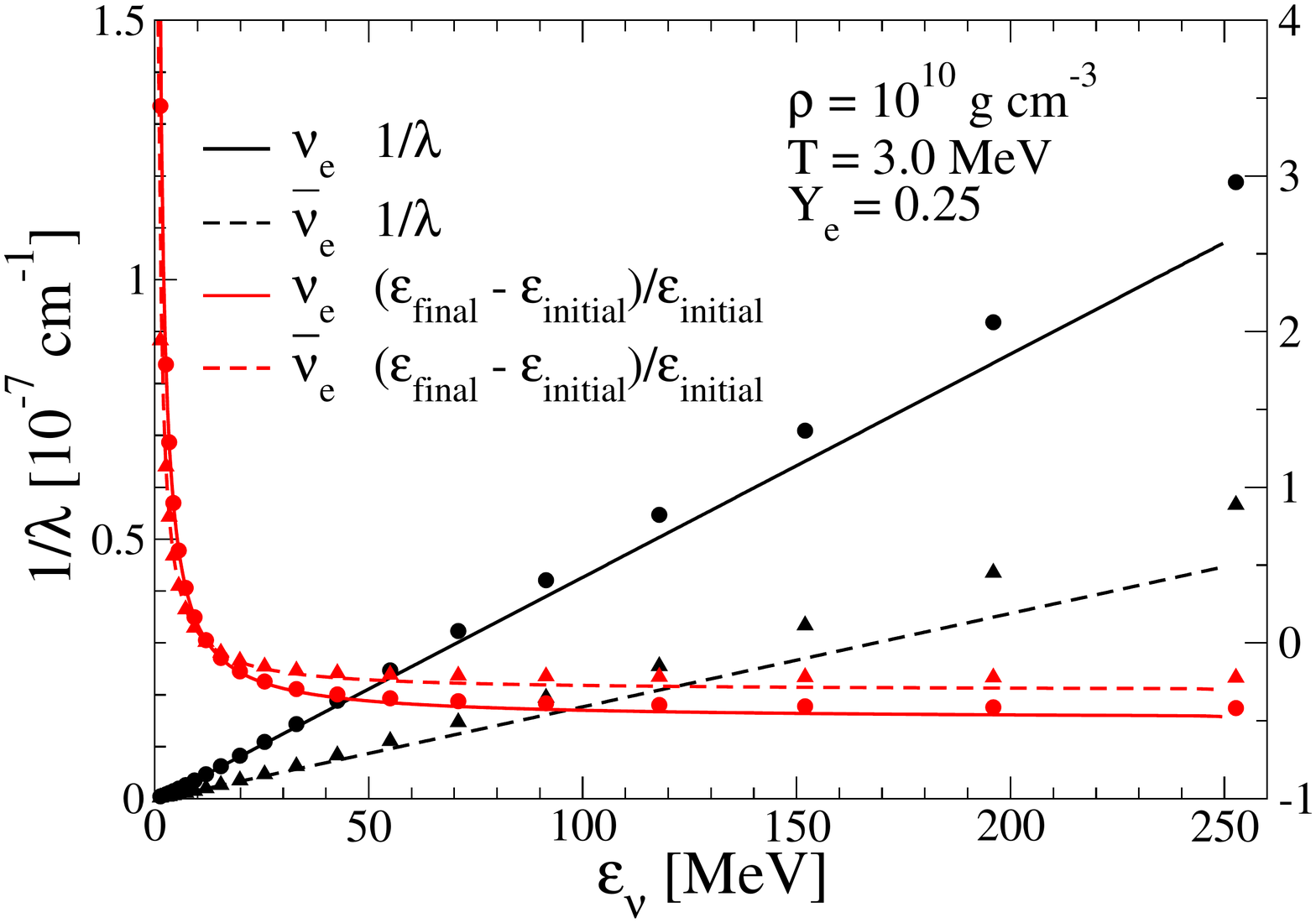}{0.5\textwidth}{(a)}
              \fig{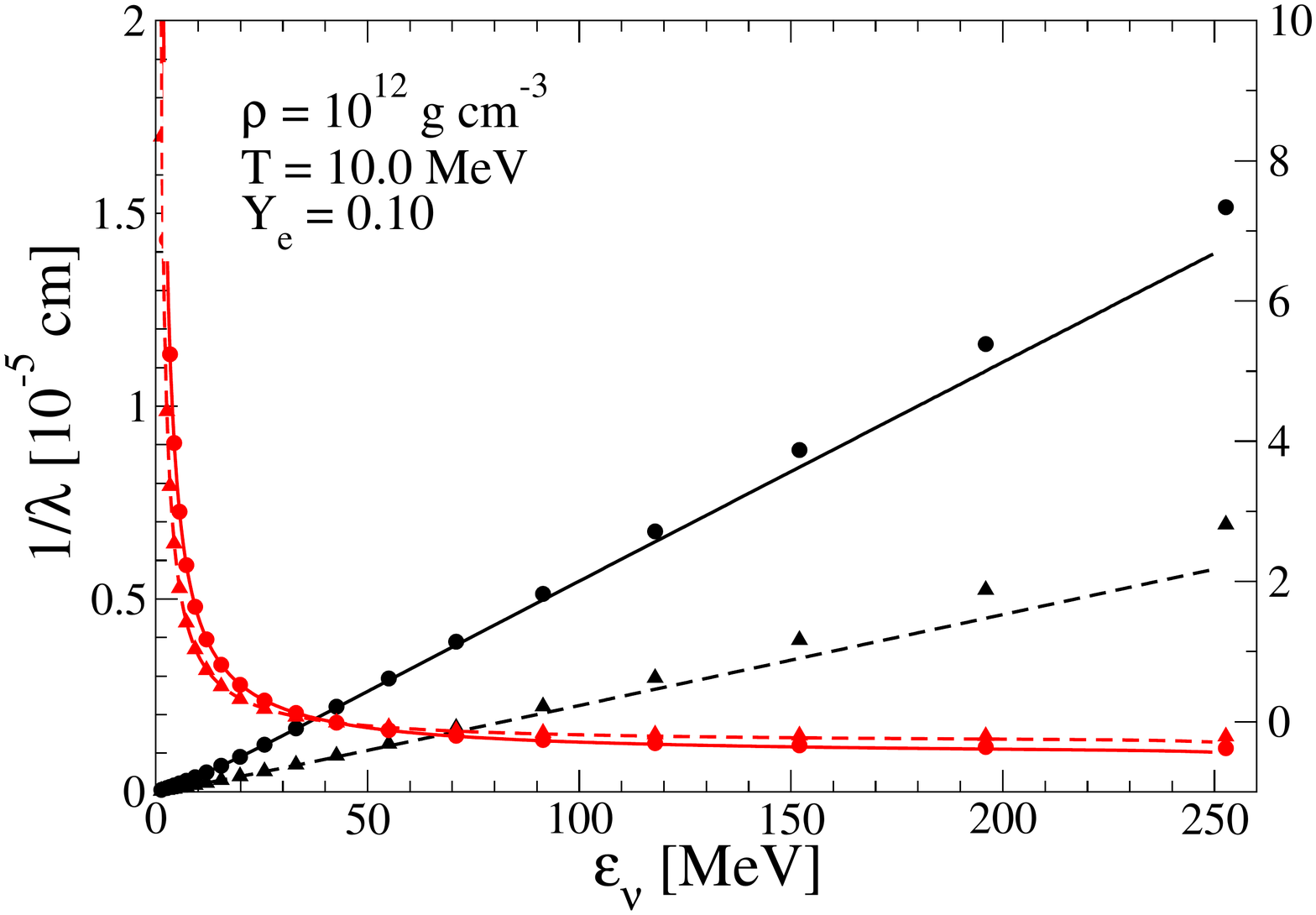}{0.5\textwidth}{(b)}
               }
\gridline{
              \fig{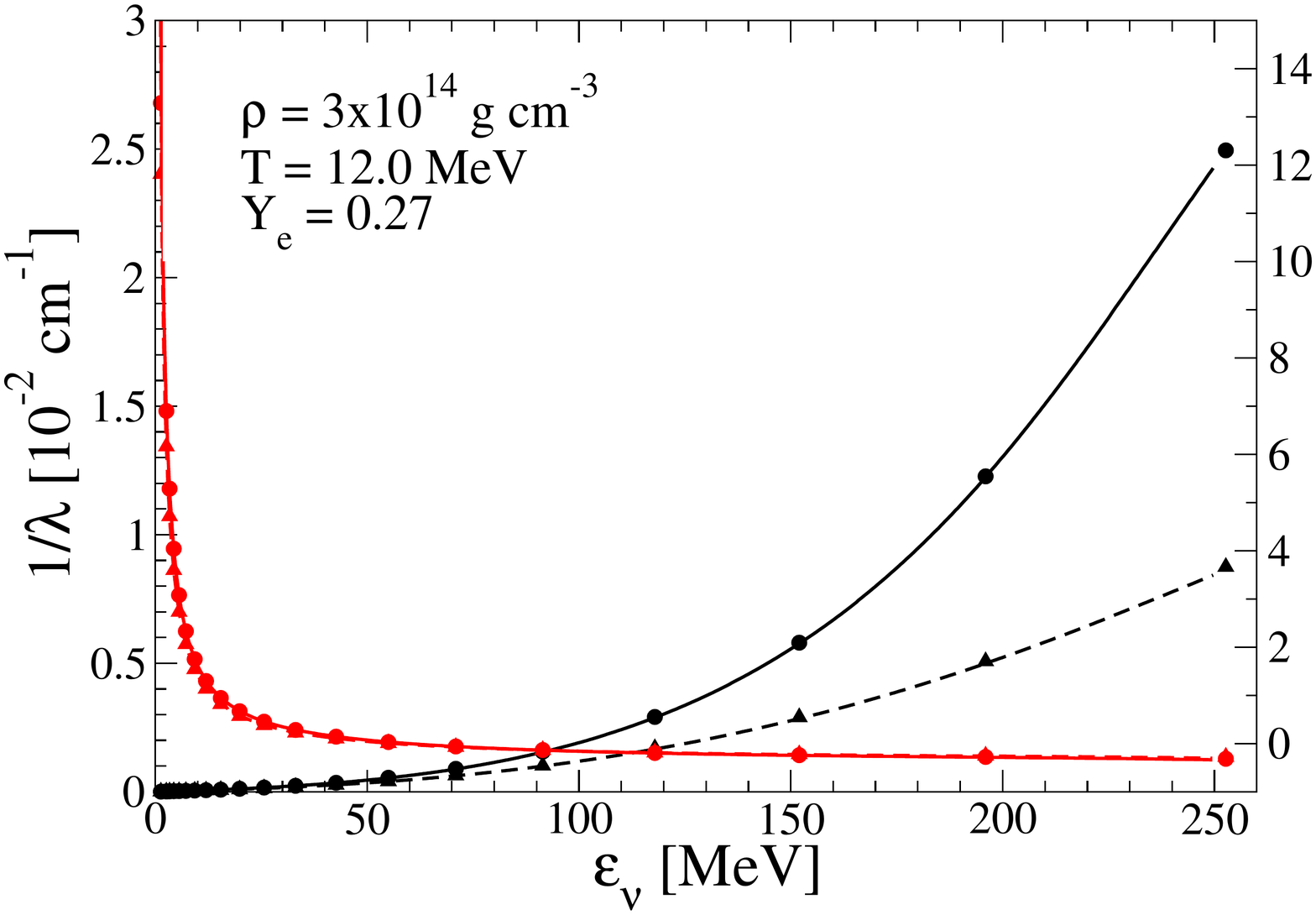}{0.5\textwidth}{(c)}
               }
\caption{\label{fig:e_sct_eos1}
Inverse mean free path (left scale, black) and fractional energy transfer, $(\epscmi{ {\rm final}} - \epscmi{ {\rm initial}})/\epscmi{ {\rm initial}}$ (right scale, red), as a function of the incident neutrino energy for \nue--electron scattering (solid) and \nuebar--electron scattering (dashed).
The symbols (\nue, circles; \nuebar, triangles) show the results for the typical \Chimera\ grid of 20 energy zones geometrically spaced from 4--250~MeV.
The solid lines are the results of a reference calculation using 600 zones evenly spaced from 1--300 MeV.
The thermodynamic conditions are  the same as those in \citet{BuRaJa06}, for comparison.}
\end{figure}

\subsection{Neutrino--Electron Scattering}
\label{scatnu_em}

For neutrino--electron scattering (NES), the scattering functions $\Phi_{0, {\rm NES}}^{\rm in}(\epscm,\epscm')$ and $\Phi_{1,{\rm XX}}^{\rm out}(\epscm,\epscm')$ are taken from Equation~(C50) of \citet {Brue85}.
The scattering functions for down-scattering and iso-energetic scattering are computed directly, using Equation~(\ref{eq:d7down}) where possible, and up-scattering is computed by use of the Equation~(\ref{eq:d7up}).
Figure~\ref{fig:e_sct_eos1} shows some representative inverse mean free paths and fractional energy transfers computed with thermodynamic states chosen to be the same as those of \citet{BuRaJa06}, for comparison, assuming empty neutrino final states for both a typical \Chimera\ geometrically-spaced energy grid of 20 zone-centered energies from 4--250~MeV and a much finer evenly-spaced grid from 1--300~MeV.

\begin{figure}
\gridline{\fig{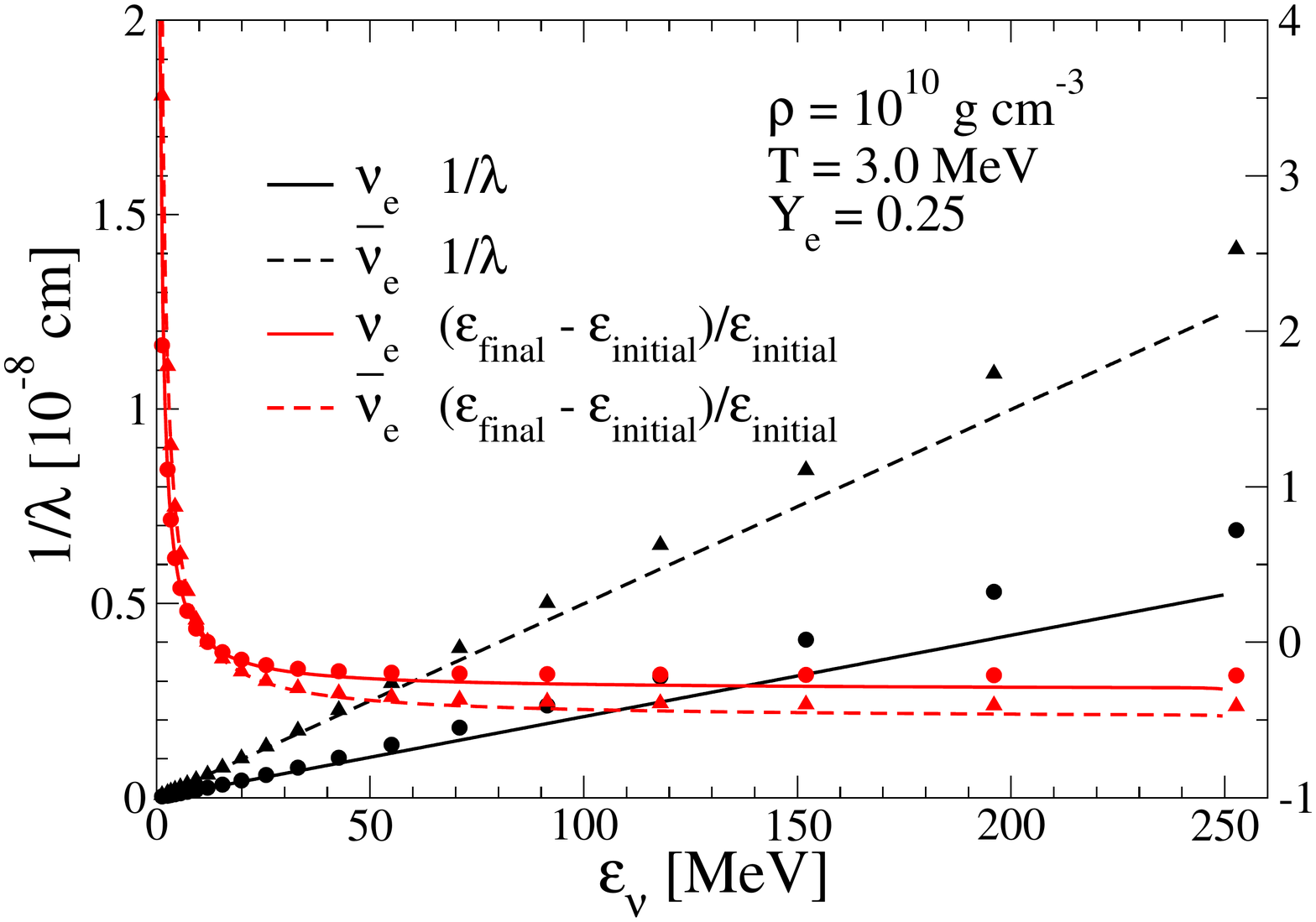}{0.5\textwidth}{(a)}
              \fig{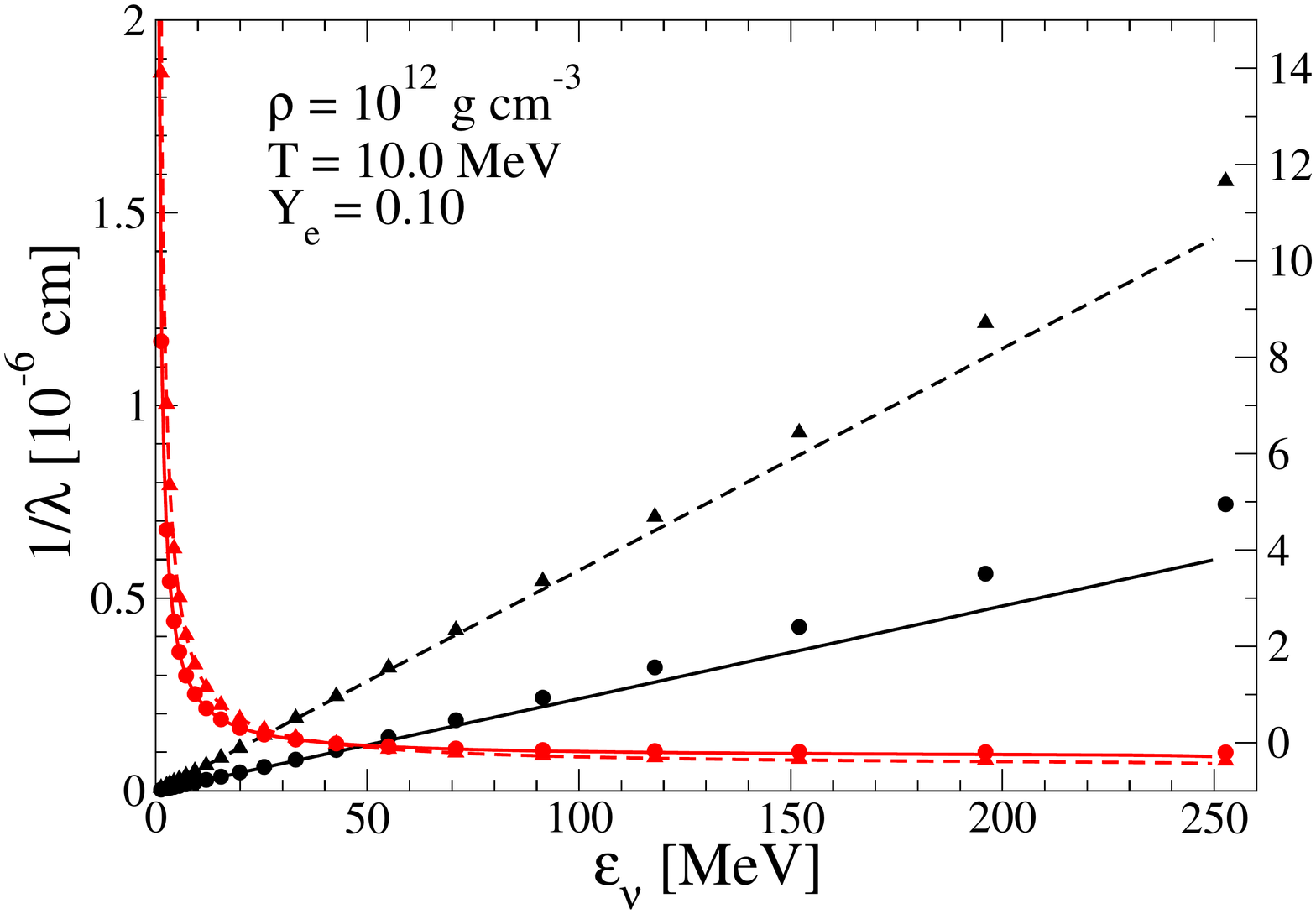}{0.5\textwidth}{(b)}
               }
\gridline{
              \fig{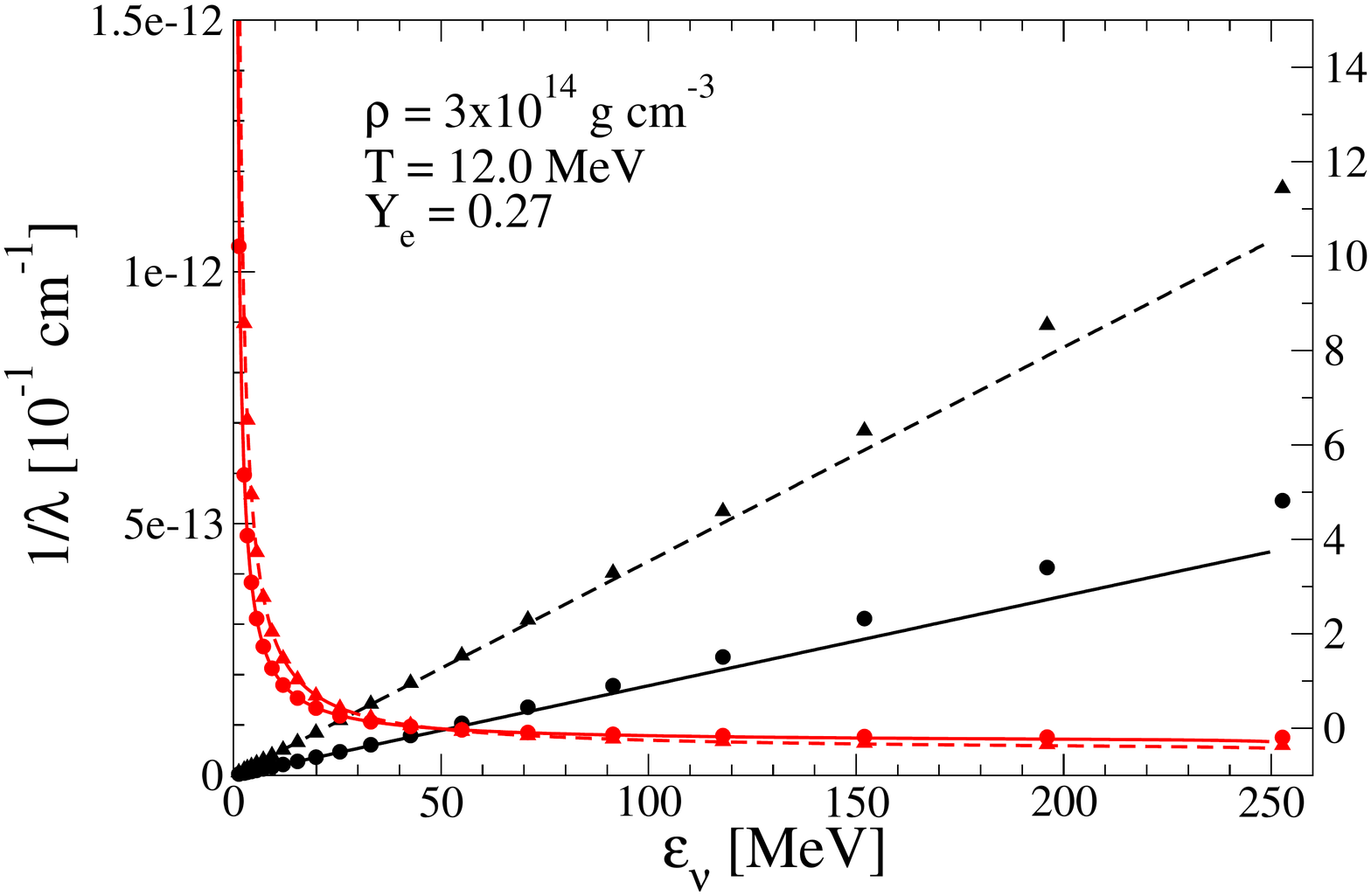}{0.5\textwidth}{(c)}
               }
\caption{\label{fig:p_sct_eos1}
Same as Figure~\ref{fig:e_sct_eos1} but for \nue--positron scattering and \nuebar--positron scattering.
}
\end{figure}

\newpage
\subsection{Neutrino--Positron Scattering}
\label{scatnu_ep}

Neutrino--positron scattering is computed using the same scattering functions as neutrino--electron scattering but with the coefficient function of $C_{V}$ and $C_{A}$ interchanged, $(C_{V} + C_{A} )^{2} \leftrightharpoons (C_{V} - C_{A} )^{2}$, and the chemical potentials in the Fermi functions replaced by their negatives, $\mu_{e^+} = - \mu_{e^-}$. 
Figure~\ref{fig:p_sct_eos1} shows the inverse mean free paths and fractional neutrino energy transfers due to neutrino--positron scattering using the same reference states as for neutrino--electron scattering (Figure~\ref{fig:e_sct_eos1}).

\subsection{Neutrino--Nucleus Scattering}
\label{scatnu_A}

\begin{figure}
 \fig{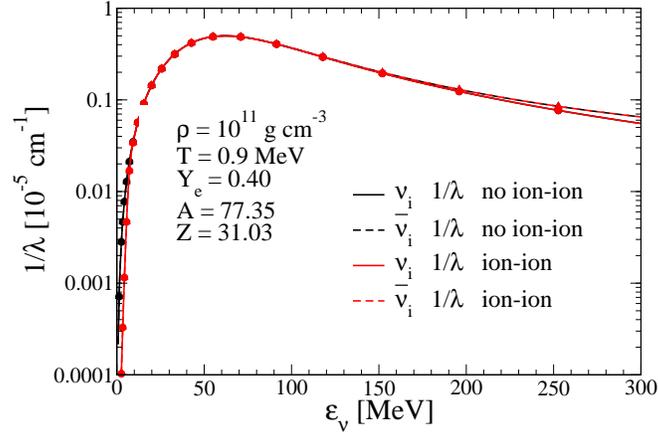}{0.5\textwidth}{}
\caption{\label{fig:a_sct_eos1}
Neutrino--nucleus neutral-current scattering inverse mean free paths for neutrinos (solid lines) and anti-neutrinos (dashed lines), with the nuclear form factors included for the indicated thermodynamic conditions. Black lines and symbols show neutrino--nucleus scattering without ion--ion correlation corrections. Red lines and symbols show the scattering with the ion--ion correlations applied. 
Filled circles (neutrinos) and triangles (anti-neutrinos) show the results for the typical \Chimera\ neutrino energy grid of 20 
geometrically-spaced energy zones. 
The solid lines show the results using a 600-zone energy grid evenly-spaced from 1-300 MeV. 
The ordinate is log scaled to show the effect of the ion--ion correlation corrections, which for these conditions, mainly affects low-energy neutrinos, which have small inverse mean free paths. 
}
\end{figure}

Neutrino--nucleus (${\nu \rm A}$) scattering is treated as iso-energetic such that 
\begin{equation}
R_{\nu \rm A}^{\rm in}(\epscm, \epscm',\cos\theta) = R_{\nu \rm A}^{\rm out}(\epscm, \epscm',\cos\theta) = R_{\nu \rm A}(\epscm, \epscm',\cos\theta) \delta(\epscm - \epscm') ,
\label{eq:d8a}
\end{equation}
with the moments 
\begin{eqnarray}
\left( \frac{ df(\mucm, \epscm) }{ dt } \right)_{ \! {\rm S, \, scat} }^{(0)} &=&0 , \nonumber \\
\left( \frac{ df(\mucm, \epscm) }{ dt } \right)_{ \! {\rm S, \, scat} }^{(1)} &=& \frac{2\pi}{(hc)^{3}} \psimomcm{1}(\epscm) \left[ \Phi_{1,{\nu \rm A}}^{\rm out}(\epscm) - \Phi_{0,{\nu \rm A}}^{\rm out}(\epscm) \right] ,
\label{eq:d8b}
\end{eqnarray}
and therefore
\begin{equation}
 \left(1/\lambda \right)_{{\nu \rm A}} = \left[ \Phi_{0, {\nu \rm A}}^{\rm out}(\epscm) - \Phi_{1,{\nu \rm A}}^{\rm out}(\epscm) \right] .
\label{eq:d8c}
\end{equation}
The calculations of $\Phi_{0, {\nu \rm A}}^{\rm out}(\epscm)$ and $\Phi_{1,{\nu \rm A}}^{\rm out}(\epscm)$ are performed using the formalism of \citet{BrMe97}, and results for the inverse mean free path are plotted in Figure~\ref{fig:a_sct_eos1} for the indicated thermodynamic conditions.
The symbols show the results using the typical \Chimera\ neutrino energy grid of 20 energy zones geometrically spaced from 4--250 MeV, the solid lines show the results obtained using a 600-zone energy grid evenly-spaced from 1 to 300 MeV. 
These thermodynamic conditions were chosen as representative of the infall epoch when material entropy is low and nuclei dominate the composition. 
As expected for an iso-energetic scattering process, the fineness of the neutrino energy grid does not affect the inverse mean free paths, as they depend only on the incident neutrino energy.
The nuclear form factor correction of the scattering rates that reduces the inverse mean free paths at high energies, where the incident neutrino wavelengths become comparable to or smaller than the inter-nucleon distances in the nuclei, is included. 
The black lines and symbols show the inverse mean free paths uncorrected for the liquid structure function (ion--ion correlations). The red lines and symbols show inverse mean free paths with the liquid structure function included, computed as described by \citet{Horo97}.
The liquid structure function has the effect of substantially reducing the inverse mean free paths at low energies, as can be seen by comparing the black and red lines for $\epscmi{ \nu} \le 10$ MeV.

\subsection{Neutrino--Nucleon Scattering}
\label{scatnu_np}

\begin{figure}
 \fig{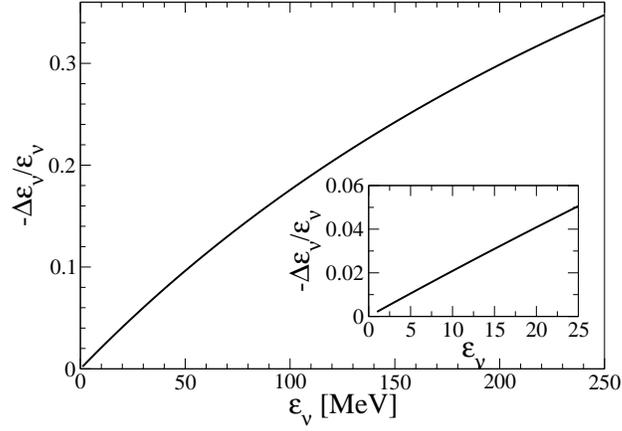}{0.5\textwidth}{}
\caption{\label{fig:denu_enu}
Maximum relative change in the energy of a neutrino, scattering on a stationary nucleon, as a function of the incident neutrino energy. Insert shows detail for incident energies typical of neutrinos emerging from the neutrinosphere.
}
\end{figure}

Differential rates for neutrino elastic scattering on free nucleons are calculated using the formalism of \citet{RePrLa98}. 
These rates include nucleon recoil and degeneracy effects, as well as special relativity, and the implementation of these rates in \Chimera\ is similar to that  of neutrino--electron scattering but differs in several important respects. 
First, analytic expressions for the Legendre moments of these rates are not available, necessitating their calculation by numerical integration. 
Second, energy transfer between a neutrino and a nucleon is small but an important component of the energy transfer between neutrinos and matter, as the scattering rates are relatively large. 
The smallness of the energy transfer is illustrated in Figure~\ref{fig:denu_enu}, which shows the maximum relative energy transfer of a neutrino scattering from a stationary nucleon. 
In the important energy range 5--30~MeV, typical of the energies of neutrinos emerging from the neutrinosphere, the maximum relative change in neutrino energy is only 0.01 to 0.05. 
The actual change in energy depends on the angle between the scattered and incident neutrino and tends to zero as this angle tends to zero. 
As the relative width of the neutrino energy zoning typically used in \Chimera\ (viz. 20 energy zones geometrically spaced from 4--250~MeV) is $\sim$0.26, and therefore considerably larger than the neutrino--nucleon energy exchange in a scattering except for neutrino energies larger than $\sim$200~MeV, a much finer energy grid is needed to adequately resolve the energetics of this process.

In the case of neutrino--electron or neutrino--positron scattering, for example, where the energy transfer is not small in comparison with the widths of our energy grid, Equation (\ref{eq:d1}) is differenced using zone-centered values of energy in both the neutrino distribution function and the scattering kernels; i.e.,
\begin{eqnarray}
\left( \frac{ df(\mucm, \epscmi{k}) }{ dt } \right)_{ \! {\rm S, \, scat \, XX}}&= &[ 1 - f(\mucm, \epscmi{k}) ] \frac{1}{(hc)^{3}} \sum_{k' = 1}^{N_{k}} \epscmi{k'}^{2} \Delta \epscmi{k'} \int_{-1}^{1} d\mucm' f(\mucm', \epscmi{k'}) \int_{0}^{2\pi} d\beta' R_{\rm XX}^{\rm in}(\epscmi{k}, \epscmi{k'}, \cos\theta ) \nonumber \\ 
&& - f(\mucm, \epscmi{k}) \frac{1}{(hc)^{3}} \sum_{k' = 1}^{N_{k}} \epscmi{k'}^{2} \Delta \epscmi{k'} \int_{-1}^{1} d\mucm' [1 - f(\mucm', \epscmi{k'}) ] \int_{0}^{2\pi} d\beta' R_{\rm XX}^{\rm out}(\epscmi{k}, \epscmi{k'}, \cos\theta ) .
\label{eq:d40}
\end{eqnarray}
However, for small energy transfers compared with the energy grid width, the scattering kernel $R_{\rm XX}^{\rm a}(\epscmi{k}, \epscmi{k'}, \cos\theta )$ will be effectively zero if $\epscmi{k} \ne \epscmi{k'}$, and the scattering will become essentially isoenergetic, with negligible energy transfer. To develop a better approximation for the energy transfer via smaller energy transfer scatterings, we continue to evaluate the neutrino distribution function at the energy zone centers, but evaluate the scattering kernels on a refined energy grid. 
To accomplish this we regard the scattering kernels as functions of \epscm\ and $\epscm''$ and operate on Equation (\ref{eq:d40}) by the unity operator
\begin{equation}
\frac{1}{\epscmi{k}^{3} \Delta \epscmi{k} } \int_{\epscmi{k-\frac{1}{2}}}^{\epscmi{k+\frac{1}{2}}} \epscm^3 d\epscm
\frac{1}{\epscmi{k'}^{2} \Delta \epscmi{k'}} \int_{\epscmi{k'-\frac{1}{2}}}^{\epscmi{k'+\frac{1}{2}}} \epscm''^{2} d\epscm'' .
\label{eq:d41}
\end{equation}
The left-hand side, being a constant operand, is unchanged by this operation. However, by integrating over $\epscm^3$ the right-hand side of Equation~(\ref{eq:d40}) becomes proportional to $dE(\mucm, \epscmi{k})/dt$, which we desire to compute accurately.
Operating on the right-hand side of Equation~(\ref{eq:d40}) by the right-hand side of Equation~(\ref{eq:d41}) gives
\begin{eqnarray}
\left( \frac{ df(\mucm, \epscmi{k}) }{ dt } \right)_{ \! {\rm S, \, scat \, XX}}&= & [ 1 - f(\mucm, \epscmi{k}) ] \frac{1}{(hc)^{3}} \sum_{k' = 1}^{N_{k}} \epscmi{k'}^{2} \Delta \epscmi{k'} \sum_{k' = 1}^{N_{k}}  \int_{-1}^{1} d\mucm' f(\mucm', \epscmi{k'}) \int_{0}^{2\pi} d\beta'  \nonumber \\
&& \hspace{-2 cm} \times \frac{1}{ \epscmi{k}^{3} \Delta \epscmi{k} } \int_{\epscmi{k-\frac{1}{2}} }^{\epscmi{k+\frac{1}{2}} } \epscm^{3} d\epscm 
\frac{1}{ \epscmi{k'}^{2} \Delta \epscmi{k'} } \int_{\epscmi{k'-\frac{1}{2}} }^{\epscmi{k'+\frac{1}{2}} } \epscmi{k''}^{2} d\epscmi{k''} R_{\rm XX}^{\rm in}(\epscm, \epscmi{k''}, \cos\theta ) \nonumber \\ 
&& \hspace{-2 cm} - f(\mucm, \epscmi{k}) \frac{1}{(hc)^{3}} \sum_{k' = 1}^{N_{k}} \epscmi{k'}^{2} \Delta \epscmi{k'} \sum_{k' = 1}^{N_{k}} \int_{-1}^{1} d\mucm' [1 - f(\mucm', \epscmi{k'}) ] \int_{0}^{2\pi} d\beta'  \nonumber \\
&& \hspace{-2 cm} \times \frac{1}{ \epscmi{k}^{3} \Delta \epscmi{k} } \int_{\epscmi{k-\frac{1}{2}} }^{\epscmi{k+\frac{1}{2}} } \epscm^{3} d\epscm 
\frac{1}{ \epscmi{k'}^{2} \Delta \epscmi{k'} } \int_{\epscmi{k'-\frac{1}{2}} }^{\epscmi{k'+\frac{1}{2}} } \epscmi{k''}^{2} d\epscmi{k''} R_{\rm XX}^{\rm out}(\epscmi{k}, \epscmi{k''}, \cos\theta ) .
\label{eq:d42}
\end{eqnarray}
The integrals in Equation~(\ref{eq:d42}) over \epscm\ and $\epscm''$ can now be replaced by summations over a refined energy grid. 

The Legendre coefficients $\Phi_{\ell,{\rm XX}}^{\rm in/out}(\epscm,\epscm')$ of the scattering kernels are given by Equation (\ref{eq:d6}) and used in the collision integrals of the transport equations as given by Equation (\ref{eq:d8}). 
In the computation of the Legendre coefficients for neutrino--nucleon scattering, the summations of the scattering kernels over refined energy grids in Equation~(\ref{eq:d42}) are directly incorporated by defining
\begin{eqnarray}
  \Phi_{\ell,{\rm n,p}}^{\rm out}(\epscmi{ k},\epscmi{ k'}) &= & \int_{-1}^{1} d(\cos\theta) P_{\ell}( \cos\theta ) \frac{1}{\epscmi{k}^{3} \Delta \epscmi{k}} \int_{\epscmi{ {\rm L,i}}(\cos\theta)}^{\epscmi{ {\rm U,i}}(\cos\theta)} \epscm^3 d\epscm \nonumber \\
&& \times \frac{1}{\epscmi{k'}^{2} \Delta \epscmi{k'}} \int_{\epscmi{ {\rm L,f}}(\cos\theta)}^{\epscmi{ {\rm U,f}}(\cos\theta)} \epscm'^2 d\epscm' R_{\rm n,p}^{\rm out}(\epscm,\epscm',\cos\theta) ,
  \label{eq:d43}
\end{eqnarray}
where $\Phi_{\ell,{\rm n,p}}^{\rm in}(\epscmi{ k},\epscmi{ k'})$ is obtained from $\Phi_{\ell,{\rm n,p}}^{\rm out}(\epscmi{ k},\epscmi{ k'})$ by Equation (\ref{eq:d4}), and where, as in Equation (\ref{eq:d42}), the integrals in Equation (\ref{eq:d43}) can be replaced by summations over a refined energy grid. 

Technically, the lower and upper limits of the energy integrals in Equation (\ref{eq:d43}) should be the zone-edged energies $\epscmi{k-\frac{1}{2}}$ and $\epscmi{k+\frac{1}{2}}$ in the initial energy integral, and $\epscmi{k'-\frac{1}{2}}$ and $\epscmi{k'+\frac{1}{2}}$ in the final. 
However, the scattering kernels can become so narrow in energy that restricting the integration range to a narrower energy width does not change the result and allows a given energy grid to be spread more finely over the smaller energy integration width. 
A useful integration range for energy as given by the widths of the scattering kernels is angle dependent and within the interval $\epscmi{ \rm L}$ to \epscmi{  \rm U} where
\begin{equation}
  [ \epscmi{ \rm L}, \epscmi{  \rm U} ] (\epscmi{r},\cos\theta) = \frac{ \epscmi{r} }{ 1 - \alpha^{2} \xi } \left[ \left( 1 - \alpha^{2} \xi \cos\theta \right) [-,+] \sqrt{ \alpha^{2} \xi ( 1 - \cos\theta ) \left( 2 - \alpha^{2} \xi ( 1 + \cos\theta ) \right) }   \right] ,
\label{eq:d8e}
\end{equation}
where
\begin{equation}
  \xi = \frac{2kT}{ m_{\rm B} c^{2} } ,
\label{eq:d8f}
\end{equation}
and where $\alpha$ is a free parameter.
Experiments found that the choice of $\alpha^{2} = 5$ gives good results in the sense that the scattering kernel is essentially negligible outside these limits. 
The derivation of Equations~(\ref{eq:d8e}) and (\ref{eq:d8f}) is given by Equations (\ref{eq:d30})--(\ref{eq:d34}) below.
Whether the zone-edged energies of the \chimera\ energy grid or the energies given by Equation (\ref{eq:d8e}) are used as integration limits is case-dependent and described below.

Three cases are considered depending on the relative values of $\epscm^{\rm in} = \epscmi{ k}$ and $\epscm^{\rm out} = \epscmi{ k'}$, where $\epscmi{ k}$ and $\epscmi{ k'}$ are the zone-centered, incoming and outgoing neutrino energy group. 
\begin{enumerate}
\item If $k = k'$, the initial energy integration over \epscm\ as given by the right-hand side of Equation (\ref{eq:d41}) is omitted, the zone centered energy \epscmi{k}\ is used as input, and the final energy integration is performed with a 32-point Gauss--Legendre integration from $\epscmi{ \rm L,f}$ to $\epscmi{  \rm U,f}$, where
\begin{equation}
  \epscmi{  \rm L,f} = \max( \epscmi{  \rm L}, \epscmi{  k-\frac{1}{2}} ), \qquad \epscmi{ \rm U,f} = \min( \epscmi{  \rm U}, \epscmi{  k+\frac{1}{2}} ) .
\label{eq:d8d} 
\end{equation}
The first two Legendre moments $\Phi^{\rm out}_{\ell,\rm n,p}(\epscmi{ k}, \epscmi{ k})$ are then obtained by a 32-point 
Gauss--Legendre angular quadrature appropriately weighted by the Legendre polynomials $P_{\ell}(\cos\theta)$.

\item  If $k' = k-1$, the zero and first Legendre moments are then obtained from Equation (\ref{eq:d43}) by first executing 8-point Gauss--Legendre energy quadratures for each of the energy integrations. 
The integration limits for the initial energy integration are from $\epscmi{ {\rm L,i}}$ to $\epscmi{ {\rm U,i}}$, given by 
\begin{equation}
 \epscmi{ {\rm L,i}} = \epscmi{  k-\frac{1}{2}}, \qquad \epscmi{ {\rm U,i}} = \min( \epscmi{  \rm U}, \epscmi{  k+\frac{1}{2}} )
\label{eq:d8g}
\end{equation}
where $\epscmi{  \rm U}$ is given by Equation~(\ref{eq:d8e}) evaluated at the boundary value \epscmi{  k+\frac{1}{2}}.
The limits of integration for the final energy are from $\epscmi{ {\rm L,f}}$ to $\epscmi{ {\rm U,f}}$, where
\begin{equation}
 \epscmi{ {\rm L,f}} =\max( \epscmi{  \rm L}, \; \epscmi{  k-\frac{3}{2}} ), \qquad \epscmi{ {\rm U,f}} = \epscmi{  k- \frac{1}{2}}
\label{eq:d8h}
\end{equation}
where $\epscmi{  \rm L}$ is given by Equation~(\ref{eq:d8e}) evaluated at \epscmi{  k+\frac{1}{2}}.
A 32-point Gauss--Legendre angular quadrature appropriately weighted by the Legendre polynomials $P_{\ell}(\cos\theta)$ is then executed to obtain the Legendre coefficients.

\item If $k' < k-1$, the differences between initial and final energy falls outside the small energy scattering range, so the necessity of refining the energy grids is less critical.
The zero and first Legendre moments are obtained from Equation (\ref{eq:d43}) by first integrating the scattering functions over the initial energy from $\epscmi{ k-\frac{1}{2}}$ to $\epscmi{ k+\frac{1}{2}}$ and the final energy grid from $\epscmi{ k'-\frac{1}{2}}$ to $\epscmi{  k'+\frac{1}{2}}$ using 4-point Gauss--Legendre energy quadratures.
The Legendre moments are then obtained by a 4-point Gauss--Legendre angular quadrature.
\end{enumerate}

\begin{figure}
\gridline{\fig{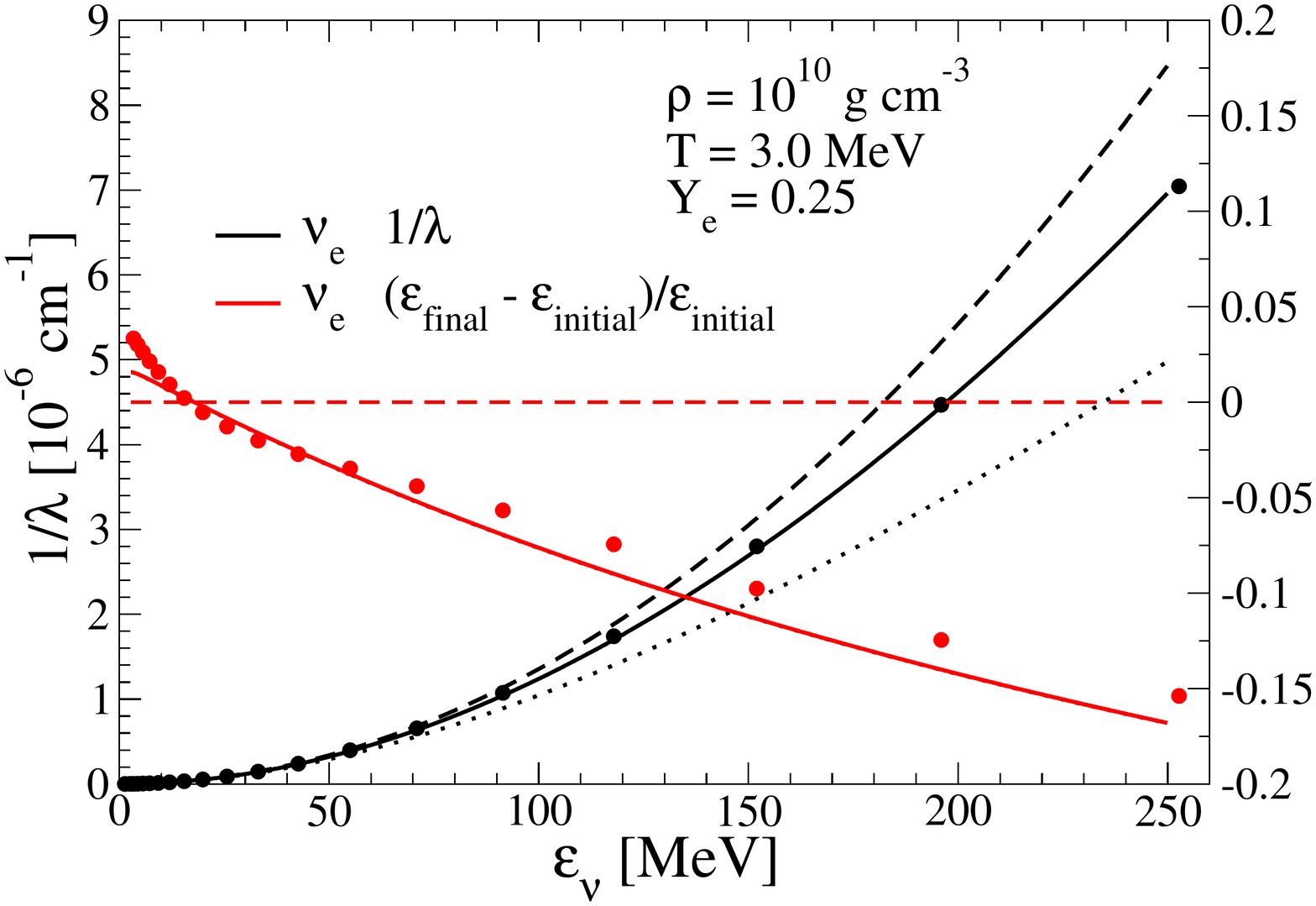}{0.5\textwidth}{(a)}
              \fig{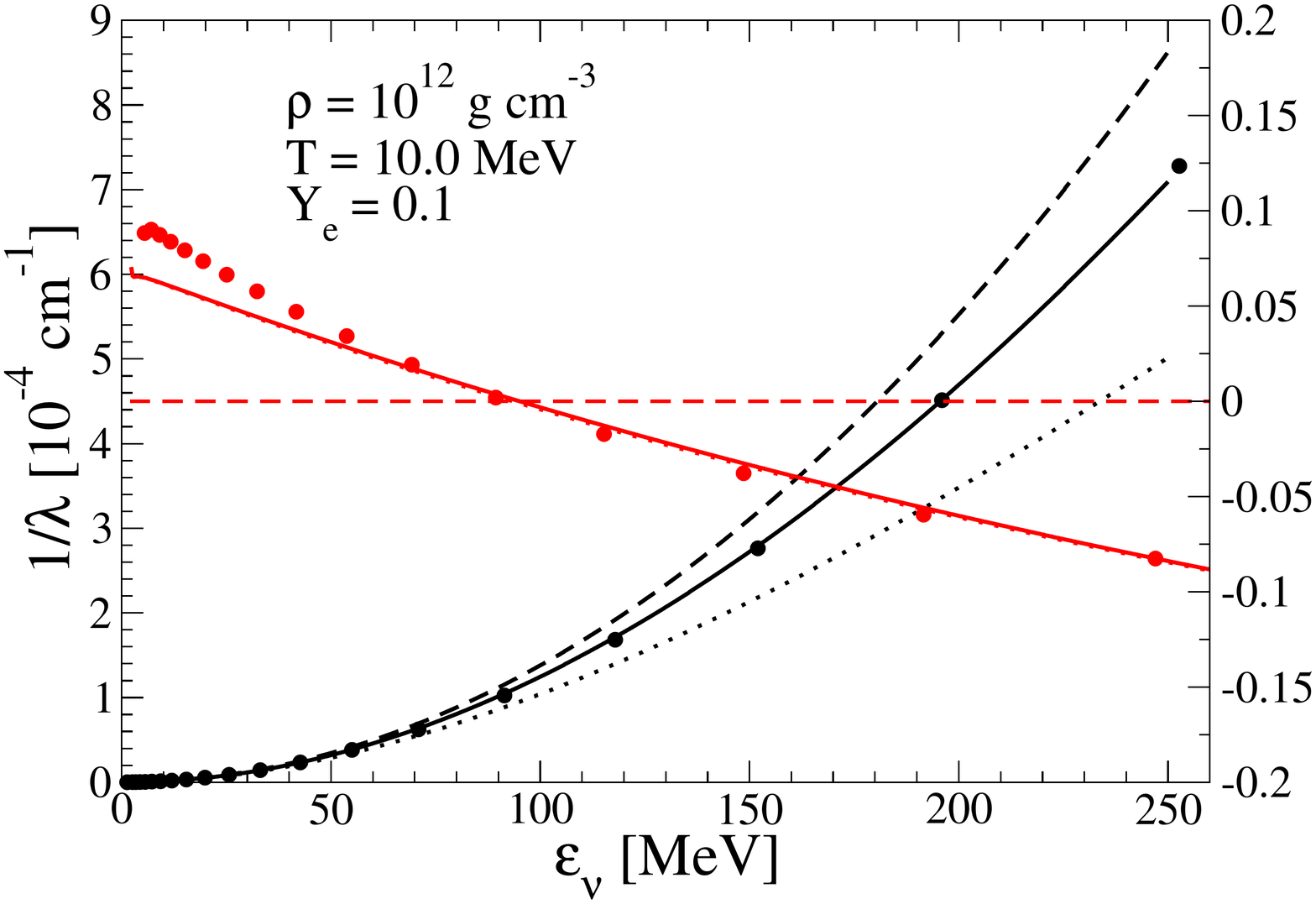}{0.5\textwidth}{(b)}
               }
\gridline{
              \fig{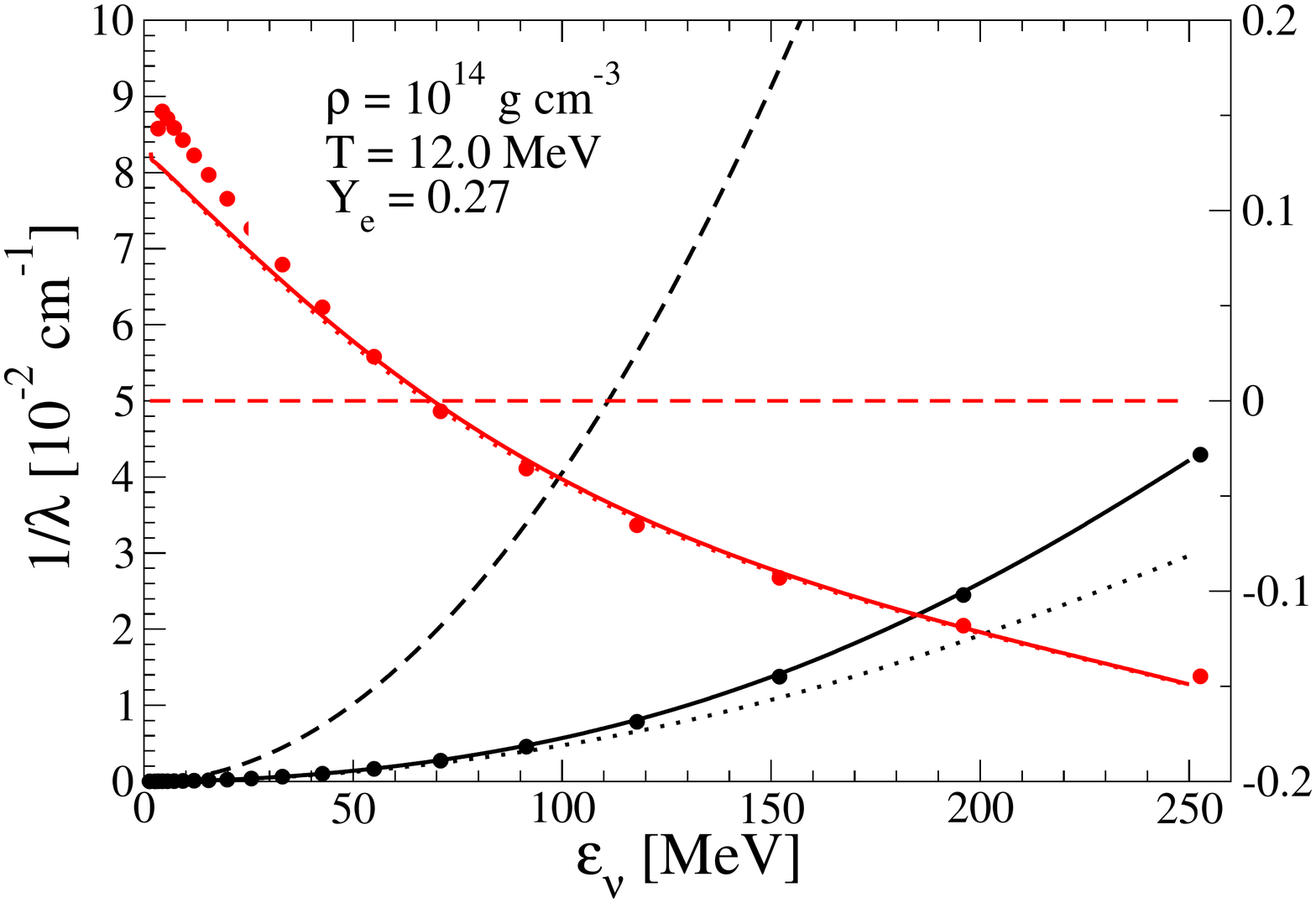}{0.5\textwidth}{(c)}
               }
\caption{\label{fig:nueN_sct_eos}
Inverse mean free paths for scattering on free nucleons by \nue-neutrinos (black, scale on left) and the relative energy transfer to the neutrino (red, scale on right), for the thermodynamic conditions listed on the upper left of the plots. The dashed lines give the inverse mean free paths using the iso-energetic approximation of \citet{Brue85} uncorrected for weak magnetism, the dotted lines give the inverse mean free paths as given by the formalism of \citep{RePrLa98}, which include recoil, nucleon final state blocking, and special relativity but not weak magnetism corrections, and the solid lines are the latter inverse mean free paths corrected for weak magnetism. The inverse mean free paths plotted by the solid, dashed, and dotted lines were computed using a 1500-zone neutrino energy grid evenly spaced between 1 and 300 MeV. The symbols show the inverse mean free paths computed with recoil, etc. and weak magnetism corrections, using the typical  \Chimera\ energy grid of 20 zones geometrically spaced between 4 and 250 MeV.
}
\end{figure}

We have also included corrections for weak magnetism to the scattering rates by the procedure outlined by \citet{BuRaJa06}, which disentangles the weak magnetism correction from the corrections given by \citet{Horo02} for both weak magnetism and recoil, $\chi_{\rm WM, Rec}^{\rm nc, n,p}$, and the corrections given for recoil only, $\chi_{\rm Rec}^{\rm nc, n,p}$. The resulting weak magnetism correction factor, $\xi^{\rm nc}(\epscm)$, is given by the ratio
\begin{equation}
  \xi^{\rm nc}(\epscm) =  \frac{ n_{\rm n} \, \chi_{\rm WM, Rec}^{\rm nc, n} + n_{\rm p} \, \chi_{\rm WM, Rec}^{\rm nc, p} }{ n_{\rm n} \, \chi_{\rm Rec}^{\rm nc, n} + n_{\rm p} \, \chi_{\rm Rec}^{\rm nc, p} },
\label{eq:d9}
\end{equation}
where $n_{\rm n}$ and $n_{\rm p}$ are the number densities of free neutrons and protons, respectively.
The Reddy rates, which do not include weak magnetism, are corrected for weak magnetism by multiplying the scattering Legendre moments by $\xi^{\rm nc}(\epscm)$.

\begin{figure}
\gridline{\fig{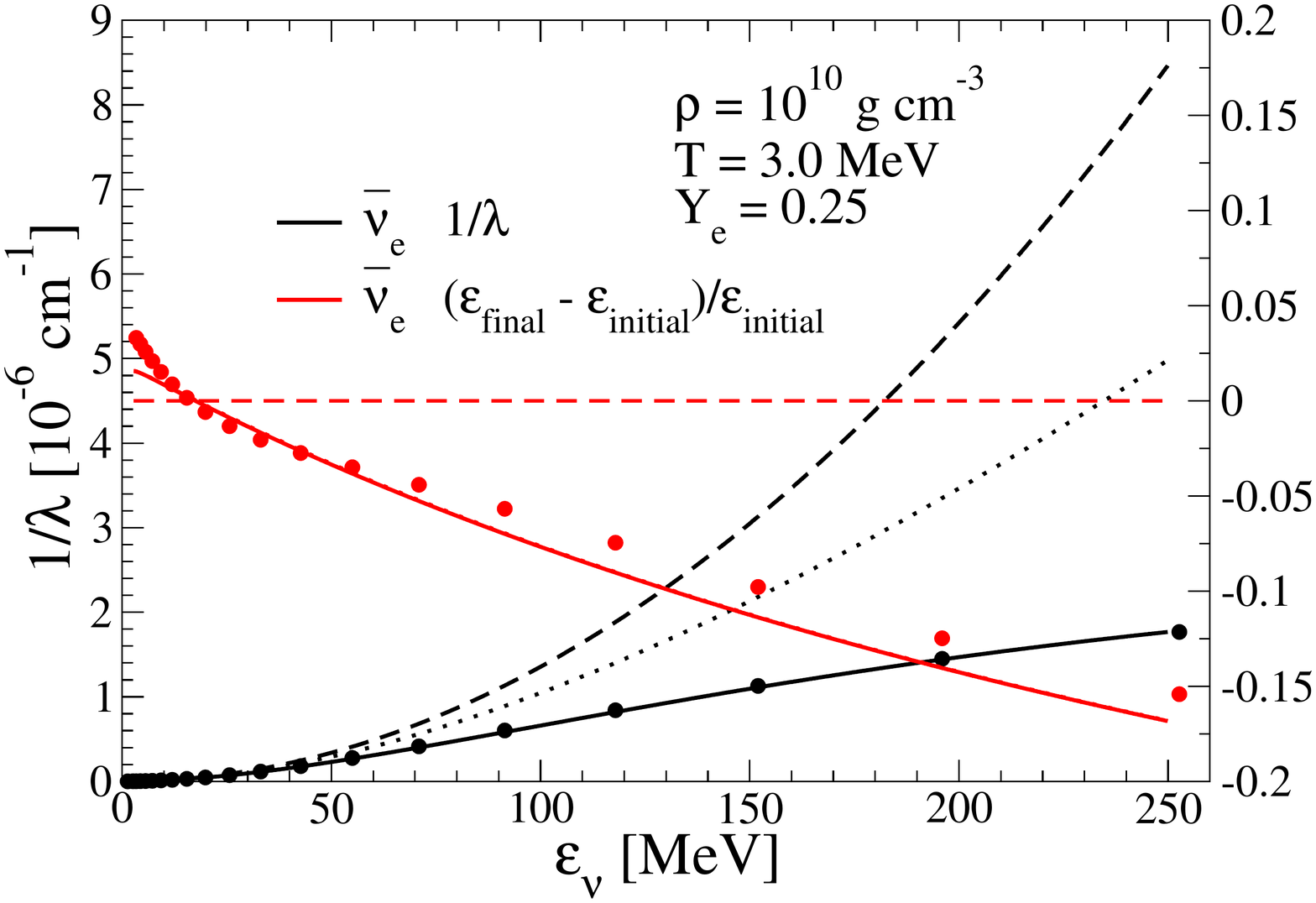}{0.5\textwidth}{(a)}
              \fig{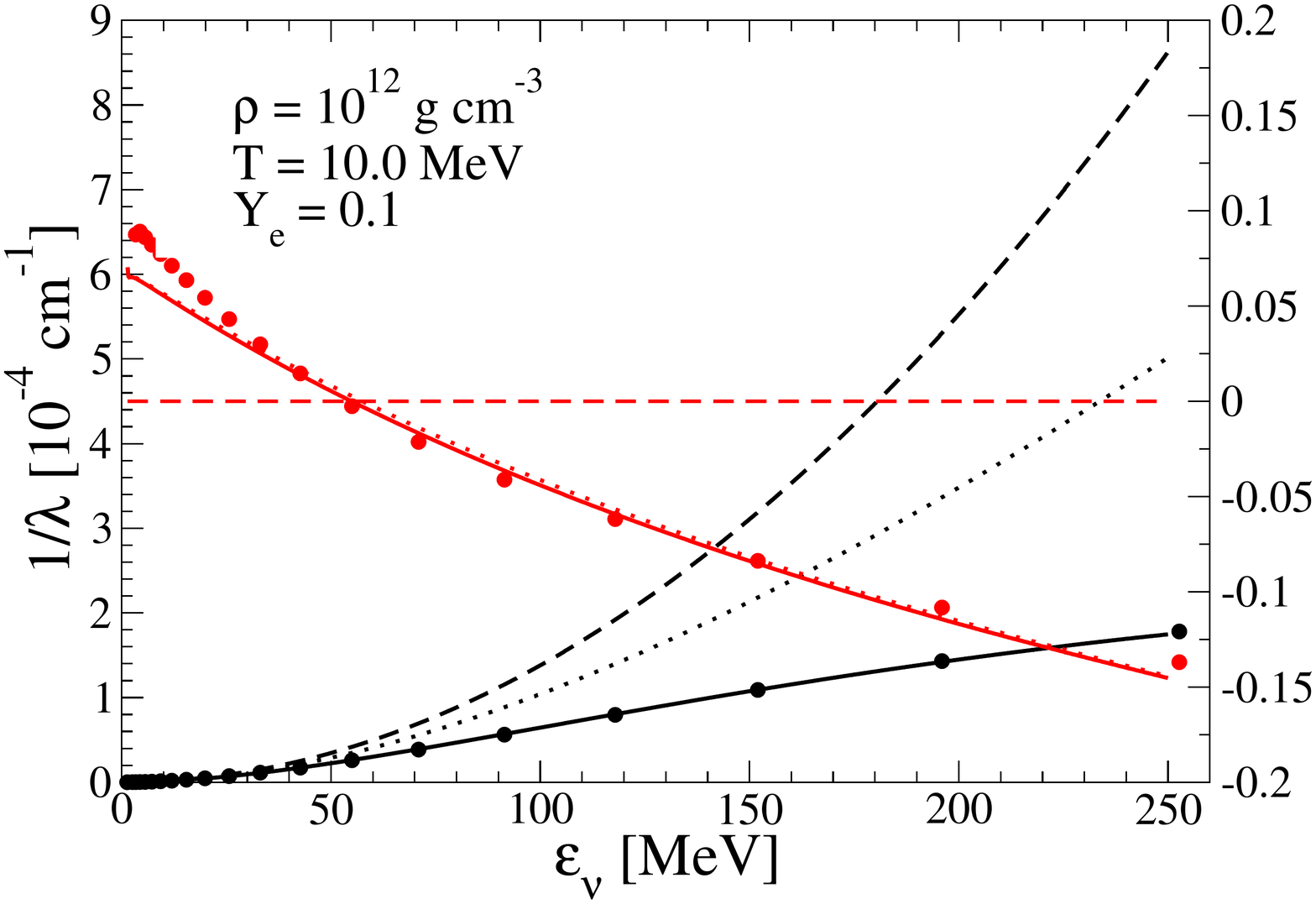}{0.5\textwidth}{(b)}
               }
\gridline{
              \fig{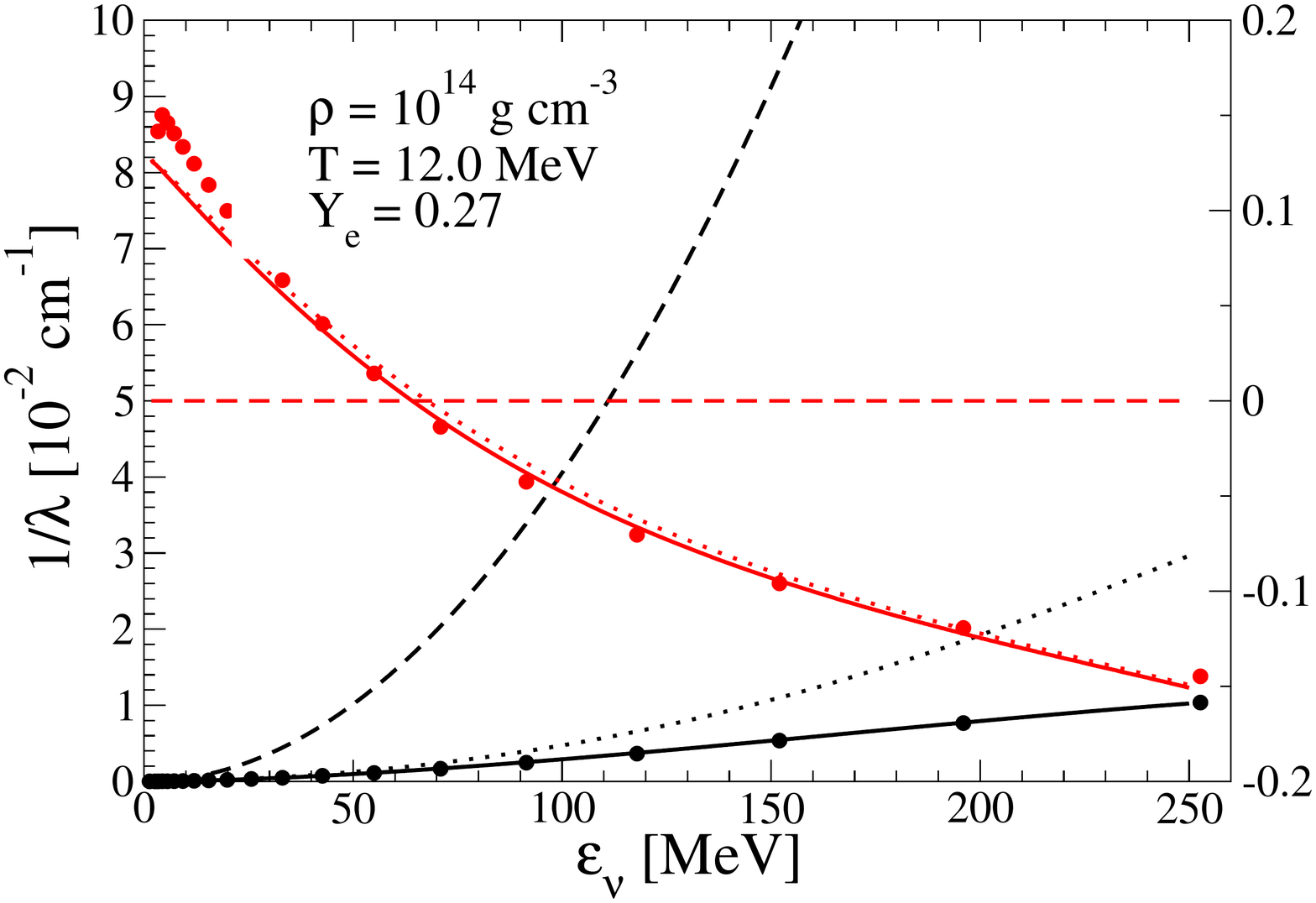}{0.5\textwidth}{(c)}
               }
\caption{\label{fig:nuebarN_sct_eos}
Same as Figure~\ref{fig:nueN_sct_eos} but for \nuebar--nucleon scattering.
}
\end{figure}

The inverse mean free paths and mean relative energy transfers are plotted for \nue\ in Figure~\ref{fig:nueN_sct_eos} and for \nuebar\ in Figure~\ref{fig:nuebarN_sct_eos}. 
The figures compare the scattering opacities computed with the procedure described above (solid lines) with those computed without including weak magnetism (dotted lines) and with those computed without including weak magnetism and with the iso-energetic approximation described by \citet{Brue85} (dashed lines). 
The effect of including recoil, nucleon final state blocking, and special relativity \citep{RePrLa98} is to reduce the rates compared with the iso-energetic approximation. 
This is mainly because the iso-energetic approximation is equivalent to assuming that the nucleon has infinite mass. 
Including recoil takes into account the finite nucleon mass and, therefore, reduces the center of mass energy of the colliding system, thus reducing the opacity for a given incident neutrino energy. 
Final-state blocking and special relativity work in the same direction but are less important for the conditions considered here. 
The effect of weak magnetism is to increase the opacities for neutrinos and decrease them for antineutrinos. 
The net effect of including recoil, blocking, and relativistic corrections and weak magnetism corrections on the magnitude of the opacities is thus considerably more pronounced for antineutrinos than for neutrinos. 
The effect of including the above corrections is essential for including the neutrino--nucleon energy transfer. 
The mean relative energy transfers are shown in Figures~\ref{fig:nueN_sct_eos} and \ref{fig:nuebarN_sct_eos}. 
They are obviously zero when computed using the iso-energetic approximation, but are nonzero and can play a significant role when recoil is taken into account. 
The effect of weak magnetism is to modify the magnitude of the cross sections but has little if any effect on the relative energy transfer.

As a reference, the dotted, dashed, and solid lines showing the inverse mean free paths and relative energy transfers were all obtained using an energy grid of 1500 zones, evenly spaced from 1 to 300 MeV. 
An energy grid of the same number of zones but geometrically spaced in the same energy range, rather than evenly spaced, gave very similar results except for very low incident neutrino energies ($\le 2$ MeV). 
The filled circles show the inverse mean free paths and relative energy transfers computed with the above corrections using the typical \Chimera\ energy grid of 20 energy zones (with sub-grids as described above) increasing geometrically between 4 and 250 MeV. 
The inverse mean free paths computed with the \Chimera\ energy grid reproduce nicely those computed with the much more refined grid. 
The relative energy transfers, however, are somewhat overestimated at low energies. 
The reason, most likely, is that \Chimera\ stores the lowest two moments of the scattering functions, which are computed using the sub-grid and from which the inverse mean free paths are directly related, while the relative energy transfers are not stored but computed from the relatively coarse 20-energy-zone \Chimera\ grid.

To derive Equations~(\ref{eq:d8e}) and (\ref{eq:d8f}), we begin with the dynamic structure function $S(q_{0}, \bar{q})$ for neutrino--nucleon scattering by setting $\hat{\mu} = \mu_{2} - \mu_{4} = 0$ and $ z = q_{0}/T$ in the expression for neutrino--nucleon absorption \citep[Equations~(21) and (22) of][]{RePrLa98} to obtain
\begin{equation}
S(q_{0}, \bar{q}) = \frac{M^{2}T}{\pi q} \left[ \frac{ z }
{ 1 - \exp \left( - z \right) } \left( 1 + \frac{ 1 }{  z } \ln \frac{ 1 + \exp[( \epsilon_{-} - \mu_{2})/T ] }{ 1 + \exp[( \epsilon_{-} + q_{0} - \mu_{4})/T ] } \right) \right],
\label{eq:d30}
\end{equation}
where
\begin{equation}
z = \frac{q_{0}}{T}, \quad \epsilon_{-} = \frac{1}{4} \frac{( q_{0} - q_{k} )^{2} }{ q_{k} }, \quad q_{k} = \frac{q^{2}}{2M}, \quad q_{0} = \epscm^{\rm in} - \epscm^{\rm out}, \quad 
q^{2} = (\epscm^{\rm in})^{2} + (\epscm^{\rm out})^{2} - 2 \epscm^{\rm in} \epscm^{\rm out} \cos \theta ,
\label{eq:d31}
\end{equation}
and where $q_{0} = \epscm^{\rm in} - \epscm^{\rm out}$, $\bar{q} = \bar{p}^{\rm in} - \bar{p}^{\rm out}$ are, respectively, the energy and momentum transferred by the neutrino, the subscripts ``2" and ``4" refer to the incident and final nucleon, respectively, $\mu$ is the nucleon chemical potential, $M$ the nucleon mass, and $\theta$ is the angle between the incident and scattered neutrino directions. In the nondegenerate limit $(\mu \ll 0)$
\begin{eqnarray}
 \ln \frac{ 1 + \exp[( \epsilon_{-} - \mu)/T ] }{ 1 + \exp[( \epsilon_{-} + q_{0} - \mu)/T ] }  &=& \frac{ - q_{0} }{T} + \ln \frac{  \left\{ 1 + \exp[-( \epsilon_{-} - \mu)/T ] \right\} }{ \left\{ 1 + \exp[-( \epsilon_{-} + q_{0} - \mu)/T ] \right\}}  \nonumber \\ 
&\simeq &- \frac{ q_{0} }{T} + \exp[-( \epsilon_{-} - \mu)/T ] - \exp[-( \epsilon_{-} + q_{0} - \mu)/T ]   \nonumber \\ 
&=& - \frac{ q_{0}  }{T} + \exp[( \mu - \epsilon_{-})/T ] \left\{ 1 - \exp[-( q_{0} )/T ]\right\} \nonumber \\ 
& = &- z + \exp[( \mu - \epsilon_{-})/T ] \left\{ 1 - \exp( -z ) \right\}  ,
 \label{eq:d32}
\end{eqnarray}
where we have set $\mu_{2} = \mu_{4} = \mu$. Using Equation (\ref{eq:d32}) in Equation (\ref{eq:d30}), we get
\begin{eqnarray}
S(q_{0}, \bar{q}) &=& \frac{M^{2}T}{\pi q} \left\{ \frac{ z }
{ 1 - \exp \left( - z \right) } \left[ 1 + \frac{ 1 }{  z } \left( - z + \exp[( \mu - \epsilon_{-})/T ] \left\{ 1 - \exp( -z ) \right\} \right) \right] \right\}  \nonumber \\ 
&=& \frac{M^{2}T}{\pi q} \left\{ \frac{ z }
{ 1 - \exp \left( - z \right) } \left[ 1 -1 + \exp[( \mu - \epsilon_{-})/T ] \frac{ \left\{ 1 - \exp( -z ) \right\} }{z} \right] \right\}  \nonumber \\ 
&=& \frac{M^{2}T}{\pi q} \exp[( \mu - \epsilon_{-})/T ] \nonumber \\ 
&=& \frac{M^{2}T}{\pi q} \exp( \mu/T) \exp \left( -\frac{1}{4T} \frac{ \left( q_{0} - q_{k} \right)^{2} }{ q_{k} } \right) \nonumber \\ 
&=& \frac{M^{2}T}{\pi q} \exp( \mu/T) \exp \left( -\frac{M}{2T} \frac{ \left( \epsilon_{\rm in} - \epsilon_{\rm out} - \left( \epsilon_{\rm in}^{2} + \epsilon_{\rm out}^{2} - 2 \epsilon_{\rm in} \epsilon_{\rm out} \cos \theta \right)/2M  \right)^{2} }{ \left( \epsilon_{\rm in}^{2} + \epsilon_{\rm out}^{2} - 2 \epsilon_{\rm in} \epsilon_{\rm out} \cos \theta \right) } \right) .  \label{eq:d33}
\end{eqnarray}
We wish to find upper and lower values, $\epscmi{ \rm U}$ and $\epscmi{ \rm L}$, such that
\begin{equation}
\exp \left( -\frac{M}{2T} \frac{ \left( \epsilon_{\rm in} - \epsilon_{\rm out} - \left( \epsilon_{\rm in}^{2} + \epsilon_{\rm out}^{2} - 2 \epsilon_{\rm in} \epsilon_{\rm out} \cos \theta \right)/2M  \right)^{2} }{ \left( \epsilon_{\rm in}^{2} + \epsilon_{\rm out}^{2} - 2 \epsilon_{\rm in} \epsilon_{\rm out} \cos \theta \right) } \right) = e^{-\alpha^{2}},
\label{eq:d34}
\end{equation}
Neglecting the third term in the numerator of the exponential, Equation (\ref{eq:d34}) is a quadratic in $\epsilon_{\rm out}$ with the two solutions given by Equation (\ref{eq:d8e}) above.

\subsection{Neutrino Absorption and Emission on Free Nucleons}
\label{abem_np}

\begin{figure}
\gridline{\fig{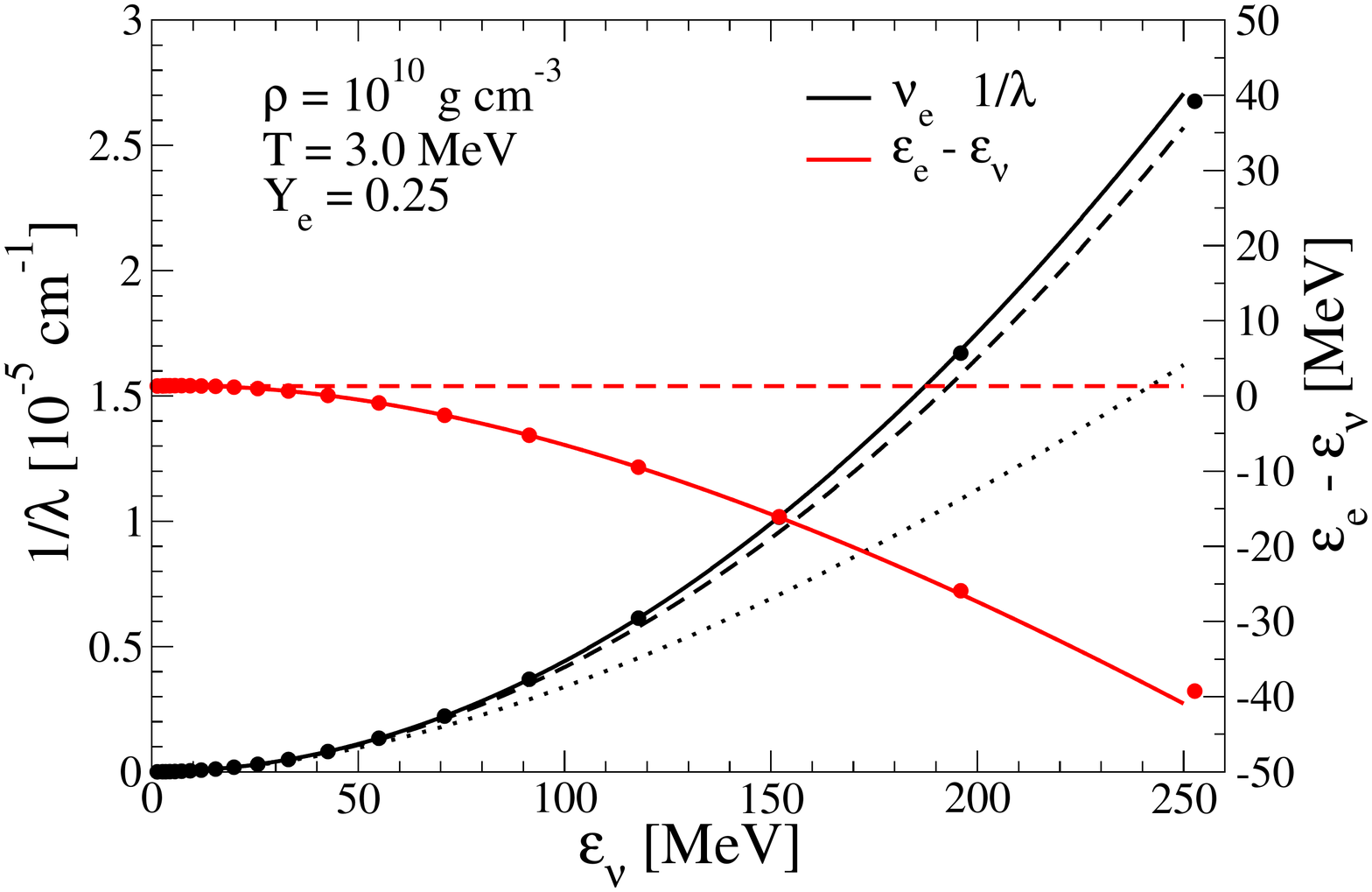}{0.5\textwidth}{(a)}
              \fig{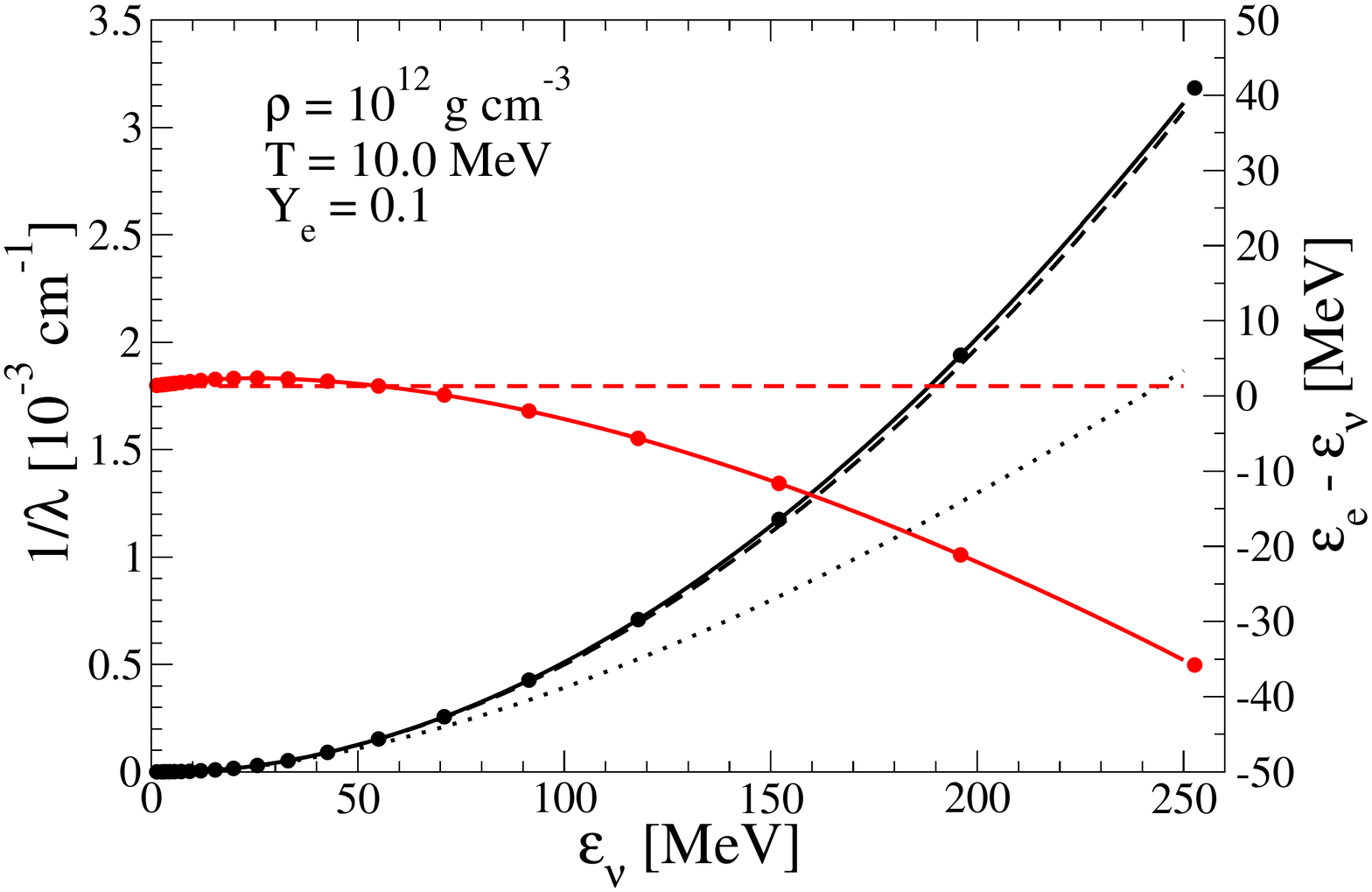}{0.5\textwidth}{(b)}
               }
\gridline{
              \fig{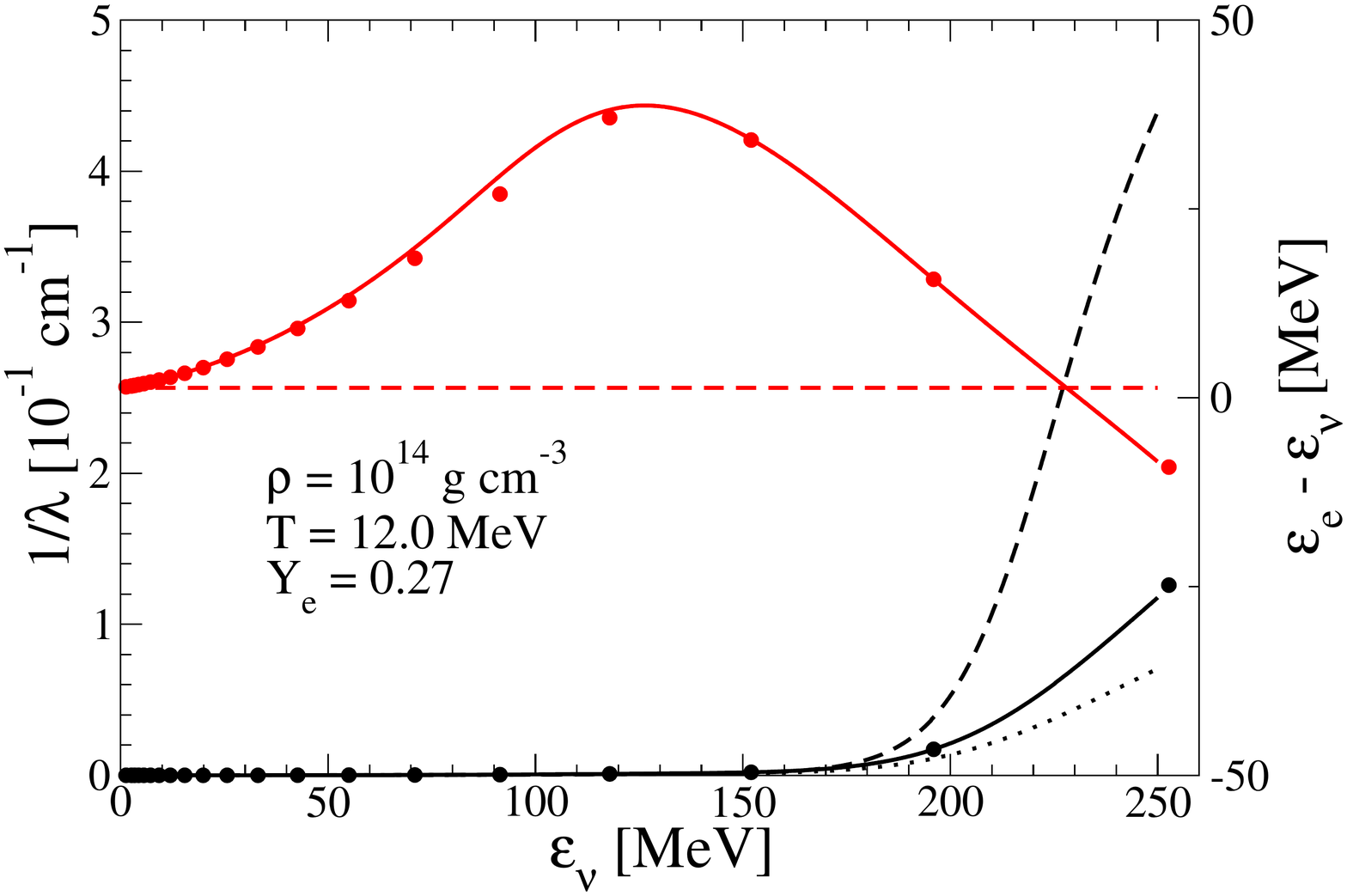}{0.5\textwidth}{(c)}
               }
\caption{\label{fig:AbEm_nu_eos}
Inverse mean free paths for the absorption/emission on free nucleons of \nue-neutrinos (black, scale on left) and the difference in energy between the absorbed neutrino and emitted electron (red, scale on right), for the thermodynamic conditions listed on the upper left of the plots. 
The dashed lines give the inverse mean free paths for the iso-energetic approximation of \citet{Brue85}, uncorrected for weak magnetism, the dotted lines give the inverse mean free paths as given by \citep{RePrLa98}, which include recoil, nucleon final state blocking, and special relativity but not weak magnetism corrections, and the solid lines are the latter inverse mean free paths corrected for weak magnetism. 
The inverse mean free paths plotted by the solid, dashed, and dotted lines were computed using a neutrino energy grid of 1500 zones spaced between 1 and 300 MeV. 
The symbols (circles for \nue, triangle for \nuebar) show the inverse mean free paths computed with the typical \Chimera\ energy grid of  20 zones geometrically spaced between 4 and 250 MeV.
}
\end{figure}

\begin{figure}
\gridline{\fig{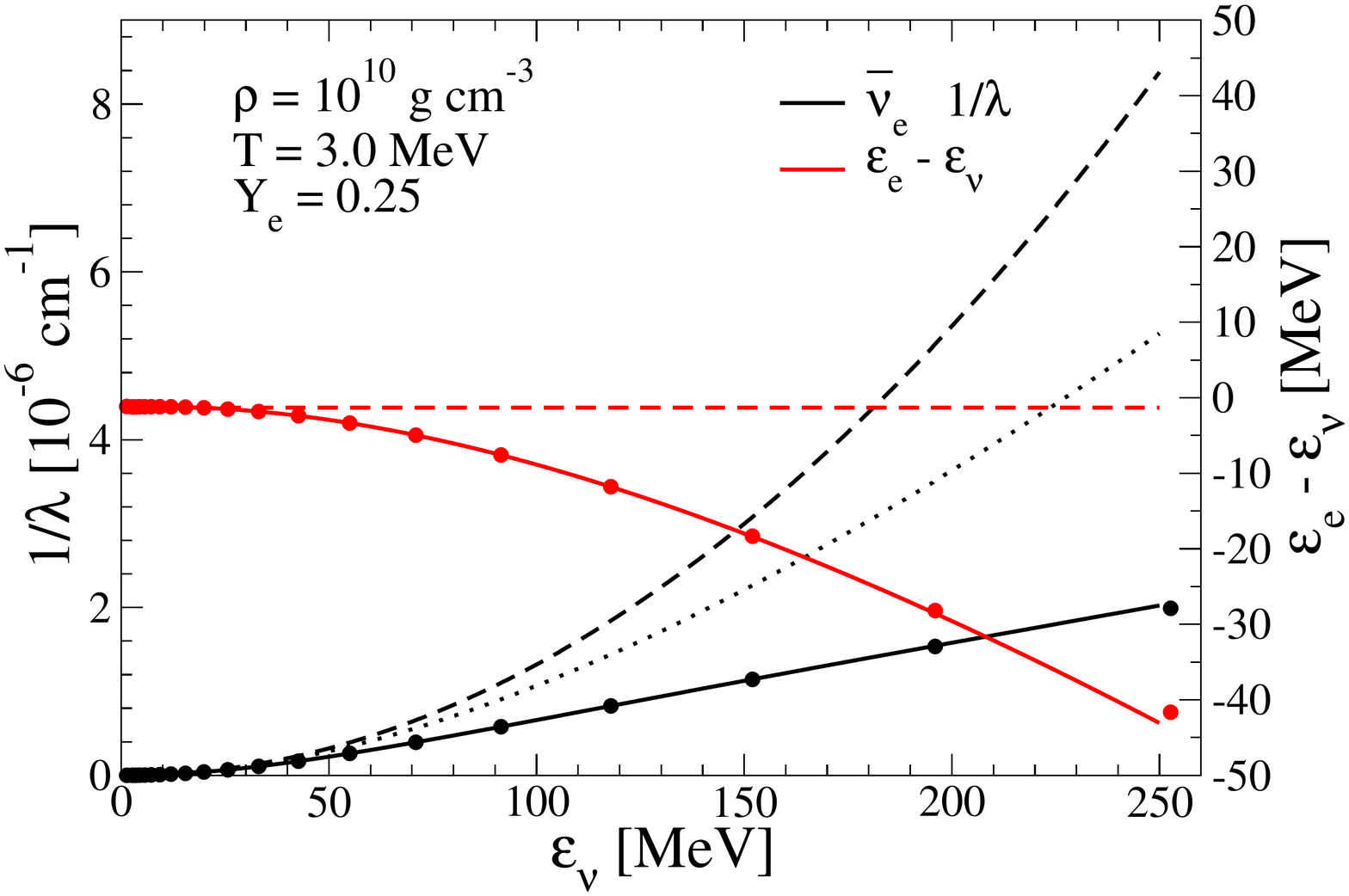}{0.5\textwidth}{(a)}
              \fig{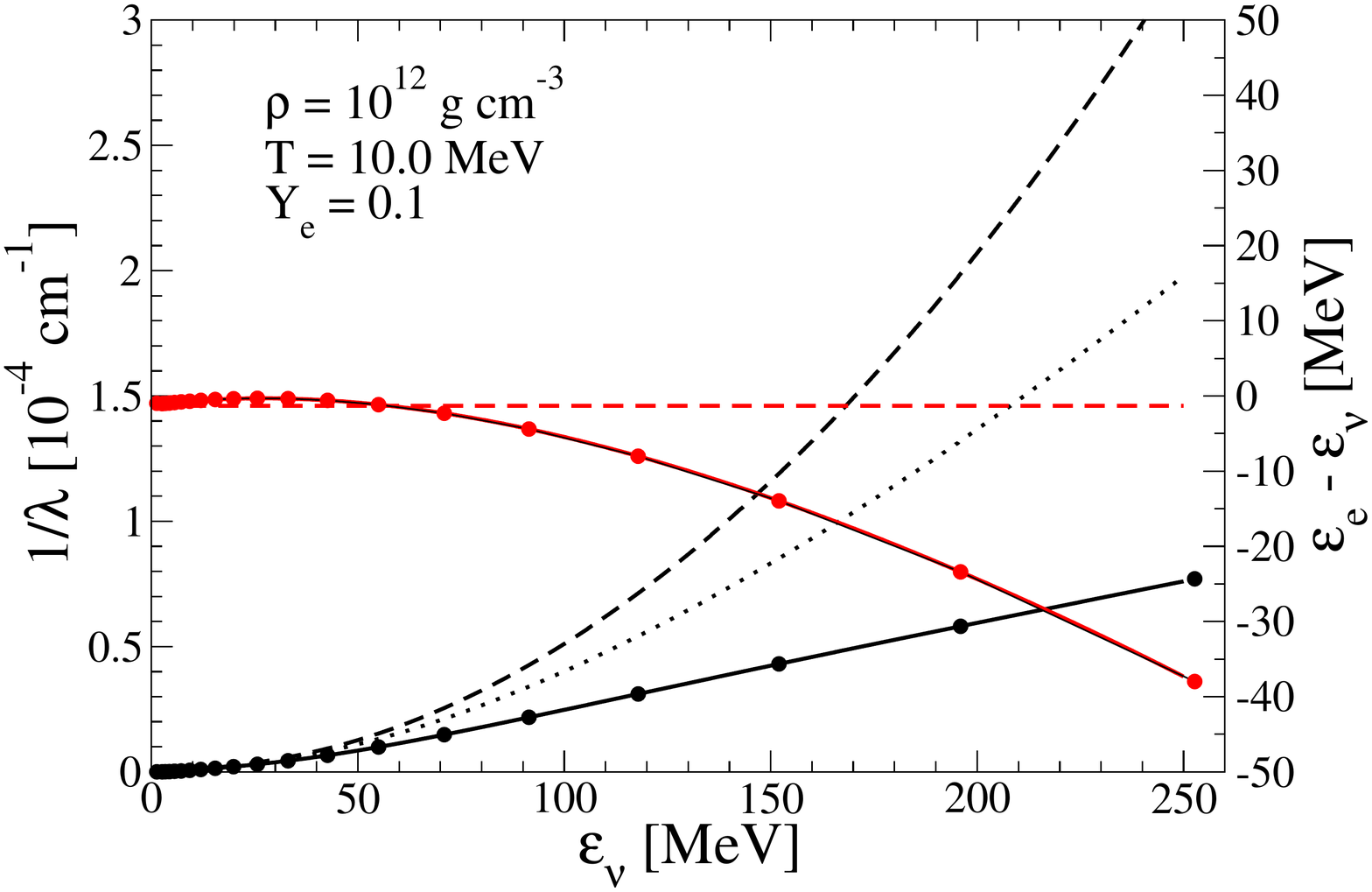}{0.5\textwidth}{(b)}
               }
\gridline{
              \fig{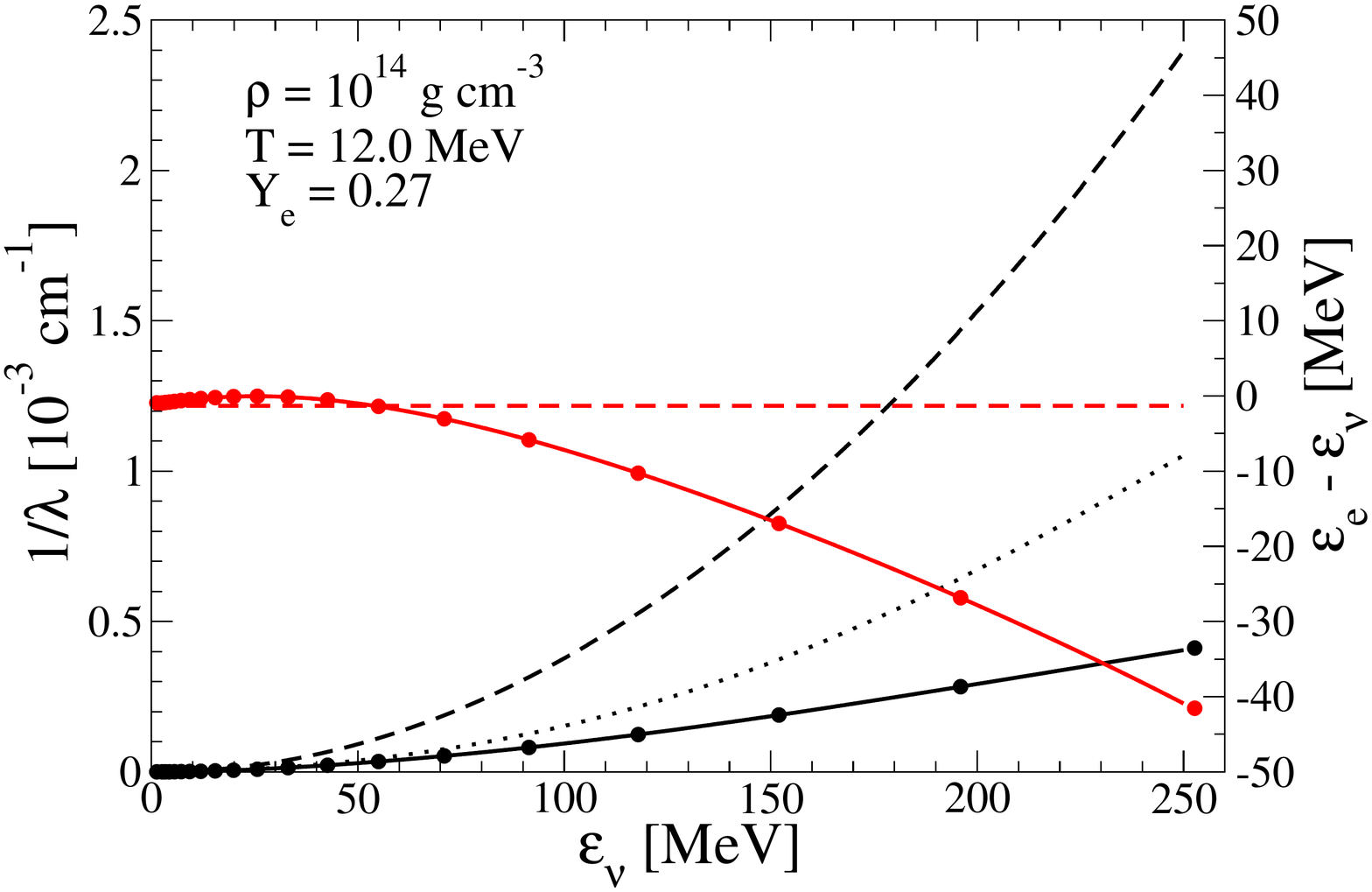}{0.5\textwidth}{(c)}
               }
\caption{\label{fig:AbEm_nubar_eos} Same as Figure~\ref{fig:AbEm_nu_eos} but for \nuebar\ absorption/emission on free nucleons.
}
\end{figure}

Emission and absorption of neutrinos is an important process, for which we plot in Figure~\ref{fig:AbEm_nu_eos},  for \nue\, the inverse mean free paths (black lines), given by
\begin{equation}
  1/\lambda(\epscm) = {\chi_{\rm a}}^{(0)}(\epscm) + {\chi_{\rm e}}^{(0)}(\epscm),
\label{eq:d10}
\end{equation}
where  ${\chi_{\rm a}}^{(0)}$ and ${\chi_{\rm e}}^{(0)}$ are the absorption and emission inverse mean free paths, respectively, and $\epscmi{ {\rm e}} - \epscmi{ \nu}$ (red lines) is the difference between the energies of the emitted electron and the absorbed neutrino.
Figure~\ref{fig:AbEm_nubar_eos} shows the same for \nuebar. 
The dashed lines give the inverse mean free paths for the iso-energetic approximation of \citet{Brue85}, uncorrected for weak magnetism. The dotted lines give the inverse mean free paths as given by \citet{RePrLa98}, which include recoil, nucleon final state blocking, and special relativity but not weak magnetism corrections. The solid lines give the inverse mean free paths for the latter but corrected for weak magnetism. 
Like neutrino scattering on free nucleons, the inverse mean free path for absorption and emission on free nucleons is reduced when recoil is taken into account by the reduced center of mass energy of the collision. 
At very low incident energies the emitted electron energy, $\epscmi{  {\rm e}}$, tends to be greater than the incident \nue\ energy, $\epscmi{ \nu}$, because of the neutron--proton mass difference and the thermal motions of the nucleons. 
At high energies,  $\epscmi{ {\rm e}}$ decreases below $\epscmi{ \nu}$ due to part of the incident collision energy being taken up by the final nucleon. 
The same is true for  \nuebar\ except that the neutron--proton mass difference is negative in this case. 
In the iso-energetic approximation, $\epscmi{ {\rm e}} - \epscmi{ \nu}$ is just the neutron--proton mass difference, positive for \nue, negative for \nuebar.

As a reference, the $1/\lambda$ and $\epscmi{ {\rm e}} - \epscmi{ \nu}$ shown by the dashed, dotted, and solid lines were computed for a given angle $\theta$ between the incident neutrino and emitted electron, by integrating the final electron energy using an energy grid of 1500 zones of equal width spaced between 1 and 300 MeV. 
The first two Legendre moments of the absorption and emission kernels were  then computed with a 64-point Gauss--Legendre angular quadrature. 
The filled circles show $1/\lambda$ and $\epscmi{ {\rm e}} - \epscmi{  \nu}$ as computed by \Chimera\ with the typical energy grid of 20 zones geometrically spaced between 4 and 250 MeV. 
To compute $1/\lambda$, \chimera\ uses a 64-point Gauss--Legendre quadrature to integrate over the final electron energy, with limits given by Equation~(\ref{eq:d8e}) above with the term $dm \, c^{2} = (M_{1} - M_{2})c^{2}$ added, where $M_{1}$ and $M_{2}$ are the initial and final nucleon masses, respectively.
A 64-point Gauss--Legendre angular quadrature is then used to compute the first two Legendre moments of the absorption and emission inverse mean free paths.

\begin{figure}
\gridline{\fig{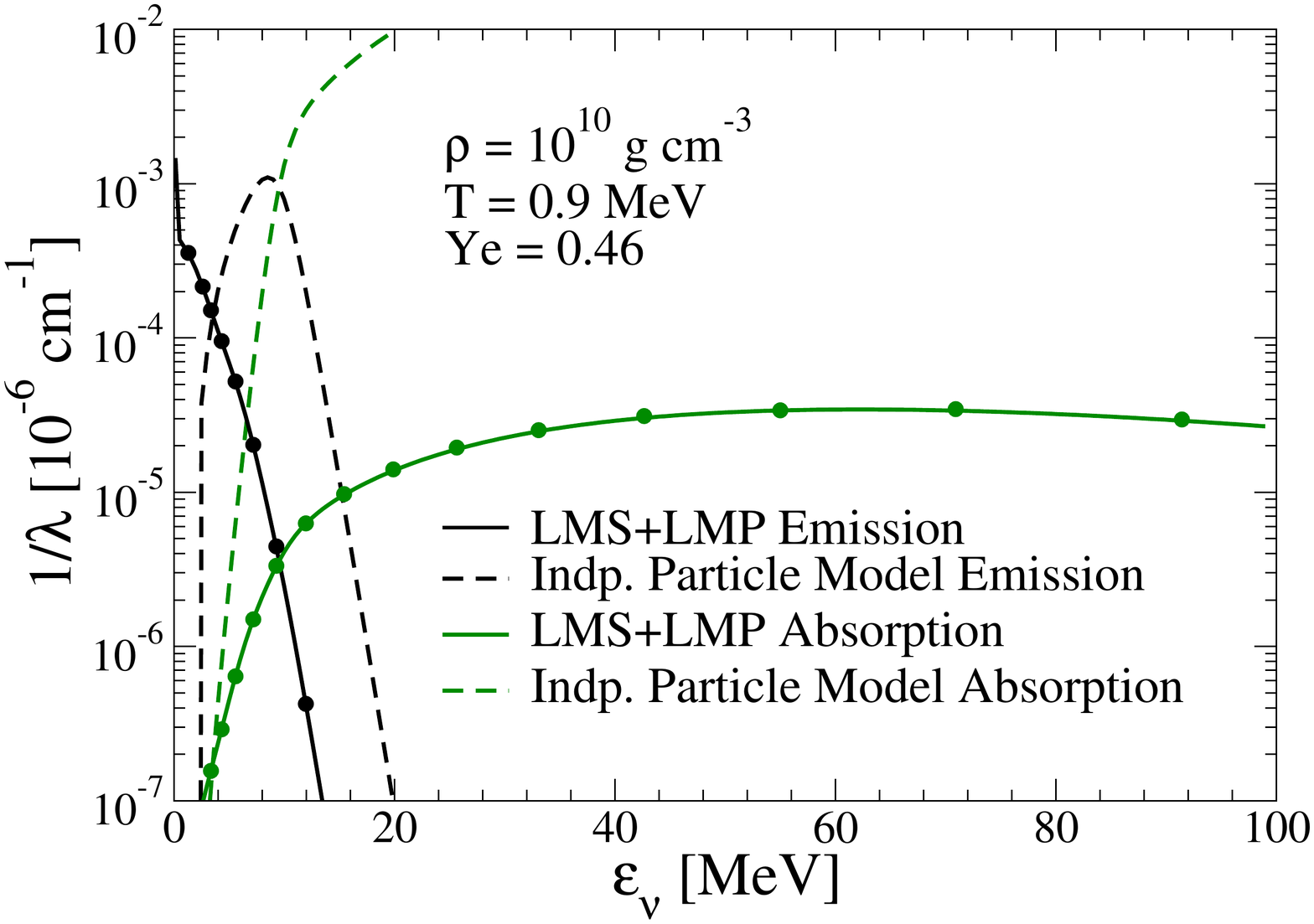}{0.5\textwidth}{(a)}
              \fig{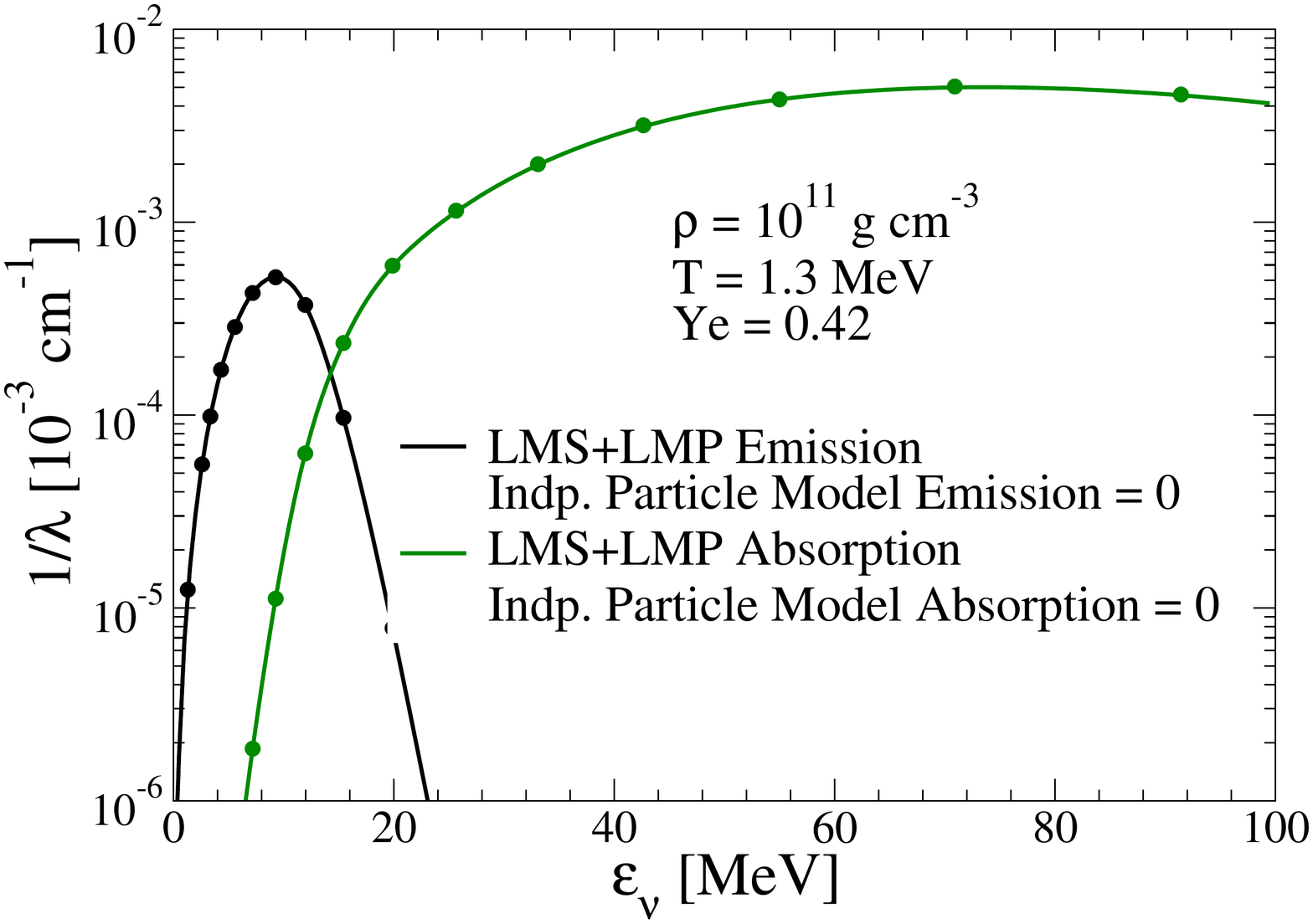}{0.5\textwidth}{(b)}
               }
\gridline{
              \fig{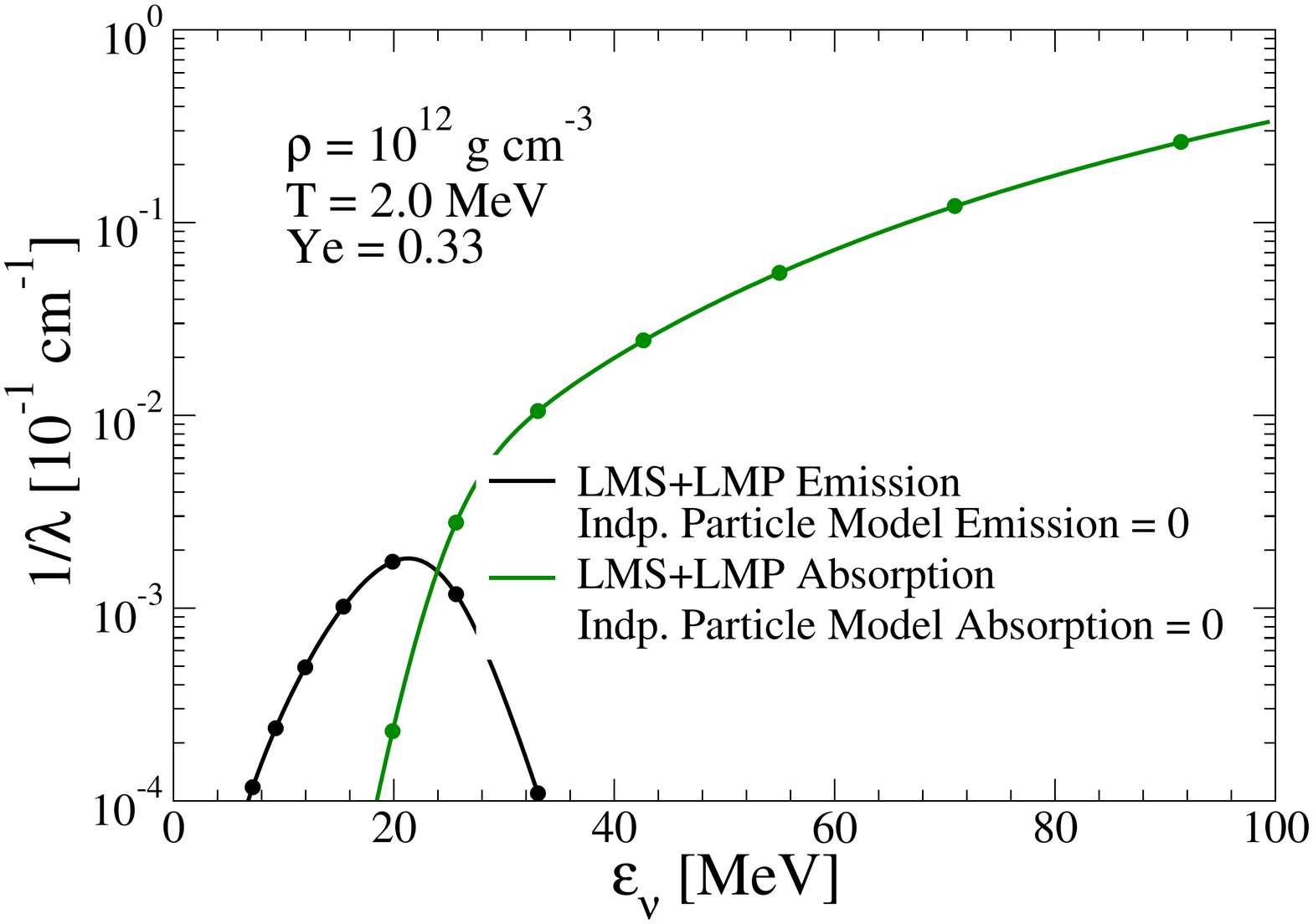}{0.5\textwidth}{(c)}
               }
\caption{\label{fig:AbEm_A_eos}
Inverse mean free paths for the \nue\ emission (black lines and circles) and \nue\ absorption (green lines and circles) on nuclei, for the thermodynamic conditions listed on the upper left of the plots. 
The dashed lines give the inverse mean free paths calculated from the independent particle model (IPM) as formulated by \citet{Brue85}.
Solid lines give the inverse mean free paths computed from tables based on the LMS-LMP formulation.
Data plotted by both the solid and dashed lines were computed using a linear energy grid of 200 zones from 0 to 100 MeV, exactly matching the energy grid of the LMS-LMP electron capture table.
Filled circles show the inverse mean free paths computed with the typical \Chimera\ energy grid of 20 zones geometrically spaced between 4 and 250 MeV.
}
\end{figure}

\subsection{Neutrino Absorption and Emission on Nuclei}
\label{abem_nuc}

Calculation of the rate of electron capture on heavy nuclei and the resulting neutrino emission in the collapsing core requires three components: the appropriate electron capture reaction rates, the spectra of emitted neutrinos, and knowledge of the nuclear composition.
In simulations of the collapsing stellar iron core, the composition is calculated by the equation of state assuming nuclear statistical equilibrium (NSE), instead of being tracked in detail via a reaction network.
As discussed in Section~\ref{StellarEoS}, the information on the nuclear composition typically provided by the equation of state is limited to the mass fractions of free neutrons and protons, $\alpha$-particles, and
the sum of all heavy nuclei, as well as the identity of an average heavy nucleus.
In \chimera, we use a prescription for nuclear electron capture first utilized by \citet{LaMaSa03} and \citet{HiMeMe03}.
This treatment is based on shell model electron capture rates from \citet[][LMP]{LaMa00} for $45<A\leq65$ and 80 reaction rates from a hybrid 
Shell-Model-Monte-Carlo (SMMC) -- Random-Phase-Approximation (RPA) calculation \citep[][LMS]{LaKoDe01,LaMaSa03} for a sample of nuclei with $66\leq A \leq 112$.
The approximation of \citet{LaMaSa01} is used for the distribution of emitted neutrinos.
To calculate the needed abundances of the heavy nuclei, a Saha-like NSE is assumed, including Coulomb corrections to the nuclear binding energy \citep{HiTh96,BrGa99}, but neglecting the effects of degenerate nucleons \citep{ElHi80}.
The combined set of LMP and LMS model rates is used to calculate an average neutrino emissivity per heavy nucleus.
The full neutrino emissivity is then the product of this average and the number density of heavy nuclei calculated by the equation of state.
With the limited coverage of rates for $A>65$, this approach provides a reasonable estimate of what the total electron capture rate would be if rates for all nuclei were available.
This averaging approach also makes the rate of electron capture consistent with the composition returned by the equation of state, while minimizing the impact of the limitations of our NSE treatment. 
A public version of an updated version of this rate tabulation is available from \citet{JuLaHi10}.
This tabulation, which is planned for inclusion in a future version of \chimera, includes more extensive coverage of heavier, more neutron-rich nuclei via a Fermi--Dirac (FD) parameterization of level occupation in place of the more costly SMMC approach.

In Figures~\ref{fig:AbEm_A_eos},  we plot the \nue\ inverse mean free paths for absorption and emission on nuclei for both the IPM as formulated by \citet{Brue85} and the more sophisticated LMP--LMS formulations. 
Both the solid and dashed lines are the results of computations using a linear energy grid of 200 zones aligned with the energy grid of the electron capture table on which the LMP--LMS inverse mean free paths were tabulated. 
The filled circles are the results obtained using the typical \Chimera\ energy grid.
At densities below a few times 10$^{10}$ \gcc\, the inverse mean free paths given by the IPM dominate (top frame of Figure \ref{fig:AbEm_A_eos}), but at higher densities electron capture reduces \Ye\ and drives the nuclear abundances toward neutron richness, including nuclei with neutron numbers $N > 40$. 
The IPM results in the vanishing of the electron capture inverse mean free paths for $N \ge 40$ nuclei due to the filling of the neutron $^{1}f_{5/3}$ orbital. 
In the LMS treatment of electron captures on nuclei, it was shown that Pauli blocking due to the filling of the neutron $^{1}f_{5/3}$ orbital is overcome by correlations and temperature effects.
Consequently, at densities above a few times 10$^{10}$ \gcc\, the LMS--LMP inverse mean free paths remain finite, whereas those given by the IPM vanish (see Figure~\ref{fig:AbEm_A_eos}(b) and (c)). 

\subsection{Pair Production: General}
\label{pairgen}

In spherical symmetry, the rate of change of the neutrino (antineutrino) occupation probability, $f(\mucm,\epscm)$, due to pair production process ``XX'' is given by
\begin{eqnarray}
\left( \frac{ df(\mucm, \epscm) }{ dt } \right)_{ \! {\rm S, \, pair, \, XX} } &=& \left[ 1 - f(\mucm, \epscm) \right] \frac{1}{(hc)^{3}} \int_{0}^{\infty} \epscm'^{2} d\epscm' \int_{-1}^{1} d\mucm' [ 1 - \bar{f}(\mucm', \epscm') ] \int_{0}^{2\pi} d\beta' R_{\rm XX}^{\rm p}(\epscm, \epscm', \cos\theta ) \nonumber \\  
&& - f(\mucm, \epscm) \frac{1}{(hc)^{3}} \int_{0}^{\infty} \epscm'^{2} d\epscm' \int_{-1}^{1} d\mucm' \bar{f}(\mucm', \epscm')  \int_{0}^{2\pi} d\beta' R_{\rm XX}^{\rm a}(\epscm, \epscm', \cos\theta )  , 
\label{eq:d11} 
\end{eqnarray} 
where $\bar{f}(\mucm', \epscm')$ is the antineutrino (neutrino) occupation probability, $R_{\rm XX}^{\rm a}(\epscm, \epscm', \cos\theta )$ and $R_{\rm XX}^{\rm p}(\epscm, \epscm', \cos\theta )$ are, respectively, the neutrino--antineutrino annihilation and neutrino--antineutrino unblocked-creation rates per neutrino--antineutrino states for process XX, $\epscm$ is the neutrino energy, $\mucm$ is the cosine of the neutrino propagation direction with respect to the radial direction, $\phi$ is the azimuthal propagation directions, and $\cos \theta$ is defined in Equation~(\ref{eq:d2}). 
Unprimed quantities refer to neutrinos (antineutrinos), and primed quantities refer to antineutrinos (neutrinos).
$R_{\rm XX}^{\rm a}$ and $R_{\rm XX}^{\rm p}$ are related by
\begin{equation}
  R_{\rm XX}^{\rm p}(\epscm, \epscm', \cos\theta ) = \exp \left( - ( E_{\rm e^{-}} + E_{\rm e^{+}} )/kT \right) R_{\rm XX}^{\rm a}(\epscm, \epscm', \cos\theta ) ,
\label{eq:d12}
\end{equation}
where $E_{\rm e^{-}}$ and $E_{\rm e^{+}}$ are the associated electron and positron energies, respectively.
Expanding the annihilation and creation kernels in a Legendre expansion and keeping the first two terms, as was done in Equation~(\ref{eq:d5}) for the scattering kernels, gives 
\begin{eqnarray}
R_{\rm XX}^{\rm a/p}(\epscm,\epscm',\cos\theta) &=& \frac{1}{2} \sum_{\ell=0}^{\infty} (2\ell + 1) \Phi_{\ell, {\rm XX}}^{\rm a/p}(\epscm,\epscm') P_{\ell}( \cos\theta ) \nonumber \\ 
& \simeq& \frac{1}{2} \Phi_{0, {\rm XX}}^{\rm a/p}(\epscm,\epscm') + \frac{3}{2} \Phi_{1,{\rm XX}}^{\rm a/p}(\epscm,\epscm') \cos\theta \label{eq:d13}.
\end{eqnarray}
Applying the moment operators $(4\pi)^{-1} \int d\Omega$ and $(4\pi)^{-1} \int \mucm d\Omega$ to Equation~(\ref{eq:d11}) and using the definitions in Equations~(\ref{eq:d2}),  (\ref{eq:d13}), (\ref{eq:d11}), and (\ref{eq:a10}), gives the moments of the pair interaction terms:
\begin{eqnarray}
 \left( \frac{ df(\mucm, \epscm) }{ dt } \right)_{ \! {\rm S, \, pair} }^{(0,1)} &=& \frac{2\pi}{(hc)^{3}} \left[ 1 - \psimomcm{0}(\epscm), - \psimomcm{1}(\epscm) \right] \int_{0}^{\infty} \epscm'^{2} d\epscm' \Phi_{0, {\rm XX}}^{\rm p}(\epscm,\epscm') \left[ 1 - \psibarmomcm{0}(\epscm') \right] \nonumber \\ 
&& - \frac{2\pi}{(hc)^{3}} \left[ 3 \psimomcm{1}(\epscm), 1 -3 \psimomcm{2}(\epscm) \right] \int_{0}^{\infty} \epscm'^{2} d\epscm' \Phi_{1,{\rm XX}}^{\rm p}(\epscm,\epscm') \psibarmomcm{1}(\epscm')  \nonumber \\ 
&& - \frac{2\pi}{(hc)^{3}} \left[ \psimomcm{0}(\epscm), \psimomcm{1}(\epscm) \right] \int_{0}^{\infty} \epscm'^{2} d\epscm' \Phi_{0, {\rm XX}}^{\rm a}(\epscm,\epscm') \psibarmomcm{0}(\epscm') \chi(\epscm,\epscm') \nonumber \\ 
&& - \frac{2\pi}{(hc)^{3}} \left[ 3 \psimomcm{1}(\epscm), 3 \psimomcm{2}(\epscm) \right] \int_{0}^{\infty} \epscm'^{2} d\epscm' \Phi_{1,{\rm XX}}^{\rm a}(\epscm,\epscm') \psibarmomcm{1}(\epscm') \chi(\epscm,\epscm') . \label{eq:d14}
\end{eqnarray}

The pair annihilation kernels in Equation~(\ref{eq:d14}) have been multiplied by angular cutoff correction factors, $\chi(\epscm,\epscm')$, to account for the nonisotropic nature of the neutrino flow \citep[e.g.,][]{GoDaNu87, CovaBa87}. 
For \Chimera, these angular correction factors are derived as follows. 
The pair neutrino--antineutrino annihilation rate is proportional to the square of the center of mass energy, or to $(1 - {\bf n_{\nu}} \cdot {\bf n_{\nubar}})^{2}$, where ${\bf n_{\nu}}$ and ${\bf n_{\nubar}}$ are unit vectors in the propagation direction of  $\nu$ and \nubar, respectively.
Inside the neutrinosphere the anisotropy of the neutrino distributions are small, and the occupation distributions $f_{\nu}$ and $f_{\nubar}$ can be approximated by 
\begin{equation}
  f_{\nu}(\epscm, \mucm) = \psimomcmi{0}{\nu}(\epscm) + 3 \mucm \psimomcmi{1}{\nu}(\epscm), \qquad f_{\nubar}(\epscm', \mucm') = \psimomcmi{0}{\nubar}(\epscm') + 3 \mucm' \psimomcmi{1}{\nubar}(\epscm'),
\label{eq:d15}
\end{equation}
which follows from the definitions in Equation~(\ref{eq:a10}). 
Using Equation~(\ref{eq:d2}) for ${\bf n_{\nu}} \cdot {\bf n_{\nubar}}$, we find that the ratio, $\chi_{1}$, of $(1 - {\bf n_{\nu}} \cdot {\bf n_{\nubar}})^{2}$ evaluated with an anisotropic neutrino distribution in the numerator and an isotropic neutrino distribution in the denominator, having the same number and energy spectrum, is given by
\begin{eqnarray}
 \chi_{1}(\epscm, \epscm') &=& \frac{ (4\pi)^{-1} \int_{4\pi} d\Omega \, (4\pi)^{-1} \int_{4\pi} d\Omega' \left( \psimomcmi{0}{\nu}(\epscm) + 3 \mucm \psimomcmi{1}{\nu}(\epscm) \right) \left( \psimomcmi{0}{\nubar}(\epscm') + 3 \mucm' \psimomcmi{1}{\nubar}(\epscm') \right) \left( 1 - {\bf n_{\nu}} \cdot {\bf n_{\nubar}} \right)^{2} }{ (4\pi)^{-1} \int_{4\pi} d\Omega \, (4\pi)^{-1} \int_{4\pi} d\Omega' \psimomcmi{0}{\nu}(\epscm) \psibarmomcmi{0}{\nubar}(\epscm') \left( 1 - {\bf n_{\nu}} \cdot {\bf n_{\nubar}} \right)^{2}} \nonumber \\ 
& =& 1 - \frac{1}{6} \frac{ \psimomcm{1}(\epscm) }{ \psimomcm{0}(\epscm) } \frac{\psibarmomcm{1}(\epscm') }{\psibarmomcm{0}(\epscm') } . \label{eq:d16}
\end{eqnarray}

\begin{figure}
\gridline{\fig{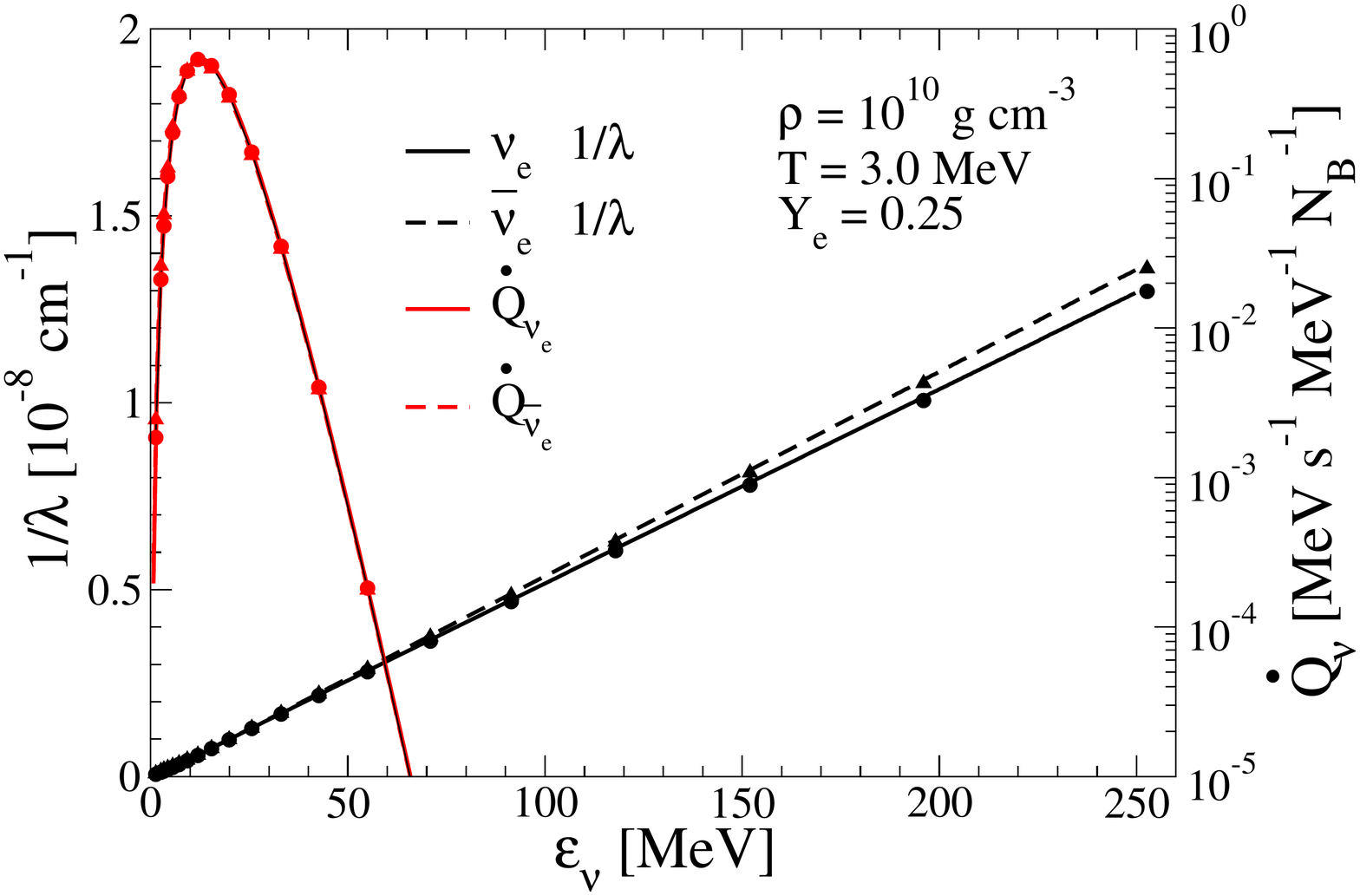}{0.5\textwidth}{(a)}
              \fig{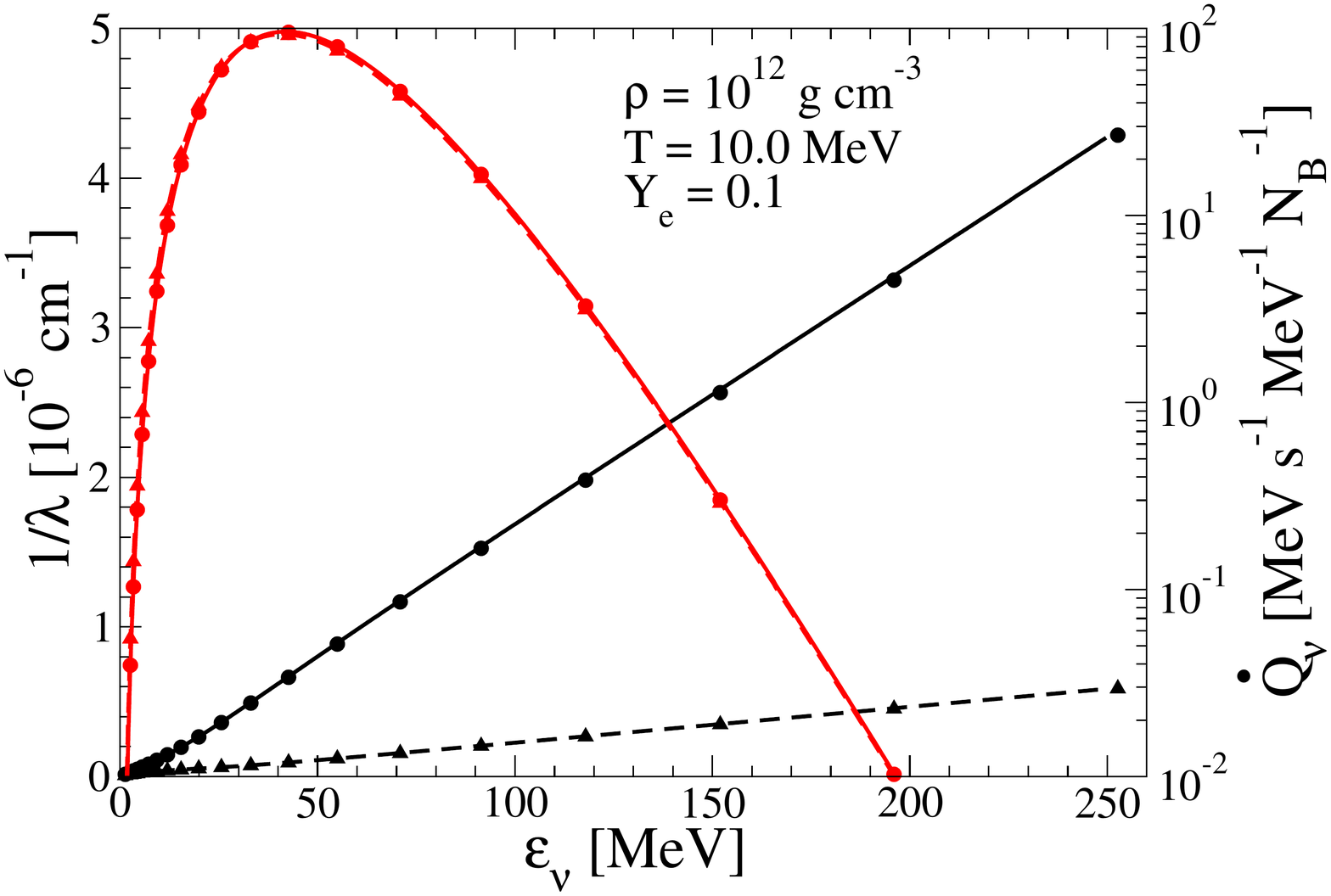}{0.5\textwidth}{(b)}
               }
\gridline{
              \fig{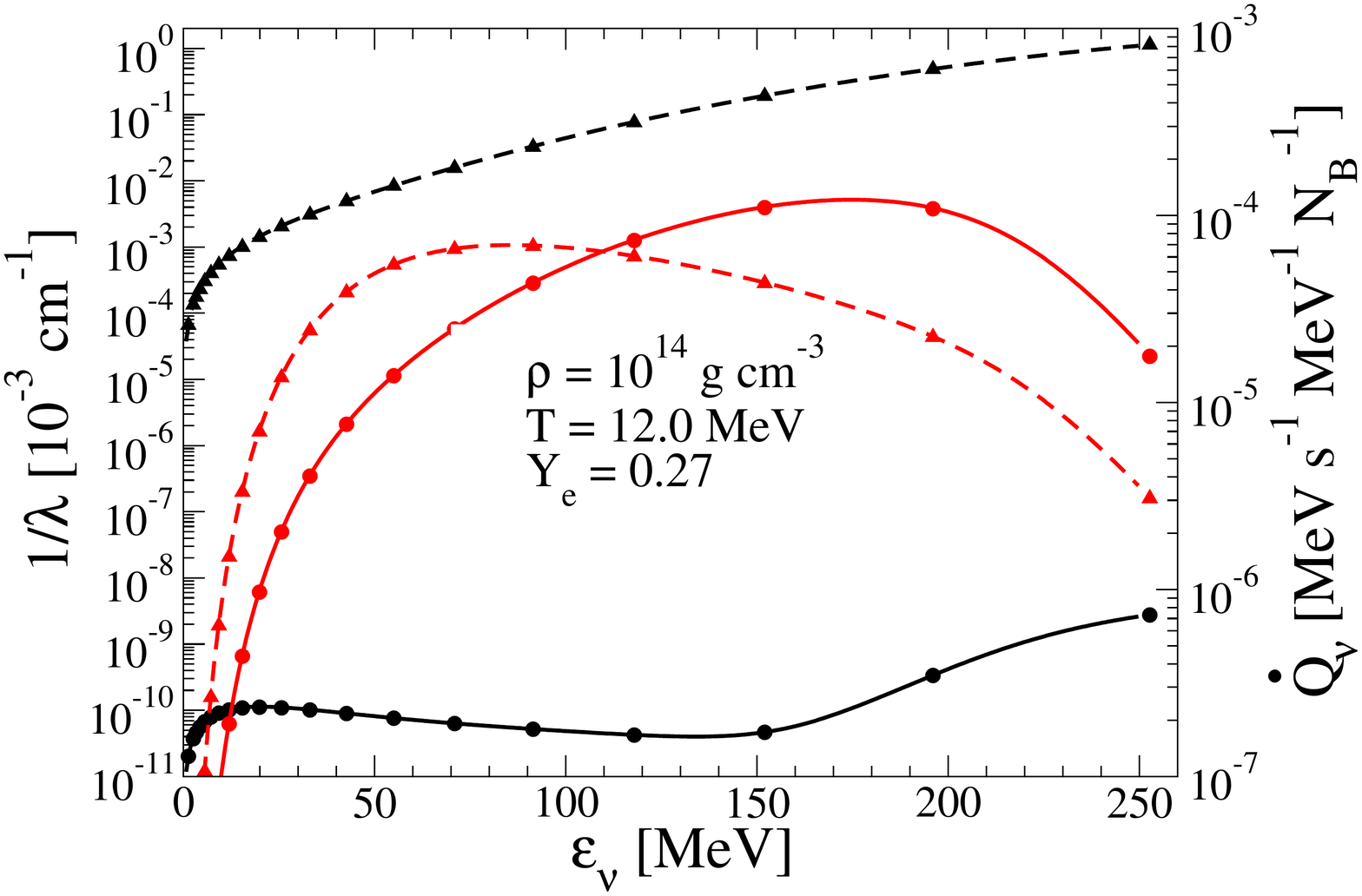}{0.5\textwidth}{(c)}
               }
\caption{\label{fig:Pair_eos}
Inverse mean free path (black) for electron--positron annihilation and pair production for \nue\ (solid)  and \nuebar\ (dashed), computed from Equation~(\ref{eq:d23}) on a 600-point evenly-spaced energy grid from 0--300~MeV, for listed thermodynamic conditions. Inverse mean free paths were computed with the \nue (\nuebar) occupation distribution set to zero and the \nuebar (\nue) distribution equilibrated with the matter and energy production rates per baryon for \nue\ (solid) and \nuebar\ (dashed) computed from Equation~(\ref{eq:d24}) with both the \nue\ and \nuebar occupation distributions set to zero. Filled circles (\nue) and triangles (\nuebar) are for values computed with a typical \Chimera\ energy grid of 20 logarithmic energy zones. 
}
\end{figure}

Outside the neutrinospheres, the neutrino occupation distributions become increasingly anisotropic and the neutrino--antineutrino center of mass collision energy becomes increasingly smaller as the beaming becomes radial and more nearly collinear.
Assuming that the neutrinosphere emits isotropically and that $f$ is constant along rays leading back to the neutrinosphere, outside the neutrinosphere $f$ is given by
\begin{equation}
  f_{\nu}(\epscm,\mucm) = \left\{
 \begin{array}{ccl}
 \ds \frac{2}{ 1 - \mucmi{  i} } \psimomcmi{0}{\nu}(\epscm)  &  \mucmi{  i}(\epscm) \le \mucm \le 1 & \mucmi{  i}(\epscm) = \sqrt{ 1 - \left( R_{\nu}(\epscm)/R \right)^{2} } \quad R > R_{\nu} \\
  0 & \mucm < \mucmi{  i}(\epscm) &
  \end{array}
  \right. ,
  \label{eq:d17}
\end{equation}
\begin{equation}
   f_{\nubar}(\epscm',\mucm') = \left\{
 \begin{array}{ccl}
 \ds \frac{2}{ 1 - \mucmi{i}' } \psimomcmi{0}{\nubar}(\epscm')  &  \mucmi{i}'(\epscm) \le \mucm' \le 1   &  \mucmi{i}'(\epscm) = \sqrt{ 1 - \left( R_{\nubar}(\epscm')/R \right)^{2} } \quad R > R_{\nubar} \\
  0 &   \mucm' <  \mucmi{i}'(\epscm) &
  \end{array}
  \right. ,
\label{eq:d18}
\end{equation}
where $R_{\nu}(\epscm)$ and $R_{\nubar}(\epscm')$ are the radii of the neutrinospheres of the neutrinos and antineutrinos, respectively, and $R$ is the radial coordinate at which the correction is being applied.
Using Equations ~(\ref{eq:d17}) and (\ref{eq:d18}) for the anisotropic neutrino radiation, we find that the ratio, $\chi_{2}$, of $(1 - {\bf n_{\nu}} \cdot {\bf n_{\nubar}})^{2}$ evaluated, as before, with an anisotropic neutrino distribution in the numerator and an isotropic neutrino distribution in the denominator, having the same number and energy spectrum, is given by
\begin{eqnarray}
\chi_{2}(\epscm, \epscm') &=& \frac{ \ds \frac{1}{4\pi} \int_{ \mucmi{  i}(\epscm) }^{1} d\mucm \int_{0}^{2\pi} d\phi \, \frac{1}{4\pi}\int_{ \mucmi{i}'(\epscm) }^{1} d\mucm' \int_{0}^{2\pi} d\phi' \frac{2}{ 1 - \mucmi{  i}(\epscm) } \psimomcmi{0}{\nu}(\epscm) \frac{2}{ 1 - \mucmi{i}'(\epscm) } \psimomcmi{0}{\nubar}(\epscm') \left( 1 - {\bf n_{\nu}} \cdot {\bf n_{\nubar}} \right)^{2}} { \ds \frac{1}{4\pi} \int_{-1}^{1} d\mucm \int_{0}^{2\pi} d\phi \, \frac{1}{4\pi} \int_{ -1}^{1} d\mucm' \int_{0}^{2\pi} d\phi' \psimomcmi{0}{\nu}(\epscm) \psimomcmi{0}{\nubar}(\epscm') \left( 1 - {\bf n_{\nu}} \cdot {\bf n_{\nubar}} \right)^{2} } \nonumber \\ 
& =& \frac{1}{8} \left\{ 8 + [ \mucmi{  i}(\epscm) \mucmi{i}'(\epscm') - 3 ] [ 1 +  \mucmi{  i}(\epscm) ] [ 1 + \mucmi{i}'(\epscm') ] \right\}.  \label{eq:d19}
\end{eqnarray}
We note that in the case $\mucmi{i}'(\epscm) = \mucmi{  i}(\epscm)$, $\chi_{2}(\epscm, \epscm)$ reduces to
\begin{equation}
  \chi_{2}(\epscm, \epscm) = \frac{1}{8} [ 1 -  \mucmi{  i}(\epscm) ]^{2} \left[ \mucmi{  i}^{2}(\epscm) + 4 \mucmi{  i}(\epscm) + 5 \right] .
\label{eq:d20}
\end{equation}
To smoothly transition between $\chi_{1}$ and $\chi_{2}$, we define the neutrino--antineutrino pair annihilation angular corrections, $\chi$, to be given by
\begin{equation}
  \chi = \min( \chi_{1}, \chi_{2} )
\label{eq:d21}
\end{equation}
and extend the definitions of \mucmi{ i}\ (and $\mucmi{i}'$) in Equation~(\ref{eq:d17}) (and (\ref{eq:d18})) to
\begin{equation}
 \mucmi{  i} = \max\left( {\rm sign}\left[ \sqrt{ \left| 1 - \left( R_{\nu}(\epscm)/R \right)^{2} \right| },  1 - \left( R_{\nu}(\epscm)/R \right)^{2} \right], -1 \right), \quad \mbox{any } R
\label{eq:d22}
\end{equation}
(similarly for $\mucmi{i}'$.)
Equation~(\ref{eq:d22}) will cause $\mucmi{  i}$ to rapidly approach $-1$ and $\chi_{2}$ to approach 1 as $R$ decreases below $R_{\nu}$, and Equation~(\ref{eq:d21}) will then cause $\chi$ to be governed by $\chi_{1}$.

\begin{figure}
 \fig{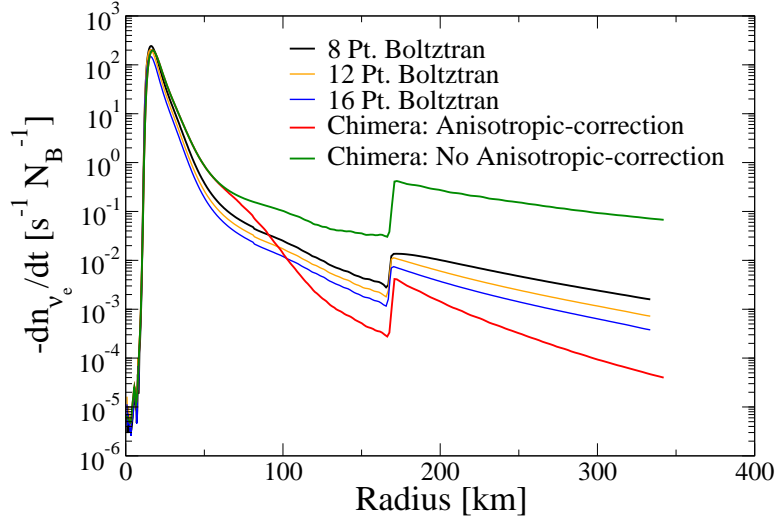}{0.6\textwidth}{}
\caption{\label{fig:ang_cor}
Pair annihilation rates computed by \chimera\ without (green) and with (red) angular cutoff correction factors, compared with the rates computed by {\sc Boltztran} with 8-point (black), 12-point (orange), and 16-point (blue) angular quadrature.
}
\end{figure}

To test the above angular cutoff correction factor, we compare in Figure \ref{fig:ang_cor} the pair annihilation rates computed by \chimera\ without and with the angular cutoff correction factor to the rates computed by the Boltzman solver {\sc Boltztran} with three different beam resolutions. 
Because the angular dependence of the transport has not been integrated over in {\sc Boltztran}, the angular resolution is limited only by the number of beams employed, and the  {\sc Boltztran} rates can therefore be regarded as a reference modulo the number of beams employed.
The core configuration used in this comparison was obtained by evolving the progenitor s15s7b2 \citep{WoWe95} in 1D to 100 ms post-bounce, at which point the core configuration was frozen and the neutrino transport was continued until a steady state was achieved. 
Without the angular cutoff correction factor, the pair annihilation rates computed by \chimera\ become significantly larger than the {\sc Boltztran} rates beyond 70 km, which is 10 km beyond the mean radius of the neutrinosphere. 
They become several orders of magnitude larger beyond 170 km.
With the angular cutoff correction factor included, the behavior of the pair annihilation rates computed by \chimera\ more closely approximates those of {\sc Boltztran}, though differences of the order of 5 are seen beyond 120 km.

\subsection{Neutrino--Antineutrino Pair Annihilation and Production from Electron--Positron Pairs}
\label{eepairgen}

The zero and first moments of the kernels for neutrino--antineutrino pair annihilation into electron--positron pairs and the inverse process are taken from the analytic expressions of \citet[][Equations~(C62)--(C74), with a typo corrected by removing the term $-a_{_{\! 0}}$ from the bracket in Equation~(C68)]{Brue85}.
The integration over the electron energy, $E_{\rm e}$, was performed for the case that $\epscm > \epscm'$ by a 24-point Gauss--Legendre energy quadrature for each of the intervals from 0 to $\epscm'$, from $\epscm'$ to $\epscm$, and from $\epscm$ to $\epscm + \epscm'$, with a similar set of integrations for the case that $\epscm' > \epscm$.

In Figure~\ref{fig:Pair_eos} we plot the inverse mean free path, 
\begin{equation}
 1/\lambda_{t}(\epscm) = \frac{2\pi}{(hc)^{3}} \left\{ \int_{0}^{\infty} \epscm'^{2} d\epscm' \Phi_{0, {\rm pair}}^{\rm p}(\epscm,\epscm') \left[ 1 - \psibarmomcm{0}(\epscm') \right] + \int_{0}^{\infty} \epscm'^{2} d\epscm' \Phi_{0, {\rm pair}}^{\rm a}(\epscm,\epscm') \psibarmomcm{0}(\epscm') \chi(\epscm,\epscm') \right\},
\label{eq:d23}
\end{equation}
where for neutrinos (antineutrinos) we have taken $\psimomcmi{0}{\nu}(\epscm)$ for the neutrinos (antineutrinos) to be zero and $\psibarmomcm{0}(\epscm')$ for the antineutrinos (neutrinos) to be in thermal equilibrium with the matter at the stated thermodynamic conditions, and plotted is the neutrino (antineutrino) spectral energy production rate per baryon,
\begin{equation}
 \dot{Q}_{\nu}(\epscm) = c \frac{m_{\rm B}}{\rho} \frac{4 \pi}{(hc)^{3}} \epscm^{3} \frac{2\pi}{(hc)^{3}} \int_{0}^{\infty} \epscm'^{2} d\epscm' \Phi_{0,{\rm pair}}^{\rm p}(\epscm,\epscm'),
\label{eq:d24}
\end{equation}
where we have set the neutrino--antineutrino blocking factors to zero.

\begin{figure}
\gridline{\fig{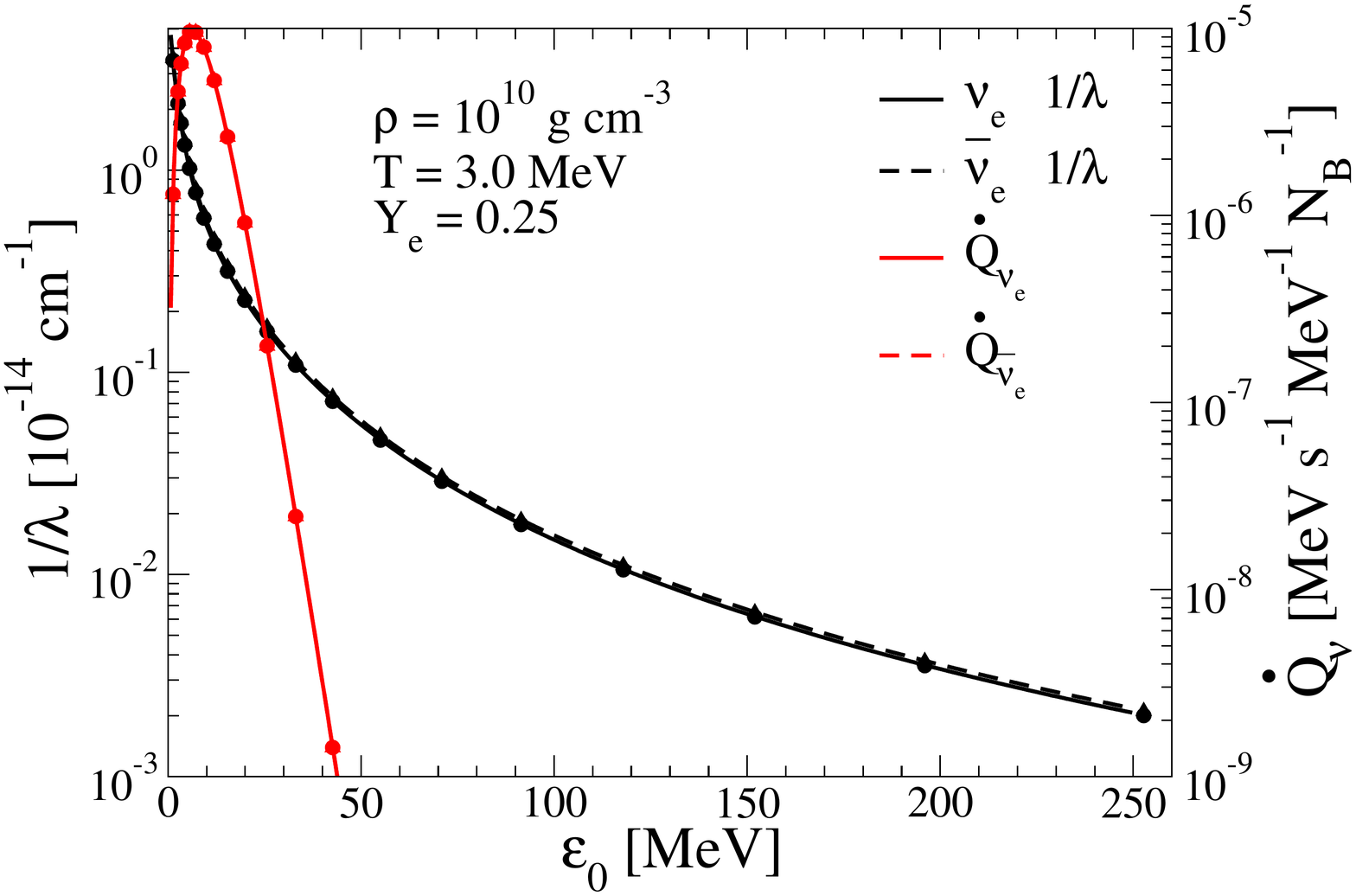}{0.5\textwidth}{(a)}
              \fig{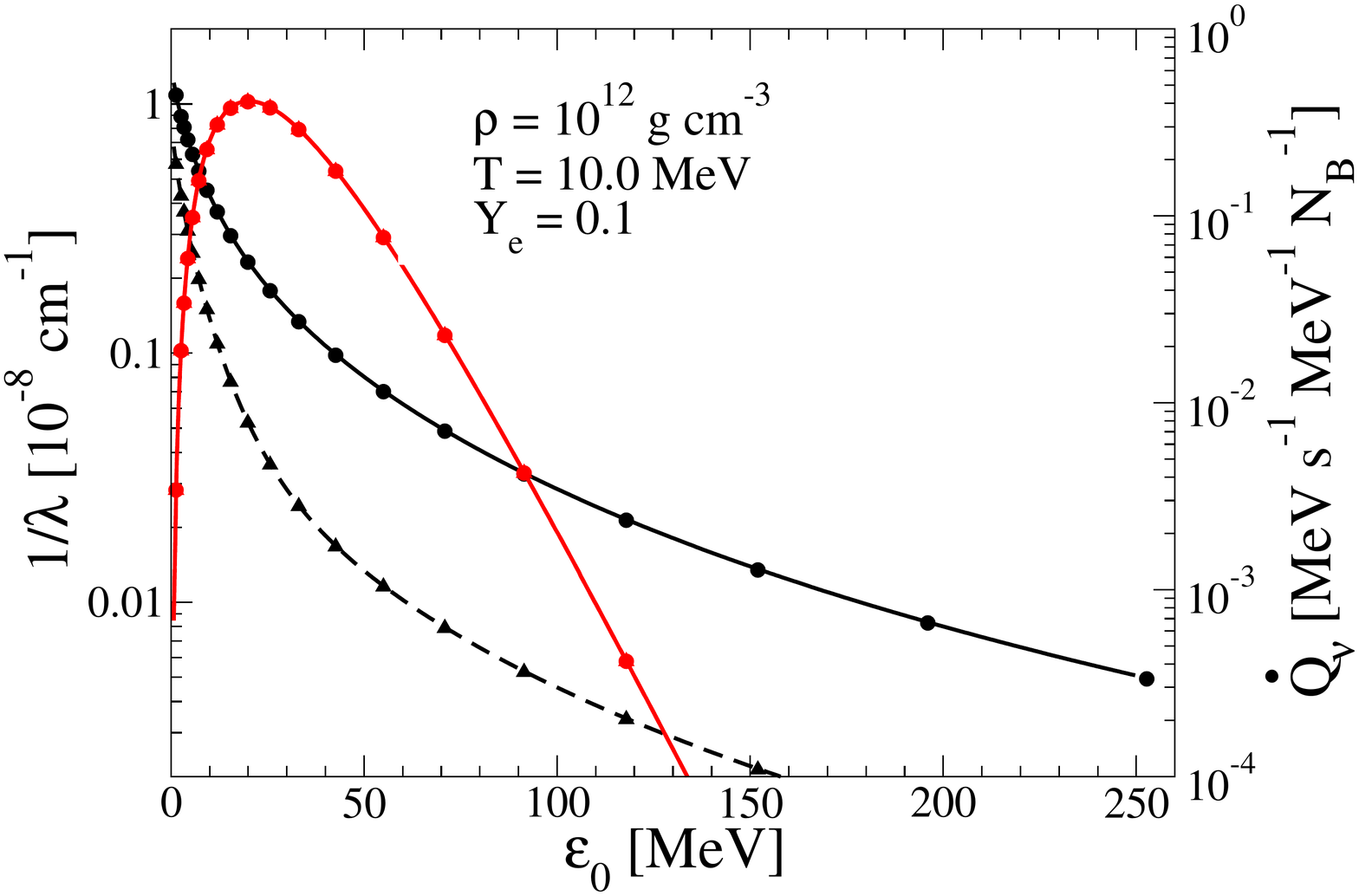}{0.5\textwidth}{(b)}
               }
\gridline{
              \fig{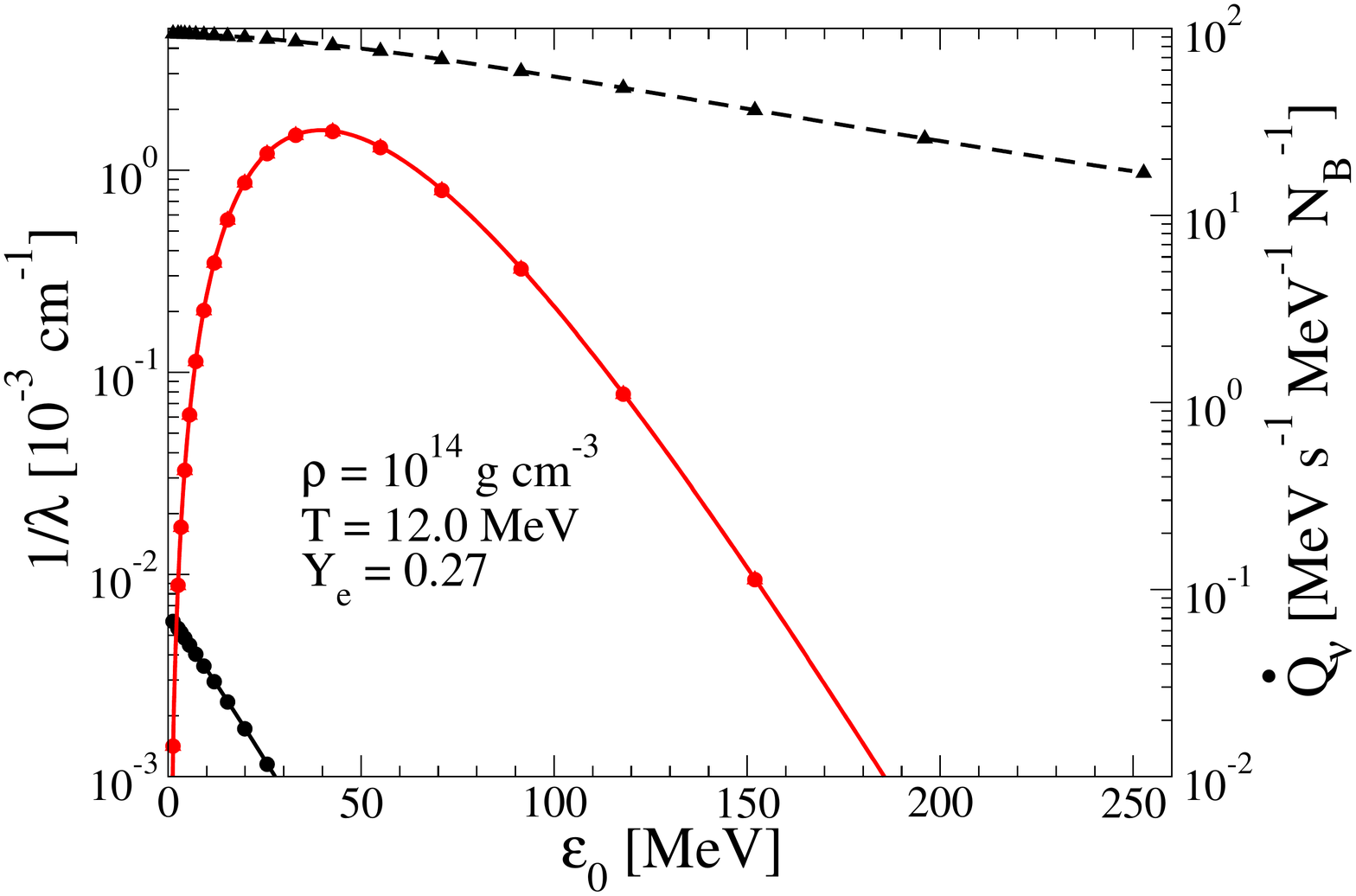}{0.5\textwidth}{(c)}
               }
\caption{\label{fig:Brem_eos}
Inverse mean free paths for \nue\ (black solid lines) and \nuebar\ (black dashed lines) computed from Equation~(\ref{eq:d23}) for nucleon--nucleon-bremsstrahlung kernels on an energy grid of 600 evenly-spaced zones from 0 to 300 MeV, for the thermodynamic conditions listed. The \nue (\nuebar) inverse mean free paths were computed with the \nue (\nuebar) occupation distribution set to zero and the \nuebar (\nue) distribution equilibrated with the matter.
Inverse mean free paths were computed with the \Chimera\ energy grid of 20 energy zones for \nue\ (black circles) and \nuebar\ (black triangles).
Energy production rates per baryon from \nue\ (solid red lines) and \nuebar\ (dashed red lines) were computed from Equation~(\ref{eq:d24}) for the  nucleon--nucleon-bremsstrahlung kernels on the above-described 600 zone energy grid, with both the \nue\ and \nuebar\ occupation distributions set to zero.
Spectral energy production from \nue\ (red circles) and \nuebar\ (red triangles) were computed with the 20 energy zone \Chimera\ energy grid. 
}
\end{figure}

\subsection{Neutrino--Antineutrino Pair Annihilation and Production from Nucleon--Nucleon Bremsstrahlung}
\label{brem}

Neutrino--antineutrino pair annihilation and production from nucleon--nucleon bremsstrahlung has been computed at various levels of approximation  \citep{FrMa79,RaSe95,HaRa98}.
The last paper examined the process and provided an interpolation scheme that interpolates the nucleon--nucleon bremsstrahlung kernel between the nucleon non-degenerate limit treated by \citet{RaSe95} and the degenerate limit treated (for the case of axion emission) by \citet{IsYo90}.
We use Equation~(35) in Equation~(23) of \citet{HaRa98}, which naturally breaks up into  $\Phi_{0,{\rm brem}}^{\rm a/p}(\epscm,\epscm')$ and  $\Phi_{0, {\rm brem}}^{\rm a/p}(\epscm,\epscm')$ terms.
We also use Equation~(\ref{eq:d21}) for the neutrino--antineutrino pair annihilation correction.
In Figure~\ref{fig:Brem_eos} we plot the inverse mean free paths, given by Equation~(\ref{eq:d23}), and the spectral energy production rates per baryon, given by Equation~(\ref{eq:d24}), with the nucleon--nucleon bremsstrahlung kernels substituted for the electron--positron pair annihilation kernels for select thermodynamic states.

Mathematically, inelastic neutrino scattering on nucleons is the bremsstrahlung process with a final-state neutrino crossed into the initial state and could be included with the same kernel.
Moreover, this process should reduce to elastic neutrino--nucleon scattering when the nucleon--nucleon interaction tends to zero.
However, the formulation of \citet{HaRa98} neglects nucleon recoil, and we have used instead of the \citet{HaRa98} formalism for inelastic neutrino--nucleon scattering the formalism of \citet{RePrLa98}, described earlier, for elastic neutrino scattering on nucleons, which includes recoil effects as well as nucleon degeneracy and relativistic effects. 
At very high densities, where inelastic neutrino scattering on nucleons becomes important, we have therefore neglected this mode of energy exchange between the neutrinos and the matter, with the hope of recovering a significant part of it through the inclusion of nucleon recoil.

\subsection{Opacity Interpolation}
\label{opac_interp}

As noted earlier in this section, the \chimera\ method for building and interpolating neutrino interactions in \lrtye\ uses a local cube of points in \rtye\ that bracket the \rtye-value of each spatial cell and that lie on a regular mesh in \lrtye-space \citep[cf.][]{MeBr93b}.
The derivatives computed within the local \rtye-cube are fully consistent with the interpolated value.
This method has a disadvantage with the on-the-fly re-computation of opacities to fill the local \rtye-cubes when the values no longer bracket the cell values. Computation of some of the opacities is expensive, notably the opacities with the sub-grid integration described in Section~\ref{scatnu_np}, and could consume up to $\sim$25--40\% of the simulation time.
This cost was compounded by computational load imbalance from the irregular number of local \rtye-cubes updates required for each transport time step, which typically numbered between zero and a few new \rtye-cubes. 

Starting with the C-series models, we implemented the scheme described below, which greatly reduced the on-the-fly computations and is better suited to future use of pre-computed tables.
Specifically, we chose to implement a sparse `local pool' of \rtye-tuples. 
Using such a `local pool', rather than retaining the existing individual \rtye-cubes for each cell with the addition of a `reuse' algorithm, has lower memory requirements, as many cells use the same logical cube of \rtye-points.
The memory savings is potentially larger when computing transport on multiple adjacent `rays' that are similar in \rtye-space. 
The reduced memory usage was particularly helpful when using machines with smaller memory footprints ($\sim$1--2~GB of memory per core).

\subsubsection{Energy Interpolation}

The energy grid used in the \chimera\ neutrino transport solver is not the comoving observer's energy, \epscm, but the energy of a comoving observer outside the gravitational well of the supernova, $a \epscm \equiv \Ecm$ (Equation~\ref{eq:a8a}), which depends on radius and thus complicates the sharing of opacity \rtye-points.
To account for this difference, we must interpolate the energy grid of the opacity table into the specific points needed for the \aeps\ grid in each group, from a reference grid, without changing the total interaction rate.

The energy grid, both $\epscmi{ k}$ and $\Ecmi{ k}$, are logarithmically spaced at the group centers
\begin{equation}
\log \zeta = \log \left(\frac{\epscmi{ k+1}}{\epscmi{k}} \right) = \log \left(\frac{\Ecmi{ k+1}}{\Ecmi{ k}}\right)
\end{equation}
(and group edges, with $k$ replaced by $k+1/2$ above).
Likewise, the difference between the two energy grids for a non-unity lapse can be written as a logarithmic shift
\begin{equation}
\log \left(\frac{1}{a} \right) = \log \left(\frac{\epscm}{\Ecm} \right),
\end{equation}
for any value of the energy.

\subsubsection{A Simple Integral Scheme}

Informed by the requirements of the \chimera\ energy grid, we can construct a more general interpolation scheme for any function, $f$, tabulated on a logarithmically spaced grid, $\epsp{\imath}$, to another logarithmically spaced grid, $\eps{i}$, that differs by a multiplicative constant that changes with time, $\epsp{\imath} = \eps{i}/\alpha$, where both grids have the same logarithmic spacing
\begin{equation}
\log \zeta = \log \left(\frac{\eps{i+1}}{\eps{i}} \right) = \log \left(\frac{\epspo{\imath}}{\epsp{\imath} } \right).
\end{equation}
The integral scheme comes from the need to evaluate integrals in the form $\int f(\eps{}) d\eps{}$ numerically by summation of the terms  $\Sigma  f_{i} \deps{i}$, where $ \deps{i} = \eps{i+1} - \eps{i}$ and $\deps{i} f_{i}$ is the integration of $f$ over that same interval, $\deps{i}$.
When the shift, $\log (1/\alpha)$, is smaller than the grid spacing, $\log \zeta$, we can define the overlap between the \deps{i} and \depsp{\imath} zones as
\begin{equation}
\beta = \frac{\eps{i+1} - \epsp{\imath}}{\delta \eps{i}} = \frac {\eps{i}}{\delta \eps{i}} \left( \zeta - \frac{1}{\alpha} \right) = \frac{1}{\zeta - 1}  \left( \zeta - \frac{1}{\alpha} \right),
\end{equation}
which is the same for all zones, $i$. 
The integral over $\bar\imath$ is then
\begin{equation}
f_{\bar\imath} \, \depsp{\imath} = \beta \, f_i \, \delta\eps{i} + (1 - \beta) f_{i+1}\, \delta\eps{i+1} . \label{eqn:interp}
\end{equation}
If the left-most zone has a coordinate $\eps{1} = 0$, we must modify the above equation to include the whole of the first zone value
\begin{equation}
f_{\bar 1} \, \depsp{1} =  f_1 \, \delta\eps{1} + (1 - \beta) f_2 \, \delta\eps{2} \label{eqn:interp0}.
\end{equation}

For grid shifts that are larger than the single-zone spacing, $\log (1/\alpha) > \log \zeta$, a small modification is required.
First we define a `shift index'
\begin{equation}
j = \tt{Int} \left[ \frac{\log \left( {1}/{\alpha} \right) }   { \log \zeta}  \right].
\end{equation}
The $\beta$ value must now reflect the larger shift, so
\begin{equation}
\beta = \frac{\eps{i+j+1} - \epsp{\imath}}{\delta \eps{i+j}} = \frac {\eps{i+j}}{\delta \eps{i+j}} \left( \zeta - \frac{1}{\alpha \zeta^j} \right) = \frac{1}{\zeta - 1}  \left( \zeta - \frac{1}{\alpha \zeta^j} \right), 
\end{equation}
 where $\zeta^j$ refers to $\zeta$ to the $j$-th power, that can be used in an extended version of Equation~(\ref{eqn:interp}):
\begin{equation}
f_{\bar\imath} \, \depsp{\imath} = \beta \, \delta\eps{i+j} \, f_{i+j} + (1 - \beta) \delta\eps{i+j+1} \, f_{i+j+1}.
\end{equation}
The additional zones shifted for $\eps{1} = 0$ grids must be added to Equation~(\ref{eqn:interp0}):
\begin{equation}
f_{\bar 1} \depsp{1} =\sum_{k=0}^{j} f_{1+k} \, \delta\eps{1+k}   + (1 - \beta) f_{2+j} \, \delta\eps{2+j}.
\end{equation}
These last two equations reduce to the earlier forms when $j=0$.

\subsubsection{Interpolation of Two-Variable Functions}

If we have a function, $f(\eps{i},\eps{\imath'})$ which needs to be remapped to a shifted grid, $f(\epsp{\imath},\epsp{\imath'})$, we can extend the single-variable method given above.
We assume that both the $\eps{i}$ and $\eps{\imath'}$ grids are the same (i.e., $\eps{i}=\eps{\imath'}$ if $i = \imath'$) and therefore have the same shift and spacing parameters, $(\zeta, \beta, 1/\alpha,j)$.
We start with a partially shifted version of Equation~(\ref{eqn:interp}) where $i$ is shifted, but $\imath'$ is not, then apply the shift to $\imath'$ to get the shifted double-grid formula:
\begin{eqnarray}
f_{\bar\imath,\bar\imath'} \, \depsp{\imath} \, \depsp{\imath'}&=& \beta \, \delta\eps{i} \, \depsp{\imath'} \, f_{i,\bar\imath'} + (1 - \beta) \delta\eps{i+1} \, \depsp{\imath'} \, f_{i+1,\bar\imath'} \nonumber \\
&=&\beta \, \deps{\imath'} \left[ \beta \, \delta\eps{i}  \, f_{i,\imath'} + (1 - \beta) \delta\eps{i+1}  \, f_{i+1,\imath'} \right]  \nonumber\\
&& \,+ (1-\beta)  \deps{\imath'+1} \left[ \beta \, \delta\eps{i}  \, f_{i,\imath'+1} + (1 - \beta) \delta\eps{i+1}  \, f_{i+1,\imath'+1} \right]. \label{eqn:double}
\end{eqnarray}
For grids shifted such that the shift index $j > 0$, we can generalize Equation~(\ref{eqn:double}) to
\begin{eqnarray}
f_{\bar\imath,\bar\imath'} \, \depsp{\imath} \, \depsp{\imath'}&=& \beta \, \deps{\imath'+j} \left[ \beta \, \delta\eps{i+j}  \, f_{i+j,\imath'+j} + (1 - \beta) \delta\eps{i+j+1}  \, f_{i+j+1,\imath'+j} \right]  \nonumber\\
&& \,+ (1-\beta)  \deps{\imath'+j+1} \left[ \beta \, \delta\eps{i+j}  \, f_{i+j,\imath'+j+1} + (1 - \beta) \delta\eps{i+j+1}  \, f_{i+j+1,\imath'+j+1} \right]
\end{eqnarray}
for cases when $i > 1, \imath' > 1$.
When either $i =1$ or $\imath' = 1$, we need to generalize the equations above:
\begin{eqnarray}
f_{\bar 1,\bar\imath'} \, \depsp{1} \, \depsp{\imath'}&=&\beta \, \deps{\imath'+j} \left[\sum_{k=0}^{j} \delta\eps{1+k}  \, f_{1+k,\imath'+j} + (1 - \beta) \delta\eps{2+j}  \, f_{2+j,\imath'+j} \right] \nonumber \\
&& \,+ (1-\beta)  \deps{\imath'+j+1} \left[ \sum_{k=0}^{j} \delta\eps{1+k}  \, f_{1+k,\imath'+j+1} + (1 - \beta) \delta\eps{2+j}  \, f_{2+j,\imath'+j+1} \right] , \\
f_{\bar\imath,\bar 1'} \, \depsp{\imath} \, \depsp{1'}&=&\sum_{k'=0}^{j}\deps{1'+k'} \left[ \beta \, \delta\eps{i+j}  \, f_{i+j,1'+k'} + (1 - \beta) \delta\eps{i+j+1}  \, f_{i+j+1,1'+k'} \right] \nonumber \\
&& \,+ (1-\beta)  \deps{2'+j} \left[ \beta \, \delta\eps{i+j}  \, f_{i+j,2'+j} + (1 - \beta) \delta\eps{i+j+1}  \, f_{i+j+1,2'+j} \right] , \\
f_{\bar 1,\bar 1'} \, \depsp{1} \, \depsp{1'}&=& \sum_{k'=0}^{j}\deps{1'+k'} \left[  \sum_{k=0}^{j}\delta\eps{1+k}  \, f_{1+k,1'+k'} + (1 - \beta) \delta\eps{2+j}  \, f_{2+j,1'+k'} \right] \nonumber \\
&& \,+ (1-\beta)  \deps{2'+j} \left[\sum_{k=0}^{j} \delta\eps{1+k}  \, f_{1+k,2'+j} + (1 - \beta) \delta\eps{2+j}  \, f_{2+j,2'+j} \right].
\end{eqnarray}

\subsubsection{Interpolation of \chimera\ Opacities}

For the interpolation of opacities in \Chimera, we choose to store the raw opacities on a grid, \eps{}, that numerically matches the specified $E_0$ grid, which makes $\alpha = a$.
Note that this is not a frame transform, just a convenient choice of variables, as all opacities are evaluated in the $\epsilon_0$ frame of the moving fluid.
For single-energy opacities (absorption and emission), $f$ is the inverse mean free path multiplied by  $\epsilon_0^2$.
For two-energy opacities (scattering and pair processes), $f$ is the Legendre coefficients of the kernels multiplied by $\epsilon_0^2 \epsilon_0'^2$.
The $\epsilon_0'^2$ term arises from within the collisions integrals, while the $\epsilon_0^2$ arises from integrating the collision integral with the operator $\int \epsilon_0^2 d\epsilon_0$ to conserve the total integral interaction rate.

\subsubsection{Local Pool Algorithm}
\label{sec:oppool}

The local pool is implemented for each opacity by first identifying the eight \rtye-points needed for each zone using that opacity on each MPI rank by generating an integer hash value that uniquely maps to the potential grid of \rtye-points, removing duplicates, and sorting them.
The sorted list is then compared to the points in the existing pool to generate a list of points that need to be added to the pool.
The list of additions is added to the pool by calling the opacity generation routines for each missing point.
The integer hash values for the eight \rtye-points for each zone are then regenerated and cross-linked to their pool index numbers.
During interpolation, the eight points indicated by the pool index numbers are used to interpolate the kernels for specific cells.

Implementation of this sharing algorithm reduced the opacity generation and management costs to a few percent of the total run time.
It had the extra benefit of reducing the opacity interpolation costs by 5--10\%, as the \rtye-points needed for interpolating the opacity for one cell frequently fully, or partially, overlap with those needed for the next cell, allowing data stored in processor cache to be reused. This was not possible under the old scheme, as every zone had its own eight points even if they were duplicates of a neighboring cell's points.
Implementation was checked with tests that were computed using a 1D reference simulation, with comparisons of critical transport and hydrodynamics variables made at core bounce and at several other times.

\section{Comparison with other Codes}
\label{comparisons}

Supernova codes are complex entities, involving the numerical solution of hydrodynamics supplemented by one or more equations of state, neutrino transport with multiple sources of absorption and scattering opacities, nuclear transmutations, and relativistic gravity, with fluid densities ranging over more than ten orders of magnitude. 
Different numerical techniques have their individual strengths and weaknesses, reinforcing the importance of code validation for a particular choice of techniques and their implementation. 
In this section we make a detailed comparison of results of a spherically symmetric simulation performed by \chimera\ using the code versions utilized in our B-series and C-series simulations with the results of two other codes that have been compared in \citet{LiRaJa05}:  \agileboltztran\ and \prometheusvertex.
A recent comparison of CCSN codes was presented in \citet{OCBoBu18}, using updated physics, particularly the EoS, and a comparison of \Chimera\ with those results will be reported in the future using a later version  of \chimera\ than described here.

\subsection{Description of Comparison Simulations and Codes}

\agileboltztran\ was developed by an Oak Ridge--Basel collaboration and is a code that solves the general relativistic hydrodynamics equations in a time-implicit fashion, with a dynamical adaptive grid, coupled to a solver for the general relativistic Boltzmann equations for the neutrino distribution functions based on a discrete-ordinates ($S_{N}$) in angle and finite-difference in space and neutrino energy discretization, all for spherically symmetric spacetimes \citep{MeBr93b, MeMe99, LiRoTh02,LiMeMe04}. 

\prometheusvertex\ is a code for multidimensional neutrino radiation hydrodynamics, here used in the context of studies assuming spherical symmetry, developed by the Garching group.
It consists of a code to solve the hydrodynamics equations, based on a finite-volume discretization \prometheus\ \citep{FrMuAr89bb},
coupled in an operator-split fashion to a code that solves the equations for two-moment neutrino transport, closed using a variable Eddington factor. This factor is derived from a formal solution of a model Boltzmann equation for the counterpart spherically-averaged matter configuration corresponding to the actual multidimensional matter configuration for a given time step \citep{RaJa02, BuRaJa06}.

From \chimera\ we include runs from two versions of the code.
The first, \chimera-B, uses the same code as the B-series models reported in \citet{BrMeHi13, BrLeHi16} but with microphysics (EoS and opacities) and progenitor \citep{WoWe95} to match the \citet{LiRaJa05} comparison models.
The second, dubbed \chimera-C, is the code used for the C-series models \citep{LeBrHi15,LeKeCa18} and differs from \chimera-B in some code refinements, particularly the modification of the transport scheme in the vicinity of large velocity discontinuities (i.e., shocks) to properly account for the large changes in the comoving reference frames described in Section~\ref{trans_shock}.

The \agileboltztran\ and \prometheusvertex\  comparisons have been documented in \citet{LiRaJa05}, as noted above, and in \citet{MuJaDi10}.
The progenitor and choice of physics we will use for our comparisons is referred to as model G15 in \citet{LiRaJa05}, and we also will refer to it as G15.
Of the two models used for the comparisons in that work, it is the model that implements the more extensive choice of physics, and thus more closely approximates the set of physics included in current state-of-the-art \ccsn\ simulations.
The progenitor used to initiate the G15 simulations is the 15-\msun\ progenitor of \citet{WoWe95}, widely used in the literature, and representative of the star in the middle to upper mass range likely to end its life as a supernova. 
From the highest densities to a density of \den{6}{7}\ the EoS used in all codes is the compressible liquid drop model of \citet{LaSw91} with incompressibility modulus $K=180$~MeV.
It assumes a composition in NSE consisting of neutrons, protons, $\alpha$-particles, a representative heavy nucleus, electrons, positrons, and photons.
At densities below \den{6}{7}\, each code switches to an EoS consisting of a composition of electrons, positrons, photons, nucleons, and nuclei, with the latter two treated as an ideal gas.
The detailed treatment of the EoS in the lower-density regime differs among the codes; therefore, our comparisons will be limited to important phenomena occurring above \den{6}{7}.
\agileboltztran\ is fully general relativistic, while both \prometheusvertex\ and \chimera\ approximate the gravitational potential by including corrections terms due to pressure and energy of the stellar medium and neutrinos, as described in \citet{MaDiJa06}.
We plot the results of two \prometheusvertex\ simulations, which we label as \vertex-1 and \vertex-2. 
The \vertex-1 simulation was performed with gravitational potential `R' of \citet{MaDiJa06} and is described in \citet{LiRaJa05}, while the \vertex-2 simulation was performed with gravitational potential `A' and is described in \citet{MuJaDi10}.
Both \prometheusvertex\ and \chimera\ include gravitational redshifting and time dilation in the transport solution but ignore the difference between coordinate and proper radial distances, which has little effect on the transport \citep{BrDeMe01}.
The hydrodynamics in both codes in Newtonian.

\begin{deluxetable*}{ll}
\tabletypesize{\scriptsize}
\tablecaption{Summary of Neutrino Opacities in Model G15\label{tab:G15opacities}}
\tablecolumns{3}
\tablewidth{0pt}
\tablehead{
\colhead{Interaction} & \colhead{References}
}
\startdata
$\nu + e^{\pm} \leftrightharpoons \nu + e^{\pm}$ & \citet{Brue85, MeBr93c} \\
$\nu + A \leftrightharpoons \nu + A$ & \citet{BrMe97, Horo97} \\
$\nu + n, p \leftrightharpoons \nu + n, p$ & \citet{Brue85, MeBr93c} \\
$\nue + n \leftrightharpoons e^{-} + p$ & \citet{Brue85, MeBr93c} \\
$\nuebar + p \leftrightharpoons e^{+} + n$ & \citet{Brue85, MeBr93c} \\
$\nue + A'  \leftrightharpoons e^{-} + A$ & \citet{Brue85, MeBr93c} \\
$\nu + \nubar  \leftrightharpoons e^{-} + e^{+}$ & \citet{Brue85, MeBr93c} \\
$N + N  \leftrightharpoons N + N + \nu + \nubar$ & \citet{HaRa98} \\
ion--ion correlations & \citet{Itoh75, Horo97, BrMe97} \\
\enddata
\end{deluxetable*}

Neutrinos of all flavors are included in model G15, with the neutrino--matter interactions shown in Table~\ref{tab:G15opacities}.
The \agileboltztran\ simulation was performed using 103 adaptive spatial zones from the center to the edge of the included progenitor, which was about 7,000 km, and a constant-pressure boundary condition was applied at the outer surface.
The neutrino energy grid was resolved with 20 geometrically spaced bins, the first centered at 3~MeV and the last at 300~MeV. The propagation directions were discretized into six angles suitable for Gaussian quadrature.
Neutrino flavors were divided into four independently transported species, \nue, \nuebar, \numt\ (\numu\ and \nutau), and \numtbar\ (\numubar\ and  \nutaubar).
The \prometheusvertex\ simulation was performed on separate radial fluid and transport grids.
The fluid grid consisted of 400 zones that moved with the fluid during collapse, and were rezoned shortly after core bounce such that inside a radius of 400 km the fluid grid  coincides with the transport grid.
The latter consisted of an Eulerian radial grid of 235 radial zones spaced logarithmically between 0 and 10,000 km.
Forty additional radial zones were added to both grids after 200~ms post-bounce to resolve the steepening density gradient at the surface of the nascent neutron star. The neutrino spectrum was discretized with 19 energy zones between 0 and 380 MeV.
Neutrino flavors were divided into three independently transported species, \nue, \nuebar, and \nux\ (\numt\ and \numtbar).
The \chimera\ simulations were performed with 512 adaptive radial zones. The zones moved with the fluid during collapse and thereafter adjusted to maintain an approximately constant $\Delta r/r$ while maintaining a maximum value of $\Delta \rho/\rho$ to maintain good resolution in the vicinity of the steepening density gradient at the surface of the nascent neutron star as the simulation progresses past bounce.
The neutrino spectrum was resolved with 20 energy zones between 0 and 279 MeV with mid-energies of 2.57 and 250 MeV in the first and last zone for the four species: \nue, \nuebar, \numt, and \numtbar.
Because the \chimera\ neutrino energy grid is defined at infinity and the actual grid is radially dependent (the local grid energies are the energies at infinity divided by the local value of the lapse function), the grid expands toward the core center as the latter contracts to higher densities. At bounce the mid-zone energies at the core center range from 2.92 to 285 MeV, and at 100 ms post-bounce the mid-zone energies at the core center range from 3.20 to 313 MeV.

\begin{figure*}
\gridline{\fig{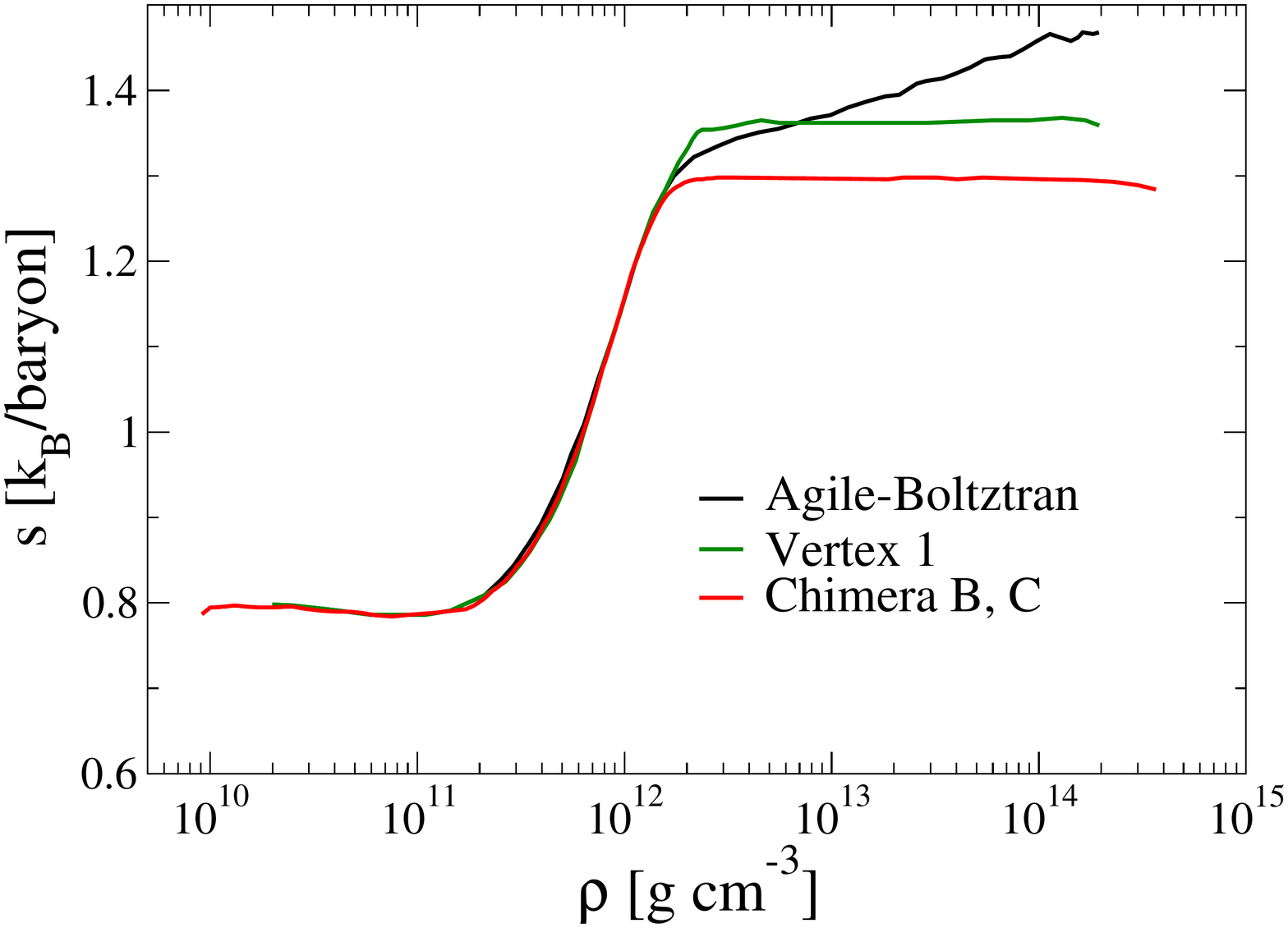}{0.5\textwidth}{(a)}
              \fig{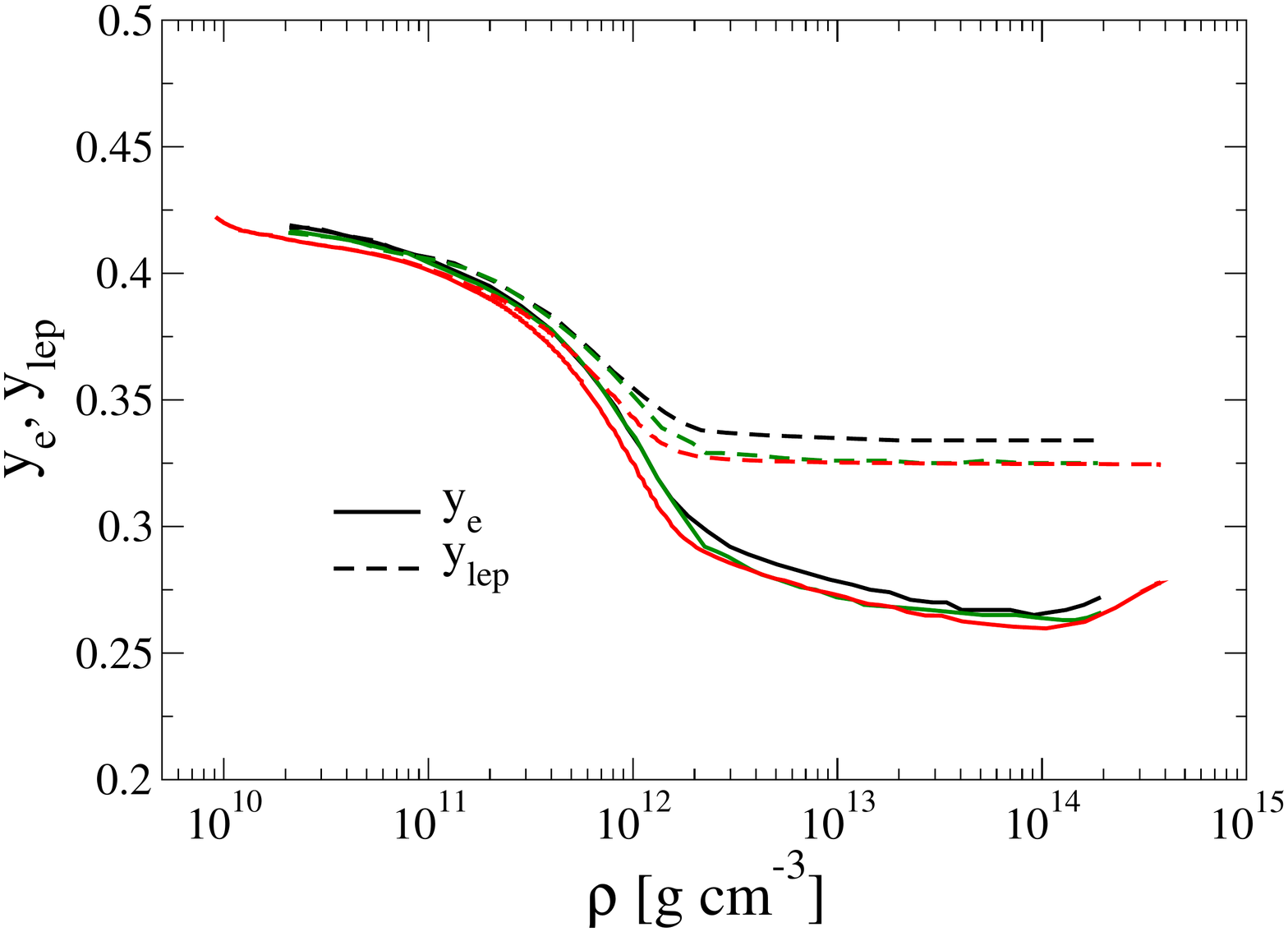}{0.5\textwidth}{(b)}
               }
\caption{\label{fig:infall}
Evolution of entropy (a), and electron (solid) and lepton (dashed) fractions (b) versus density in the central zone during infall, for \chimera\ (red),  \agileboltztran\ (black), and \vertex-1 (green) simulations. 
}
\end{figure*}

\begin{figure*}
\gridline{\fig{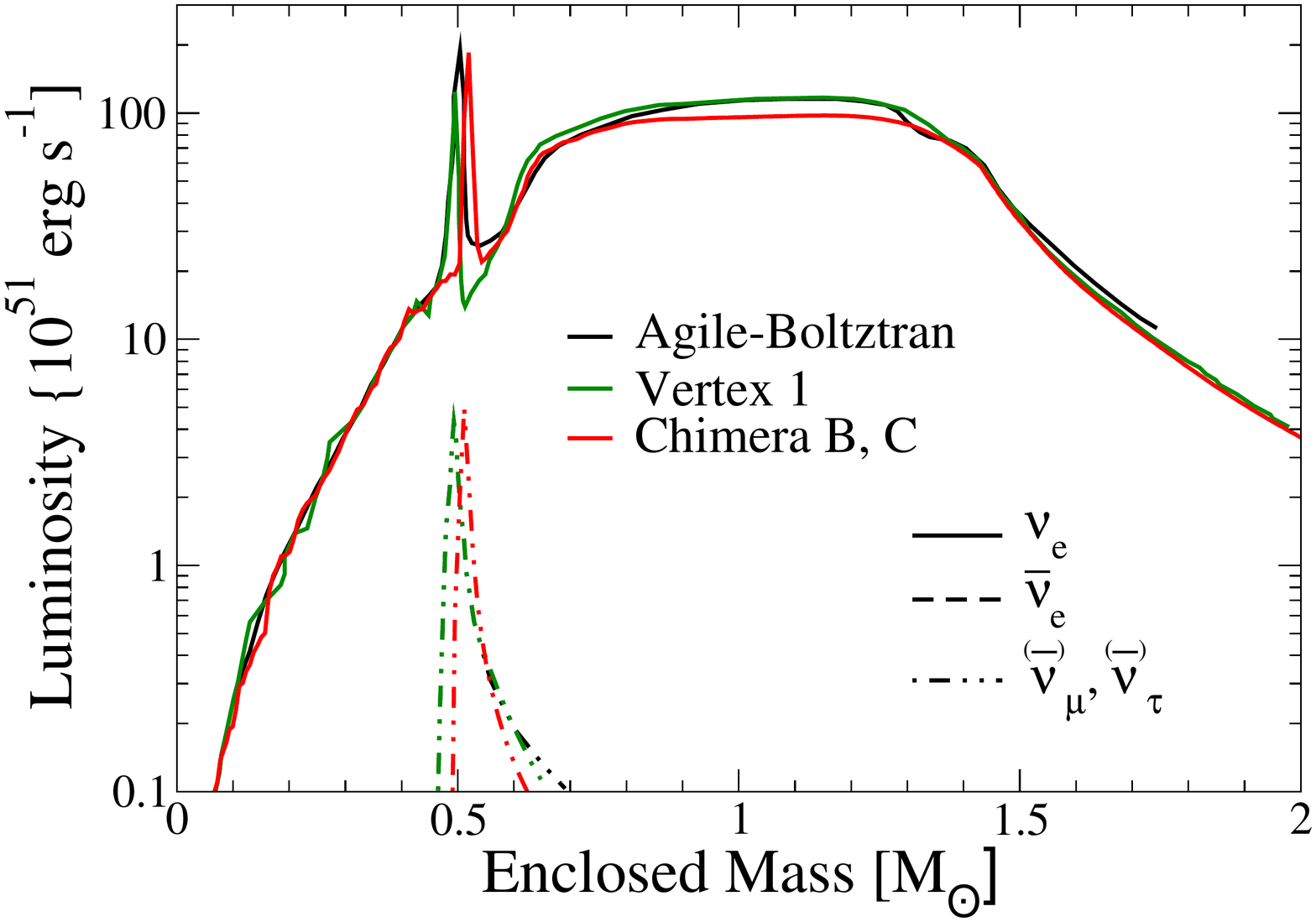}{0.5\textwidth}{(a)}
              \fig{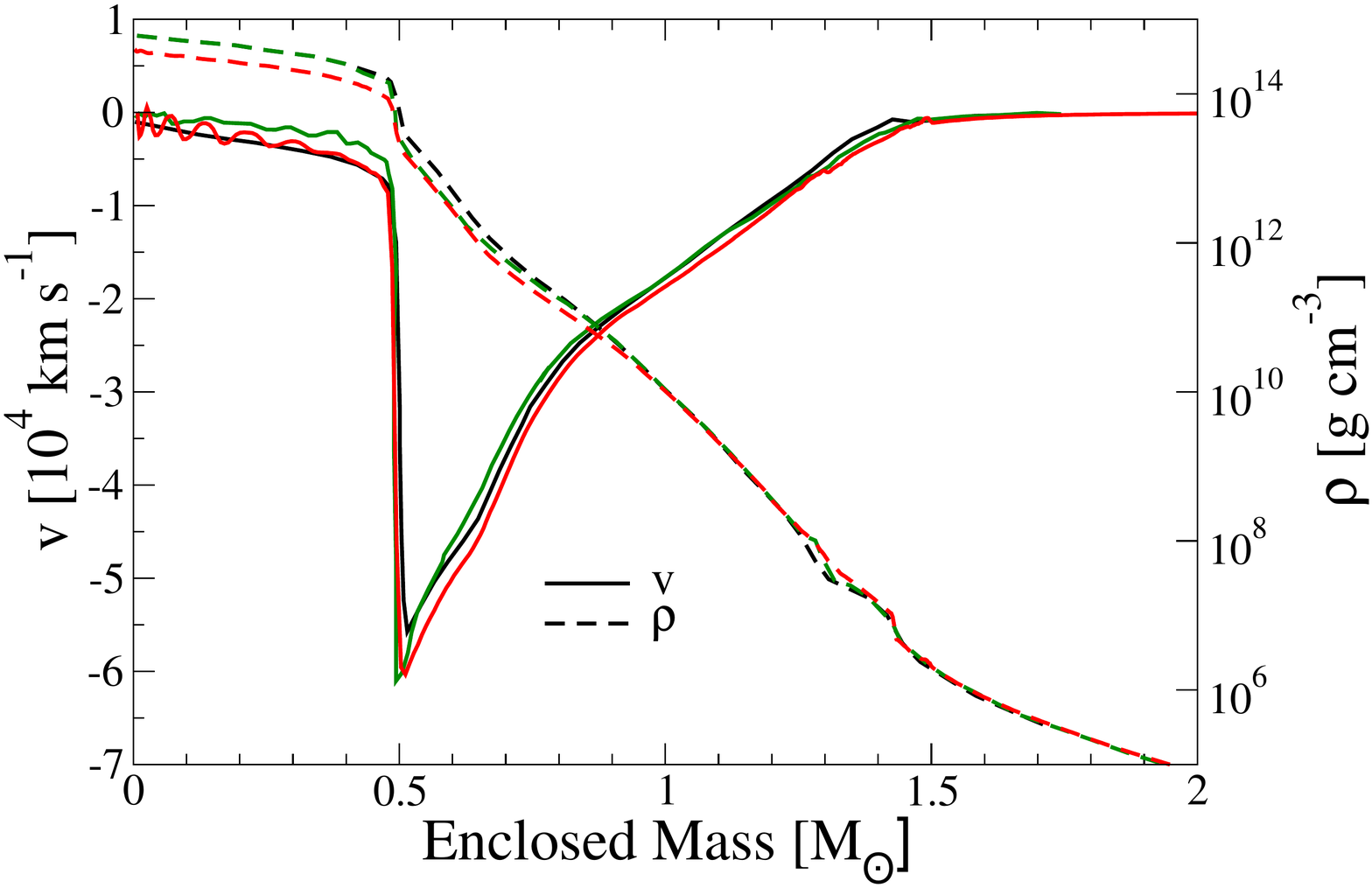}{0.5\textwidth}{(b)}
               }
\gridline{
              \fig{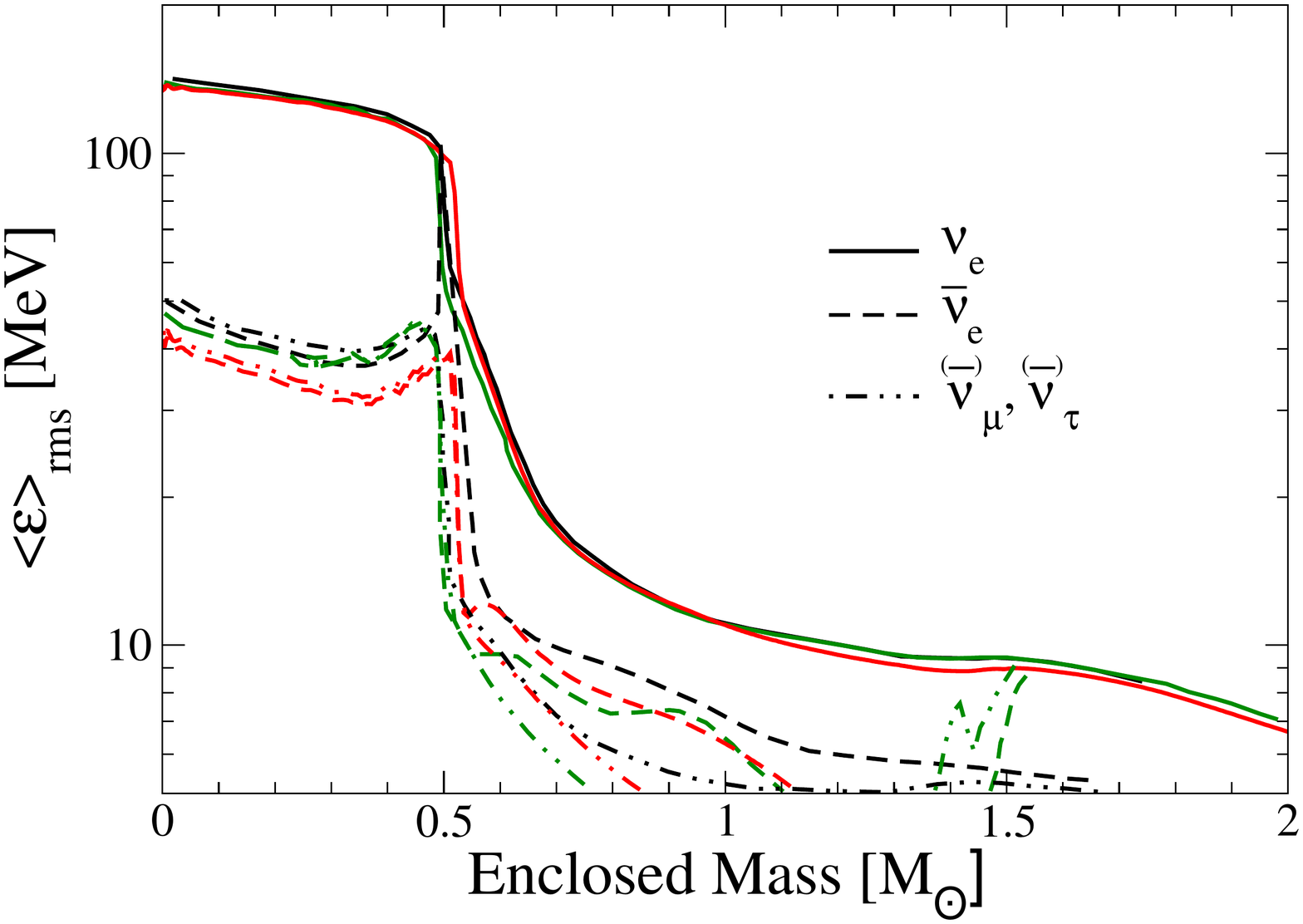}{0.5\textwidth}{(c)}
              \fig{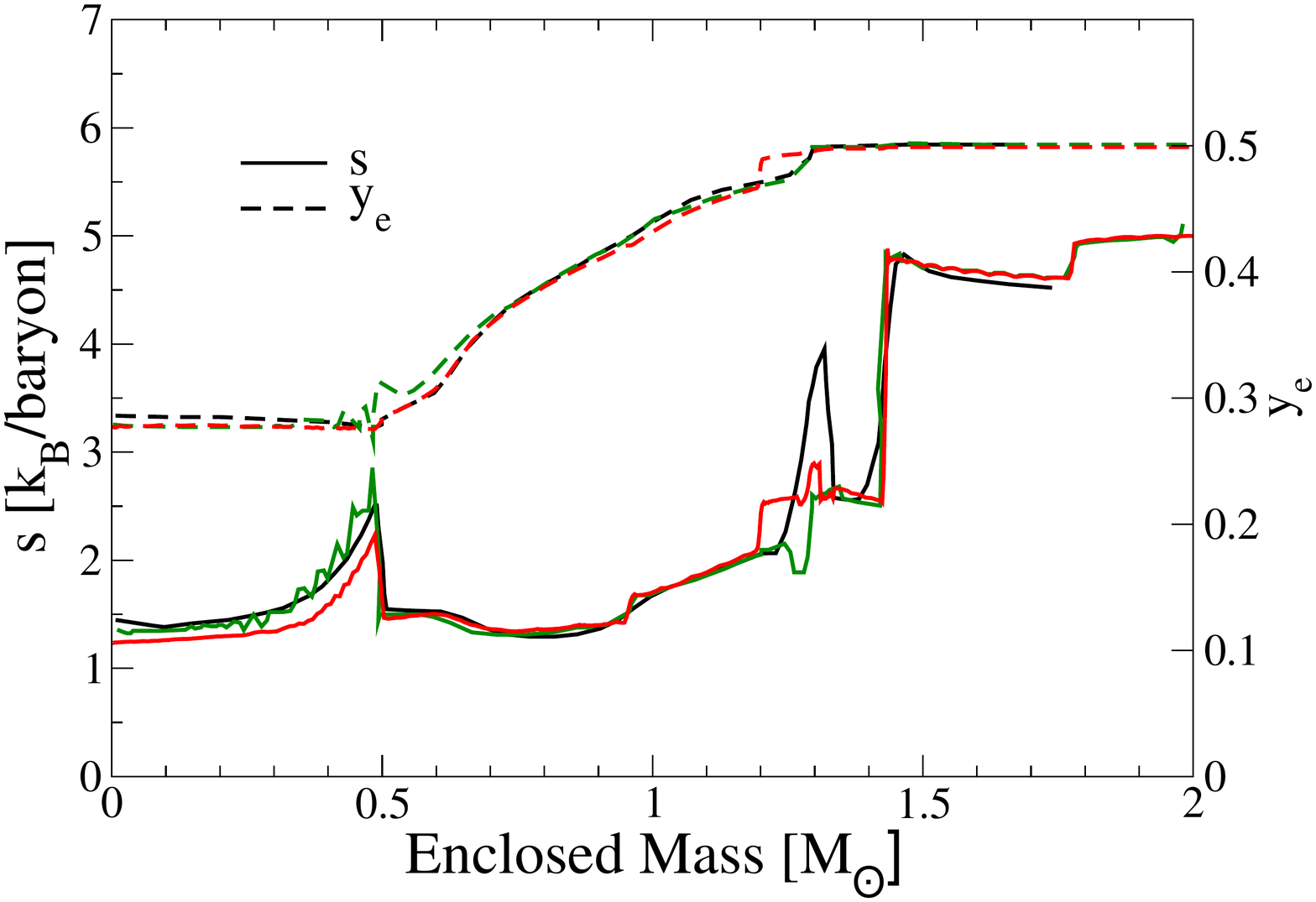}{0.5\textwidth}{(d)}
               }
\caption{\label{fig:bounce}
Snapshots at bounce for (a) luminosity, (b) density and velocity, (c) rms-energies, and (d) entropy and \Ye\ of the G15 profile, for the \agileboltztran\ (black), \vertex-1 (green), and \Chimera\  (red) simulations.
}
\end{figure*}

\subsection{Infall}
\label{infall}

The evolution of the entropy, electron fraction, and lepton fraction during infall is shown in Figure~\ref{fig:infall}. 
Prior to the $\sim$10~ms after shock formation, the \chimera-B and \chimera-C results are essentially identical, and are shown simply as ``\chimera'' in graphs until such times as the differences between them become significant.
Before trapping, the entropy evolution is almost identical for all three simulations. Trapping occurs slightly later for \vertex-1 compared with \chimera. 
Trapping seems to occur for \agileboltztran\ near that of \chimera, but the entropy continues to slowly increase thereafter, likely because of zones moving away from the center to resolve the forming shock.
This causes the central zone to encompassing more and more mass of higher entropy, causing the zone-average entropy to rise.

\begin{figure*}
\gridline{\fig{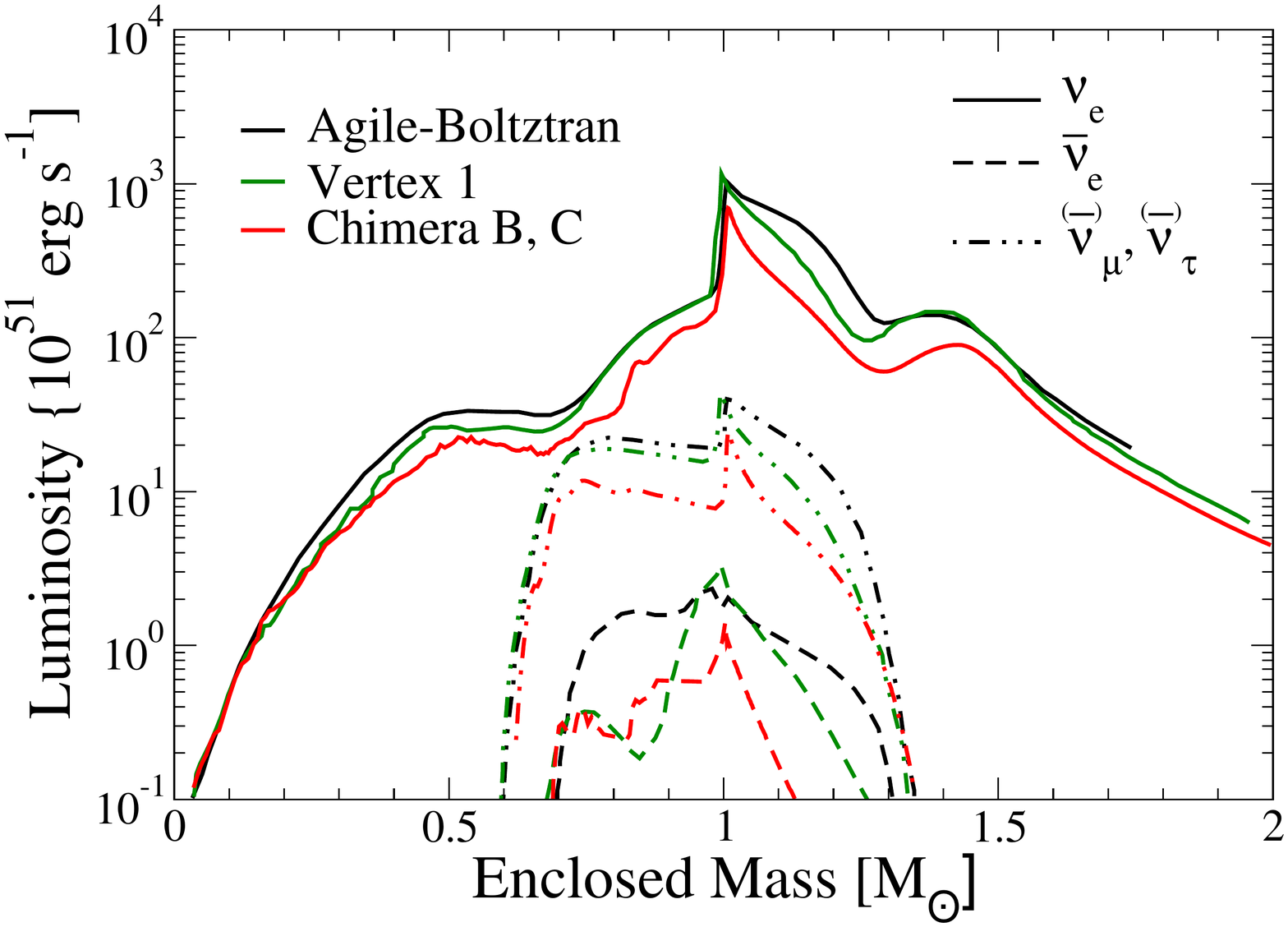}{0.5\textwidth}{(a)}
              \fig{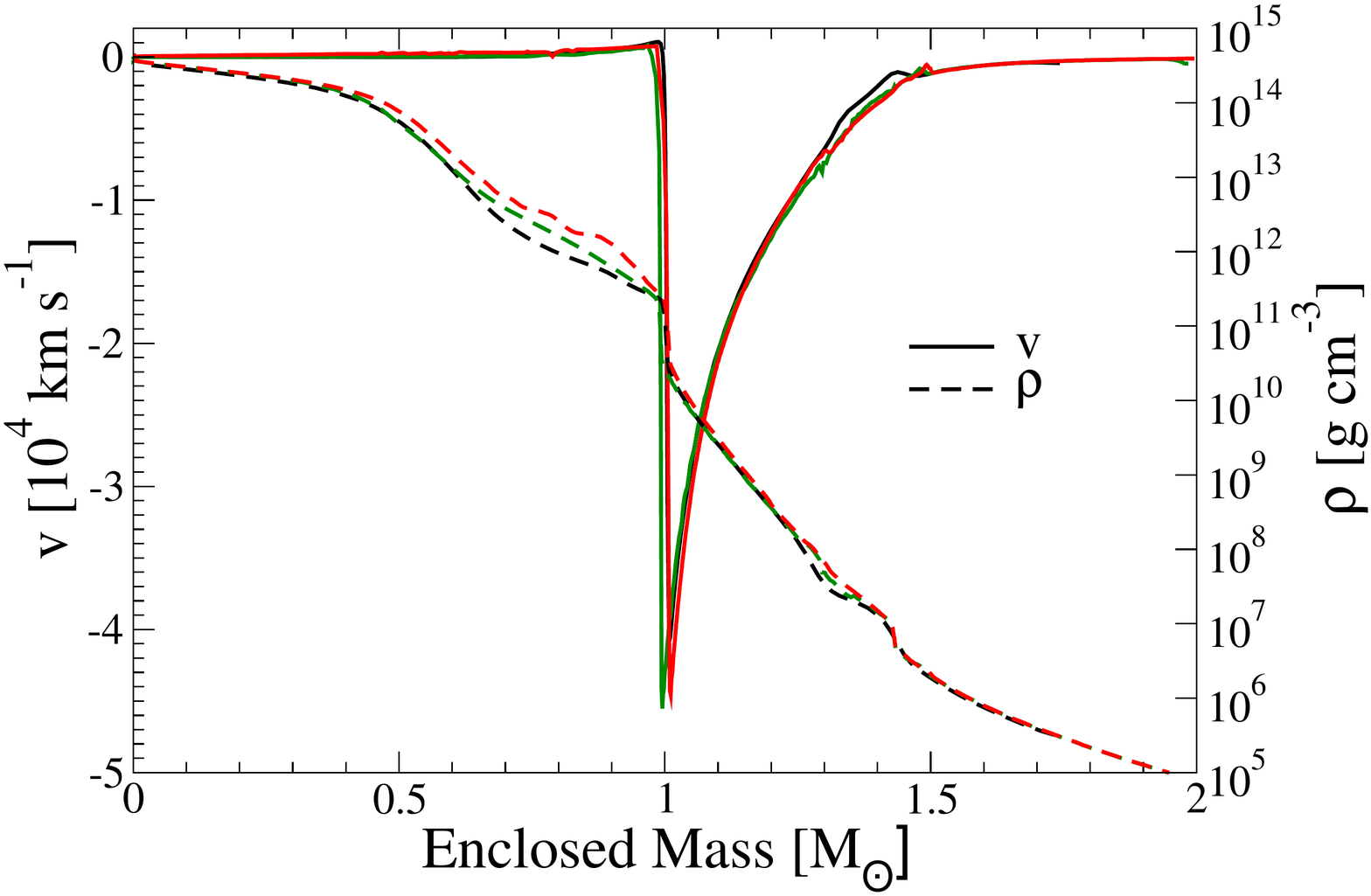}{0.5\textwidth}{(b)}
               }
\gridline{
              \fig{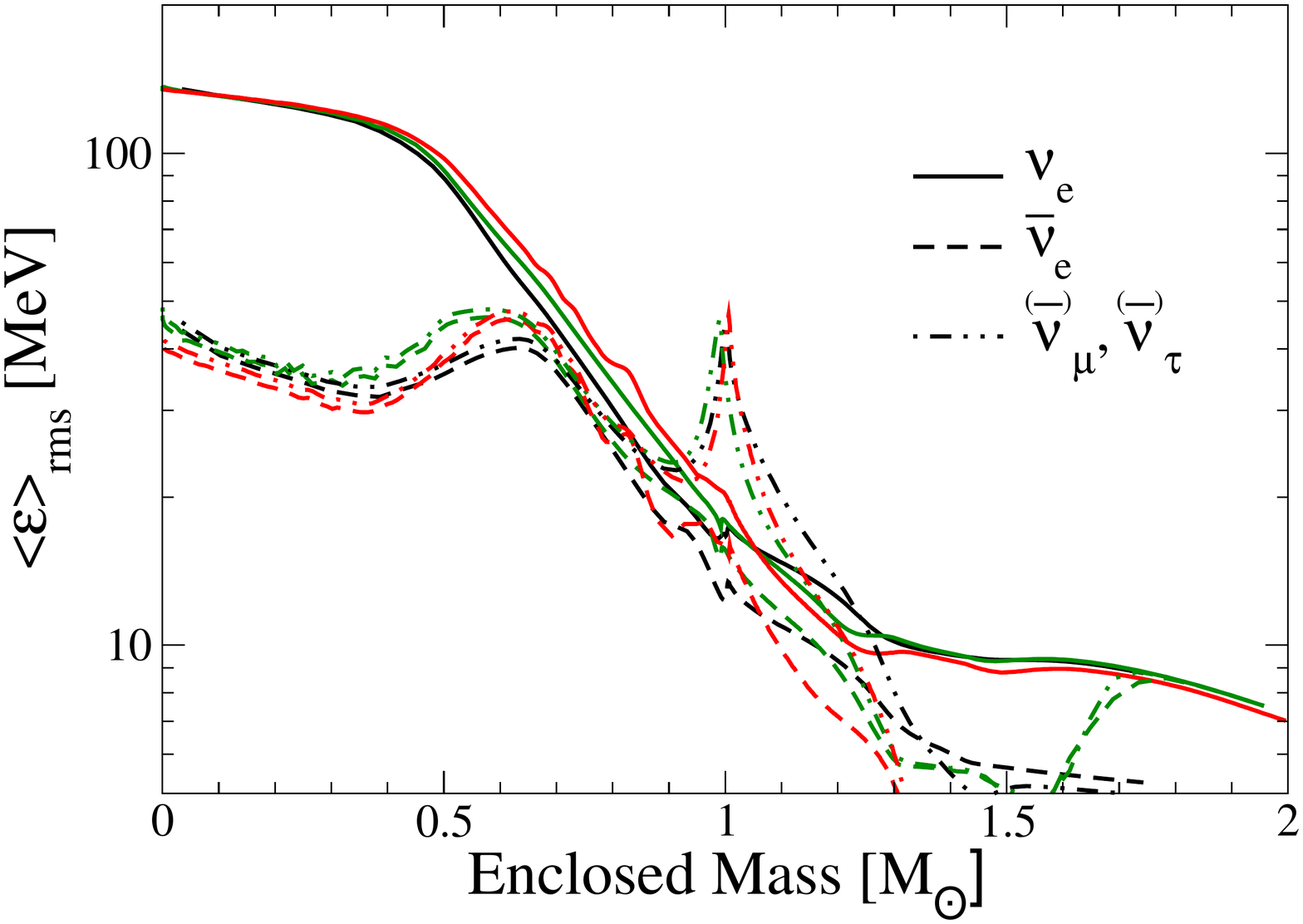}{0.5\textwidth}{(c)}
              \fig{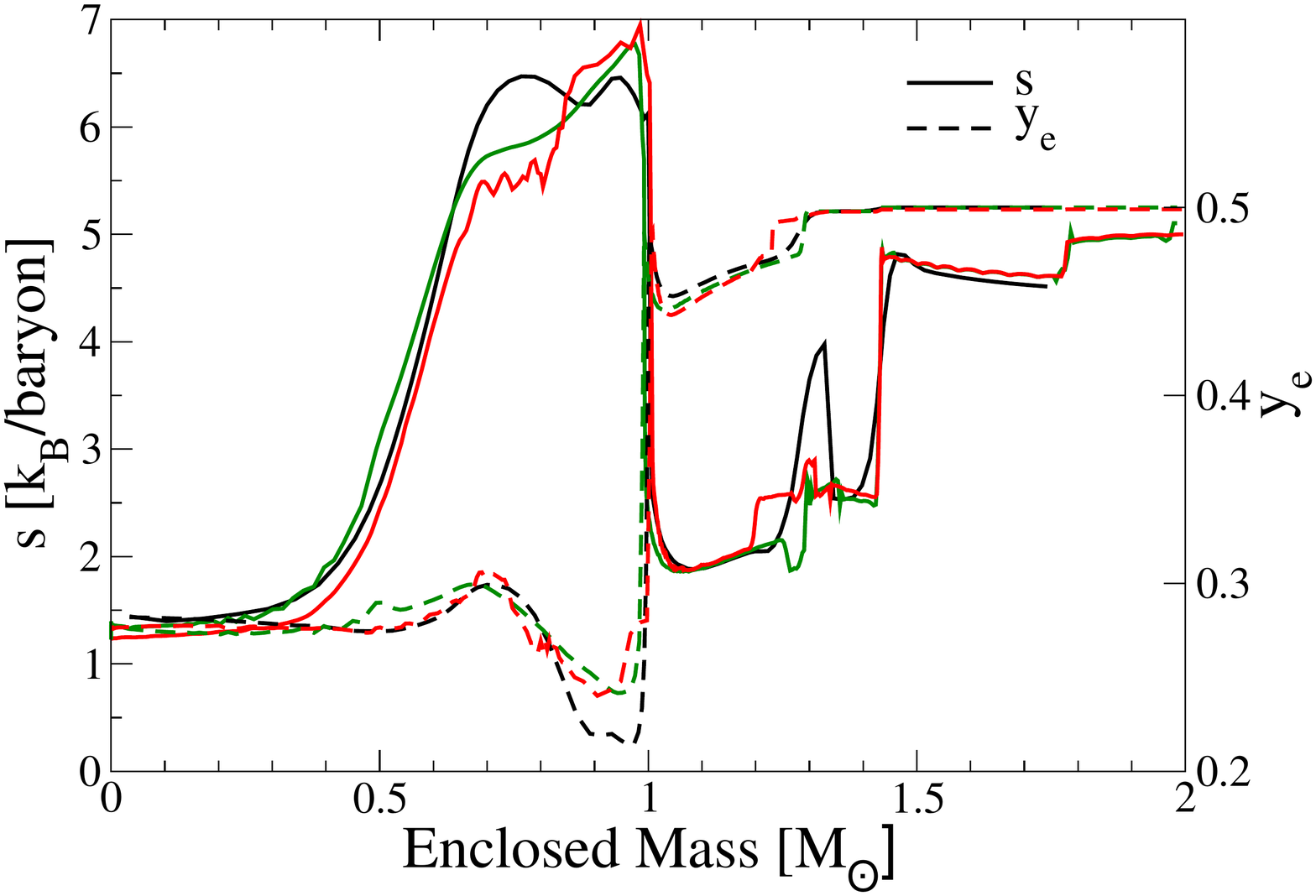}{0.5\textwidth}{(d)}
               }
\caption{\label{fig:postbounce3ms}
As in Figure~\ref{fig:bounce} but for 3~ms after bounce
}
\end{figure*}

\subsection{Bounce}
\label{bounce}

Figure \ref{fig:bounce} displays the material and neutrino configurations of model G15 at bounce. 
The results of all simulations are quite similar at this time.
Figure~\ref{fig:bounce}(a) plots the neutrino luminosities, with the \nue-luminosity for \chimera\ between 0.8 and 1.3~\msun\ slightly lower than that for the others --- a difference which may be related to the slightly lower core densities (Figure~\ref{fig:bounce}(b)), and slightly lower trapped entropy (Figure~\ref{fig:bounce}(d) and Figure \ref{fig:infall}).
The \nuebar-luminosities are still too small to be shown, and the \numt-luminosities are just beginning to develop.
The \nue-rms energies (Figure~\ref{fig:bounce}(c)) are almost the same for all models, while the \numt-rms energies are slightly smaller in the core in the \chimera\ simulation.
The jump in the entropy and electron fraction at an enclosed mass of 1.18~\msun\ in the \chimera\ simulation (Figure~\ref{fig:bounce}(d)) occurs at about 1.28~\msun\ in the other two simulations.
These jumps in the \chimera\ simulation appear in the initial model at the same enclosed mass, so it is not a feature that has evolved during infall by \chimera.

\subsection{Comparisons at 3 ms after Bounce}
\label{postbounce3ms}

Figure~\ref{fig:postbounce3ms} displays the material and neutrino configurations of Model~G15 3~ms after bounce. 
Again, the results of all simulations are quite similar at this time, but small differences can be observed. 
The velocity profiles (Figure~\ref{fig:postbounce3ms}(b)) are almost identical, with the \chimera\ and \agileboltztran\ shock being slightly farther out in enclosed mass. 
The \chimera\ density in the region from 0.5~\msun\ to the shock is slightly higher than the \vertex-1 density, which in turn is slightly higher than the \agileboltztran\ density. 
This probably accounts for the similar hierarchy in the \nue-rms energy in that region (Figure~\ref{fig:postbounce3ms}(c)). 
The \chimera\ shock at formation is slightly weaker than that of \vertex-1, which is slightly weaker than that of \agileboltztran, as inferred by the entropy profile behind the shock (Figure~\ref{fig:postbounce3ms}(d)), which likely accounts for the differences in the neutrino luminosities at this time (Figure~\ref{fig:postbounce3ms}(a)). 

\begin{figure*}
\gridline{\fig{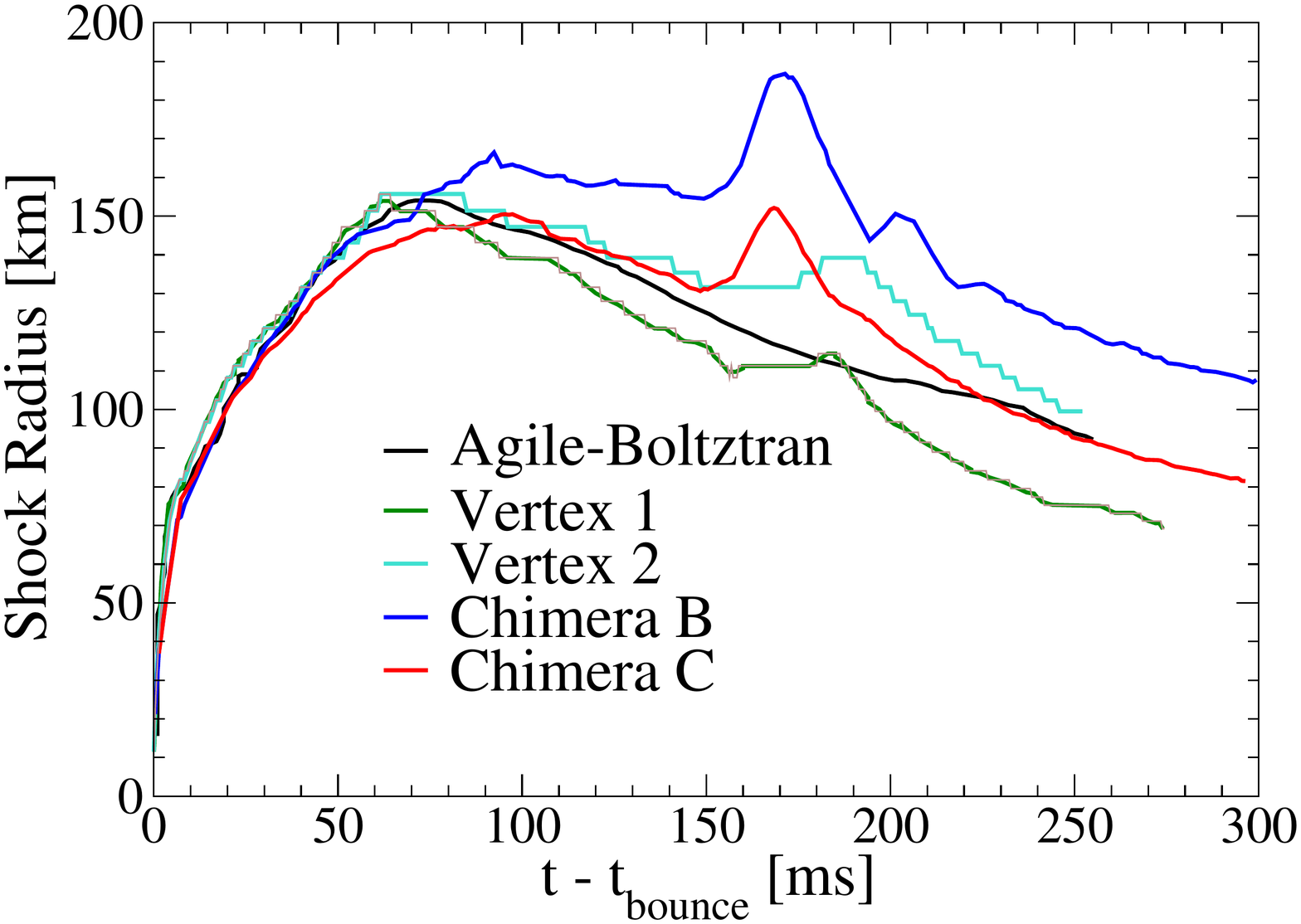}{0.5\textwidth}{(a)}
              \fig{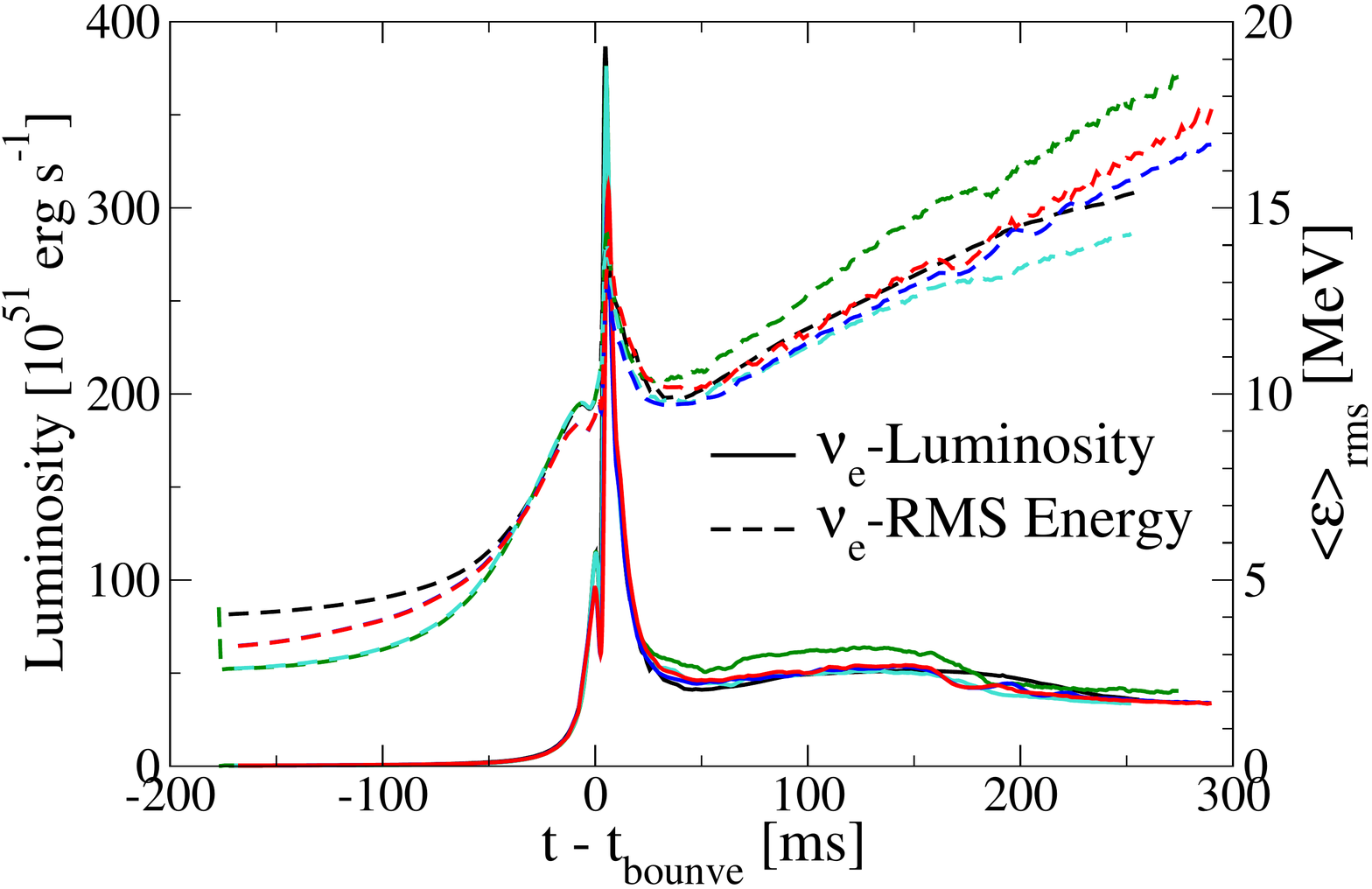}{0.5\textwidth}{(b)}
               }
\gridline{
              \fig{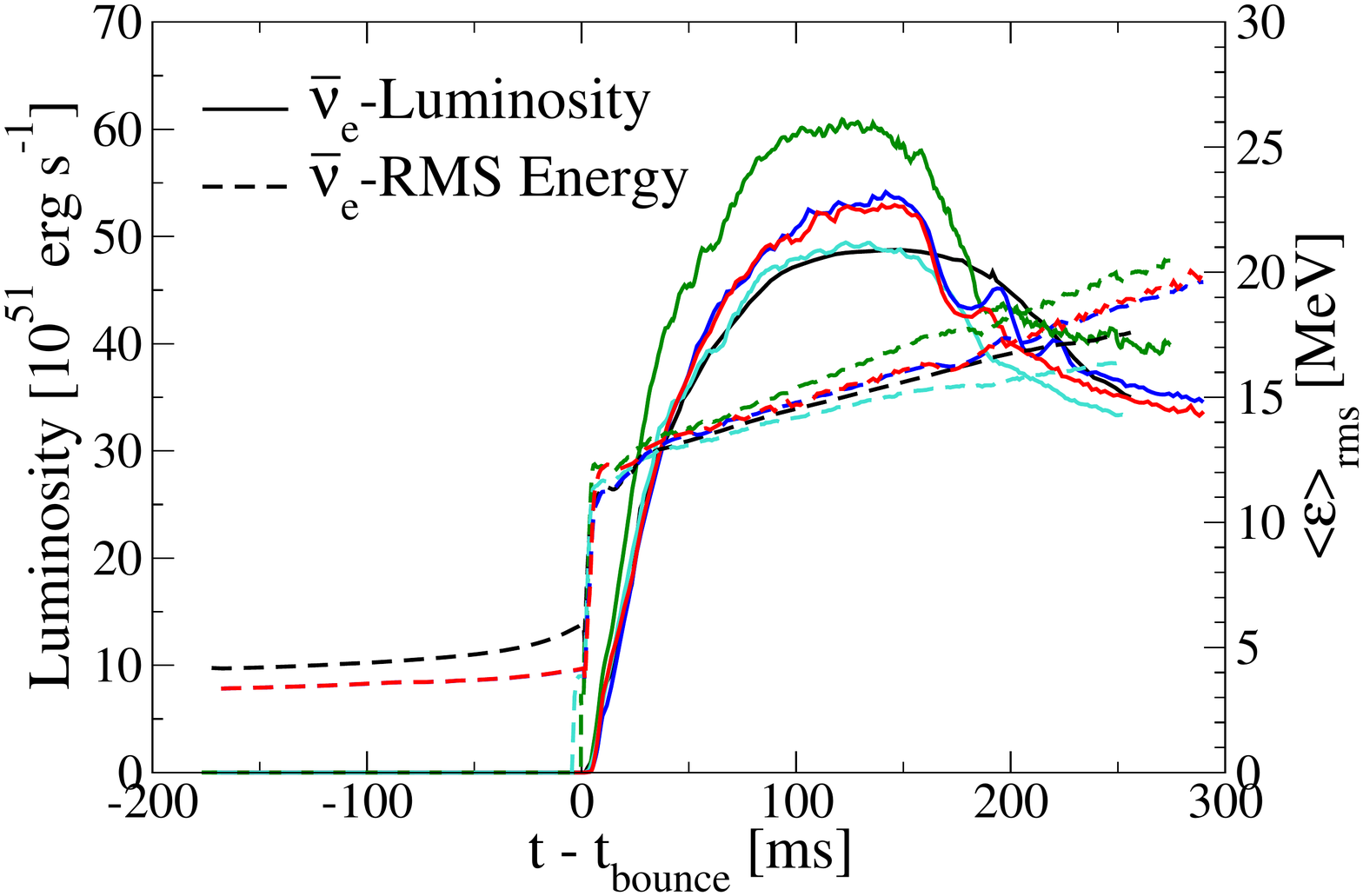}{0.5\textwidth}{(c)}
              \fig{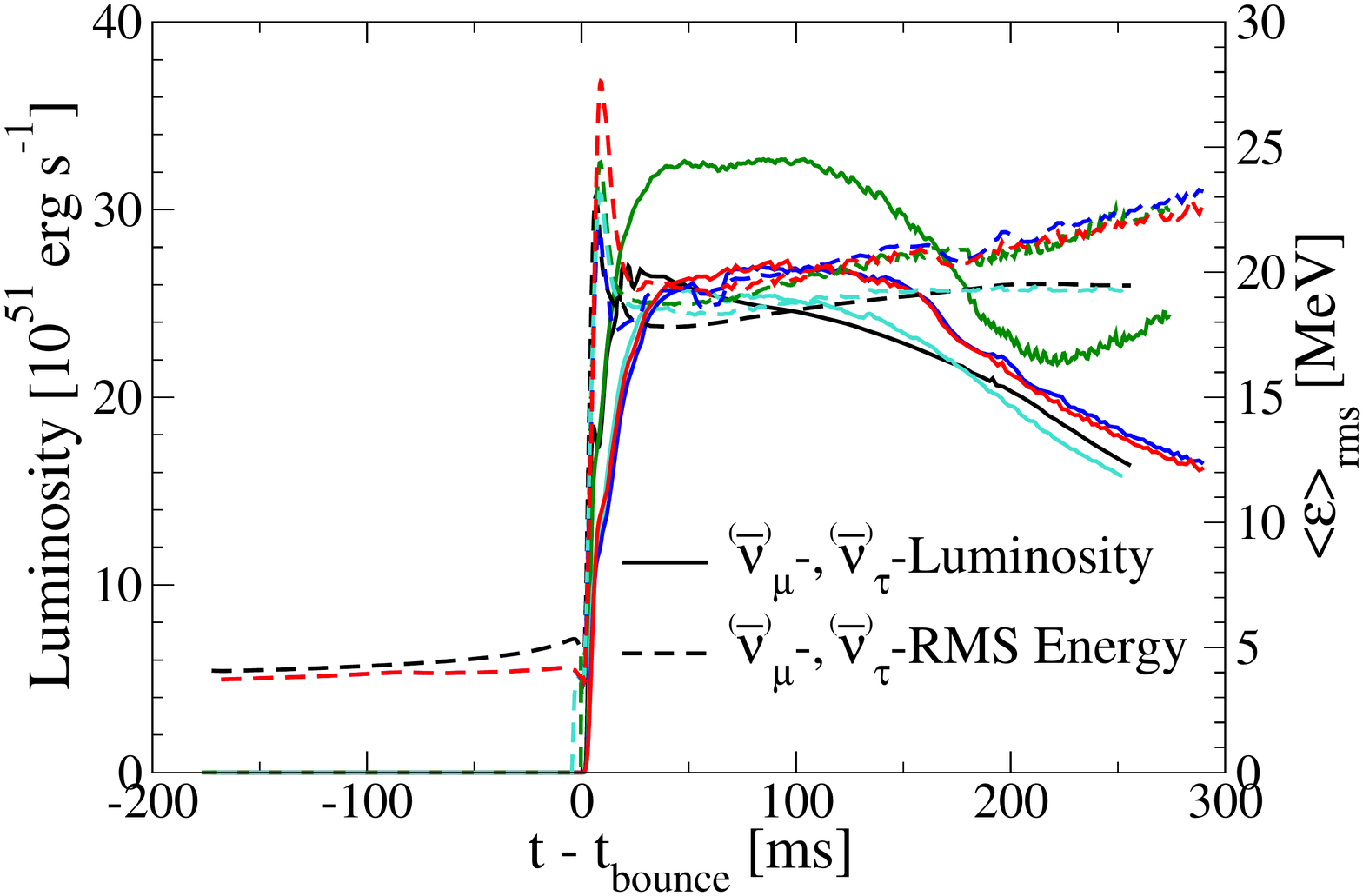}{0.5\textwidth}{(d)}
               }
\caption{\label{fig:config_vs_time}
Evolution of the (a) model G15 shock radius and the (b) \nue, (c) \nuebar, and (d) \nux\ neutrino luminosities and rms energies, for \agileboltztran\ (black),  \vertex-1 (green) and  \vertex-2 (turquoise), \chimera-B (blue), and  \chimera-C (red) simulations.
}
\end{figure*}

\subsection{Comparisons as a Function of Time}
\label{compvtime}

\added{Figures \ref{fig:config_vs_time}(a-d) compare the shock trajectories, neutrino luminosities and neutrino rms energies computed by the codes as a function of time from bounce to 300~ms post-bounce. }
Figure \ref{fig:config_vs_time}(a) plots the position of the shocks as a function of time.
\added{The outwardly directed ``hump'' in the shock trajectories at about 170~ms, not seen in the \agileboltztran\ results, is a feature that is due to the passage through the shocks of the silicon layer, with its associated drop in density and consequent reduction in the inwardly directed ram pressure on the shocks.}
It should be noted that after about 150~ms the shock exits the region covered by the Lattimer-Swesty EoS and enters a region where the EoS is treated differently by each code, making comparisons problematic. 
The \chimera-B shock trajectory is very close to that of \agileboltztran\  for the first 60~ms post-bounce, then, after that, rises some 20~km above it. 
The \chimera-C shock trajectory is initially close to that of \agileboltztran, falls below it by up to 5~km from 30 to 80~ms, then stays within 2~km of it from 90~ms to the end of the plot. 
The \vertex-1 shock trajectory is initially within a few km of that of \agileboltztran, then retreats more rapidly after 70~ ms post-bounce, falling below it by about 5~km by 100~ms post-bounce. 
The \vertex-2 shock trajectory follows closely but slightly above that of \agileboltztran. 
Both the \vertex-1 and \vertex-2 shocks exhibit the outwardly directed hump due to the passage through the shock of the silicon layer about 20-30 ms after those of the \chimera\ shocks. 

Figures \ref{fig:config_vs_time}(b)--(d) plot, respectively, the \nue-, \nuebar-, and one of the \numt-, \numtbar-luminosities and rms energies as a function of time. 
The \chimera-B and \chimera-C \nue-luminosities exhibit a lower peak at bounce than those of \agileboltztran\ or \vertex-1, 2, but track the \agileboltztran\ luminosities very closely thereafter, except for a slight decline at 170~ms when the shock reaches the large density decrement in the progenitor, which \agileboltztran\ fails to adequately resolve. 
The \vertex-1 luminosities are above those of the other simulations, reflecting its more rapid shock retraction, while the \vertex-2 \nue-luminosities track \agileboltztran\ closely until about 170 ms at which point they, like the \chimera\ luminosities, fall below.
The \nuebar-luminosities of \chimera-B and \chimera-C fall between those of \agileboltztran\ and \vertex-1 after bounce, rising about $5 \times 10^{51}$ ergs s$^{-1}$ above those of \agileboltztran\ after bounce and falling below \agileboltztran\ after 150 ms.
The \vertex-2 \nuebar-luminosities track those of \agileboltztran\ until about 150 ms at which point they too fall below.
The \numt-, \numtbar-luminosities of \chimera-B and \chimera-C also fall between those of \agileboltztran\ and \vertex-1 after bounce, rising about $5 \times 10^{51}$ ergs s$^{-1}$ above \agileboltztran\ until about 170 ms. 
\vertex-2 tracks \agileboltztran\ closely until about 170 ms at which point it falls below.

The rms energies exhibit a similar pattern after bounce. 
The \nue- and \nuebar-rms energies of both \chimera-B and \chimera-C are between those of \agileboltztran\ and \vertex-1 before bounce, tracking closely those of \agileboltztran\ 
after bounce and rising slightly above \agileboltztran\ after 190 ms. 
The \nue- and \nuebar-rms energies of \vertex-1 veer above those of \agileboltztran\ after bounce while those of \vertex-2 track \agileboltztran\ closely after bounce veering slightly below \agileboltztran\ after 50--100~ms.
The \numt-, \numtbar-rms energies of \chimera-B, \chimera-C, and \vertex-1 track each other within 1~MeV, all three of which lie about 2~MeV higher than those of \agileboltztran, while those of \vertex-2 and \agileboltztran\ lie within 1~MeV of each other.

\begin{figure*}
\gridline{\fig{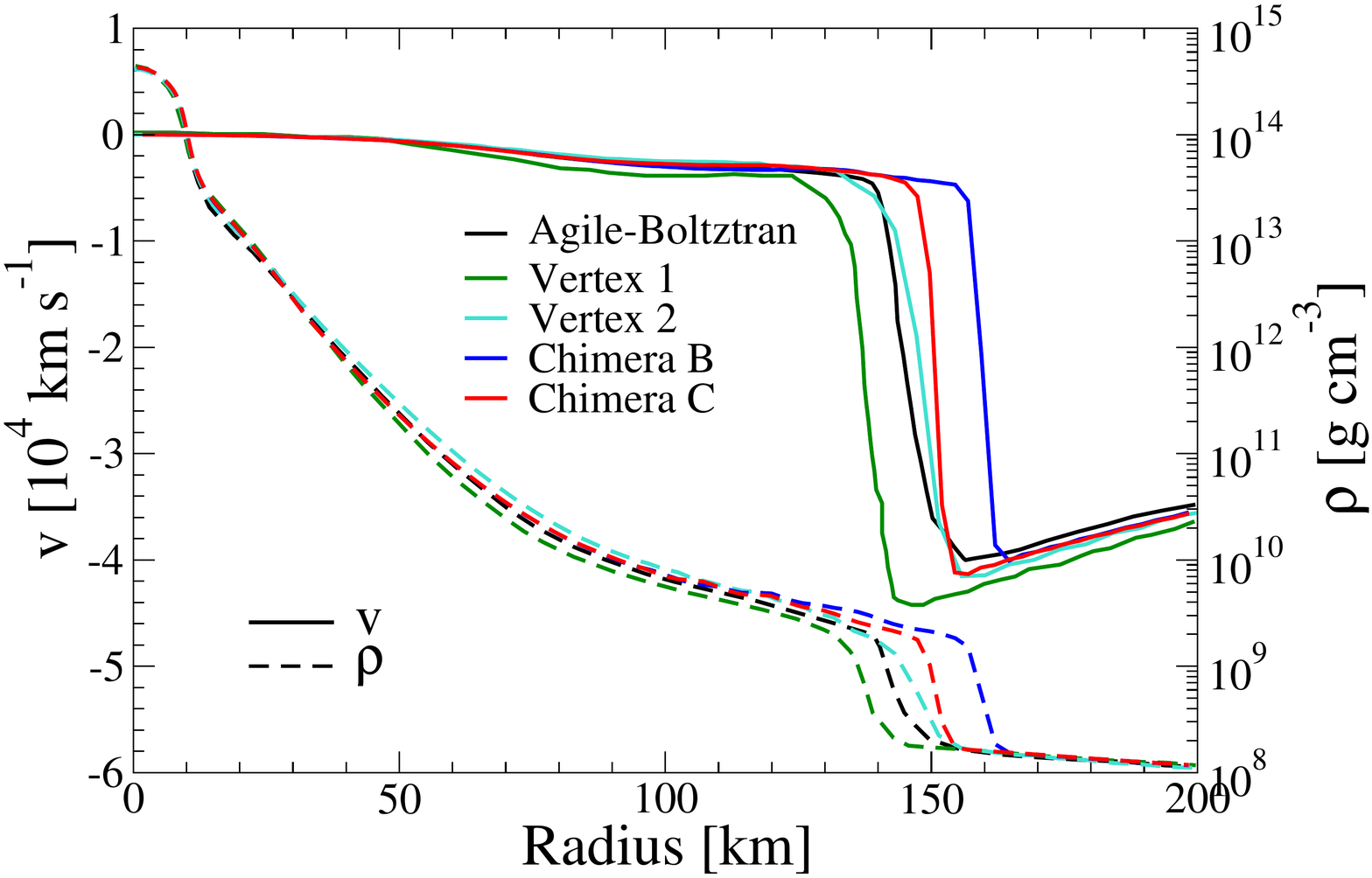}{0.5\textwidth}{(a)}
              \fig{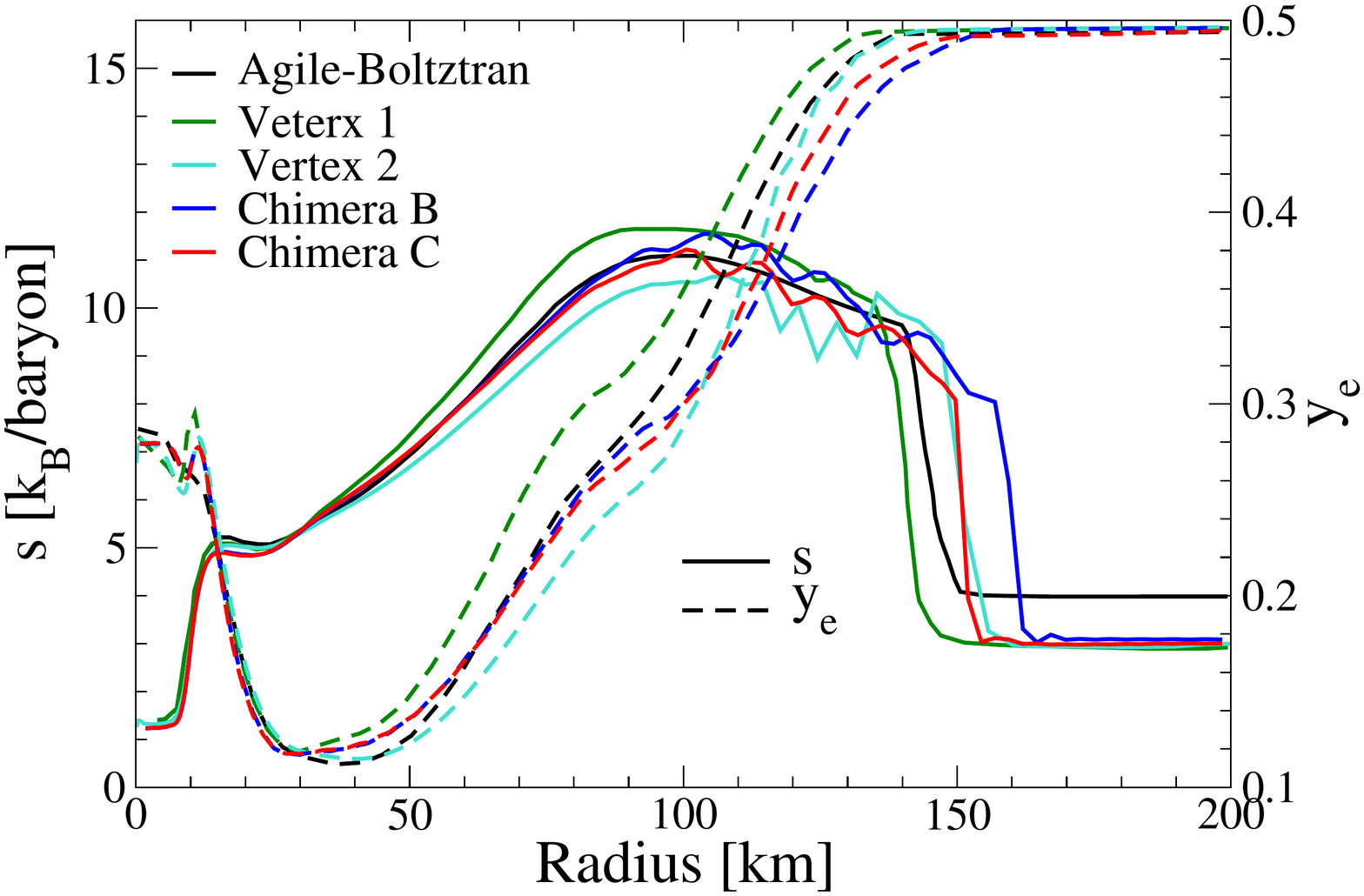}{0.5\textwidth}{(b)}
               }
\caption{\label{fig:config100ms}
Snapshots of model G15 at 100~ms after bounce for (a) velocity and density, and (b) entropy and \Ye\, for the \agileboltztran\ (black), \vertex-1 (green),  \vertex-2 (turquoise), and  \chimera-B (blue), \Chimera-C (red) simulations.
}
\end{figure*}

\subsection{Comparisons at 100 ms After Bounce}
\label{postbounce100ms}

At 100 ms post-bounce (Figure~\ref{fig:config100ms}), the \chimera-B and \chimera-C shocks are, respectively, about 20 km and 8 km farther out than that of \agileboltztran, the front of the \chimera-C shock being close to that of \agileboltztran. 
The \vertex-2 shock radius is close to that of \agileboltztran\, while the \vertex-1 shock has retreated about 10 km inward.
Aside from the positions of the shocks, the densities and velocities as a function of radius, shown in Figure~\ref{fig:config100ms}(a), agree with each other quite closely. 
The entropy profiles, shown in Figure~\ref{fig:config100ms}, also agree with each other modulo the position of the shock. 
Those of \chimera-B and \chimera-C exhibit entropy wiggles behind the shock. 
These are not due to the computation of the effective index, $\Gamma_{\rm e}$ \citep{BuRaJa06}, but the use of the \citet{CoWo84} suggested parameters for supplying dissipation in the vicinity of strong shocks. 
These wiggles disappear with a somewhat more aggressive parameter choice for shock dissipation, and this choice is now used in current versions of \chimera.
The electron fraction profiles are quite similar up to 30~km and beyond 150~km, but are displaced horizontally relative to each other in between these distances. 
This displacement is also reflected in the entropy profiles, but is less obvious, as the plots themselves are more horizontal. 
The origin of these differences is unclear, but may reflect the slight differences in the shock trajectories as a function of time, and the somewhat higher infall velocities in the case of \vertex-1.

\begin{figure*}
\gridline{\fig{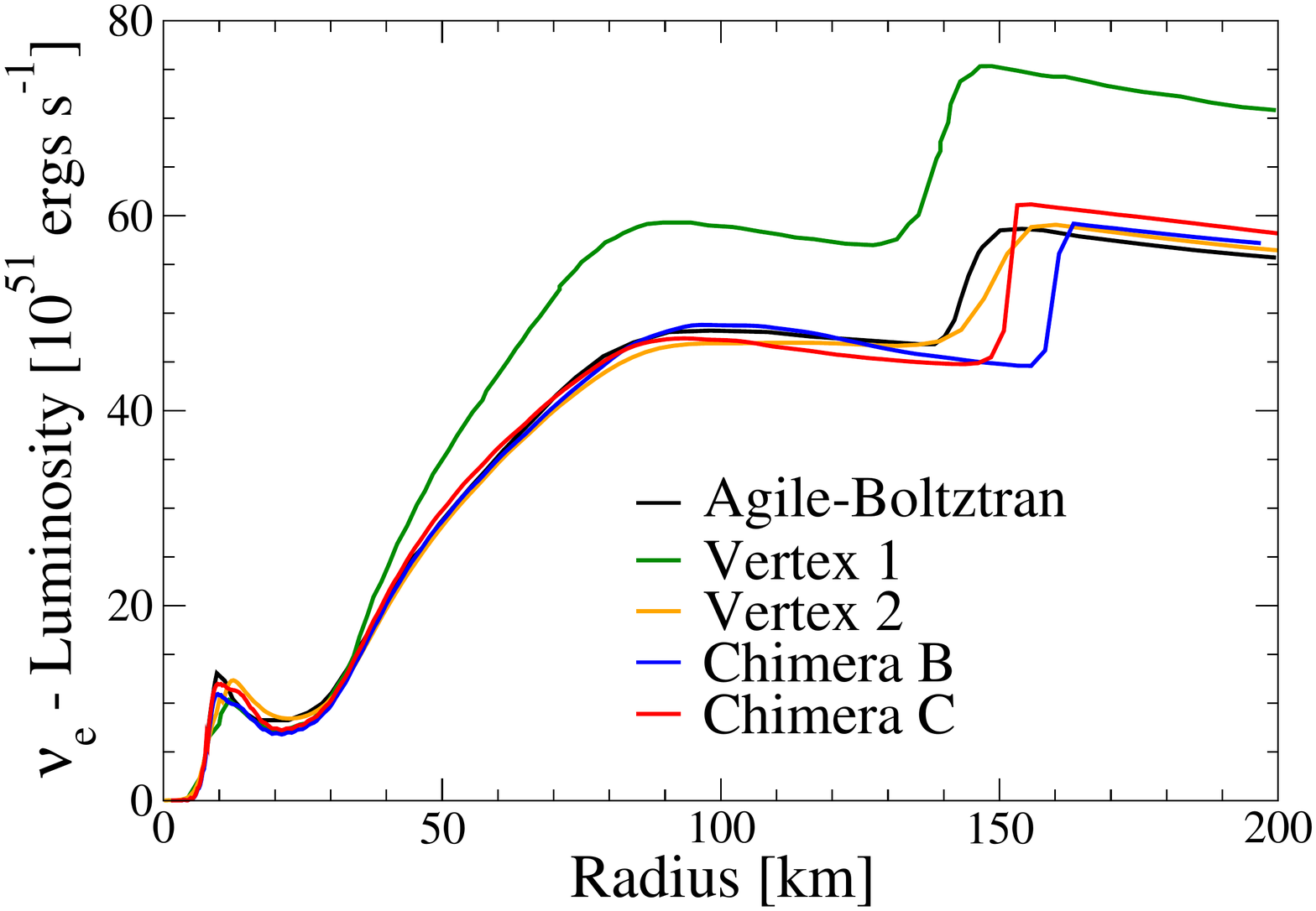}{0.5\textwidth}{(a)}
              \fig{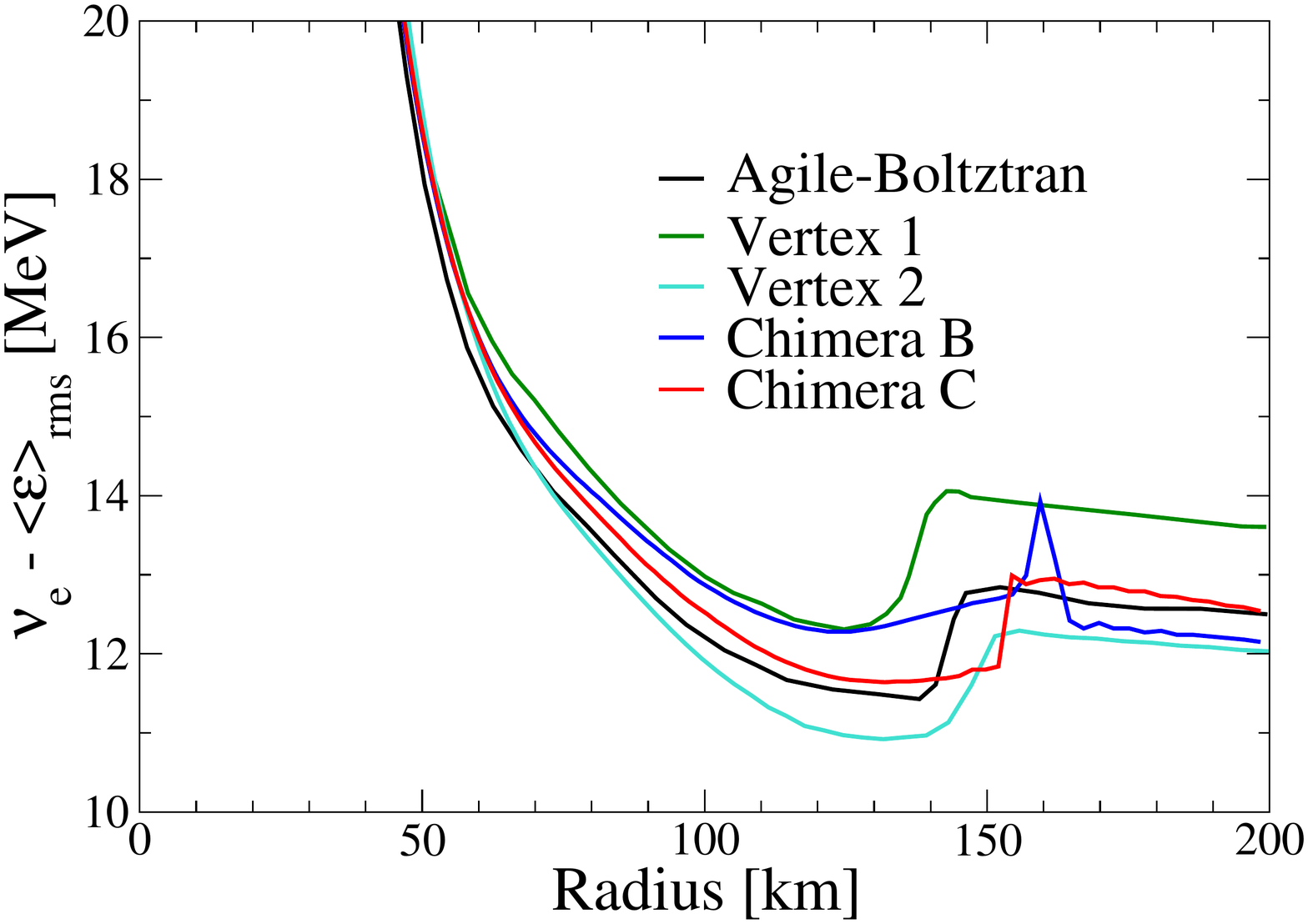}{0.5\textwidth}{(b)}
               }
\gridline{
              \fig{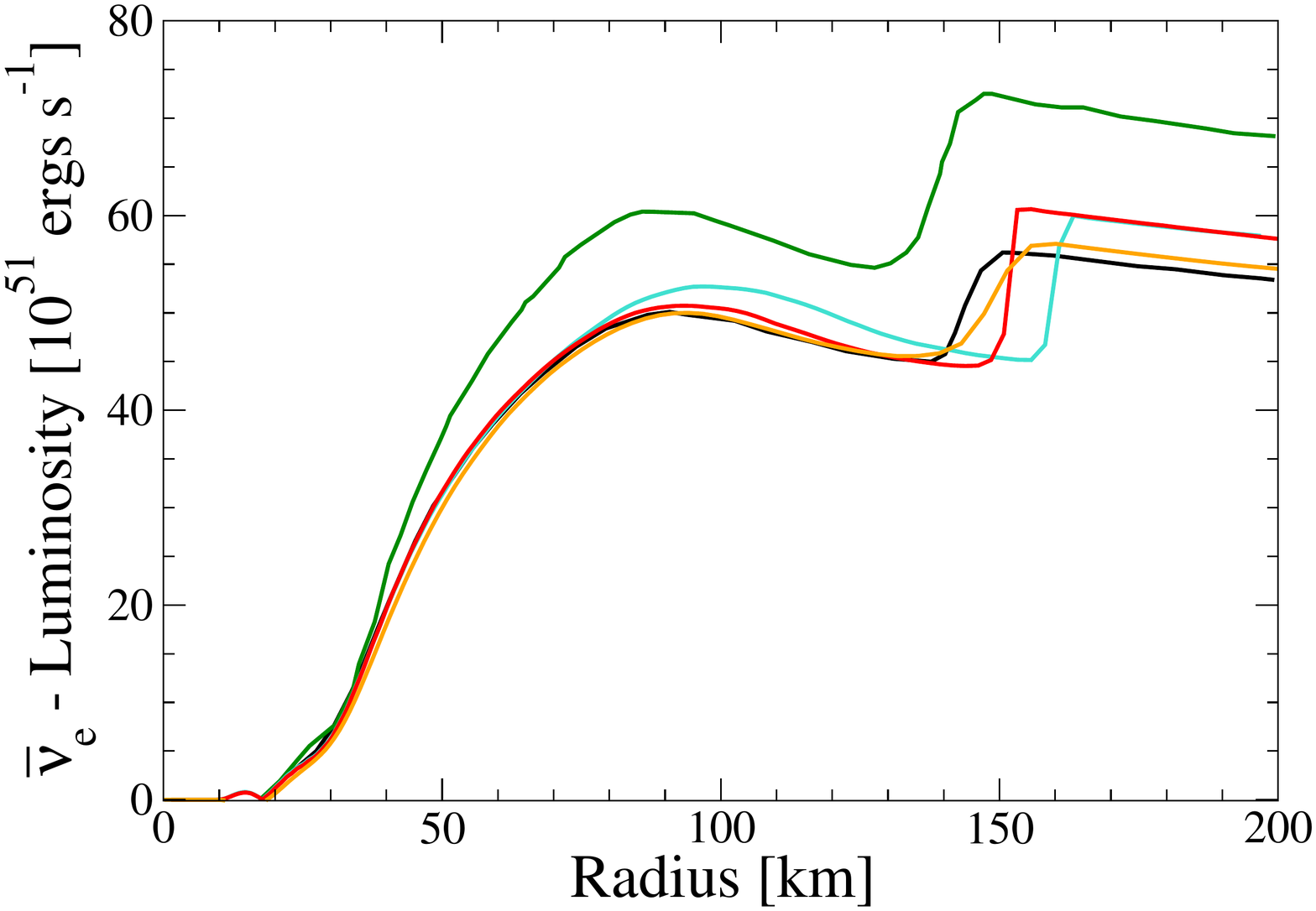}{0.5\textwidth}{(c)}
              \fig{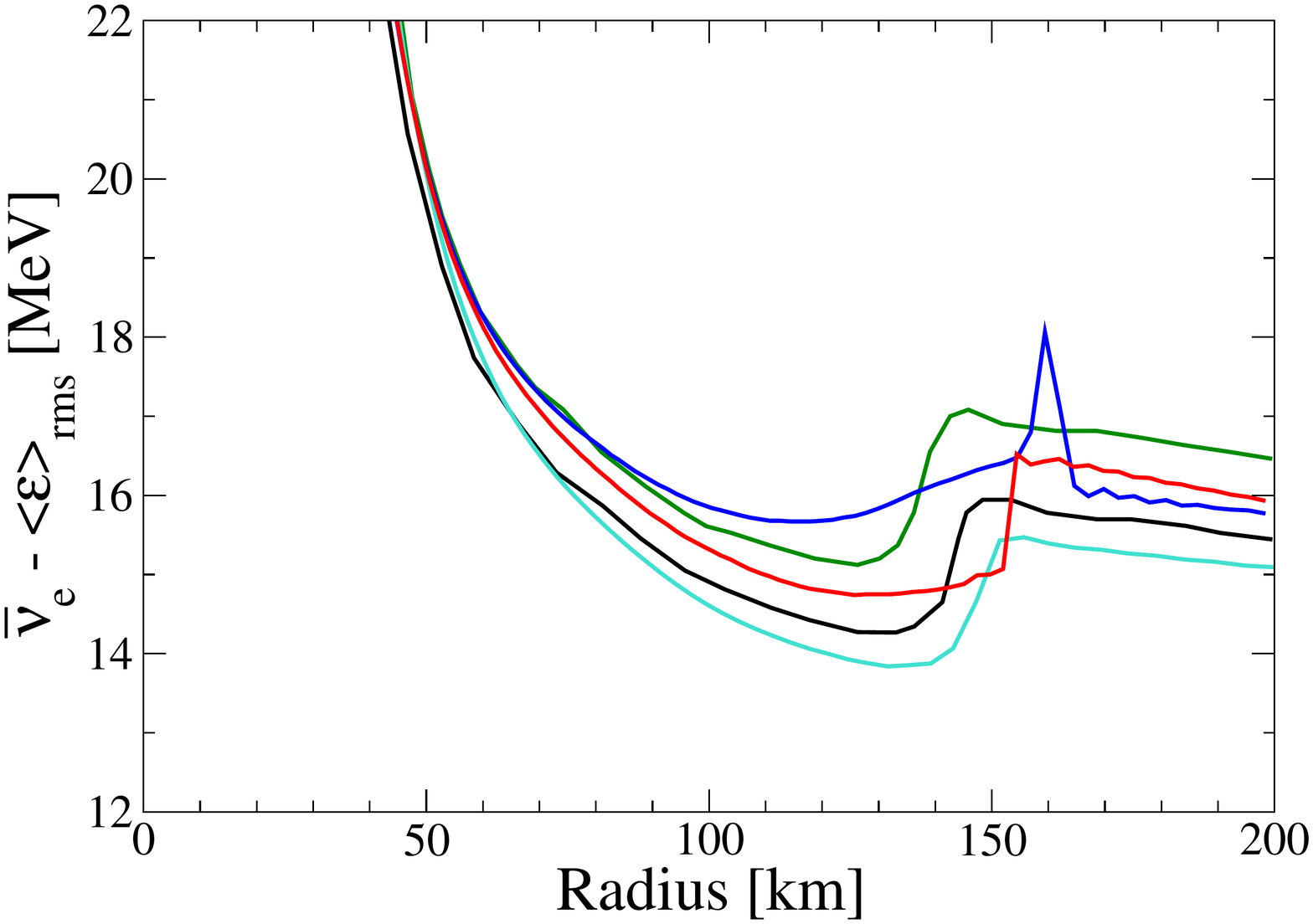}{0.5\textwidth}{(d)}
               }
\gridline{\fig{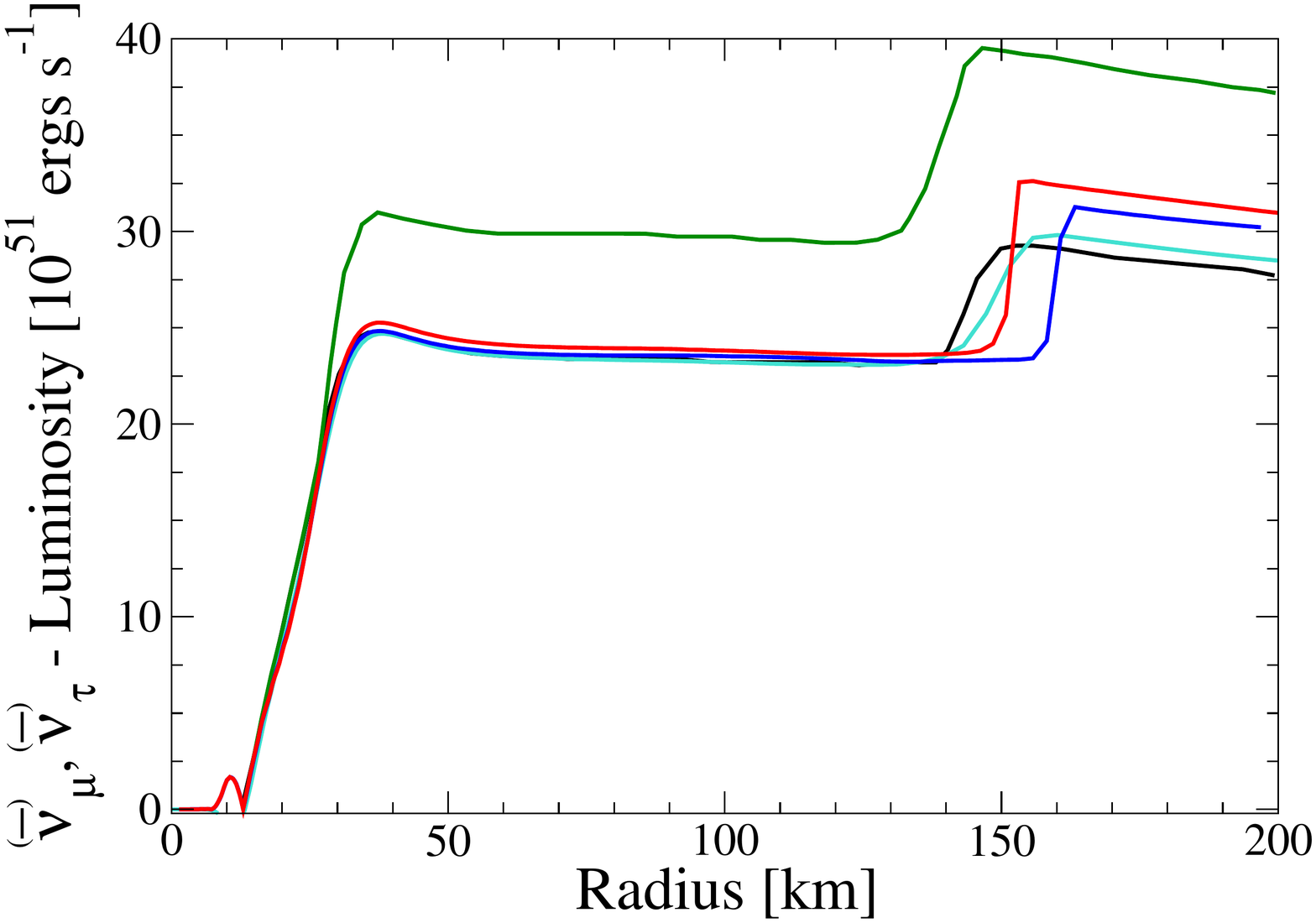}{0.5\textwidth}{(e)}
              \fig{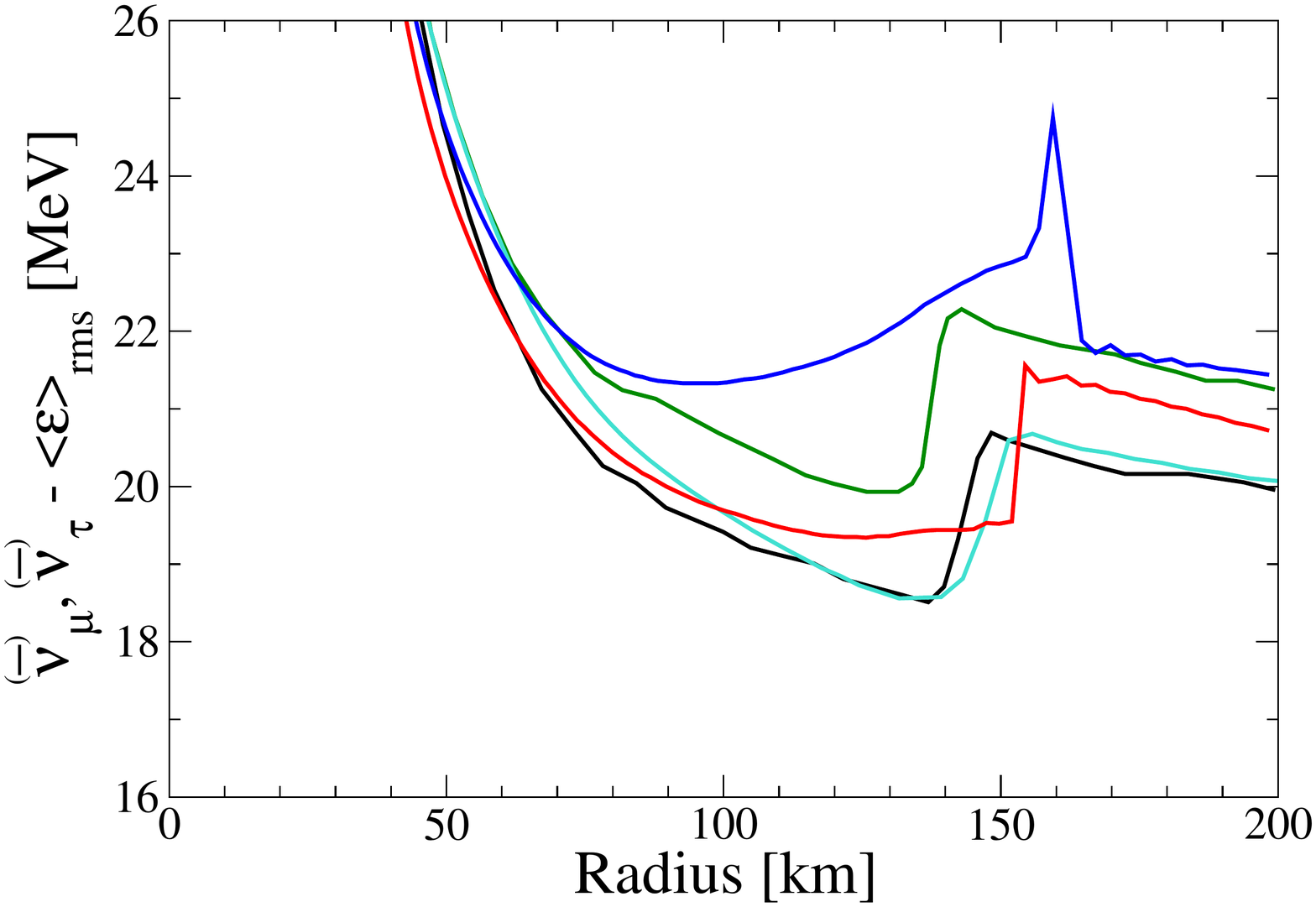}{0.5\textwidth}{(f)}
               }
\caption{\label{fig:nulumrms100ms}
Snapshot of model G15 neutrino properties at 100 ms after bounce, for  the \agileboltztran\ (black),  \vertex-1 (green),  \vertex-2 (turquoise), \chimera-B (blue), and  \chimera-C (red) simulations.
Neutrino luminosities are plotted on the left in Panels (a), (c), and (e) and rms-energies on the right in Panels (b), (d), and (f), with \nue\ in Panels (a) and (b), \nuebar\ in Panels (c) and (d), and \nux\ (or \numt) in Panels (e) and (f).
}
\end{figure*}

Figure~\ref{fig:nulumrms100ms}  compares the \nue, \nuebar, and \nux\ luminosity and rms neutrino energy profiles at 100 ms post-bounce.
Compared with the \agileboltztran\ luminosities, \chimera-B and \vertex-1 tend to overestimate these quantities in the region behind the shock, while the agreement between the results from \chimera-C, \vertex-2, and \agileboltztran\ for these quantities is extremely good. 
\chimera-B and \chimera-C tend to compute larger shock jumps than does \agileboltztran, while the shock jumps computed by \vertex-2 and \agileboltztran\ are in good agreement.
Both \vertex-1 and \chimera-B compute rms energies larger than those of \agileboltztran.
The \nue\ rms energies computed by \chimera-C are quite close to those of \agileboltztran, while the \nux\ rms energies computed by \vertex-2 and \agileboltztran\ are in excellent agreement.
In other cases, the rms neutrino energies computed by \chimera-C, \vertex-2, and \agileboltztran\ are typically within an MeV of each other. 

\section{Summary}

This report  has documented the development and construction of the \chimera\ code through its C-series implementations. 
\chimera\  has been designed to simulate core collapse supernovae throughout the entire neutrino-driven explosion phase, with outputs that can be used to extract important associated observables, such as element synthesis and dispersal, neutrino signatures, and gravitational radiation. 
The code couples a multidimensional, PPM-plus-remap, Newtonian hydrodynamics module with radial-ray-plus, multi-group flux-limited diffusion neutrino transport and a multi-species nuclear reaction network. 
The transport module stems from a general relativistic treatment and currently retains the most important element of general relativity, namely the lapse function, which ensures proper red-shifting of neutrinos as they propagate out of the gravitational well. General relativity enters the computation of self gravity through a monopole correction to the Newtonian gravitational potential.
\chimera\ evolves all six neutrino and antineutrino distributions and includes an extensive suite of neutrino weak interactions, including sophisticated treatments of nuclear electron capture and non-isoenergetic scattering on nucleons, as well as industry-standard equations of state.

We have subjected \chimera\ to a suite of test problems and made comparisons with two other sophisticated neutrino radiation-hydrodynamics codes, \agileboltztran, a code based on  general relativistic gravity, hydrodynamics, and six-species Boltzmann neutrino kinetics, and \prometheusvertex, a code similar in many respects to \chimera\ but based on a more sophisticated, two-moment neutrino transport scheme closed by a variable Eddington factor computed from approximate Boltzmann kinetics.

Development of \chimera\ continues to include new features and other enhancements beyond those described herein and will be reported, as appropriately, with the associated results.

\acknowledgments

The authors would like to thank Bernard MŸ\"{u}ller for the \vertex-1, 2 data, and Mathias Liebend\"{o}šrfer for posting the \agileboltztran\ data online. We also would like to express our appreciation to an anonymous referee for a careful reading of this paper and for suggestions of some tests that were not included in the original draft of this paper.
This work has been supported by a wide range of programs in the 15 years since \chimera's inception.
\chimera\ development has been supported by the NSF PetaApps program (OCI-0749242, OCI-0749248, and OCI-0749204), by the NSF Gravitational Physics Theory program (PHY-0555644, PHY-0652874, PHY-0855315, PHY-1505933, PHY-1806692), by the NSF Nuclear Theory Program (PHY-0244783, PHY-1516197), and by the NSF Stellar Astronomy and Astrophysics program (AST-0653376).
NASA supported \chimera\ development under the Astrophysics Theory Program (NNH08AH71I,  NNH11AQ72I, 07-ATFP07-0011). \chimera\  development was also supported by the U.S. Department of Energy Offices of Nuclear Physics and Advanced Scientific Computing Research.

This research used resources of the Oak Ridge Leadership Computing Facility, which is a DOE Office of Science User Facility supported under Contract DE-AC05-00OR22725, and the National Energy Research Scientific Computing Center, which is supported by the U.S. DOE Office of Science under Contract No. DE-AC02-05CH11231.
This research also utilized resources of the NSF TeraGrid provided by the National Institute for Computational Sciences under grant number TG-MCA08X010.

P.M. is supported by the National Science Foundation through its employee IR/D program. The opinions and conclusions expressed herein are those of the authors and do not represent the National Science Foundation.

We wish to thank John C. Hayes for his assistance early in the development of \chimera\, including its cognomination, as well as our many colleagues in the core collapse supernova simulation community, and other of our colleagues, from whom we have benefited greatly through discussions related to supernova simulation.

\bibliographystyle{aasjournal}

\end{document}